\documentstyle[aps,multicol]{revtex}

\begin{document}
\draft

\def\BCMO{\rm {Bi_{1-x} Ca_x Mn O_3}}
\def\PSMO{\rm {Pr_{1-x} Sr_x Mn O_3}}
\def\NSMO{\rm {Nd_{1-x} Sr_x Mn O_3}}
\def\PCMO{\rm {Pr_{1-x} Ca_x Mn O_3}}
\def\LCMOhalf{\rm {La_{0.5} Ca_{0.5} Mn O_3}}
\def\LCMO{\rm {La_{1-x} Ca_x Mn O_3}}
\def\LSMO{\rm {La_{1-x} Sr_x Mn O_3}}
\def\bilayered(1.8){\rm {La_{1.2} Sr_{1.8} Mn_2 O_7 }}
\def\bilayer{\rm {La_{2-2x} Sr_{1+2x} Mn_2 O_7 }}
\def\single{\rm {La_{1-x} Sr_{1+x} Mn O_4}}
\def\densi{$\langle$$n$$\rangle$}

\title{COLOSSAL MAGNETORESISTANT MATERIALS: \\
  THE KEY ROLE OF PHASE SEPARATION}

\author{Elbio Dagotto$^1$, Takashi Hotta$^2$, Adriana Moreo$^1$}

\address{$^1$National High Magnetic Field Lab and Department of
  Physics, Florida State University, Tallahassee, FL 32306, USA} 

\address{$^2$Institute for Solid State Physics, University of Tokyo, 
  5-1-5 Kashiwa-no-ha, Kashiwa, Chiba 277-8581, Japan}

\date{\today}

\maketitle

\begin{abstract}

The study of the manganese oxides, widely known as manganites, that 
exhibit the ``Colossal'' Magnetoresistance (CMR) effect is among the
main areas of research within the area of Strongly Correlated
Electrons. After considerable theoretical effort in recent years,
mainly guided by computational and mean-field studies of realistic
models, considerable progress has been achieved in understanding
the curious properties of these compounds.
These recent studies suggest that the ground states of manganite
models tend to be intrinsically inhomogeneous due to the presence of
strong tendencies toward phase separation, typically involving
ferromagnetic metallic and antiferromagnetic charge and orbital
ordered insulating domains.
Calculations of the resistivity versus temperature using mixed states
lead to a good agreement with experiments.
The mixed-phase tendencies have two origins:
(i) electronic phase separation between phases with different
densities that lead to nanometer scale coexisting clusters, and
(ii) disorder-induced phase separation with percolative
characteristics between equal-density phases, driven by disorder near
first-order metal-insulator transitions.
The coexisting clusters in the latter can be as large as a micrometer in
size. It is argued that a large variety of experiments reviewed in
detail here contain results compatible with the theoretical
predictions. The main phenomenology of mixed-phase states appears to
be independent of the fine details of the model employed, since the
microscopic origin of the competing phases does not influence the
results at the phenomenological level. 
However, it is quite important to clarify the electronic properties of
the various manganite phases based on microscopic Hamiltonians,
including strong electron-phonon Jahn-Teller and/or Coulomb
interactions.
Thus, several issues are discussed here 
from the microscopic viewpoint as well, 
including the phase diagrams of manganite models, 
the stabilization of the charge/orbital/spin ordered
half-doped CE-states, the importance of the
naively small Heisenberg coupling among localized spins, the setup of
accurate mean-field approximations, the existence of a new temperature
scale $T^*$ where clusters start forming above the Curie temperature,
the presence of stripes in the system, and many others. However, much
work remains to be carried out, and a list of open questions is 
included here. It is also argued that the mixed-phase phenomenology of
manganites may appear in a large variety of compounds as well,
including ruthenates, diluted magnetic semiconductors, and others.
It is concluded that manganites reveal such a wide variety of
interesting physical phenomena that their detailed study is quite
important for progress in the field of Correlated Electrons.

\end{abstract}


\begin{multicols}{2}
\narrowtext

\section{INTRODUCTION}

This is a review of theoretical and experimental work in the context 
of the manganese oxides widely known as manganites. 
These materials are currently being investigated by a sizable fraction 
of the Condensed Matter community, and their popularity is reaching
levels comparable to that of the high temperature superconducting
cuprates. From this review hopefully the reader will be able to
understand the reasons behind this wide interest in manganites, the
problems that have been solved in this context, and those that remain
to be investigated.
The authors have made a considerable effort in trying to include in
this review the majority of what they consider to be the most relevant 
literature on the subject.
However, clearly it is not possible to cover all aspects of the
problem in a single manuscript.
Here the main focus has been directed into recent theoretical
calculations that address the complex spin, charge, and/or orbital
ordered phases of manganites, which have important and prominent
intrinsic inhomogeneities, and also on the recent experimental results 
against which those calculations can be compared.
Due to the complexity of the models needed to address manganites, it
is natural that the most robust results have been obtained with
computational tools, and those are the calculations that will be
emphasized in the text.
The continuous growth of available computer power has allowed
simulations that were simply impossible not long ago.
In addition, the physics of manganites appears dominated by intrinsic
inhomogeneities and its description is quite difficult in purely
analytic frameworks that usually assume uniform states.
However, several calculations, notably some mean-field approximations, 
have also reached a high accuracy level and they are important in
deciding which are the phases of relevance in manganites.
These calculations are also discussed in detail below.
Finally, it is reassuring for the success of manganite investigations
that a variety of experiments, reviewed here, appear to be in qualitatively
good agreement with the most recent theoretical calculations.
Even quantitative agreement is slowly starting to emerge, although
there are still many aspects of the problem that require further
investigation.
At a more general level, from this review it is expected that the
readers will understand the richness of manganite physics and how it
challenges aspects of our present understanding of Condensed Matter
systems.
The effort to fully unveil the behavior of electrons in manganites
should continue at its current fast pace in the near future.

The field of manganites started with the seminal paper of Jonker and
van Santen (1950) where the existence of ferromagnetism in mixed
crystals of $\rm LaMnO_3$-$\rm CaMnO_3$, $\rm LaMnO_3$-$\rm SrMnO_3$,
and $\rm LaMnO_3$-$\rm BaMnO_3$ was reported. 
The general chemical formula for the manganese oxides described in
Jonker and van Santen's paper (1950), and many other compounds
investigated later on, is $\rm T_{1-x} D_x Mn O_3$, with T a trivalent 
rare earth or Bi$^{3+}$ cation, and D a divalent alkaline or Pb$^{2+}$ 
cation. Oxygen is in a O$^{2-}$ state, and the relative fraction of
Mn$^{4+}$ and Mn$^{3+}$ is regulated by ``x''.
The perovskite lattice structure of these materials is illustrated in 
Fig.I.1a. 
Jonker and van Santen (1950) adopted the terminology ``manganites'' 
to refer to these mixed compounds, although it is not strictly
correct, as they emphasized in a footnote, since the term manganite
should in principle apply only to the 100\% $\rm Mn^{4+}$ compound.
 
More detailed information about $\LCMO$ using neutron scattering
techniques was obtained later by Wollan and Koehler (1955).
In their study the antiferromagnetic (AF or AFM) and ferromagnetic
(FM) phases were characterized, and the former was found to contain
nontrivial arrangements of charge at particular hole densities.
Wollan and Koehler (1955) noticed the mixture of C-type and E-type
magnetic unit cells in the structure at x=0.5, and labeled the
insulating state at this density as a ``CE-state'' (the seven possible 
arrangements A, B, C, D, E, F, and G for the spin in the unit-cell are 
shown in Fig.I.1b, with the spins of relevance being those located in
the manganese ions).
Theoretical work at approximately the same time, to be reviewed below, 
explained the ferromagnetic phase as caused by an effect called
``double-exchange'' (DE), and thus one of the most interesting
properties of these materials appeared to have found a good
rationalization in the early studies of these compounds.
Perhaps as a consequence of the apparent initial theoretical success, 
studies of the manganites continued in
subsequent years at a slow pace.

The renewed surge of interest in manganites in the 1990's started with
the experimental observation of large magnetoresistance (MR) effects
in $\rm Nd_{0.5} Pb_{0.5} Mn O_3$ by Kusters et al. (1989) and in
$\rm La_{2/3} Ba_{1/3} Mn O_x$ by von Helmolt et al. (1993)
(actually Searle and Wang (1967) were the first to report MR studies
in manganites, which were carried out using $\rm La_{1-x} Pb_x Mn O_3$ 
single crystals).
Resistivity vs. temperature results for the (La,Ba) compound are shown 
in Fig.I.2a, reproduced from von Helmolt et al. (1993).
The MR effect was found to be as high as 60\% at room temperature
using thin-films, and it was exciting to observe that this value was
higher than found in artificial magnetic/nonmagnetic multilayers,
allowing for potential applications in magnetic recording.
However, as discussed extensively below, while a large body of
subsequent experimental work has shown that the MR factor can actually
be made very close to 100\% (for a definition of the MR ratio see
below), this occurs unfortunately at the cost of reducing the Curie
temperature $T_{\rm C}$, which jeopardizes those possible technological 
applications.
Consider for instance in Fig.I.2b the results for 
$\rm Nd_{0.5} Pb_{0.5} Mn O_3$ reproduced from Kusters et al. (1989). 
In this case the change in resistivity is larger than for the (La,Ba)
compound, however its Curie temperature is reduced to 184K. 
Also complicating possible applications, it is known that giant MR
multilayer structures present their appreciable changes in resistivity
with fields as small as 0.01 Tesla (Helman and Abeles, 1976; Fert and
Campbell, 1976), while manganites typically need larger fields of
about 1 Tesla or more for equivalent resistivity changes, which appear
too large for potential use in magnetic recording.
Although progress in the development of applications is frequently
reported (for recent references see Chen et al. (2000), Venimadhav et
al. (2000), Kida and Tomouchi (2000b)), 
in this review the manganites will be mainly considered as
an interesting basic-physics problem, with emphasis focused on 
understanding the microscopic origin of the large MR effect which
challenges our current knowledge of strongly Correlated Electron
systems.
The discussion of possible applications of manganites is
left for future reviews.

The big boost to the field of manganites that led to the present
explosion of interest in the subject was produced by the discovery of 
the so-called ``Colossal'' Magnetoresistance (CMR) effect.
In studies of thin films of $\rm La_{0.67} Ca_{0.33} Mn O_x$,
a MR effect three orders of magnitude larger than the
typical ``giant'' MR of superlattice films was observed (the name colossal was
coined mainly to distinguish the effect from this previously found
giant MR effect).   
Defining the MR ratio as $\Delta R/R(H)$=$(R(0)-R(H))/R(H)$, where
$R(0)$ and $R(H)$ are the resistances without and with a magnetic
field $H$, respectively, and expressing the result as a percentage
(i.e., multiplying by an additional factor 100) it has been shown that
MR ratios as large as 127,000\% near 77K can be obtained
(Jin et al., 1994). 
This corresponds to more than a thousand-fold change in resistivity.
Alternatively, expressed in terms of
$\Delta R/R(0)$=$(R(0)-R(H))/R(0)$ the MR ratio in this case is as
large as 99.92\%. 
Xiong et al. (1995) reported thin-films studies of 
$\rm Nd_{0.7} Sr_{0.3} Mn O_{\delta}$ and in this case 
$\Delta R/R(H)$ was as high as $10^6$\%, a truly colossal factor.
Triggered by such huge numbers, the experimental and theoretical study
of manganites reignited, and is presently carried out by dozens of
groups around the world. 
Early work tended to focus on the x=0.3 doping due to its large
$T_{\rm C}$.
However, more recently the attention has shifted towards other
densities such as x$<$0.2 or x$>$0.5, where the competition between
the various states of manganites can be better analyzed. In fact, one
of the main results of recent investigations is that in order to
understand the CMR effect, knowledge of the ferromagnetic metallic
phase is not sufficient. The competing phases must be understood as
well. This issue will be discussed extensively below.

Previous reviews on manganites are already available, but they have
mainly focused on experiments.
For instance, the reader can consult the reviews of Ramirez (1997),
Coey, Viret, and von Molnar (1998), as well as the books recently
edited by Rao and Raveau (1998), Kaplan and Mahanti (1999) and Tokura
(1999). See also Loktev and Pogorelov (2000).
The present review differs from previous ones in several aspects: 
(i) It addresses theoretical work in detail, especially regarding the 
stabilization in a variety of calculations of the many nontrivial
charge/spin/orbital phases found in experiments; 
(ii) it highlights the importance of {\it phase separation} tendencies in 
models for manganites and the potential considerable influence of
disorder on transitions that would be first-order in the clean limit,
leading to percolative processes; and 
(iii) it emphasizes the experimental results that have recently
reported the presence of intrinsic inhomogeneities in manganites,
results that appear in excellent agreement with the theoretical
developments.
If the review could be summarized in just a few words, the overall
conclusion would be that theoretical and experimental work is
rapidly converging to a unified picture pointing toward a physics of
manganites in the CMR regimes clearly dominated by $inhomogeneities$
in the form of coexisting competing phases.
This is an $intrinsic$ feature of single crystals, unrelated to grain
boundary effects of polycrystals, and its theoretical understanding
and experimental control is a challenge that should be strongly
pursued.
In fact, in spite of the considerably progress in recent years
reviewed here, it is clear that the analysis of mixed-phase systems 
is at its early stages, and considerable more work should be devoted
to the detailed study of manganese oxides and related compounds in
such a regime.

In this review, it is assumed that the reader is familiar with some
basic phenomenology involving the $d$-orbitals of relevance in
manganese oxides.
In the cubic lattice environment, the five-fold degenerate
$3d$-orbitals of an isolated atom or ion are split into a manifold of
three lower energy levels ($d_{xy}$, $d_{yz}$, and $d_{zx}$), usually
referred to as ``$t_{\rm 2g}$'', and two higher energy  states
($d_{x^2-y^2}$ and $d_{3z^2-r^2}$)  called ``$e_{\rm g}$''.
The valence of the Mn-ions in this context is either four (Mn$^{3+}$)
or three (Mn$^{4+}$), and their relative fraction is controlled through
chemical doping. 
The large Hund coupling favors the population of the $t_{\rm 2g}$
levels with  three electrons forming a spin 3/2 state, and the 
$e_{\rm g}$ level either contains one electron or none.
A sketch with these results is shown in Fig.I.3.

The organization is the following.
In Section II, the most basic properties of manganites will be
reviewed from the experimental viewpoint.
Emphasis will be given to the phase diagrams and magnitude of the CMR 
effect in various manganites.
For this section, the manganites will be divided into large,
intermediate, and small bandwidth $W$ compounds, a slightly unorthodox 
classification since previous work simply labeled them as either large
or small $W$.
In Section III, the theoretical aspects are presented, starting with 
the early developments in the subject.
Models and approximations will be discussed in detail, and results
will be described. Especially, the key importance of the recently
found phase separation tendencies will be remarked.
In Section IV, the experimental work that have reported evidence of 
intrinsic inhomogeneities in manganites compatible with the
theoretical calculations will be reviewed. 
Finally, conclusions are presented in Section V including the problems
still open.

\section{BASIC PROPERTIES, PHASE DIAGRAMS, AND CMR EFFECT IN
  MANGANITES} 

\begin{center}
  {\bf II.a Large-Bandwidth Manganites: \\
    The case of $\bf {La_{1-x} Sr_x Mn O_3}$}
\end{center}

In the renaissance of the study of manganites during the 90's, 
a considerable emphasis has been given to the analysis of $\LSMO$,
a material considered to be representative of the ``large'' bandwidth
subset of manganese oxides.
It is believed that in this compound the hopping amplitude for
electrons in the $e_{\rm g}$-band is larger than in other manganites,
as a consequence of the sizes of the ions involved in the chemical
composition, as discussed below.
Its Curie temperature $T_{\rm C}$ as a function of hole doping is
relatively high, increasing its chances for future practical
applications.
It has also been found that $\LSMO$ presents a complex behavior in the 
vicinity of x=1/8 (see below), with a potential phenomenology as rich
as found in the low-bandwidth manganites described later in this
section. 
Resistivities vs. temperature for this compound at several densities
are shown in Fig.II.a.1a (taken from Urushibara et al. (1995)). From
these transport measurements, the phase diagram of this compound can
be determined and it is shown in Fig.II.a.1b (Y. Tokura
and Y. Tomioka, prepared with data from Urushibara et al. (1995) and
Fujishiro et al. (1998). See also Y. Moritomo et al. (1998)).
At hole concentrations such as x=0.4, the system is metallic 
(defined straightforwardly as regions where
${\rm d}\rho_{\rm dc}/{\rm d}T$$>$0) even above $T_{\rm C}$.
At densities above x=0.5 an interesting A-type antiferromagnetic
metallic state is stabilized, with ferromagnetism in planes and
antiferromagnetism between those planes.
This phase was actually first observed in another compound
$\NSMO$ (Kawano et al., 1997) and is believed to have
$d_{x^2-y^2}$-type uniform orbital order within the ferromagnetic
planes (for a visual representation of this state see Fig.4 of
Kajimoto et al., 1999).
Considering now lower hole densities, at concentrations slightly below
x=0.30 the state above $T_{\rm C}$ becomes an insulator, which is an
unexpected property of a paramagnetic state that transforms into a
metal upon reducing the temperature.
This curious effect is present in all the intermediate and low
bandwidth manganites, and it is a key property of this family of
compounds.
At densities x$\lesssim$0.17, an insulating state is found even at low
temperatures.
As reviewed in more detail below, currently considerable work in the
context of $\LSMO$ is being focused on the x$\sim$0.12 region, simply
labeled ``ferromagnetic insulator'' in Fig.II.a.1b.
In this region, indications of charge-ordering have been found even in
this putative large-bandwidth material, establishing an
interesting connection with intermediate and low bandwidth manganites
where charge-ordering tendencies are very prominent.
A revised phase diagram of $\LSMO$ will be presented later when the
experimental evidence of inhomogeneous states in this compound is
discussed.

Studies by Tokura et al. (1994) showed that the MR effect is maximized
in the density region separating the insulating from metallic states
at low temperature, namely x=0.175. The MR effect here is shown in
Fig.II.a.2a, taken from Tokura et al. (1994).
The Curie temperature is still substantially large in this regime,
which makes it attractive from the viewpoint of potential
applications.
A very important qualitative aspect of the results shown in
Fig.II.a.2a is that the MR effect is maximized at the
smallest $T_{\rm C}$ which leads to a metallic ferromagnetic state;
this observation let the authors to conclude that the large MR regime can
only be clearly understood when the various effects which are
in competition with DE are considered.
This point will be a recurrent conclusion emerging from the
theoretical calculations reviewed below, namely in order to achieve a
large MR effect the insulating phase is at least as important as
the metallic phase, and the region of the most interest should be the
$boundary$ between the metal and the insulator (Moreo, Yunoki and
Dagotto, 1999a; Moreo et al., 2000).
These insulating properties occur at low temperature by changing the 
density, or, at fixed density, by increasing the temperature, at least in
some density windows. It is at the metal-insulator boundary where the
tendencies to form coexisting clusters and percolative transitions are
the most important. This point of view is qualitatively different from
the approach followed
in previous theories based on polaronic formation, Anderson
localization, or on modifications of the double-exchange ideas, but it
is crucial in the phase-separation based approaches described here.
Note, however, that the metallic phase of $\LSMO$ at sufficiently
large hole density seems quite properly described by double-exchange
approaches, namely there is a simple relation between the resistivity
and the magnetization in the metallic ferromagnetic phase
(Tokura et al., 1994; Furukawa, 1998).
Then, it is important to distinguish the theoretical understanding of
the individual phases, far from others in parameter space, from the
understanding of the $competition$ among them.
It is the latest issue that is the most important for the explanation
of the MR effect according to recent calculations (Moreo, Yunoki, and
Dagotto, 1999a; Moreo et al., 2000).

It is also interesting to point out that the low temperature
ferromagnetic metallic state that appears prominently in Fig.II.a.1b
is actually ``unconventional'' in many respects.
For instance, in Fig.II.a.2b, taken from Tokura (1999), it is shown
that the total low energy spectral weight of the optical conductivity
$N_{\rm eff}$ is still changing even in the low temperature region,
where the spin is already almost fully polarized. Clearly there is
another scattering mechanism in the system besides the spin.
Actually, even within the FM-state the carrier motion is mostly
incoherent since the Drude weight is only 1/5 of the total low-energy
weight. The conventional Drude model is not applicable to the FM-state
of manganites. Probably the orbital degrees of freedom are important
to account for this effect.

\begin{center}
  {\bf II.b Intermediate-Bandwidth Manganites: \\ 
    The case of $\bf {La_{1-x} Ca_x Mn O_3}$}
\end{center}

Currently, a large fraction of the work in manganites focuses on
intermediate to low bandwidth materials since these are the ones that 
present the largest CMR effects, which are associated with the
presence of charge ordering tendencies.
Unfortunately, as discussed in the Introduction, this comes at a
price: The Curie temperature decreases as the magnitude of the MR
effects increases.
In this section, the properties of $\LCMO$ will be discussed in
detail.
This compound presents some characteristics of large bandwidth
manganites, such as the presence of a robust ferromagnetic phase.
However, it also has features that indicate strong deviations from
double-exchange behavior, including the existence of
charge/orbital-ordered phases.
For this reason, the authors consider that this compound should be 
labeled as of ``intermediate bandwidth'', to distinguish it from the truly
low bandwidth compound $\PCMO$ where a metallic ferromagnetic phase
can only be stabilized by the application of magnetic fields.

$\LCMO$ has been analyzed since the early days of manganite studies
(Jonker and van Santen, 1950; Wollan and Koehler, 1955; Matsumoto,
1970b), but it is only recently that it has been systematically
scrutinized as a function of density and temperature.
In particular, it has been observed that $\LCMO$ has a large MR
effect.
For example, Fig.II.b.1 reproduces results from Schiffer et al. (1995)
at x=0.25 showing the magnetization and resistivity as a function of
temperature, and the existence of a robust MR, larger than 80\%.
The drop in $\rho_{\rm dc}(T)$ with decreasing temperature and the
peak in MR are correlated with the ferromagnetic transition in the 
magnetization.
The insulating behavior above the Curie temperature is very prominent
and the explanation of its origin is among the most important issues
to be addressed in theories of manganites, as already discussed in the 
previous subsection.
Below $T_{\rm C}$ the presence of ferromagnetism was 
tentatively attributed to the double exchange mechanism, but further
work reviewed in Sec.~IV has actually revealed a far more complex
structure with coexisting phases even in this metallic regime.
In fact, hints of this behavior may already be present in Fig.II.b.1
which already reveals a MR effect as large as 30\% well below
$T_{\rm C}$.
In addition, it is also interesting to observe that hydrostatic
pressure leads to large changes in resistivity comparable to those
found using magnetic fields [see for instance Fig.II.b.2 where
$\rho_{\rm dc}(T)$ is shown parametric with pressure at x=0.21,
taken from Neumeier et al. (1995). See also Hwang et al. (1995b)].

The qualitative features of Fig.II.b.1 contribute to the
understanding, at least in part, of the CMR effect found in 
thin-films of this same 
compound (Jin et al., 1994. For references on
thin film work in manganites, see Ramirez, 1997, page 8182. See also
Kanki, Tanaka and Kawai, 2000).
In the work of Jin et al. (1994), $T_{\rm C}$ was suppressed by
substrate-induced strain, and, as a consequence, the $\rho_{\rm dc}$
was much higher immediately above the transition than in crystals
since the system was still in the insulating state, inducing an
enormous change in resistivity.
Thus, it appears that to understand the large MR values the insulating
state is actually $more$ important than the metallic state, and the
relevance of the DE ideas is limited to the partial explanation of the
low temperature phase in a narrow density window, as explained in more
detail later.
Clearly the DE framework is insufficient for describing the physics of
the manganites.

The complete phase diagram of $\LCMO$, based on magnetization and
resistivity measurements, is reproduced in Fig.II.b.3, taken from
Cheong and Hwang (1999).
Note that the FM phase actually occupies just a fraction of the whole
diagram, illustrating once again that DE does not provide a full
understanding of the manganites.
For instance, equally prominent are the charge ordered (CO) states
between x=0.50 and 0.87.
The CO state at x=0.50 was already described by Wollan and Koehler
(1955) as a CE-state, and the characteristics at other densities are
discussed below.
In the regime of CO-states, studies by Ramirez et al. (1996) of the
sound velocity, specific heat, and electron diffraction were
attributed to strong electron-phonon coupling, in
agreement with the predictions of Millis, Shraiman and Littlewood
(1995). The ``canted'' state at x close to 1 could be a mixed-phase
state with coexisting FM-AF characteristics based on recent
theoretical calculations (see below Sec.III), but the issue is still
under discussion.
The low hole-density regime is quite unusual and nontrivial, and it
appears to involve a charge-ordered phase, and a curious ferromagnetic
insulator.
Actually at x=0.10, there is no large MR effect using fields of 12T,
according to Fig.~6 of Ibarra and De Teresa (1998c). 
Figure 25 of the same reference shows that at x=0.65, well inside the
charge-ordered state, a 12T field is also not sufficient to
destabilize the insulating state into a metallic one.
Thus, to search for a large MR effect, the density must be closer to
that leading to the FM metallic regime, as emphasized before.

In Fig.II.b.3 note also the presence of well-defined features at
commensurate carrier concentrations x=$N$/8 ($N$=1,3,4,5 and 7).
The Curie temperature is maximized at x=3/8 according to Cheong and
Hwang (1999), contrary to the x=0.30 believed by many to be the most
optimal density for ferromagnetism.
Cheong and Hwang (1999) also remarked that in the large-bandwidth
compound $\LSMO$, $T_{\rm C}$ is also maximized at the same x=3/8
concentration, implying that this phenomenon is universal.
It is important to realize that within a simple one-orbital
double-exchange model, as described later, the optimal density for
ferromagnetism should be x=0.50.
The fact that this is not observed is already indicative of the
problems faced by a double-exchange description of manganites. 
Note also that Zhao et al. (1996,1999) found a giant oxygen isotope
shift in $T_{\rm C}$ of about 20K at x=0.2, showing the relevance of 
electron-phonon couplings in manganites, a recurrent result of many
papers in this context.

The charge-ordering temperature $T_{\rm CO}$ peaks at x=5/8 (the same
occurs in (Bi,Ca)-based compounds), while at x=4/8=1/2 there is a
sharp change from ferromagnetic to antiferromagnetic ground states.
The whole phase diagram has a pronounced electron-hole $asymmetry$,
showing again that simple double-exchange models with only one orbital
are not realistic.
At x=1/8 the low density charge-ordered state appears to have the
largest strength, while on the other side at x=7/8 charge ordering
disappears into a mixed FM-AF state.
Finally, at x=0 the ground state is an A-type antiferromagnet (see
also Matsumoto, 1970a) with ferromagnetic spin correlations on a plane 
and antiferromagnetism between planes, while at x=1 it is a G-type
antiferromagnet (AF in all directions), both of them insulating.

The pattern of charge- and orbital-order in the CO states of
Fig.II.b.3 is highly nontrivial and at several densities still under
discussion (for early work in the context of orbital ordering see
Kugel and Khomskii (1974), and Eremin and Kalinenkov (1978 and 1981)).
Some of the arrangements that have been identified are those shown in
Fig.II.b.4, reproduced from Hwang and Cheong (1999).
At x=0, the A-type spin state is orbitally-ordered as it appears in
Fig.II.b.4a. At x=0.5 the famous CE-type arrangement (Fig.II.b.4b)
already found in early studies of manganites is certainly stabilized.
This state has been recently observed experimentally using resonant
x-ray scattering (Zimmermann et al., 1999. See also Zimmermann et al., 2000).
At x=2/3, and also x=3/4, a novel ``bi-stripe'' arrangement is found 
(Mori, Chen, and Cheong, 1998a).
The x=0.65 state is very stable upon the application of a magnetic
field (Fig.25 of Ibarra and De Teresa, 1998c).
The origin of the term bi-stripe is obvious from Fig.II.b.4c.
However, theoretical work (Hotta et al., 2000 and references therein)
has shown that it is more appropriate to visualize this arrangement as
formed by FM zigzag chains running in the direction $perpendicular$ to 
those of the charge stripes of Fig.II.b.4c. 
This issue will be discussed in more detail later in the review when
the theoretical aspects are addressed.
Based on electron microscopy techniques Hwang and Cheong (1999)
believe that at the, e.g., x=5/8 concentration a mixture of the x=1/2
and x=2/3 configurations forms the ground state.
The size of the coexisting clusters is approximately 100 \AA.
Once again, it appears that phase separation tendencies are at work in
manganese oxides.

However, note that studies by Radaelli et al. (1999) on the x=2/3 
compound arrived to the conclusion that a ``Wigner
crystal'' charge arrangement is stable at this density, with the
charge ordered but spread as far from each other as possible.
It appears that bi-stripes and Wigner crystal states must be very
close in energy.
While the results at x=0.0 and 0.5 have been reproduced in recent 
theoretical studies of manganite models, the more complex arrangements 
at other densities are still under analysis (Hotta et al., 2000) and
will be discussed in more detail below.

Finally, there is an interesting observation that is related with 
some theoretical developments to be presented later in the review.
In Fig.II.b.5, the resistivity at 300K and 100K vs. hole density is
shown, reproduced from Cheong and Hwang (1999).
Note at 300K the smooth behavior as x grows from 0, only interrupted
close to x=1 when the G-type AF insulating state is reached.
Then, at 300K there is no $precursor$ of the drastically different
physics found at, e.g., 100K where for x$<$0.5 a FM-state is found
while for x$>$0.5 the state is CO and AF. 
This lack of precursors is also in agreement with neutron scattering 
results that reported FM fluctuations above the CO and N\'eel
temperatures in the large x regime (Dai et al., 1996), similar to those
observed at lower hole densities.
These results are consistent with an $abrupt$ first-order-like
transition from the state with FM fluctuations 
to the CO/AF-state, as observed in other 
experiments detailed in later sections (for very recent work see
Ramos et al., 2000).
These two states are so different that a smooth transition between
them is not possible.
In addition, recent theoretical developments assign considerable
importance to the influence of disorder on this type of first-order
transitions to explain the large MR effect in manganites 
(Moreo et al., 2000), as shown elsewhere in this review.
Then, the sudden character of the transition from ferromagnetism to
charge-ordered antiferromagnetism appears to play a $key$ role in the
physics of manganites, and its importance is emphasized in many places
in the text that follows.

\begin{center}
  {\bf II.c Low-Bandwidth Manganites: \\ 
    The case of $\bf {Pr_{1-x} Ca_x Mn O_3}$}
\end{center}

As explained before, in perovskite manganites, such as $\LCMO$ and
the compound $\PCMO$ described here, the bandwidth $W$ is smaller than 
in other compounds that have a behavior more in line with the standard
double-exchange ideas. 
In the low bandwidth compounds, a charge-order state is stabilized in
the vicinity of x=0.5, while manganites with a large $W$ ($\LSMO$ as 
example) present a metallic phase at this hole density.
Let us focus in this subsection on $\PCMO$ which presents a
particularly stable CO-state in a broad density region between x=0.30
and x=0.75, as Jirak et al. (1985) showed.
Part of the phase diagram of this compound is in Fig.II.c.1, reproduced
from Tomioka et al. (1996) (see also Tomioka and Tokura (1999)).
Note that a metallic ferromagnetic phase is $not$ stabilized at zero
magnetic field and ambient pressure in this low-bandwidth compound.
Instead, a ferromagnetic insulating (FI) state exists in the range
from x=0.15 to 0.30. This FI state has not been fully explored to the
best of our knowledge, and it may itself present charge-ordering as
some recent theoretical studies have suggested (Hotta and Dagotto,
2000).
For x$\geq$0.30, an antiferromagnetic CO-state is stabilized.
Neutron diffraction studies (Jirak et al., 1985) showed that at $all$
densities between 0.30 and 0.75, the arrangement of
charge/spin/orbital order of this state is similar to the CE-state
(see Fig.II.b.4b) already discussed in the context of x=0.5
(La,Ca)-based manganites.
However, certainly the hole density is changing with x, and as a
consequence the CE-state cannot be ``perfect'' at all densities
but electrons have to be added or removed from the structure.
Jirak et al.(1985) discussed a ``pseudo''-CE-type structure for x=0.4 
that has the proper density. Other authors simply refer to the
x$\neq$0.5 CO-states as made out of the x=0.5 structure plus
``defects''. 
Hotta and Dagotto (2000) proposed an ordered state for x=3/8 based on 
mean-field and numerical approximations.
Neutron diffraction studies have shown that the coupling along the
$c$-axis changes from AF at x=0.5 to FM at x=0.3 (Yoshizawa et al.,
1995) and a canted state has also been proposed to model this
behavior.
Certainly more work is needed to fully understand the distribution 
of charge in the ground-state away from x=0.5.

The effect of magnetic fields on the CO-state of $\PCMO$ is
remarkable.
In Fig.II.c.2 the resistivity vs. temperature is shown parametric with 
magnetic fields of a few Teslas, which are small in typical electronic
units.
At low temperatures, changes in $\rho_{\rm dc}$ by several orders of 
magnitude can be observed.
Note the stabilization of a metallic state upon the application of the
field.
This state is ferromagnetic according to magnetization measurements,
and thus it is curious to observe that a state not present at zero
field in the phase diagram, is nevertheless stabilized at finite
fields, a  puzzling result that is
certainly difficult to understand.
The shapes of the curves in Fig.II.c.2 resemble similar measurements 
carried out in other manganites which also present a large MR effect,
and a possible origin based on percolation between the CO- and
FM-phases will be discussed later in the review.
First-order characteristics of the metal-insulator transitions in this
context are very prominent, and they have been reviewed by Tomioka and
Tokura (1999).
It is interesting to observe that pressure leads to a colossal MR
effect quite similar to that found upon the application of magnetic
fields (see for example Fig.II.c.3, where results at x=0.30 from
Moritomo et al. (1997) are reproduced).

The abrupt metal-insulator transition at small magnetic fields found
in $\PCMO$ at x=0.30 appears at other densities as well, as
exemplified in Fig.II.c.4, which shows the resistivity vs temperature
at x=0.35, 0.4 and 0.5, reproduced from Tomioka et al. (1996). 
Figure II.c.5 (from Tomioka and Tokura, 1999) shows that as x grows
away from x=0.30, larger fields are needed to
destabilize the charge-ordered state at low temperatures  
(e.g., 27T at x=0.50 compared
with 4T at x=0.30).
It is also interesting to observe that the replacement of Ca by Sr 
at x=0.35 also leads to a metal-insulator transition, as shown in
Fig.II.c.6 taken from Tomioka et al. (1997).
Clearly $\PCMO$ presents a highly nontrivial behavior that challenges
our theoretical understanding of the manganese oxide materials. 
The raw huge magnitude of the CMR effect in this compound highlights
the relevance of the CO-FM competition. 

\begin{center}
  {\bf II.d Other Perovskite Manganite Compounds}
\end{center}

Another interesting perovskite manganite compound is $\NSMO$, and its
phase diagram is reproduced in Fig.II.d.1 (from Kajimoto et al., 1999).
This material could be labelled as ``intermediate-bandwidth'' 
due to the presence of a stable CO phase at x=0.50, state which can be 
easily destroyed by a magnetic field in a first-order transition
(Kuwahara et al., 1995). 
However, this phase appears only in a tiny range of densities and at
low temperature.
In fact, aside from this CO-phase, the rest of the phase diagram is
very similar to the one of $\LSMO$.
In particular, it is interesting to observe the presence of an A-type
antiferromagnetic metallic structure which is believed to have
ferromagnetic planes with $uniform$ $d_{x^2-y^2}$-type orbital order
(Kawano et al., 1997), making the system effectively anisotropic
(Yoshizawa et al., 1998). 
A compound that behaves similarly to $\NSMO$ is $\PSMO$, with the
exception of x=0.5: In $\PSMO$(x=0.5) the CO-state is not stable. 
Actually, Tokura clarified to the authors (private communication) that
the polycrystal results that appeared in Tomioka et al. (1995) showing
a CO-phase in this compound were later proven incorrect after the
preparation of single crystals. 
Nevertheless such a result illustrates the fragile
stability of the CO-phase in materials where the bandwidth is not
sufficiently small.
Results for this compound reporting mixed-phase tendencies were
recently presented by Zvyagin et al. (2000).
The corresponding phase diagram for a mixture 
$\rm (La_{1-z} Nd_z)_{1-x} Sr_x Mn O_3$ can be found in 
Akimoto et al. (1998) and it shows that the CO-phase at x=0.5 of the
pure (Nd,Sr) compound disappears for z smaller than $\sim$0.5.
The phase diagram of $\rm (La_{1-z} Nd_z)_{1-x} Ca_x Mn O_3$
investigated by Moritomo (1999) also shows 
a competition between FM and CO, with
phase separation characteristics in between.

Other manganites present CO-phases at x=0.5 as well, and the compound 
where this phase seems to be the strongest is 
$\rm Sm_{0.5} Ca_{0.5} Mn O_3$, as exemplified in Fig.II.d.2, where
the effect of magnetic fields  on several low-bandwidth manganites is
shown.
An interesting way to visualize the relative tendencies of manganite 
compounds to form a CO-state at x=0.5 can be found in Fig.II.d.3, taken 
from Tomioka and Tokura (1999).
As discussed in more detail at the end of this section, the key
ingredient determining the FM vs CO character of a state at a fixed
density is the size of the ions involved in the chemical composition.
In Fig.II.d.3 the radius of the trivalent and divalent ions, as well
as their average radius at x=0.5, appear in the horizontal axes.
The Curie and CO temperature are shown below in part (b).
As an example, for the extreme case of (La,Sr) based manganites, a
metallic ferromagnetic state is observed at x=0.5, while (Pr,Ca)
compounds are charge-ordered.
As a byproduct of Fig.II.d.2, it is quite interesting to note the
similarities between the actual values of the critical temperatures
$T_{\rm C}$ and $T_{\rm CO}$.
Being two rather different states, there is no obvious reason why
their critical temperatures are similar. 
A successful theory must certainly address this curious fact.

\begin{center}
  {\bf II.e Double-Layer Compound}
\end{center}

Not only three-dimensional perovskite-type structures are present in
the family of manganite compounds, but layered ones as well.
In fact, Moritomo et al. (1996) showed that it is possible to prepare
double-layer compounds with a composition $\bilayer$. 
Single-layer manganites can also be synthesized, as will be discussed
in the next subsection.
In fact these are just special cases of the Ruddlesden-Popper series
$\rm (T_{1-x} D_x )_{n+1} Mn_n O_{3n+1}$, with T a trivalent cation, 
D a divalent cation, and n=1 corresponding to the single layer, n=2 to
the double layer, and n=$\infty$ to the cubic perovskite structure
(see Fig.II.e.1 for the actual structure).
The temperature dependence of the resistivity in representative
multilayer structures is shown in Fig.II.e.2 for the n=1, n=2, and
n=$\infty$ (3D perovskite) compounds at a hole concentration of  
x=0.4. In the regime where the single-layer is insulating and the
n=$\infty$ layer is metallic, the double-layer has an intermediate
behavior, with insulating properties above a critical temperature and
metallic below. A large MR effect is observed in this double-layer 
system as shown in Fig.II.e.3, larger than the one found for $\LSMO$.
The full phase diagram of this compound will be discussed later in
this review (Section IV) in connection with the presence of
inhomogeneities and clustering tendencies.

\begin{center} 
  {\bf II.f Single-Layer Compound}
\end{center}

As mentioned in the previous subsection, the single-layer manganite
has also been synthesized (see Rao et al., 1988).
Its chemical formula is $\single$.
This compound does not have a ferromagnetic phase in the range
from x=0.0 to x=0.7, which is curious. Note that other manganites
have not presented a ferromagnetic metallic phase also, but they had
at least a ferro-insulating regime (e.g., $\PCMO$).
A schematic phase diagram of the one-layer compound is given in
Fig.II.f.1, reproduced from Moritomo et al. (1995). 
At all the densities shown, insulating behavior has been found.
Note the prominent CO-phase near x=0.5, and especially the
``spin-glass'' phase in a wide range of densities between 
x=0.2 and x$\sim$0.5.
The x=0.5 charge-ordered phase is of the CE-type (Sternlieb et al.,
1996; Y. Murakami et al., 1998).
The large x regime has phase separation according to Bao et al. (1996),
as discussed in more detail below.
The actual microscopic arrangement of charge and spin in the
intermediate spin-glass regime has not been experimentally studied in
detail, to the best of our knowledge, but it certainly deserves more
attention since the two-dimensionality of the system makes possible
reliable theoretical studies and simulations.

\begin{center}
  {\bf II.g Importance of Tolerance Factor}
\end{center}

It has been clearly shown experimentally that working at a fixed
hole density the properties of manganites strongly depend on a
geometrical quantity called the ``tolerance factor'',
defined as $\Gamma$=$d_{\rm A-O}/(\sqrt{2} d_{\rm Mn-O})$.
Here $d_{\rm A-O}$ is the distance between the A site, where the
trivalent or divalent non-Mn ions are located, to the nearest oxygen.
Remember that the A ion is at the center of a cube with Mn in the
vertices and O in between the Mn's.
$d_{\rm Mn-O}$ is the Mn-O shortest distance.
Since for an undistorted cube with a straight Mn-O-Mn link, 
$d_{\rm A-O}$=$\sqrt{2}$ and $d_{\rm Mn-O}$=1 in units of the Mn-O
distance, then $\Gamma$=1 in this perfect system. 
However, sometimes the A ions are too small to fill the space in the
cube centers and for this reason the oxygens tend to move toward that
center, reducing $d_{\rm A-O}$. 
In general $d_{\rm Mn-O}$ also changes at the same time.
For these reasons, the tolerance factor becomes less than unity, 
$\Gamma$$<$1, as the A radius is reduced, and the Mn-O-Mn angle
$\theta$ becomes smaller than 180$^{\circ}$.
The hopping amplitude for carriers to move from Mn to Mn naturally 
decreases as $\theta$ becomes smaller than 180$^{\circ}$ 
(remember that for a 90$^{\circ}$ bond the hopping involving a
$p$-orbital at the oxygen simply cancels, as explained in more detail 
below).
As a consequence, as the tolerance factor decreases, the tendencies to
charge localization increase due to the reduction in the mobility of
the carriers.
Since in the general chemical composition for perovskite manganites 
$\rm A_{1-x} A'_{x} Mn O_3$ there are two possible ions at the ``A''
site, then the tolerance factor for a given compound can be defined as
a density-weighted average of the individual tolerance factors.
In Fig.II.d.3 the reader can find  some of the ionic radius in 
$\rm \AA$, for some of the most important elements in the manganite
composition.

Note that the distance Mn-Mn is actually reduced in the situation 
described so far ($\Gamma$$<$1), while the tolerance factor
(monotonically related with the hopping amplitude) is also reduced,
which is somewhat counterintuitive since it would be expected that
having closer Mn-ions would increase the electron hopping between
them. 
However, the hopping amplitude is not only proportional to
$1/(d_{\rm Mn-O})^{\alpha}$, where $\alpha$$>$1 (see Harrison, 1989)
but also to $\cos \theta$ due to the fact that it is the
$p$-orbital of oxygen that is involved in the process and if this
orbital points toward one of the manganese ions, it can not point
toward the other one simultaneously for $\theta$$\neq$180$^{\circ}$.

Hwang et al. (1995a) carried out a detailed study of the 
$\rm A_{0.7} A'_{0.3} Mn O_3$ compound for a variety of A and A$'$ ions.
Figure II.g.1a summarizes this effort, and it shows the presence of 
three dominant regimes: a paramagnetic insulator at high temperature,
a low temperature ferromagnetic metal at large tolerance factor,
and a low temperature charge-ordered ferromagnetic-insulator at small
tolerance factor.  
This figure clearly illustrates the drastic dependence with the
tolerance factor of the properties of doped manganites.
These same results will be discussed in more detail below in this
review, when issues related with the presence of coexisting phases are
addressed. In particular, experimental work have shown that the
``FMI'' regime may actually correspond to coexisting CO and FM large
clusters. 
The CO phase has both charge and orbital order.

An example upon which Fig.II.g.1a has been constructed is shown in
Fig.II.g.1b that mainly corresponds to results obtained for
$\rm La_{0.7-x} Pr_x Ca_{0.3} Mn O_3$.
The temperature dependence of $\rho_{\rm dc}(T)$ presents hysteresis 
effects, suggesting that the PMI-FMM transition has some $first$ order
characteristics, a
feature that is of crucial importance in recent theoretical
developments to be discussed later
(Yunoki, Hotta, Dagotto, 2000; Moreo et al., 2000).
Note the huge MR ratios found in these compounds and the general trend
that this ratio dramatically increases as $T_{\rm C}$ is reduced,
mainly as a consequence of the rapid increase of the resistivity of
the PM insulating state as the temperature is reduced. 
Certainly the state above $T_{\rm C}$ is not a simple metal where
ferromagnetic correlations slowly build up with decreasing temperature
as in a second order transition, and as expected in the DE mechanism.
A new theory is needed to explain these results.

\section{THEORY OF MANGANITES}

\begin{center}
  {\bf III.a Early Studies} 
\end{center}

\noindent{\bf Double Exchange:}

Most of the early theoretical work on manganites focused on the
qualitative aspects of the experimentally discovered relation between
transport and magnetic properties, namely the increase in conductivity
upon the polarization of the spins. 
Not much work was devoted to the magnitude of the magnetoresistance
effect itself.
The formation of coexisting clusters of competing phases was not
included in the early considerations. 
The states of manganites were assumed to be uniform, and 
``Double Exchange'' (DE) was proposed by Zener (1951b) as a way to 
allow for charge to move in manganites by the generation of 
a spin polarized state.
The DE process has been historically explained in two somewhat
different ways.
Originally, Zener (1951b) considered the explicit movement of
electrons schematically written (Cieplak, 1973) as 
$\rm Mn^{3+}_{1\uparrow}$$\rm O_{2\uparrow,3\downarrow}$$\rm Mn^{4+}$
$\rightarrow$
$\rm Mn^{4+}$$\rm O_{1\uparrow,3\downarrow}$$\rm Mn^{3+}_{2\uparrow}$
where 1, 2, and 3 label electrons that belong either to the oxygen
between manganese, or to the $e_{\rm g}$-level of the Mn-ions.
In this process there are two $simultaneous$ motions (thus the name
double-exchange) involving electron 2 moving from the oxygen to the
right Mn-ion, and electron 1 from the left Mn-ion to the oxygen
(see Fig.III.a.1a). 
The second way to visualize DE processes was presented in detail by
Anderson and Hasegawa (1955) and it involves a second-order process in
which the two states described above go from one to the other using an
intermediate state $\rm Mn^{3+}_{1\uparrow}$$\rm O_{3\downarrow}$
$\rm Mn^{3+}_{2\uparrow}$.
In this context the effective hopping for the electron to move from
one Mn-site to the next is proportional to the square of the hopping
involving the $p$-oxygen and $d$-manganese orbitals ($t_{\rm pd}$). 
In addition, if the localized spins are considered classical and with
an angle $\theta$ between nearest-neighbor ones, the effective hopping
becomes proportional to $\rm cos(\theta/2)$, as shown by Anderson and
Hasegawa (1955). 
If $\theta$=0 the hopping is the largest, while if $\theta$=$\pi$,
corresponding to an antiferromagnetic background, then the hopping
cancels. The quantum version of this process has been described by
Kubo and Ohata (1972).  

Note that the oxygen linking the Mn-ions is crucial to understand the
origin of the word ``double'' in this process. 
Nevertheless, the majority of the theoretical work carried out in the
context of manganites simply forgets the
presence of the oxygen and uses a manganese-only Hamiltonian.
It is interesting to observe that ferromagnetic states appear in this
context even $without$ the oxygen.
It is clear that the electrons simply need a polarized background to
improve their kinetic energy, in similar ways as the Nagaoka phase is
generated in the one-band Hubbard model at large $U/t$ 
(for a high-$T_{\rm c}$ review, see Dagotto, 1994). 
This tendency to optimize the kinetic energy is at work in a variety
of models and the term double-exchange appears unnecessary.
However, in spite of this fact it has become customary to refer to
virtually any ferromagnetic phase found in manganese models as 
``DE induced'' or ``DE generated'', forgetting the historical origin
of the term. 
In this review a similar convention will be followed, namely the
credit for the appearance of FM phases will be given to the DE
mechanism, although a more general and simple kinetic-energy
optimization is certainly at work. 

\medskip
\noindent{\bf Ferromagnetism due to a Large Hund Coupling:}

Regarding the stabilization of ferromagnetism, computer simulations
(Yunoki et al., 1998a) and a variety of other approximations have
clearly shown that models $without$ the oxygen degrees of freedom (to
be reviewed below) can also produce FM phases, as long as the Hund
coupling is large enough.
In this situation, when the $e_{\rm g}$ electrons directly jump from
Mn to Mn their kinetic energy is minimized if all spins are aligned
(see Fig.III.a.1b). 
As explained in the previous subsection, this procedure to obtain
ferromagnetism is usually also called double-exchange and even the
models from where it emerges are called double-exchange models.
However, there is little resemble of these models and physical process
with the original DE ideas (Zener, 1951b) where two electrons were
involved in the actual hopping. 
Actually the FM phases recently generated in computer simulations and
a variety of mean-field approximations resemble more closely the
predictions of another work of Zener (1951a), where indeed a large
Hund coupling is invoked as the main reason for ferromagnetism in some
compounds.

In addition, it has been questioned whether double-exchange or the
large $J_{\rm H}$ mechanism are sufficient to indeed produce the
ferromagnetic phase of manganites. 
An alternative idea (Zhao, 2000) relies on the fact that holes are
located mostly in the oxygens due to the charge-transfer character of
manganites and these holes are linked antiferromagnetically with the
spins in the adjacent Mn ions due to the standard exchange coupling,
leading to an effective Mn-Mn ferromagnetic interaction.
In this context the movement of holes would be improved if all Mn
spins are aligned leading to a FM phase, although many-body
calculations are needed to prove that this is indeed the case for
realistic couplings.
A comment about this idea: in the context of the cuprates a similar 
concept was discussed time ago (for references see Dagotto, 1994)
and after considerable discussion it was concluded that this process
has to be contrasted against the so-called Zhang-Rice singlet
formation where the spin of the hole at the oxygen couples in a
singlet state with the spin at the Cu. 
As explained in Riera, Hallberg and Dagotto (1997) the analogous
process in manganites would lead to the formation of effective 
$S$=3/2 ``hole'' states, between the $S$=2 of manganese (3+) and 
the $S$=1/2 of the oxygen hole. 
Thus, the competition between these two tendencies
should be addressed in detail, similarly as done for the cuprates
to clarify the proposal of Zhao (2000).

\medskip
\noindent{\bf Spin-Canted State:}

At this point it is useful to discuss the well-known proposed
 ``spin canted'' state for manganites. Work by de Gennes (1960) using 
mean-field approximations suggested that the interpolation between the 
antiferromagnetic state of the undoped limit and the ferromagnetic
state at  finite hole density, where the DE mechanism works, occurs
through a ``canted state'', similar as the state produced by a
magnetic field acting over an antiferromagnet (Fig.III.a.1c).
In this state the spins develop a moment in one direction, while
being mostly antiparallel within the plane perpendicular to that
moment. The coexistence of FM and AF features in several experiments
carried out at low hole doping (some of them reviewed below) led to
the widely spread belief until recently that this spin canted state
was indeed found in real materials.
However, a plethora of recent theoretical work (also discussed below)
has shown that the canted state is actually $not$ realized in the
model of manganites studied by de Gennes (i.e., the simple one-orbital
model). Instead phase separation occurs between the AF and FM states,
as extensively reviewed below. 
Nevertheless, a spin canted state is certainly still a possibility in
real low-doped manganites but its origin, if their presence is
confirmed, needs to be revised.
It may occur that substantial Dzyaloshinskii-Moriya (DM) interactions
appear in manganese oxides, but the authors are not aware of
experimental papers confirming or denying their relevance, although
some estimations (Solovyev, Hamada, and Terakura, 1996;
Lyanda-Geller et al., 1999, Chun et al., 1999a)
appear to indicate that the DM couplings are small.
However,  even if the DM coupling were large there are still subtle
issues to be addressed. For instance, it is widely believed that DM
interactions lead to canting.
However, Coffey, Bedell and Trugman (1990) showed that in the simple
case where the $D_{ij}$ factor for the DM interaction 
is a constant, as in most early work on
the subject, the DM term leads to a spiral state rather than a truly
canted state. In addition, the authors of this review are not aware of
reliable calculations showing that a canted state can indeed be
stabilized in a model without terms added that break explicitly the
invariance under rotations of the system in such a way that a given
direction, along which the moment develops, is made by hand different
from the others.
For all these reasons and from the discussion below it may appear that
the simplest way to explain the experimental data at low doping is to
assume an AF-FM phase coexistence instead of a canted state. 
However, the issue is still open and more experimental
work should be devoted to its clarification.

\medskip
\noindent{\bf Charge-Ordered State at x=0.5:}

Early theoretical work on manganites carried out by Goodenough (1955)
explained many of the features observed in the neutron scattering
$\LCMO$ experiments by Wollan and Koehler (1955), notably the
appearance of the A-type AF phase at x=0 and the CE-type phase at
x=0.5. The approach of Goodenough (1955) was based on the notion of
``semicovalent exchange'' and the main idea can be roughly explained
as follows.
Suppose one considers a Mn-O bond directed, say, along the $x$-axis,
and let us assume that the Mn-cation has an occupied orbital pointing
along $y$ or $z$ instead of $x$ (in other words, there is an empty
orbital along $x$).
The oxygen, being in a (2$-$) state, will try to move towards this Mn
site since it does not have a negative cloud of Mn electrons to fight
against. This process shortens the distance Mn-O and makes this bond
quite stable. This is a semicovalent bond.
Suppose now the occupied Mn orbital has an electron with an up spin.
Of the two relevant electrons of oxygen, the one with spin up will
feel the exchange force toward the Mn electron, i.e., if both
electrons involved have the same spin, the space part of their common
wave function has nodes which reduce the electron-electron repulsion
(as in the Hund's rules).
Then, effectively the considered Mn-O bond becomes ferromagnetic
between the Mn electron and one of the oxygen electrons. 

Consider now the left-side O-Mn portion of the Mn-O-Mn bond.
In the example under consideration, the second electron of oxygen must
be down and it spends most of the time away from the left Mn-ion
(rather than close to it as the oxygen spin-up electron does).
If the Mn-ion on the right of the link Mn-O-Mn also has an occupied
orbital pointing perpendicular to the $x$-axis, then O-Mn and Mn-O
behave similarly (individually FM) but with pairs of spins pointing in
opposite directions. As a consequence an effective $antiferromagnetic$ 
Mn-Mn interaction has been generated (see Fig.III.a.2a), and both Mn-O
and O-Mn are shorten in length.
However, if the right Mn has an electron in an orbital pointing along 
$x$, namely along the relevant $p$-orbital of the oxygen, the
Hund-rule-like argument does not apply anymore since now a simple
direct exchange is more important, leading to an AF O-Mn bond.
In this case, the overall Mn-Mn interaction is $ferromagnetic$, as
sketched in Fig.III.a.2b. 
Then, simple arguments lead to the notion that both AF and FM
couplings among the Mn-ions can be effectively generated, depending on 
the orientation of the orbitals involved.

Analyzing the various possibilities for the orbital directions and
generalizing to the case where Mn$^{4+}$ ions are also present,
Goodenough (1955) arrived to the A- and CE-type phases of manganites
very early in the theoretical study of these compounds 
(the shape of these states was shown in Fig.II.b.4).
In this line of reasoning, note that the Coulomb interactions are
important to generate Hund-like rules and the oxygen is also important 
to produce the covalent bonds. The lattice distortions are also quite 
relevant in deciding which of the many possible states minimizes the
energy. However, it is interesting to observe that in more recent
theoretical work described below in this review, both the A- and
CE-type phases can be generated $without$ the explicit appearance of
oxygens in the models and also without including long-range Coulombic
terms.
 
Summarizing, there appears to be three mechanisms to produce effective
FM interactions: (i) double exchange, where electrons are mobile,
which is valid for non charge-ordered states and where the oxygen 
plays a key role, (ii) Goodenough's approach where covalent bonds are
important (here the electrons do not have mobility in spite of the FM
effective coupling), and it mainly applies to charge-ordered states,
and (iii) the approach already described in this subsection based on
purely Mn models (no oxygens) which leads to FM interactions mainly as
a consequence of the large Hund coupling in the system.
If phonons are introduced in the model it can be shown that the A-type
and CE-type states are generated, as reviewed later in this section.
In the remaining theoretical part of the review most of the emphasis
will be given to approach (iii) to induce FM bonds since a large
number of experimental results can be reproduced by this procedure,
but it is important to keep in mind the historical path followed in
the understanding of manganites.

Based on all this discussion, it is clear that reasonable proposals to
understand the stabilization of AF and FM phases in manganites have
been around since the early theoretical studies of manganese oxides.
However, these approaches (double exchange, ferromagnetic covalent
bonds, and large Hund coupling) are still $not$ sufficient to handle
the very complex phase diagram of manganites.
For instance, there are compounds such as $\LSMO$ that actually 
do not have the CE-phase at x=0.5, while others do. 
There are compounds that are never metallic, while others have a
paramagnetic state with standard metallic characteristics. 
And even more important, in the early studies of manganites there was
no proper rationalization for the large MR effect. 
It is only with the use of state-of-the-art many-body tools that the
large magnetotransport effects are starting to be understood, thanks to 
theoretical developments in recent years that can address the competition
among the different phases of manganites, their clustering and
mixed-phase tendencies, and dynamical Jahn-Teller polaron
formation.

\begin{center}
  {\bf III.b More Recent Theories}
\end{center}

The prevailing ideas to explain the curious magnetotransport behavior 
of manganites changed in the mid-90's from the simple double-exchange
scenario to a more elaborated picture where a large Jahn-Teller (JT)
effect, which occurs in the Mn$^{3+}$ ions, produces a strong 
electron-phonon coupling that persists even at densities where a
ferromagnetic ground-state is observed.
In fact, in the undoped limit x=0, and even at finite but small x, it
is well-known that a robust static structural distortion is present in
the manganites (see Goodenough, 1955, and Elemans et al., 1971).
In this context it is natural to imagine the existence of small
lattice polarons in the paramagnetic phase above $T_{\rm C}$, and it
was believed that these polarons lead to the insulating behavior of
this regime. Actually, the term polaron (see Holstein (1959))
is somewhat ambiguous.
In the context of manganites it is usually associated with a local
distortion of the lattice around the charge, sometimes
together with a magnetic cloud or region with ferromagnetic 
correlations (magneto polaron or lattice-magneto polaron).

\medskip
\noindent{\bf Double-Exchange is not Enough:}

The fact that double-exchange cannot be enough to understand the
physics of manganites is clear from several different points of view.
For instance, Millis, Littlewood and Shraiman (1995) arrived at this
conclusion by presenting estimations of the critical Curie temperature
and of the resistivity using the DE framework.
Regarding ferromagnetism, their calculations for a model having as an
interaction only a large Hund coupling between $e_{\rm g}$ and
$t_{\rm 2g}$ electrons led to a $T_{\rm C}$ prediction between 0.1eV
and 0.3eV, namely of the order of the bare hopping amplitude and 
considerably higher than the experimental results.
Thus, it was argued that DE produces the wrong $T_{\rm C}$ by a large
factor.
However, note that computational work 
led to a much smaller estimation of the Curie
temperature of the order of 0.1$t$ for the double-exchange model
($t$ is the $e_{\rm g}$-electron hopping amplitude), and compatible
with experiments (Yunoki et al., 1998a; Calderon and Brey, 1998;
Yi, Hur and Yu, 1999; Motome and Furukawa, 1999; Motome and Furukawa,
2000b; some of which will be reviewed in more detail later.
Results for $S$=1/2 localized spins can be found in R\"oder, Singh,
and Zang, 1997).
For this reason arguments based on the value of $T_{\rm C}$ are $not$
sufficient to exclude the double-exchange model. 
Regarding the resistance, using the memory function method (in
principle valid at large frequency) to estimate the dc component,
Millis, Littlewood and Shraiman (1995) found a resistivity that grows 
with reducing temperature (insulating behavior) even below $T_{\rm C}$.
For this reason Millis, Littlewood and Shraiman (1995) concluded that
the model based only on a large $J_{\rm H}$ is not adequate for the
manganites, and instead the relevance of the Jahn-Teller phonons was
invoked.
These results have to be contrasted with computer-based calculations
of the resistivity for the one-orbital model at $J_{\rm H}$=$\infty$
by Calderon, Verg\'es and Brey (1999) that reported instead a metallic
behavior for the double-exchange model, actually both above and below 
$T_{\rm C}$.
Paradoxically, this behavior also leads to the same conclusion, namely
that double-exchange is not sufficient to explain the manganite
behavior of, e.g., $\LCMO$ which has insulating characteristics above
$T_{\rm C}$ but it is metallic below.
However, both lines of attack to the DE model may need further
revision, since the computational work of Yunoki et al. (1998a) at a
large but not infinite Hund coupling has established that the simple
one-orbital double-exchange model has regions with mixed-phase
tendencies, presenting an insulating resistivity (Moreo, Yunoki 
and Dagotto, 1999a) at and near $n$=1 ($n$ is the $e_{\rm g}$
electron number per site), which becomes metallic as the electronic
density is further reduced.
The existence of a metal-insulator transition in this model opens the
possibility that the one-orbital system may still present physics
$qualitatively$ similar to that found experimentally, where such a 
transition is crucial in manganites. For this reason, using the
one-orbital model as a toy model for manganites is still quite
acceptable, as long as the region of study is close to the
metal-insulator regime.
In fact recent work reporting percolative effects in this context use
both the one- and two-orbital models with or without a strong JT
coupling (Moreo et al., 2000). 
However, it is clear that the one-orbital model is incomplete for
quantitative studies since it cannot describe, e.g., the key
orbital-ordering of manganites and the proper charge-order states at x
near 0.5, which are so important for the truly CMR effect found in
low-bandwidth manganites. 
Then, the authors of this review fully agree with the conclusions of
Millis, Littlewood and Shraiman (1995), although the arguments leading
to such conclusion are different. 
It is clear that not even a fully disordered set of classical spins
can scatter electrons as much as needed to reproduce the experiments
(again, unless large antiferromagnetic regions appear in a mixed-phase
regime).

\medskip
\noindent{\bf Jahn-Teller Phonons and Polarons:}

Millis, Shraiman and Mueller (1996) (see also Millis, Mueller, and 
Shraiman, 1996a, and Millis, 1998) argued that the physics of
manganites is dominated by the interplay between a strong
electron-phonon coupling and the large Hund coupling effect that
optimizes the electronic kinetic energy by the generation of a FM
phase. 
The large value of the electron phonon coupling is clear in the regime 
of manganites below x=0.20 where a static JT distortion plays a key
role in the physics of the material. Millis, Shraiman and Mueller
(1996) argued that a dynamical JT effect may persist at higher hole
densities, without leading to long-range order but producing important
fluctuations that localize electrons by splitting the degenerate $e_{\rm g}$
levels at a given MnO$_6$ octahedron. 
The calculations were carried out using the infinite dimensional
approximation that corresponds to a $local$ mean-field technique where
the polarons can have only a one site extension, and the classical
limit for the phonons and spins was used. 
The latter approximation is not expected to be severe unless the
temperatures are very low (for a discussion see Millis, Mueller and
Shraiman, 1996b).
The Coulomb interactions were neglected, but further work reviewed
below showed that JT and Coulombic interactions lead to very similar
results (Hotta, Malvezzi, and Dagotto, 2000), and, as a consequence,
this approximation is not severe either.
Orbital or charge ordering were not considered in the formalism of
Millis, Shraiman and Mueller (1996).
Following the work of Millis, Littlewood and Shraiman (1995), phonons 
were also argued to be of much importance in manganites by 
R\"oder, Zang and Bishop (1996), who found a tendency toward the
formation of polarons in a single-orbital DE model with quantum
phonons, treating the localized spins in the mean-field approximation
and the polaron formation with the Lang-Firsov variational
approximation. 
Coulomb interactions were later incorporated using the Gutzwiller
approximation (Zang, Bishop and R\"oder, 1996).

Millis, Shraiman and Mueller (1996) argued that the ratio 
$\lambda_{\rm eff}$=$E_{\rm JT}$/$t_{\rm eff}$ dominates the physics
of the problem. Here $E_{\rm JT}$ is the static trapping energy 
at a given octahedron, and $t_{\rm eff}$ is an effective hopping that
is temperature dependent following the standard DE discussion. 
In this context it was conjectured that when the temperature is larger
than $T_{\rm C}$ the effective coupling $\lambda_{\rm eff}$ could be
above the critical value that leads to insulating behavior due to
electron localization, while it becomes smaller than the critical
value below $T_{\rm C}$, thus inducing metallic behavior.
The calculations were carried out using classical phonons and 
$t_{\rm 2g}$ spins.
The results of Millis, Shraiman and Mueller (1996) for $T_{\rm C}$ and
the resistivity at a fixed density $n$=1 when plotted as a function of
$\lambda_{\rm eff}$ had formal similarities with experimental results
(which are produced as a function of density). 
In particular, if $\lambda_{\rm eff}$ is tuned to be very close to
the metal-insulator transition, the resistivity naturally strongly
depends on even small external magnetic fields. 
However, in order to describe the percolative nature of the transition 
found experimentally and the notorious phase separation tendencies,
calculations beyond mean-field approximations are needed, as reviewed
later in this paper.

The existence of a critical value of the electron-phonon coupling
constant $\lambda$ of order unity at $n$=1 leading to a metal-insulator
transition is natural and it was also obtained in Monte Carlo (MC)
simulations by Yunoki et al. (1998b).
However, computational studies of the conductivity led to either
insulating or metallic behavior at all temperatures, for values of 
$\lambda$ above or below the critical temperature, respectively.
A mixture of metal/insulator behavior in the resistivity at a fixed
$\lambda$ was not observed at $n$=1.

\begin{center}
  {\bf III.c Models and Parameters}
\end{center}

In the previous subsections, the theoretical work on manganites has
been reviewed mainly in a historical order. 
In this section, the first steps toward a description of the latest 
theoretical developments in this context are taken. 
First, it is important to clearly write down the model Hamiltonian for 
manganites.
For complex material such as the Mn-oxides, unfortunately, the full\
Hamiltonian
includes several competing tendencies
and couplings.
However, as shown below, the essential physics can be obtained using
relatively simple models, deduced from the complicated full
Hamiltonian.

\medskip
\noindent{\bf Effect of crystal field:}

In order to construct the model Hamiltonian for manganites, let us
start our discussion at the level of the atomic problem, in which just 
one electron occupies a certain orbital in the $3d$ shell of a
manganese ion.
Although for an isolated ion a five-fold degeneracy exists for the
occupation of the $3d$ orbitals, this degeneracy is partially lifted
by the crystal field due to the six oxygen ions surrounding the
manganese forming an octahedron.
This is analyzed by the ligand field theory that shows that the
five-fold degeneracy is lifted into doubly-degenerate 
$e_{\rm g}$-orbitals ($d_{x^2-y^2}$ and $d_{3z^2-r^2}$) and
triply-degenerate $t_{\rm 2g}$-orbitals ($d_{xy}$, $d_{yz}$, and
$d_{zx}$).
The energy difference between those two levels is usually expressed as 
$10Dq$, by following the traditional notation in the ligand field
theory (see, for instance, Gerloch and Slade, 1973).

Here note that the energy level for the $t_{\rm 2g}$-orbitals is lower 
than that for $e_{\rm g}$-orbitals.
Qualitatively this can be understood as follows:
The energy difference originates in the Coulomb interaction
between the $3d$ electrons and the oxygen ions surrounding manganese.
While the wave-functions of the $e_{\rm g}$-orbitals is extended along
the direction of the bond between manganese and oxygen ions, those in
the $t_{\rm 2g}$-orbitals avoid this direction.
Thus, an electron in $t_{\rm 2g}$-orbitals is not heavily influenced
by the Coulomb repulsion due to the negatively charged oxygen ions, 
and the energy level
for $t_{\rm 2g}$-orbitals is lower than that for $e_{\rm g}$-orbitals. 

As for the value of $10Dq$, it is explicitly written as
(see Gerloch and Slade, 1973)
\begin{equation}
  10Dq=\frac{5}{3} \frac {Ze^2}{a}
  \frac{\langle r^4 \rangle}{a^4},
\end{equation}
where $Z$ is the atomic number of the ligand ion, $e$ is the electron 
charge, $a$ is the distance between manganese and oxygen ions, 
$r$ is the coordinate of the $3d$ orbital, and
${\langle \cdots \rangle}$ denotes the average value by using the
radial wavefunction of the $3d$ orbital.
Estimations by Yoshida (1998, page 29) suggest that 10$Dq$ is about
10000-15000cm$^{-1}$ (remember that 1eV = 8063 cm$^{-1}$).

\medskip

\noindent{\bf Coulomb interactions:}

Consider now a Mn$^{4+}$ ion, in which three electrons exist
in the $3d$ shells.
Although those electrons will occupy $t_{\rm 2g}$-orbitals due to the
crystalline field splitting, the configuration is not uniquely
determined.
To configure three electrons appropriately, it is necessary to take 
into account the effect of the Coulomb interactions.
In the localized ion system, the Coulomb interaction term
among $d$-electrons is generally given by
\begin{eqnarray}
  \label{eq:Coulomb0}
  H^{\rm C}_{\bf i} &=& (1/2)
  \sum_{\gamma_1 \gamma_2 \gamma'_1 \gamma'_2}
  \sum_{\sigma_1 \sigma_2 \sigma'_1 \sigma'_2}
  \langle \gamma_1 \sigma_1,\gamma_2 \sigma_2 ||
  \gamma'_1 \sigma'_1,\gamma'_2 \sigma'_2 \rangle \nonumber \\
  &\times& d_{{\bf i}\gamma_1\sigma_1}^{\dag}  
  d_{{\bf i}\gamma_2\sigma_2}^{\dag}
  d_{{\bf i}\gamma'_2\sigma'_2}
  d_{{\bf i}\gamma'_1\sigma'_1},
\end{eqnarray}
where $d_{{\bf i}\gamma\sigma}$ is the annihilation operator for a 
$d$-electron with spin $\sigma$ in the 
$\gamma$-orbital at site ${\bf i}$,
and the Coulomb matrix element is given by
\begin{eqnarray}
  \label{element}
  && \langle \gamma_1 \sigma_1,\gamma_2 \sigma_2 ||
  \gamma'_1 \sigma'_1,\gamma'_2 \sigma'_2 \rangle \nonumber \\
  && \!=\! \!\int \!  \int \! d{\bf r} d{\bf r'} \!
  \phi_{\gamma_1\sigma_1}^{*}({\bf r})
  \phi_{\gamma_2\sigma_2}^{*}({\bf r'})
  g_{{\bf r}-{\bf r'}}
  \phi_{\gamma'_1\sigma'_1}({\bf r})
  \phi_{\gamma'_2\sigma'_2}({\bf r'}).
\end{eqnarray}
Here $g_{{\bf r}-{\bf r'}}$ is the screened Coulomb potential,
and $\phi_{\gamma\sigma}({\bf r})$ is the Wannier function for an electron
with spin $\sigma$ in the $\gamma$-orbital at position ${\bf r}$.
By using the Coulomb matrix element, 
the so-called ``Kanamori parameters'', 
$U$, $U'$, $J$, and $J'$, are defined as follows
(see Kanamori, 1963; Dworin and Narath, 1970;
Castellani et al., 1978).
$U$ is the intraband Coulomb interaction, given by 
\begin{eqnarray}
  U =\langle \gamma \sigma,\gamma \sigma' ||
  \gamma \sigma,\gamma \sigma' \rangle,
\end{eqnarray}
with $\sigma \ne \sigma'$.
$U'$ is the interband Coulomb interaction, expressed by 
\begin{eqnarray}
  U'= \langle \gamma \sigma,\gamma' \sigma' ||
  \gamma \sigma,\gamma' \sigma' \rangle,
\end{eqnarray}
with $\gamma \ne \gamma'$.
$J$ is the interband exchange interaction, written as 
\begin{eqnarray}
  J= \langle \gamma \sigma,\gamma' \sigma' ||
  \gamma' \sigma,\gamma \sigma' \rangle,
\end{eqnarray}
with $\gamma \ne \gamma'$.
Finally, $J'$ is the pair-hopping amplitude
between different orbitals, given by
\begin{eqnarray}
  J'=\langle \gamma \sigma,\gamma \sigma' ||
  \gamma' \sigma,\gamma' \sigma' \rangle,
\end{eqnarray}
with $\gamma \ne \gamma'$ and $\sigma \ne \sigma'$.

Note the relation $J$=$J'$, which is simply due to the fact that each of 
the parameters above is given by an integral of the Coulomb interaction
sandwiched with appropriate orbital wave functions. 
Analyzing the form of those integrals the equality between $J$ and $J'$
can be deduced [see equation Eq.~(2.6) of Castellani et al. (1978);
See also the Appendix of Fr\'esard and Kotliar (1997)].

\medskip
\begin{center}
  \begin{tabular}{|c|c|c|c|} \hline
    \makebox[20mm]{$\gamma$} & \makebox[20mm]{$\gamma'$} &
    \makebox[15mm]{$U'$} & \makebox[15mm]{$J$} \\ \hline
    $xy,yz,zx$ & $xy,yz,zx$ &
    $A$$-$$2B$+$C$  & $3B$+$C$  \\ \hline 
    $x^2$$-$$y^2$, $3z^2$$-$$r^2$ & $x^2$$-$$y^2$, $3z^2$$-$$r^2$ &
    $A$$-$$4B$+$C$  & $4B$+$C$  \\ \hline 
    ${xy}$ & $x^2$$-$$y^2$ &
    $A$+$4B$+$C$  & $C$  \\ \hline 
    ${xy}$ & $3z^2$$-$$r^2$ &
    $A$$-$$4B$+$C$  & $4B$+$C$  \\ \hline 
    ${yz}$, ${zx}$ & $x^2$$-$$y^2$  &
    $A$$-$$2B$+$C$  & $3B$+$C$  \\ \hline 
    ${yz}$, ${zx}$ & $3z^2$$-$$r^2$  &
    $A$+$2B$+$C$  & $B$+$C$  \\ \hline 
  \end{tabular}
\end{center}

\noindent TABLE I.~ Expressions for $U'$ and $J$ by using
Racah parameters $A$, $B$, and $C$.
Note that $U$=$A$+$4B$+$3C$ for each orbital.
For more information, see Tang, Plihal, and Mills (1998).
\medskip

Using the above parameters, it is convenient to rewrite
the Coulomb interaction term in the following form:
\begin{eqnarray}
  \label{eq:coulomb}
  H^{\rm C}_{\bf i} &=&
  (U/2) \sum_{\gamma,\sigma \ne \sigma'}
   n_{{\bf i}\gamma\sigma}n_{{\bf i}\gamma\sigma'} \nonumber \\
  &+& (U'/2)  \sum_{\sigma,\sigma',\gamma \ne \gamma'} 
  n_{{\bf i}\gamma\sigma}n_{{\bf i}\gamma'\sigma'}
  \nonumber \\
  &+& (J/2)  \sum_{\sigma,\sigma',\gamma \ne \gamma'} 
  d_{{\bf i}\gamma\sigma}^{\dag}  
  d_{{\bf i}\gamma'\sigma'}^{\dag}
  d_{{\bf i}\gamma\sigma'}
  d_{{\bf i}\gamma'\sigma} \nonumber \\
  &+& (J'/2) \sum_{\sigma \ne \sigma',\gamma \ne \gamma'} 
  d_{{\bf i}\gamma\sigma}^{\dag}  
  d_{{\bf i}\gamma\sigma'}^{\dag}
  d_{{\bf i}\gamma'\sigma'}
  d_{{\bf i}\gamma'\sigma},
\end{eqnarray}
where 
$n_{{\bf i}\gamma\sigma}$=
$d_{{\bf i}\gamma\sigma}^{\dag}d_{{\bf i}\gamma\sigma}$.
Here it is important to clarify that the parameters $U$, $U'$, and $J$
in Eq.~(\ref{eq:coulomb}) are not independent (here $J$=$J'$ is used).
The relation among them in the localized ion problem
has been clarified by group theory arguments,
showing that all the above Coulomb interactions
can be expressed by the so-called ``Racah parameters'' $A$, $B$, and $C$
(for more details, see J. S. Griffith, 1961.
See also Tang, Plihal, and Mills, 1998).
Here only the main results are summarized in Table~I,
following Tang, Plihal, and Mills (1998).
Note that the values of $U'$ and $J$ depend on the combination
of orbitals, namely they take different values depending on the
orbitals used (Table I),
while $U$=$A$+$4B$+$3C$ is independent of the orbital choice.
Thus, it is easily checked that the relation
\begin{equation}
  \label{relation}
  U=U'+2J
\end{equation}
holds in any combination of orbitals.

Although Eq.~(\ref{relation}) has been clearly shown to be valid
using the Racah parameters,
the discussions in the current literature 
regarding this issue are somewhat
confusing, probably since the arguments usually rely 
directly on the Hamiltonian Eq.~(\ref{eq:coulomb}), 
rather than Eqs.~(\ref{eq:Coulomb0}) and (\ref{element}).
Thus, it is instructive to discuss the above mentioned
relation among the several 
couplings using arguments directly based on the
model Eq.~(\ref{eq:coulomb}),
without using  the Racah parameters.
First note that even using $J$=$J'$, the electron-electron interaction is
still not invariant under rotations in orbital space.
This can be easily understood simply using two orbitals as an example,
and two particles.
In the absence of hopping terms, the problem involves just one site
and it can be easily diagonalized, leading to 
four eigenenergies. The lowest one is $U'$$-$$J$, has
degeneracy three, and it corresponds to a spin-triplet and
orbital-singlet state. In order to verify that indeed this state is a
singlet in orbital space, the operators 
\begin{equation}
 \left\{
  \begin{array}{l}
  T^x_{{\bf i}} =
  (1/2)\sum_{\sigma}
  (d_{{\bf i}{\rm a}\sigma}^{\dag}d_{{\bf i}{\rm b}\sigma}
  +d_{{\bf i}{\rm b}\sigma}^{\dag}d_{{\bf i}{\rm a}\sigma}) \\
  T^y_{{\bf i}} =
  -(i/2)\sum_{\sigma}
  (d_{{\bf i}{\rm a}\sigma}^{\dag}d_{{\bf i}{\rm b}\sigma}
  -d_{{\bf i}{\rm b}\sigma}^{\dag}d_{{\bf i}{\rm a}\sigma}) \\
  T^z_{{\bf i}} =
  (1/2)\sum_{\sigma}
  (d_{{\bf i}{\rm a}\sigma}^{\dag}d_{{\bf i}{\rm a}\sigma}
  -d_{{\bf i}{\rm b}\sigma}^{\dag}d_{{\bf i}{\rm b}\sigma})
  \end{array}
  \right.
\end{equation}
are needed.
The next state is non-degenerate, it has energy $U'$+$J$ and it is a
spin-singlet.
Regarding the orbital component, it corresponds to the $T^z_{\bf i}$=0
part of an orbital triplet.
This result already suggests us that orbital invariance is not respected 
in the system unless restrictions are imposed on the couplings,
since a state of an orbital triplet is energetically separated from
another state of the same triplet
(see, for instance, Kuei and Scalettar, 1997).
The next two states have energies $U$+$J'$ and $U$$-$$J'$,
each is non-degenerate and spin singlet, 
and they are combinations of orbital triplets with 
$T^z_{\bf i}$=+1 and $-1$.
Note that the state characterized by $U$+$J'$ is invariant under rotations 
in orbital space (using a real rotation matrix parametrized by only 
one angle), while the other one with $U$$-$$J'$ is not.
Then, it is clear now how to proceed to restore rotational invariance.
It should be demanded that $U$$-$$J'$=$U'$+$J$, namely, $U$=$U'$+$J$+$J'$

In addition, following Castellani et al. (1978), it is known that
$J$=$J'$, as already discussed.
Then, Eq.~(\ref{relation}) is again obtained as a condition
for the rotational invariance in orbital space.
It should be noted that spin rotational invariance does not impose
any constraints on the parameters.
Also it should be noted that the orbital rotational invariance
achieved here is not a full SU(2) one, but a subgroup, similarly
as it occurs in 
the anisotropic Heisenberg model that has invariance under rotations
in the xy plane only. For this reason, the states are either singlets
or doblets, but not triplets, in orbital space.


For the case of three orbitals, a similar study can be carried out, 
although it is more tedious.
For two particles, the energy levels now are at $U'$$-$$J$ (degeneracy
nine, spin-triplet and orbital-triplet), $U'$+$J$ (degeneracy three,
spin singlet, contains parts of an orbital quintuplet), $U$$-$$J'$
(degeneracy two, spin singlet, contains portions of an orbital
quintuplet), and $U$+$2J'$ (non-degenerate, spin-singlet and
orbital-singlet).
In order to have the proper orbital multiplets that are characteristic of a
rotational orbital invariant system, it is necessary to require that
$U'$+$J$=$U$$-$$J'$. 
If the relation $J$=$J'$ is further used,
then the condition again becomes $U$=$U'$+$2J$.
A better proof of this condition can be carried out by
rewriting the Hamiltonian in terms of spin and orbital rotational
invariant operators such as 
$N_{\bf i}$=$\sum_{\gamma,\sigma} n_{{\bf i} \gamma \sigma}$,
${\bf S}_{\bf i}^2$=$\sum_{\gamma,\gamma'} 
{{{\bf S}_{{\bf i} \gamma}}\cdot{{\bf S}_{{\bf i} \gamma'}} }$,
and
${\bf L}_{\bf i}^2$=$\sum_{\sigma,\sigma'} 
{{{\bf L}_{{\bf i} \sigma}}\cdot{{\bf L}_{{\bf i} \sigma'}} }$,
where ${\bf S}_{\bf i}$ and ${\bf L}_{\bf i}$ are the spin
and orbital operators, respectively.
By this somewhat tedious procedure, a final expression is reached
in which only one term is not in the form of explicitly invariant
operators. To cancel that term, the condition mentioned above is needed.

Now let us move to the discussion of the configuration of 
three electrons for the Mn$^{4+}$ ion.
Since the largest energy scale among the several Coulombic
interactions is $U$, the orbitals are not doubly occupied by both up-
and down-spin electrons.
Thus, only one electron can exist in each orbital of the triply
degenerate $t_{\rm 2g}$ sector.
Furthermore, in order to take advantage of $J$, the spins of those 
three electrons point along the same direction.
This is the so-called ``Hund's rule''.

By adding one more electron to Mn$^{4+}$ with three up-spin
$t_{\rm 2g}$-electrons, let us consider the configuration for the
Mn$^{3+}$ ion.
Note here that there are two possibilities due to the balance between
the crystalline-field splitting and the Hund coupling:
One is the ``high-spin state'' in which an electron occupies the
$e_{\rm g}$-orbital with up spin if the Hund coupling is dominant.
In this case, the energy level appears at $U'$$-$$J$+10$Dq$.
Another is the ``low-spin state'' in which one of the
$t_{\rm 2g}$-orbitals is occupied with a down-spin electron, when the 
crystalline-field splitting is much larger than the Hund coupling.
In this case, the energy level occurs at $U$+$2J$.
Thus, the high spin state appears if $10Dq$$<$$5J$ holds.
Since $J$ is a few eV and $10Dq$ is about 1eV in the manganese oxide,
the inequality $10Dq$$<$$5J$ is considered to hold.
Namely, in the Mn$^{3+}$ ion, the high spin state is realized.

In order to simplify the model without loss of essential physics,
it is reasonable to treat the three spin-polarized
$t_{\rm 2g}$-electrons as a localized ``core-spin'' expressed by 
${\bf S}_{\bf i}$ at site ${\bf i}$, since the overlap integral
between $t_{\rm 2g}$ and oxygen $p\sigma$ orbital is small compared to
that between $e_{\rm g}$ and $p\sigma$ orbitals.
Moreover, due to the large value of the total spin $S$=$3/2$, it is
usually approximated by a classical spin (this approximation will be tested
later using computational techniques).
Thus, the effect of the strong Hund coupling between the 
$e_{\rm g}$-electron spin and localized $t_{\rm 2g}$-spins is
considered by introducing 
\begin{equation}
  H_{\rm Hund} = -J_{\rm H} \sum_{\bf i}
  {\bf s}_{\bf i} \cdot {\bf S}_{\bf j},
\end{equation}
where ${\bf s}_{\bf i}$=
$\sum_{\gamma\alpha\beta}d^{\dag}_{{\bf i}\gamma\alpha}
{\sigma}_{\alpha\beta}d_{{\bf i}\gamma\beta}$,
$J_{\rm H}$($>$0) is the Hund coupling between
localized $t_{\rm 2g}$-spin and mobile $e_{\rm g}$-electron, 
and ${\sigma}$=$(\sigma_x, \sigma_y, \sigma_z)$ are the Pauli
matrices. The magnitude of $J_{\rm H}$ is of the order of $J$.
Here note that ${\bf S}_{\bf i}$ is normalized as
$|{\bf S}_{\bf i}|$=1. 
Thus, the direction of the classical $t_{\rm 2g}$-spin at site 
${\bf i}$ is defined as 
\begin{equation}
  {\bf S}_{\bf i}=(\sin\theta_{\bf i}\cos\phi_{\bf i}, 
  \sin\theta_{\bf i}\sin\phi_{\bf i}, 
  \cos\theta_{\bf i}),
\end{equation}
by using the polar angle $\theta_{\bf i}$ and the azimuthal angle 
$\phi_{\bf i}$.

Unfortunately, the effect of the Coulomb interaction is not fully
taken into account only by $H_{\rm Hund}$ since there remains the 
direct electrostatic repulsion between $e_{\rm g}$-electrons, 
which will be referred to as the ``Coulomb interaction'' hereafter.
Then, the following term should be added to the Hamiltonian.
\begin{equation}
  H_{\rm el-el} = \sum_{{\bf i}} H_{\bf i}^{\rm C}
  + V \sum_{\langle {\bf i,j} \rangle}
  \rho_{\bf i} \rho_{\bf j},
\end{equation}
where $\rho_{\bf i}=\sum_{\gamma\sigma}n_{{\bf i}\gamma\sigma}$.
Note here that in this expression, the index $\gamma$ for the orbital
degree of freedom runs only in the $e_{\rm g}$-sector.
Note also that in order to consider the effect of the long-range 
Coulomb repulsion between $e_{\rm g}$-electrons, the term including
$V$ is added, where $V$ is the nearest-neighbor Coulomb interaction. 

\medskip
\noindent{\bf Electron-phonon coupling:}

Another important ingredient in manganites is the lattice distortion 
coupled to the $e_{\rm g}$-electrons.
In particular, the double degeneracy in the $e_{\rm g}$-orbitals is
lifted by the Jahn-Teller distortion of the MnO$_6$ octahedron
(Jahn and Teller, 1937).
The basic formalism for the study of electrons coupled to Jahn-Teller
modes has been set up by Kanamori (1960).
He focused on cases where the electronic orbitals are degenerate in
the undistorted crystal structure, as in the case of Mn in an 
octahedron of oxygens. 
As explained by Kanamori (1960), the Jahn-Teller effect 
(Jahn and Teller, 1937) in this context
can be simply stated as follows: when a given electronic level of a
cluster is degenerate in a structure of high symmetry, this structure
is generally unstable, and the cluster will present a
distortion toward a lower symmetry ionic arrangement.
In the case of Mn$^{3+}$, which is doubly degenerate when the crystal 
is undistorted, a splitting will occur when the crystal is distorted. 
The distortion of the MnO$_6$ octahedron is ``cooperative'' since once 
it occurs in a particular octahedron, it will affect the neighbors. 
The basic Hamiltonian to describe the interaction between electrons
and Jahn-Teller modes was written by Kanamori (1960) and it is of the
form
\begin{equation}
  H^{\rm JT}_{\bf i} = 2g (Q_{2{\bf i}} T^x_{\bf i} +
  Q_{3{\bf i}} T^z_{\bf i})
  + (k_{\rm JT}/2)(Q_{2{\bf i}}^2+Q_{3{\bf i}}^2),
\end{equation}
where 
$g$ is the coupling constant between the $e_{\rm g}$-electrons and
distortions of the MnO$_6$ octahedron,
$Q_{2{\bf i}}$ and $Q_{3{\bf i}}$ are normal modes of vibration of the 
oxygen octahedron that remove the degeneracy between the electronic
levels, and $k_{\rm JT}$ is the spring constant for the Jahn-Teller
mode distortions.
In the expression of $H^{\rm JT}_{\bf i}$, a $T^y_{\bf i}$-term does
not appear for symmetry reasons, since it belongs to 
the $A_{\rm 2u}$ representation.
The non-zero terms should correspond to the irreducible
representations included in $E_{\rm g}$$\times$$E_{\rm g}$, namely,
$E_{\rm g}$ and $A_{\rm 1g}$. 
The former representation is expressed by using the pseudo spin
operators $T^x_{\bf i}$ and $T^z_{\bf i}$ as discussed here, while the
latter, corresponding to the breathing mode, is discussed later in
this subsection. 
For more details the reader should consult Yoshida (1998, page 40).

Following Kanamori, $Q_{2{\bf i}}$ and $Q_{3{\bf i}}$ are explicitly
given by
\begin{equation}
  \label{eq:q2}
  Q_{2{\bf i}}={1 \over \sqrt{2}}(X_{1{\bf i}}-X_{4{\bf i}}
  -Y_{2{\bf i}}+Y_{5{\bf i}}),
\end{equation}
and 
\begin{equation}
  \label{eq:q3}
  Q_{3{\bf i}}\!=\! {1 \over \sqrt{6}}(2Z_{3{\bf i}}-2Z_{6{\bf i}}-
  X_{1{\bf i}}+X_{4{\bf i}}-Y_{2{\bf i}}+Y_{5{\bf i}}),
\end{equation}
where $X_{\mu j}$, $Y_{\mu j}$, and $Z_{\mu j}$ are the displacement
of oxygen ions from the equilibrium positions along the $x$-, $y$-,
and $z$-direction, respectively. The convention for the labeling $\mu$
of coordinates is shown in Fig.III.c.1. To solve this Hamiltonian, it
is convenient to scale the phononic degrees of freedom as
\begin{equation}
  Q_{2{\bf i}}=(g/k_{\rm JT})q_{2{\bf i}},~~
  Q_{3{\bf i}}=(g/k_{\rm JT})q_{3{\bf i}},
\end{equation}
where $g/k_{\rm JT}$ is the typical energy scale for the Jahn-Teller 
distortion, which is of the order of 0.1$\rm \AA$, namely, 2.5\% of
the lattice constant.
When the JT distortion is expressed in the polar coordinate as 
\begin{equation}
  \label{eq:polar}
  q_{2{\bf i}} = q_{{\bf i}} \sin \xi_{\bf i},~~
  q_{3{\bf i}} = q_{{\bf i}} \cos \xi_{\bf i},
\end{equation}
the ground state is easily obtained as
$(-\sin [\xi_{\bf i}/2] {d}_{{\bf i} a\sigma}^{\dag}
+\cos [\xi_{\bf i}/2] {d}_{{\bf i} b\sigma}^{\dag})|0\rangle$
with the use of the phase $\xi_{\bf i}$. 
The corresponding eigenenergy is given by $-E_{\rm JT}$,
where $E_{\rm JT}$ is the static Jahn-Teller energy,
defined by 
\begin{equation}
  E_{\rm JT}=g^2/(2k_{\rm JT}).
\end{equation}
Note that the ground state energy is independent of the phase 
$\xi_{\bf i}$. Namely, the shape of the deformed isolated octahedron
is not uniquely determined in this discussion.
In the Jahn-Teller crystal, the kinetic motion of $e_{\rm g}$
electrons, as well as the cooperative effect between adjacent
distortions, play a crucial role in lifting the degeneracy and fixing
the shape of the local distortion.
This point will be discussed later in detail.

To complete the electron-phonon coupling term, it is necessary to 
consider the breathing mode distortion, coupled to the local electron 
density as
\begin{equation}
  H^{\rm br}_{\bf i} = g Q_{1{\bf i}} \rho_{{\bf i}}
  + (1/2) k_{\rm br}Q_{1{\bf i}}^2,
\end{equation}
where the breathing-mode distortion $Q_{1{\bf i}}$ is given by
\begin{equation}
  \label{eq:q1}
  Q_{1{\bf i}}={1 \over \sqrt{3}}(X_{1{\bf i}}-X_{4{\bf i}}
  +Y_{2{\bf i}}-Y_{5{\bf i}}+Z_{3{\bf i}}-Z_{6{\bf i}}),
\end{equation}
and $k_{\rm br}$ is the associated spring constant.
Note that, in principle, the coupling constants of the $e_{\rm g}$ 
electrons with the $Q_1$, $Q_2$, and $Q_3$ modes could be 
different from one another. For simplicity, here it is assumed that
those coupling constants take the same value.
On the other hand, for the spring constants, a different notation for
the breathing mode is introduced, since the frequency for the
breathing mode distortion has been found experimentally to be
different from that for the Jahn-Teller mode.
This point will be briefly discussed later. 
Note also that the Jahn-Teller and breathing modes are competing with
each other. As it was  shown above, the energy gain due to the
Jahn-Teller distortion is maximized when one electron exists per
site. On the other hand, the breathing mode distortion energy is
proportional to the total number of $e_{\rm g}$ electrons per site,
since this distortion gives rise to an effective on-site attraction
between electrons. 

By combining the JT mode and breathing mode distortions, the
electron-phonon term is summarized as
\begin{equation}
 H_{\rm el-ph}= \sum_{\bf i}(H_{\bf i}^{\rm JT}+H_{\bf i}^{\rm br}).
\end{equation}
This expression depends on the parameter $\beta$=$k_{\rm br}/k_{\rm JT}$,
which regulates which distortion, the Jahn-Teller or breathing 
mode, play a more important role. This point will be discussed 
in a separate subsection.

Note again that the distortions at each site are not independent,
since all oxygens are shared by neighboring MnO$_6$ octahedra,
as easily understood by the explicit expressions of $Q_{1{\bf i}}$,
$Q_{2{\bf i}}$, and $Q_{3{\bf i}}$ presented before.
A direct and simple way to consider this cooperative effect is
to determine the oxygen positions $X_{1{\bf i}}$, $X_{4{\bf i}}$,
$Y_{2{\bf i}}$, $Y_{5{\bf i}}$, $Z_{3{\bf i}}$, and $Z_{6{\bf i}}$,
by using, for instance, the Monte Carlo simulations or numerical
relaxation methods (see Press et al., 1992, chapter 10).
To reduce the burden on the numerical calculations,
the displacements of oxygen ions are assumed to be along the bond
direction between nearest neighboring manganese ions.
In other words, the displacement of the oxygen ion perpendicular to 
the Mn-Mn bond, i.e., the buckling mode, is usually ignored. 
As shown later, even in this simplified treatment, several
interesting results haven been obtained for the spin, charge, and
orbital ordering in manganites.

Rewriting Eqs.~(\ref{eq:q2}), (\ref{eq:q3}), and (\ref{eq:q1})
in terms of the displacement of oxygens from the equilibrium
positions, it can be shown that
\begin{equation}
  Q_{1{\bf i}}=Q_{1}^{(0)}+
  {1 \over \sqrt{3}}(\Delta_{\bf xi}+\Delta_{\bf yi}+\Delta_{\bf zi}), 
\end{equation}
\begin{equation}
  Q_{2{\bf i}}=Q_{2}^{(0)}+
  {1 \over \sqrt{2}}(\Delta_{\bf xi}-\Delta_{\bf yi}),
\end{equation}
\begin{equation}
  Q_{3{\bf i}}=Q_{3}^{(0)}+
  {1 \over \sqrt{6}}(2\Delta_{\bf zi}-\Delta_{\bf xi}-\Delta_{\bf yi}),
\end{equation}
where $\Delta_{\bf ai}$ is given by
\begin{equation}
  \Delta_{\bf ai}=u_{\bf i}^{\bf a}-u_{\bf i-a}^{\bf a},
\end{equation}
with $u_{\bf i}^{\bf a}$ being the displacement of oxygen ion at site
${\bf i}$ from the equilibrium position along the ${\bf a}$-axis.
The offset values for the distortions, $Q_{1}^{(0)}$, $Q_2^{(0)}$, and
$Q_3^{(0)}$, are respectively given by
\begin{equation}
  Q_{1}^{(0)}\!=\!
  {1 \over \sqrt{3}}(\delta L_{\bf x}+\delta L_{\bf y}+\delta L_{\bf z}),
\end{equation}
\begin{equation}
  Q_{2}^{(0)}\!=\!
  {1 \over \sqrt{2}}(\delta L_{\bf x}-\delta L_{\bf y}),
\end{equation}
\begin{equation}  
  Q_{3}^{(0)}\!=\!
  {1 \over \sqrt{6}}(2\delta L_{\bf z}-\delta L_{\bf x}-\delta L_{\bf y}),
\end{equation}
where $\delta L_{\bf a}$=$L_{\bf a}-L$,
the non-distorted lattice constants are $L_{\bf a}$,
 and $L$=$(L_{\bf x}+L_{\bf y}+L_{\bf z})/3$.
In the $cooperative$ treatment,
the $\{ u \}$'s  are directly optimized in the numerical calculations
(see Allen and Perebeinos, 1999; Hotta et al. 1999).
On the other hand, in the $non$-$cooperative$ calculations,
$\{Q\}$'s are treated instead of the $\{u\}$'s.
The similarities and differences between those two treatments will be discussed
later for some particular cases.

\medskip
\noindent{\bf Hopping amplitudes:}

Although the $t_{\rm 2g}$-electrons are assumed to be localized, 
the $e_{\rm g}$-electrons can move around the system via the oxygen
$2p$ orbital. 
This hopping motion of $e_{\rm g}$-electrons is expressed as 
\begin{equation}
  H_{\rm kin} =-\sum_{{\bf ia}\gamma \gamma'\sigma}
  t^{\bf a}_{\gamma \gamma'} d_{{\bf i} \gamma \sigma}^{\dag}
  d_{{\bf i+a} \gamma' \sigma},
\end{equation}
where ${\bf a}$ is the vector connecting nearest-neighbor sites and 
$t^{\bf a}_{\gamma \gamma'}$ is the nearest-neighbor hopping amplitude
between $\gamma$- and $\gamma'$-orbitals along the 
${\bf a}$-direction.

The amplitudes are evaluated from the overlap integral between
manganese and oxygen ions by following Slater and Koster (1954).
The overlap integral between $d_{x^2-y^2}$- and $p_x$-orbitals 
is given by
\begin{equation}
  E_{{\bf x},{\rm a}}(\ell,m,n)=(\sqrt{3}/2) \ell
  (\ell^2-m^2)(pd\sigma), 
\end{equation}
where $(pd\sigma)$ is the overlap integral between the $d\sigma$- and
$p\sigma$-orbital and $(\ell,m,n)$ is the unit vector along the
direction from manganese to oxygen ions.
The overlap integral between $d_{3z^2-r^2}$- and $p_x$-orbitals is
expressed as
\begin{equation}
  E_{{\bf x},{\rm b}}(\ell,m,n)= \ell [n^2-(\ell^2+m^2)/2](pd\sigma).
\end{equation}
Thus, the hopping amplitude between adjacent manganese ions along the
$x$-axis via the oxygen $2p_x$-orbitals is evaluated as
\begin{equation}
  -t_{\gamma\gamma'}^{\bf x}=E_{{\bf x},\gamma}(1,0,0) \times 
  E_{{\bf x},\gamma'}(-1,0,0).
\end{equation}
Note here that the minus sign is due to the definition of hopping amplitude
in $H_{\rm kin}$. 
Then, $t_{\gamma\gamma'}^{\bf x}$ is explicitly given by
\begin{equation}
  t_{\rm aa}^{\bf x}
  =-\sqrt{3}t_{\rm ab}^{\bf x}
  =-\sqrt{3}t_{\rm ba}^{\bf x}
  =3t_{\rm bb}^{\bf x}=3t_0/4,
\end{equation}
where $t_0$ is defined by $t_0=(pd\sigma)^2$.
By using the same procedure, the hopping amplitude along the $y$- 
and $z$-axis are given by
\begin{equation}
  t_{\rm aa}^{\bf y}
  =\sqrt{3}t_{\rm ab}^{\bf y}
  =\sqrt{3}t_{\rm ba}^{\bf y}
  =3t_{\rm bb}^{\bf y}=3t_0/4,
\end{equation}
and
\begin{equation}
  t_{\rm bb}^{\bf z}=t_0,
  t_{\rm aa}^{\bf z}=t_{\rm ab}^{\bf z}=t_{\rm ba}^{\bf z}=0,
\end{equation}
respectively.
It should be noted that the signs in the hopping amplitudes between
different orbitals are different between the $x$- and $y$-directions,
which will be important when the charge-orbital ordered phase
in the doped manganites is considered.
Note also that in some cases, it is convenient to define 
$t_{\rm aa}^{\bf x}$ as the energy scale $t$, given as $t$=$3t_0/4$.

\medskip
\noindent{\bf Heisenberg term:}

Thus far, the role of the $e_{\rm g}$-electrons has been discussed
to characterize the manganites.
However, in the fully hole-doped manganites composed of Mn$^{4+}$ 
ions, for instance CaMnO$_3$, it is well known that a G-type 
antiferromagnetic phase appears,
and this property cannot be understood within the above discussion. 
The minimal term to reproduce this antiferromagnetic property
is the Heisenberg-like coupling between localized $t_{\rm 2g}$ spins,
given in the form 
\begin{equation}
  H_{\rm AFM} = J_{\rm AF} \sum_{\langle {\bf i,j} \rangle}
  {\bf S}_{\bf i} \cdot {\bf S}_{\bf j},
\end{equation}
where $J_{\rm AF}$ is the AFM coupling between nearest neighbor 
$t_{\rm 2g}$ spins.
The existence of this term is quite natural from the viewpoint of the
super-exchange interaction, working between neighboring localized
$t_{\rm 2g}$-electrons.
As for the magnitude of $J_{\rm AF}$, it is discussed later
in the text. 

\medskip
\noindent{\bf Full Hamiltonian:}

As discussed in the previous subsections, there are five important
ingredients that regulate the physics of electrons in manganites:
(i) $H_{\rm kin}$, the kinetic term of the $e_{\rm g}$ electrons. 
(ii) $H_{\rm Hund}$, the Hund coupling between the $e_{\rm g}$
electron spin and the localized $t_{\rm 2g}$ spin.
(iii) $H_{\rm AFM}$, the AFM Heisenberg coupling between nearest
neighbor $t_{\rm 2g}$ spins.
(iv) $H_{\rm el-ph}$, the coupling between the $e_{\rm g}$ electrons
and the local distortions of the MnO$_6$ octahedron.
(v) $H_{\rm el-el}$, the Coulomb interactions among the $e_{\rm g}$
electrons.
By unifying those five terms into one, the full Hamiltonian $H$ is
defined as
\begin{equation}
  \label{Hamiltonian}
  H = H_{\rm kin} + H_{\rm Hund} + H_{\rm AFM} 
  + H_{\rm el-ph} + H_{\rm el-el}.
\end{equation}
This expression is believed to define an appropriate starting
model for manganites, but, unfortunately, it is quite difficult to
solve such a Hamiltonian.
In order to investigate further the properties of manganites,
some simplifications are needed.

\medskip
\noindent{\bf Free $e_{\rm g}$-electron model:}

The simplest model is obtained by retaining only the kinetic term. 
Although this is certainly an oversimplification for describing the
complex nature of manganites, it can be a starting model to study the 
transport properties of these compounds, particularly in the
ferromagnetic region in which the static Jahn-Teller distortion does not
occur and the effect of the Coulomb interaction is simply renormalized 
into the quasi-particle formation.
In fact, some qualitative features of manganites can be addressed 
in the band-picture, as discussed by Shiba et al. (1997) and Gor'kov
and Kresin (1998).
The kinetic term is rewritten in momentum space as
\begin{equation}
  H_0 = \sum_{{\bf k} \gamma \gamma' \sigma}
  \varepsilon_{{\bf k}\gamma\gamma'} d_{{\bf k}\gamma \sigma}^{\dag}
  d_{{\bf k}\gamma' \sigma},
\end{equation}  
where $d_{{\bf k}\gamma\sigma}=(1/N)\sum_{\bf i}
e^{i{\bf R}_{\bf i} \cdot {\bf k}} d_{{\bf i}\gamma\sigma}$,
$\varepsilon_{{\bf k}{\rm aa}}=-(3t_0/2)(C_x +C_y)$,
$\varepsilon_{{\bf k}{\rm bb}}=-(t_0/2)(C_x + C_y+4C_z)$, 
and 
$\varepsilon_{{\bf k}{\rm ab}}=\varepsilon_{{\bf k}{\rm ba}}=
(\sqrt{3} t_0/2)(C_x - C_y)$,
with $C_{\mu}=\cos k_{\mu}$ ($\mu$=$x$, $y$, and $z$).
After the diagonalization of $\varepsilon_{{\bf k}\gamma\gamma'}$,
two bands are obtained as
\begin{eqnarray}
  E_{\bf k}^{\pm} \! &=& \! -t_0 \Bigl( C_x + C_y + C_z\nonumber \\
   \! &\pm& \! \sqrt{C_x^2  + C_y^2 + C_z^2
  - C_x C_y - C_y C_z - C_z C_x} ~ \Bigr).
\end{eqnarray}
Note that the cubic symmetry can be seen clearly in $E_{\bf k}^{\pm}$,
although the hopping amplitudes at first sight are quite anisotropic,
due to the choice of a particular basis for the $d$-orbitals. Other
basis certainly lead to the same result.
Note also that the bandwidth $W$ is given by $W=6t_0$.

\medskip
\noindent{\bf One-orbital Model:}

A simple model for manganites to illustrate the CMR effect is obtained
by neglecting the electron-phonon coupling and the Coulomb
interactions.
Usually, an extra simplification is carried out  by neglecting the
orbital degrees of freedom, leading to the FM Kondo model or
one-orbital double-exchange model, which will be simply referred as
the ``one-orbital model'' hereafter,
given as (Zener, 1951b; Furukawa, 1994) 
\begin{eqnarray}
  H_{\rm DE} &=& -t \sum_{\langle{\bf i,j} \rangle,\sigma}
  (a_{{\bf i} \sigma}^{\dag}a_{{\bf j} \sigma}+{\rm H.c.})
  -J_{\rm H} \sum_{\bf i}{\bf s}_{\bf i} \cdot {\bf S}_{\bf j} 
  \nonumber \\
  &+& J_{\rm AF} \sum_{\langle {\bf i,j} \rangle}
  {\bf S}_{\bf i} \cdot {\bf S}_{\bf j},
\end{eqnarray}
where $a_{{\bf i} \sigma}$ is the annihilation operator for an
electron with spin $\sigma$ at site ${\bf i}$, but without orbital
index.
Note that $H_{\rm DE}$ is quadratic in the electron operators,
indicating that it is reduced to a one-electron problem on the
background of localized $t_{\rm 2g}$ spins.
This is a clear advantage for the Monte Carlo simulations,
as discussed later in detail.
Neglecting the orbital degrees of freedom is clearly an
oversimplification, and important phenomena such as orbital ordering
cannot be obtained in this model.
However, the one-orbital model is still important, since it already
includes part of the essence of manganese oxides.
For example, recent computational investigations have clarified that
the very important phase separation tendencies and metal-insulator
competition exist in this model.
The result will be discussed in detail in the following subsection.

\medskip
\noindent{\bf $J_{\rm H}$=$\infty$ limit:}

Another simplification without the loss of essential physics is to
take the widely used limit $J_{\rm H}$=$\infty$,
since in the actual material $J_{\rm H}/t$ is much larger than unity.
In such a limit, the $e_{\rm g}$-electron spin perfectly aligns along 
the $t_{\rm 2g}$-spin direction, reducing the number of degrees of
freedom.
Then, in order to diagonalize the Hund term,
the ``spinless"  $e_{\rm g}$-electron
operator, $c_{{\bf i} \gamma}$, is defined as
\begin{equation}
  c_{{\bf i} \gamma} =
  \cos(\theta_{\bf i}/2)d_{{\bf i}\gamma \uparrow}
  + \sin(\theta_{\bf i}/2)e^{-i\phi_{\bf i}}d_{{\bf i}\gamma\downarrow}.
\end{equation}
In terms of the $c$-variables, the kinetic energy acquires
the simpler form
\begin{equation}
H_{\rm kin} = -\sum_{{\bf ia}\gamma \gamma'}
  S_{\bf i,i+a} t^{\bf a}_{\gamma \gamma'} 
  c_{{\bf i} \gamma}^{\dag}c_{{\bf i+a} \gamma'},
\end{equation}
where $S_{\bf i,j}$ is given by
\begin{eqnarray}
  S_{\bf i,j} &=& \cos (\theta_{\bf i}/2)\cos (\theta_{\bf j}/2)
  \nonumber \\ 
  &+& \sin (\theta_{\bf i}/2)\sin (\theta_{\bf j}/2)
  e^{-i(\phi_{\bf i}-\phi_{\bf j})}.
\end{eqnarray}
This factor denotes the change of hopping amplitude due to the
difference in angles between $t_{\rm 2g}$-spins at sites ${\bf i}$ and
${\bf j}$.
Note that the effective hopping in this case is a complex number
(Berry phase), contrary to the real number widely used in a large
number of previous investigations (for details in the case of the
one-orbital model see M\"uller-Hartmann and Dagotto, 1996).

The limit of infinite Hund coupling reduces the number of degrees of 
freedom substantially since the spin index is no longer needed.
In addition, the $U$- and $J$-terms in the electron-electron
interaction within the $e_{\rm g}$-sector are also no longer needed.
In this case, the following simplified model is obtained:
\begin{eqnarray}
  H^{\infty} &=& -\sum_{{\bf ia}\gamma \gamma'}
  S_{\bf i,i+a} t^{\bf a}_{\gamma \gamma'} 
  c_{{\bf i} \gamma}^{\dag}c_{{\bf i+a} \gamma'}
  + J_{\rm AF} \sum_{\langle {\bf i,j} \rangle}
  {\bf S}_{\bf i} \cdot {\bf S}_{\bf j} \nonumber \\
  &+& U' \sum_{\bf i} n_{{\bf i}{\rm a}} n_{{\bf i}{\rm b}}
  + V \sum_{\langle {\bf i,j} \rangle} n_{\bf i}n_{\bf j} \nonumber \\ 
  &+& E_{\rm JT} \sum_{\bf i}
  [2( q_{1{\bf i}} n_{\bf i} 
  + q_{2{\bf i}} \tau_{x{\bf i}}
  + q_{3{\bf i}} \tau_{z{\bf i}}) \nonumber \\
  &+& \beta q_{1{\bf i}}^2 + q_{2{\bf i}}^2 +q_{3{\bf i}}^2],
\end{eqnarray}
where 
$n_{{\bf i} \gamma}$=$c_{{\bf i} \gamma}^{\dag}c_{{\bf i} \gamma}$,
$n_{\bf i}$=$\sum_{\gamma}n_{{\bf i} \gamma}$,
$\tau_{x\bf i}$=
$c_{{\bf i}a}^{\dag}c_{{\bf i}b}$+$c_{{\bf i}b}^{\dag}c_{{\bf i}a}$,
and 
$\tau_{z\bf i}$=
$c_{{\bf i}a}^{\dag}c_{{\bf i}a}$$-$$c_{{\bf i}b}^{\dag}c_{{\bf i}b}$.

Considering the simplified Hamiltonian $H^{\infty}$, two other
limiting models can be obtained. One is the Jahn-Teller model 
$H^{\infty}_{\rm JT}$, defined as 
$H^{\infty}_{\rm JT}$=$H^{\infty}(U'=V=0)$,
in which the Coulomb interactions are simply ignored.
Another is the Coulombic model
$H^{\infty}_{\rm C}$, defined as 
$H^{\infty}_{\rm C}$=$H^{\infty}(E_{\rm JT}=0)$,
which denotes the two-orbital double exchange model 
influenced by the Coulomb interactions, neglecting the phonons.
Of course, the actual situation is characterized by 
$U' \ne 0$, $V \ne 0$, and $E_{\rm JT} \ne 0$,
but in the spirit of the adiabatic continuation,
it is convenient and quite meaningful to consider the 
minimal models possible to describe correctly the 
complicated properties of manganites.

\medskip
\noindent{\bf JT phononic and Coulombic models:}

Another possible simplification could have been obtained by neglecting 
the electron-electron interaction in the full Hamiltonian
but keeping the Hund coupling finite, leading to the following purely
JT-phononic model with active spin degrees of freedom:
\begin{equation}
  H_{\rm JT} = H_{\rm kin} + H_{\rm Hund} + H_{\rm AFM} 
  + H_{\rm el-ph}.
\end{equation}
Often in this review this Hamiltonian will be referred to as the
``two-orbital'' model (unless explicitly stated otherwise).
To solve $H_{\rm JT}$, numerical methods such as Monte Carlo
techniques and the relaxation method have been employed.
Qualitatively, the negligible values of the probability of double
occupancy in the strong electron-phonon coupling region with large
$J_{\rm H}$ justifies the neglect of $H_{\rm el-el}$, since the
Jahn-Teller energy is maximized when one $e_{\rm g}$ electron exists
at each site. 
Thus, the JT-phonon induced interaction will produce physics quite 
similar to that due to the on-site correlation.

It would be important to verify this last expectation by studying a
multi-orbital model with only Coulombic terms, without the extra 
approximation of using mean-field techniques for its analysis.
Of particular relevance is whether phase separation tendencies and
charge ordering appear in this case, as they do in the JT-phononic
model. 
This analysis is particularly important since, as explained before,
a mixture of phononic and Coulombic interactions is expected to be
needed for a proper quantitative description of manganites.
For this purpose, yet another simplified model has been analyzed
in the literature:
\begin{equation}
  H_{\rm C} = H_{\rm kin} + H_{\rm el-el}.
\end{equation}
Note that the Hund coupling term between $e_{\rm g}$ electrons and  
$t_{\rm 2g}$ spins is not explicitly included.
The reason for this extra simplification is that the numerical
complexity in the analysis of the model is drastically reduced by
neglecting the localized $t_{\rm 2g}$ spins.
In the FM phase, this is an excellent approximation, but not
necessarily for other magnetic arrangements. 
Nevertheless the authors believe that it is important to establish 
with accurate numerical techniques whether the PS tendencies are
already present in this simplified two-orbital models with Coulomb
interactions, even if not all degrees of freedom are incorporated from 
the outset.
Adding the $S$=3/2 quantum localized spins to the problem would
considerably increase the size of the Hilbert space of the model,
making it intractable with current computational techniques.

\medskip
\noindent{\bf Estimations of Parameters:}

In this subsection, estimations of the couplings that appear in the
models described before are provided.
However, before proceeding with the details the reader must be warned
that such estimations are actually quite difficult, for the simple
reason that in order to compare experiments with theory reliable
calculations must be carried out.
Needless to say, strong coupling many-body problems are notoriously
difficult and complex, and it is quite hard to find accurate
calculations to compare against experiments.
Then, the numbers quoted below must be taken simply as rough
estimations of orders of magnitude. The reader should consult the
cited references to analyze the reliability of the estimations mentioned 
here.
Note also that the references discussed in this subsection correspond
to only a small fraction of the vast literature on the subject.
Nevertheless, the ``sample'' cited below is representative of the
currently accepted trends in manganites.

Regarding the largest energy scales, the on-site $U$ repulsion was
estimated to be 5.2$\pm$0.3 eV and 3.5$\pm$0.3 eV, for $\rm Ca Mn O_3$ 
and $\rm La Mn O_3$, respectively, by Park et al. (1996) using
photoemission techniques.
The charge-transfer energy $\Delta$ was found to be 3.0$\pm$0.5 eV for
$\rm Ca Mn O_3$ in the same study (note that in the models described
in previous Sections, the oxygen ions were simply ignored).
In other photoemission studies, Dessau and Shen (1999) estimated the
exchange energy for flipping an $e_{\rm g}$-electron to be 2.7eV.

Okimoto et al. (1995) studying the optical spectra of $\LSMO$ with
x=0.175 estimated the value of the Hund coupling to be of the order of 
2 eV, much larger than the hopping of the one-orbital model for
manganites.
Note that in estimations of this variety care must be taken with the
actual definition of the exchange $J_{\rm H}$, which sometimes is in
front of a ferromagnetic Heisenberg interaction where classical
localized spins of module 1 are used, while in other occasions quantum 
spins of value 3/2 are employed.  
Nevertheless, the main message of Okimoto et al.'s paper is that 
$J_{\rm H}$ is larger than the hopping.
A reanalysis of Okimoto et al.'s results led Millis, Mueller and
Shraiman (1996) to conclude that the Hund coupling is actually even
larger than previously believed.
The optical data of Quijada et al. (1998) and Machida et al. (1998) 
also suggest that the Hund coupling is larger than 1 eV.
Similar conclusions were reached by Satpathy et al. (1996) using
constrained LDA calculations.

The crystal-field splitting between the $e_{\rm g}$- and 
$t_{\rm 2g}$-states was estimated to be of the order of 1 eV by
Tokura (1999) (see also Yoshida, 1998). Based on the discussion in the
previous subsection, it is clear that manganites are in high-spin
ionic states due to their large Hund coupling.

Regarding the hopping ``$t$'', Dessau and Shen (1999) reported a value
of order 1eV, which is somewhat larger than other estimations.
In fact, the results of Bocquet et al. (1992), Arima et al. (1993),
and Saitoh et al. (1995) locate its magnitude between 
0.2 eV and 0.5 eV, which is reasonable in transition metal oxides.
However, note that fair comparisons between theory and experiment require
calculations of, e.g., quasiparticle band dispersions, which are
difficult at present. Nevertheless it is widely accepted that the
hopping is just a fraction of eV.

Dessau and Shen (1999) estimated the static Jahn-Teller energy 
$E_{\rm JT}$ as 0.25eV. From the static Jahn-Teller energy and the 
hopping amplitude, it is convenient to define the dimensionless
electron-phonon coupling constant $\lambda$ as
\begin{equation}
  \lambda=\sqrt{2E_{\rm JT}/t}=g/\sqrt{k_{\rm JT}t}.
\end{equation}
By using $E_{\rm JT}$=0.25eV and $t$=0.2$\sim$0.5eV,
$\lambda$ is estimated as between 1 $\sim$ 1.6.
Actually, Millis, Mueller and Shraiman (1996) concluded
that $\lambda$ is between 1.3 and 1.5.

As for the parameter $\beta$, it is given by 
$\beta$=$k_{\rm br}/k_{\rm JT}$=
$(\omega_{\rm br}/\omega_{\rm JT})^2$,
where $\omega_{\rm br}$ and $\omega_{\rm JT}$ are the vibration
energies for manganite breathing- and JT-modes, respectively, assuming
that the reduced masses for those modes are equal. From experimental 
results and band-calculation data (see Iliev et al. 1998),
$\omega_{\rm br}$ and $\omega_{\rm JT}$ are
estimated as $\sim 700$cm$^{-1}$ and $500$-$600$cm$^{-1}$,
respectively,
leading to $\beta$$\approx$2.
However, in practice it has been observed that the main conclusions
are basically unchanged as long as $\beta$ is larger than unity.
Thus, if an explicit value for $\beta$ is not provided, the reader can
consider that $\beta$ is simply taken to be $\infty$ to suppress the
breathing mode distortion.

The value of $J_{\rm AF}$ is the smallest of the set of couplings
discussed here. 
In units of the hopping, it is believed to be of the order of 0.1$t$ 
(see Perring et al., 1997), namely about 200K. Note, however, that it
would be a bad approximation to simply neglect this parameter since in
the limit of vanishing density of $e_{\rm g}$ electrons, $J_{\rm AF}$
is crucial to induce antiferromagnetism, as it occurs in 
$\rm Ca Mn O_3$ for instance.
Its relevance, at hole densities close to 0.5 or larger, to the
formation of
antiferromagnetic charge-ordered states is remarked elsewhere in this
review. Also in mean-field approximations by Maezono, Ishihara, and
Nagaosa  (1998) the importance of $J_{\rm AF}$ has been mentioned,
even though in their work this coupling was estimated to be only
0.01$t$.

Summarizing, it appears well-established that:
(i) the largest energy scales in the Mn-oxide models studied here 
are the Coulomb repulsions between electrons in the same ion, which is 
quite reasonable.
(ii) The Hund coupling is between 1 and 2 eV, larger than the typical
hopping amplitudes, and sufficiently large to form high-spin Mn$^{4+}$
and Mn$^{3+}$ ionic states. 
As discussed elsewhere in the review, a large $J_{\rm H}$ leads
naturally to a vanishing probability of $e_{\rm g}$-electron
double-occupancy of a given orbital, thus mimicking the effect of a
strong on-site Coulomb repulsion. 
(iii) The dimensionless electron-phonon coupling constant $\lambda$
is of the order of unity, showing that the electron lattice
interaction is substantial and cannot be neglected.
(iv) The electron hopping energy is a fraction of eV.
(v) The AF-coupling among the localized spins is about a tenth of the
hopping. However, as remarked elsewhere, this apparent small coupling
can be quite important in the competition between FM and AF states.

\medskip
\noindent{\bf Monte Carlo Simulations:}


In this subsection the details related to the Monte Carlo
calculations are provided. 
For simplicity, let us focus here on one dimensional systems as an
example.
Generalizations to higher dimensions are straightforward.
Also, as a simple example, the case of the one-orbital model will be used, 
with the two-orbital case left as exercise to the readers.
Note that the one-orbital model $H_{\rm DE}$ is simply denoted by 
${\hat H}$ in this subsection.
Note also that $\beta$ indicates the inverse temperature, i.e.,
$\beta$=$1/T$, in this subsection.

As explained before, the Hamiltonian for the one-orbital model is
quadratic in the $\{ a, a^{\dagger} \}$ operators and thus, it
corresponds to a ``one electron'' problem, with a density regulated by
a chemical potential $\mu$. For the case of a chain with $L$ sites, 
the base can be considered as
$a^{\dagger}_{1,\uparrow}|0\rangle$, $\ldots$,
$a^{\dagger}_{L,\uparrow}|0\rangle$, 
$a^{\dagger}_{1,\downarrow}|0 \rangle$, $\ldots$,
$a^{\dagger}_{L,\downarrow}|0 \rangle$, 
and thus $\hat H$ is given by a $2L \times 2L$ matrix
for a fixed configuration of the classical spins.

The partition function in the grand canonical ensemble can be
written as
\begin{equation}
  \label{eq:part}
  Z=\prod_i^L(\int_0^{\pi}d\theta_i \sin \theta_i \int_0^{2 \pi}
  d\phi_i) Z_g(\{\theta_i,\phi_i\}).
\end{equation}
Here $g$ denotes conduction electrons and
$Z_g(\{\theta_i,\phi_i\})={\rm Tr}_g(e^{-\beta {\hat K}})$,
where ${\hat K}={\hat H}-\mu {\hat N}$
with ${\hat N}$ the number operator and the trace is taken for the
mobile electrons in the $e_{\rm g}$-orbital, which are created and
destroyed by the fermionic operators $a^{\dagger}$ and $a$.
It will be shown that $Z_g$ can be calculated in terms of the
eigenvalues of ${\hat K}$ denoted by $\epsilon_{\lambda}$ 
($\lambda=1, \cdots, 2L$).
The diagonalization is performed numerically using library routines.

Since ${\hat K}$ is an hermitian operator, it can be represented
in terms of a hermitian matrix which can be diagonalized
by an unitary matrix $U$ such that
\begin{equation}
U^{\dagger}KU=\pmatrix{\epsilon_1&0         &\ldots&0\cr
                       0         &\epsilon_2&\ldots&0\cr
                       \vdots    &\vdots    &\ddots&\vdots\cr
                       0         &0         &\ldots&\epsilon_{2L}\cr}.
\end{equation}
The base in which the matrix $K$ is diagonal is given by the
eigenvectors 
$u^{\dagger}_1|0\rangle$, $\ldots$, $u^{\dagger}_{2L}|0\rangle$.
Defining $u^{\dagger}_m u_m$=$\hat n_m$ and denoting by $n_m$ 
the eigenvalues of $\hat n_m$, the trace can be written
\begin{eqnarray}
 && {\rm Tr}_g(e^{-\beta {\hat K}})=\sum_{n_1,\ldots,n_{2L}} \!
 \langle n_1 \ldots n_{2L}| e^{-\beta \hat K}
 |n_1 \ldots n_{2L}\rangle \! \nonumber \\
 && = \sum_{n_1,\ldots,n_{2L}} \! \langle n_1 \ldots n_{2L}|
  e^{-\beta \sum_{\lambda=1}^{2L} \epsilon_{\lambda}n_{\lambda}}
  |n_1 \ldots n_{2L}\rangle,
\end{eqnarray}
since in the $\{u^{\dagger}_m|0\rangle \}$ basis, the operator 
$\hat K$ can be replaced by its eigenvalues.
The exponential is now a ``c'' number and it is equivalent to a 
product of exponentials given by
\begin{equation}
  \label{eq:Zg}
  \! Z_g \!=\! \sum_{n_1}\langle n_1|e^{-\beta\epsilon_1 n_1}|n_1 \rangle
  \! \ldots \!
  \sum_{n_{2L}}\langle n_{2L}|e^{-\beta\epsilon_{2L} n_{2L}}|n_{2L} \rangle,
\end{equation}
which can be written compactly as
\begin{equation}
 Z_g=\prod_{\lambda=1}^{2L} {\rm Tr}_\lambda
 (e^{-\beta\epsilon_\lambda n_\lambda}).
\end{equation}

Since the particles are fermions, the occupation numbers are either 0
or 1, and the sum in Eq.~(\ref{eq:Zg}) is restricted to those values,
\begin{equation}
  \label{eq:explic}
  Z_g= \prod_{\lambda=1}^{2L} \sum_{n=0}^1 e^{-\beta\epsilon_\lambda n}= 
  \prod_{\lambda=1}^{2L}(1+e^{-\beta\epsilon_\lambda}).
\end{equation}
Thus, combining Eq.~(\ref{eq:part}) and Eq.~(\ref{eq:explic}),
$Z$ is obtained as
\begin{eqnarray}
  Z = \prod_i^L(\int_0^{\pi}d\theta_i \sin \theta_i \int_0^{2 \pi}d\phi_i)
  \prod_{\lambda=1}^{2L}(1+e^{-\beta \epsilon_{\lambda}}).
\end{eqnarray}
Note here that the integrand is clearly positive, and thus, 
``sign problems'' are not present.
The integral over the angular variables can be performed 
using a classical Monte Carlo simulation.
The eigenvalues must be obtained for each classical spin configuration
using library subroutines. Finding the eigenvalues is the most time
consuming part of the numerical simulation.

\noindent{\it Calculation of static observables:}
The equal-time or static observables $\hat O(\{a_i,a_i^{\dagger}\})$
are given by
\begin{eqnarray}
 \langle {\hat O} \rangle
 &=& {1 \over Z} \prod_i^L(\int_0^{\pi}d\theta_i \sin \theta_i 
     \int_0^{2 \pi}d \phi_i) {\rm Tr}_g(\hat O e^{-\beta\hat K})
 \nonumber \\
 &=& {1 \over Z} \prod_i^L(\int_0^{\pi}d\theta_i \sin \theta_i 
 \int_0^{2 \pi}d \phi_i) Z_g \langle \tilde O \rangle,
\end{eqnarray}
where $\langle \tilde O \rangle$=
${\rm Tr}_g(\hat O e^{-\beta\hat K})/Z_g$.
In practice only the Green function has to be calculated, and more 
complicated operators are evaluated using Wick's theorem.
The Green function for a given configuration of classical spins are 
given by 
$G_{i,j,\sigma,\sigma'}$
=$\langle a_{i,\sigma}a_{j,\sigma'}^{\dagger}\rangle$.

Let us consider the case in which 
$\hat O=a_{i,\sigma}a_{j,\sigma'}^{\dagger}$, relevant for the Green 
function. In this case, 
\begin{equation}
 \label{eq:bg}
 G_{i,j,\sigma,\sigma'}
 = {\rm Tr}_g(a_{i,\sigma}a_{j,\sigma'}^{\dagger} e^{-\beta\hat K})/Z_g.
\end{equation}
Changing to the base in which $\hat K$ is diagonal through the
transformation
$a_{i\sigma}^{\dagger}$=
$\sum_{\mu=1}^{2L}u_{\mu}^{\dagger} U^{\dagger}_{\mu,i_{\sigma}}$,
where $i_{\sigma}=(i,\sigma)$,
it can be shown that
\begin{eqnarray}
  && {\rm Tr}_g (a_{i,\sigma}a_{j,\sigma'}^{\dagger} e^{-\beta \hat K})
  \nonumber \\
  && =\sum_{\lambda=1}^{2L}\sum_{\eta=1}^{2L}U_{i_{\sigma},\lambda}
  U^{\dagger}_{\eta,j_{\sigma'}}
  {\rm Tr}_g[u_{\lambda}u^{\dagger}_{\eta}\prod_{\nu=1}^{2L}
  e^{-\beta\epsilon_{\nu}n_{\nu}}]
  \nonumber \\
  && =\sum_{\lambda=1}^{2L}\sum_{\eta=1}^{2L}U_{i_{\sigma},\lambda}
  U^{\dagger}_{\eta, j_{\sigma'}}
  {\rm Tr}_g[\prod_{\nu=1}^{2L}
  \{1+(e^{-\beta \epsilon_{\nu}}-1)n_{\nu}\}u_{\lambda}u^{\dagger}_{\eta}]
  \nonumber \\
  && =\sum_{\lambda=1}^{2L}U_{i_{\sigma},\lambda}
  U^{\dagger}_{\lambda, j_{\sigma'}} 
  \prod_{\nu=1}^{2L}\{1+(e^{-\beta\epsilon_{\nu}}-1)n_{\nu}\}(1-n_{\lambda})
  \nonumber \\
  && =\sum_{\lambda=1}^{2L}U_{i_{\sigma}, \lambda}
  U^{\dagger}_{\lambda, j_{\sigma'}}
  \prod_{\nu=1 (\nu \ne \lambda)}^{2L}(\sum_{n_{\nu}=0}^1
  \{1+(e^{-\beta\epsilon_{\nu}}-1)n_{\nu}\})
  \nonumber \\
  && =\sum_{\lambda=1}^{2L}U_{i_{\sigma}, \lambda}
  U^{\dagger}_{\lambda, j_{\sigma'}}\prod_{\nu=1 (\nu \ne \lambda)}^{2L}
  (1+e^{-\beta\epsilon_{\nu}}).
\end{eqnarray}
Thus, the Green function is given by
\begin{eqnarray}
  \label{eq:traza}
  && G_{i,j,\sigma,\sigma'}
  \nonumber \\
  &&=\sum_{\lambda=1}^{2L}U_{i_{\sigma} \lambda}
  U^{\dagger}_{\lambda j_{\sigma'}}\prod_{\nu=1 (\nu \ne
    \lambda)}^{2L}(1+e^{-\beta\epsilon_{\nu}})/\prod_{\nu=1}^{2L}
  (1+e^{-\beta\epsilon_{\nu}})
  \nonumber \\
  &&=\sum_{\lambda=1}^{2L} U_{i_{\sigma},\lambda}
  {1\over{1+e^{-\beta\epsilon_{\lambda}}}}U_{\lambda,j_{\sigma'}}^{\dagger}.
\end{eqnarray}

Let us now consider some examples. The $e_{\rm g}$ electron number is
given by 
\begin{equation}
  \langle\hat n\rangle=
  \sum_{i,\sigma}\langle a^{\dagger}_{i,\sigma}a_{i,\sigma}\rangle
  =2L-\sum_{i,\sigma}G_{i,i,\sigma,\sigma}.
\end{equation}
More complicated operators can be written in terms of Green
functions using Wick's theorem (Mahan, 1990, page 95) which states
that 
\begin{eqnarray}
  && \langle a_{j_1,\sigma}a_{j_2,\sigma}^{\dagger}
  a_{j_3,\sigma}a_{j_4,\sigma}^{\dagger}\rangle 
  \nonumber \\
  && =\langle a_{j_1,\sigma}a_{j_2,\sigma}^{\dagger}\rangle
  \langle a_{j_3,\sigma}a_{j_4,\sigma}^{\dagger}\rangle-
  \langle a_{j_1,\sigma}a_{j_4,\sigma}^{\dagger}\rangle
  \langle a_{j_3,\sigma}a_{j_2,\sigma}^{\dagger}\rangle.
\end{eqnarray}
For example, if 
$\hat O$=$a_{j_1,\sigma}^{\dagger}a_{j_2,\sigma}
a_{j_3,\sigma}^{\dagger}a_{j_4,\sigma}$, 
a combination that appears in the calculation of spin and charge
correlations, and using Wick's theorem in combination with the fact
that
$\langle a_{i,\sigma}^{\dagger}a_{j,\sigma'} \rangle$=
$\delta_{i,j}\delta_{\sigma,\sigma'}$$-$
$\langle a_{j,\sigma'}a_{i,\sigma}^{\dagger} \rangle$=
$\delta_{i,j}\delta_{\sigma,\sigma'}$$-$
$G_{j,i,\sigma',\sigma}$,
it can be shown that
\begin{eqnarray}
  \langle\hat O \rangle &=&
  \langle a_{j_1,\sigma}^{\dagger}a_{j_2,\sigma}\rangle
  \langle a_{j_3,\sigma}^{\dagger}a_{j_4,\sigma}\rangle -
  \langle a_{j_1,\sigma}^{\dagger}a_{j_4,\sigma}\rangle
  \langle a_{j_3,\sigma}^{\dagger}a_{j_2,\sigma}\rangle
  \nonumber \\
  &=&(\delta_{j_1,j_2}-\langle G_{j_2,j_1,\sigma,\sigma}\rangle)
  (\delta_{j_3,j_4}-\langle G_{j_4,j_3,\sigma,\sigma}\rangle)
  \nonumber \\
  &&- (\delta_{j_1,j_4}-\langle G_{j_4,j_1,\sigma,\sigma}\rangle)
  (\delta_{j_3,j_2}-\langle G_{j_2,j_3,\sigma,\sigma}\rangle).
\end{eqnarray}

\noindent{\it Calculation of time-dependent observables:}
Time-dependent observables are evaluated through the time dependent
Green function which can be readily calculated numerically. 
The Green function is defined as
\begin{equation}
  \label{eq:green}
  G_{i,j,\sigma,\sigma}^>(t)=\langle
  a_{i,\sigma}(t)a_{j,\sigma}^{\dagger}(0)\rangle,
\end{equation}
where
\begin{equation}
  a_{i,\sigma}(t)=e^{i\hat Ht}a_{i,\sigma}e^{-i\hat Ht}.
\end{equation}
Note that $\hat H$ and $\hat K$ can be diagonalized by the same basis
of eigenvectors $\{ u_m^{\dagger}|0\rangle \}$, 
and the eigenvalues of $\hat H$ are denoted by $\rho_\lambda$.
Working in this basis it is
possible to write $a_{i,\sigma}(t)$ in terms of $a_{i,\sigma}$ as
\begin{eqnarray}
  \label{eq:creation}
  a_{i,\sigma}(t) &=& e^{it\sum_{\nu}\rho_{\nu}n_{\nu}}\sum_{\eta}
  U_{i_{\sigma},\eta}u_{\eta}
  e^{-it\sum_{\nu}\rho_{\nu}n_{\nu}} \nonumber \\
  &=& \sum_{\nu=1}^{2L}[\sum_{\lambda=1}^{2L}
  U_{i_{\sigma},\lambda}e^{-it\rho_{\lambda}}U^{\dagger}_{\lambda,\nu}]a_{\nu},
\end{eqnarray}
where $a_{\nu}=a_{\nu,\uparrow}$ if $\nu\le L$ and 
$a_{\nu}=a_{\nu-L,\downarrow}$ if $\nu>L$.

Replacing Eq.~(\ref{eq:creation}) in Eq.~(\ref{eq:green}),
the time dependent Green function given by 
\begin{equation}
  \label{eq:gito}
  G_{i,j,\sigma,\sigma}^>(t)
  =\sum_{\nu=1}^{2L}[\sum_{\lambda=1}^{2L}U_{i_{\sigma},\lambda}
  e^{-it\rho_{\lambda}}U^{\dagger}_{\lambda,\nu}]
  \langle a_{\nu}a_{j,\sigma}^{\dagger}\rangle.
\end{equation}
In Eq.~(\ref{eq:bg}) to Eq.~(\ref{eq:traza}),
it has been shown that
$\langle a_{\nu}a_{j,\sigma}^{\dagger}\rangle
=\sum_{\lambda=1}^{2L}U_{\nu,\lambda}{1\over{1+e^{-\beta\epsilon_{\lambda}}}}
U^{\dagger}_{\lambda,j_{\sigma}}$,
where $\epsilon_{\lambda}=\rho_{\lambda}-\mu$.
Thus, replacing Eq.~(\ref{eq:traza}) in Eq.~(\ref{eq:gito}), 
\begin{equation}
  \label{eq:git}
  G_{i,j,\sigma,\sigma}^>(t)=\sum_{\lambda=1}^{2L}
  U_{i_{\sigma},\lambda}{e^{-it\rho_{\lambda}}
    \over{1+e^{-\beta(\rho_{\lambda}-\mu)}}}U^{\dagger}_{\lambda, j_{\sigma}}.
\end{equation}

Now, as an example, let us calculate the spectral function $A(k,\omega)$,
given by
\begin{equation}
  A(k,\omega)=-{1\over{\pi}} {\rm Im} G_{\rm ret}(k,\omega),
\end{equation}
where the retarded Green function $G_{\rm ret}(k,\omega)$
is given by (see Mahan, 1990, page 135)
\begin{equation}
  \label{eq:inte}
  G_{\rm ret}(k,\omega)=\int_{-\infty}^{\infty}dt e^{i\omega t}
  G_{\rm ret}(k,t),
\end{equation}
and
\begin{eqnarray}
  \label{eq:gret}
  G_{\rm ret}(k,t) &\!=\!& -i\theta(t)\sum_{\sigma}\langle
  [a_{k,\sigma}(t)a^{\dagger}_{k,\sigma}(0) 
  +a^{\dagger}_{k,\sigma}(0)a_{k,\sigma}(t)]\rangle
  \nonumber \\
  &\!=\!& -i\theta(t)\sum_{\sigma}(G^{>}_{k,\sigma}+G^{<}_{k,\sigma}).
\end{eqnarray}
Note here that 
$G^{>}_{k,\sigma}$=
$\langle a_{k,\sigma}(t)a^{\dagger}_{k,\sigma}(0)\rangle$
and
$G^{<}_{k,\sigma}$=
$\langle a^{\dagger}_{k,\sigma}(0)a_{k,\sigma}(t)\rangle$
are implicitly defined.

Since the measurements are performed in coordinate space,
$G^{>}_{k,\sigma}$ and $G^{<}_{k,\sigma}$ must be expressed
in terms of real space operators using 
$a_{k,\sigma}={1\over{\sqrt{L}}}\sum_je^{-ikj}a_{j,\sigma}$.
Then
\begin{eqnarray}
  G^{>}_{k,\sigma} &=& {1\over{L}}\sum_{j,l}e^{-ik(j-l)}
  \langle a_{j,\sigma}(t)a^{\dagger}_{l,\sigma}(0)\rangle \nonumber \\
  &=& {1\over{L}}\sum_{j,l}e^{-ik(j-l)}G^{>}_{j,l,\sigma,\sigma},
\end{eqnarray}
and analogously an expression for $G^{<}_{k,\sigma}$ can be obtained.
Thus, Eq.~(\ref{eq:gret}) becomes
\begin{eqnarray}
  \label{eq:gretn}
  && G_{\rm ret}(k,t) \nonumber\\
  && =-i\theta(t){1\over{L}}\sum_{j,l,\sigma}[e^{ik(l-j)}
  G^{>}_{j,l,\sigma,\sigma}+ e^{-ik(l-j)}G^{<}_{j,l,\sigma,\sigma}].
\end{eqnarray}

Replacing Eq.~(\ref{eq:gretn}) in Eq.~(\ref{eq:inte}) it can be shown that
\begin{eqnarray}
  \label{eq:gkw}
  G_{\rm ret}(k,\omega) &=& {-i\over{L}}\int_0^{\infty}dt e^{i\omega t}
  \nonumber\\
  &\times& \sum_{j,l,\sigma}[e^{ik(l-j)}G^{>}_{j,l,\sigma,\sigma}+
  e^{-ik(l-j)}G^{<}_{j,l,\sigma,\sigma}].
\end{eqnarray}
The next step is to evaluate the integral. 
Using Eq.~(\ref{eq:git}), the first term in Eq.~(\ref{eq:gkw}) becomes
\begin{equation}
  {-i\over{L}}\sum_{j,l,\sigma}e^{ik(l-j)}
  \sum_{\lambda=1}^{2L}
  {U_{j_{\sigma},\lambda}U^{\dagger}_{\lambda, l_{\sigma}}
    \over{1+e^{-\beta(\rho_{\lambda}-\mu)}}} 
  \int_0^{\infty}dt e^{i(\omega-\rho_{\lambda})t}.
\end{equation}
Note that the integral is equal to $\pi\delta(\omega-\rho_{\lambda})$. 
A similar expression is obtained for the second term and finally the spectral
function can be expressed in terms of the eigenvectors and eigenvalues of 
$\hat H$ as
\begin{eqnarray}
  \label{eq:akw}
  A(k,\omega) &=& {1\over{L}}{\rm Im} \{i\sum_{j,l,\sigma,\lambda}[e^{ik(l-j)}
  {U_{j_{\sigma},\lambda}U^{\dagger}_{\lambda, l_{\sigma}} 
    \over{1+e^{-\beta(\rho_{\lambda}-\mu)}}} 
  \nonumber \\
  &+& e^{-ik(l-j)}{U_{l_{\sigma},\lambda}U^{\dagger}_{\lambda, j_{\sigma}}
    \over{1+e^{\beta(\rho_{\lambda}-\mu)}}}]\delta(\omega-\rho_{\lambda})\}
  \nonumber\\
  &=& {1\over{L}}{\rm Im} \{i\sum_{j,l,\sigma,\lambda}[e^{ik(j-l)}
  U_{l_{\sigma},\lambda}U^{\dagger}_{\lambda, j_{\sigma}}
  \delta(\omega-\rho_{\lambda}) \nonumber \\
  &\times& ({1\over{1+e^{-\beta(\rho_{\lambda}-\mu)}}}
  +{1\over{1+e^{\beta(\rho_{\lambda}-\mu)}}})]\}.
\end{eqnarray}
Noticing that the sum on the final line is equal 1, the final expression is
\begin{equation}
  A(k,\omega)={1\over{L}}{\rm Re}
  [\sum_{j,l,\sigma,\lambda}e^{ik(j-l)}
  U_{l_{\sigma},\lambda}U^{\dagger}_{\lambda, j_{\sigma}}
  \delta(\omega-\rho_{\lambda})].
\end{equation}

Similar algebraic manipulations allow to express other dynamical
observables, such as the optical conductivitity and dynamical spin
correlation functions, in terms of 
the eigenvalues and eigenvectors of the Hamiltonian matrix.

\medskip
\noindent{\bf Mean-field approximation for $H^{\infty}$:}

Even a simplified model such as $H^{\infty}$ is still difficult to
be solved exactly, except for some special cases.
Thus, in this subsection, the mean-field approximation (MFA) is
developed for $H^{\infty}$ to attempt to grasp its essential physics.
Note that even at the mean-field level, due care should be paid to the 
self-consistent treatment to lift the double degeneracy of the
$e_{\rm g}$-electrons. 
The present analytic MFA will be developed based on the following
assumptions:
(i) The background $t_{\rm 2g}$-spin structure is fixed through the
calculation by assuming that the nearest-neighbor $t_{\rm 2g}$-spins
(not to be confused with the full state) 
can only be in the FM or AF configuration.
(ii) The JT- and breathing-mode distortions are non-cooperative.
These assumptions will be discussed later in this subsection.

First, let us rewrite the electron-phonon term by applying a
standard mean-field decoupling procedure.
In this approximation, a given operator $O$ is written as
$O$=$\langle O \rangle$+$\delta O$,
where $\delta O$=$O-\langle O \rangle$.
In a product of operators $O_1 O_2$, 
terms of order $\delta O_1 \delta O_2$
are simply discarded.
Applying this trick to our case, it is shown that
\begin{equation}
  \left\{
  \begin{array}{l}
  q_{1{\bf i}} n_{\bf i} \approx 
  \langle q_{1{\bf i}} \rangle n_{\bf i} + 
  q_{1{\bf i}} \langle n_{\bf i} \rangle - 
  \langle q_{1{\bf i}} \rangle \langle n_{\bf i} \rangle, \\
  q_{2{\bf i}} \tau_{x{\bf i}} \approx 
  \langle q_{2{\bf i}} \rangle \tau_{x{\bf i}} + 
  q_{2{\bf i}} \langle \tau_{x{\bf i}} \rangle - 
  \langle q_{2{\bf i}} \rangle \langle \tau_{x{\bf i}} \rangle, \\
  q_{3{\bf i}} \tau_{z{\bf i}} \approx 
  \langle q_{3{\bf i}} \rangle \tau_{z{\bf i}} + 
  q_{3{\bf i}} \langle \tau_{z{\bf i}} \rangle - 
  \langle q_{3{\bf i}} \rangle \langle \tau_{z{\bf i}} \rangle, \\
  q_{\alpha{\bf i}}^2 \approx 
  2 \langle q_{\alpha{\bf i}} \rangle q_{\alpha{\bf i}} -
  \langle q_{\alpha{\bf i}} \rangle^2 ~(\alpha=1,2,3),
  \end{array}
  \right.
\end{equation}
where the bracket denotes the average value using the mean-field
Hamiltonian described below.
By minimizing the phonon energy, the local distortion is determined in
the MFA as 
\begin{equation}
\label{eqs:mfqs}
  q_{1{\bf i}}\!=\!-\langle n_{\bf i} \rangle /\beta,~
  q_{2{\bf i}}\!=\!-\langle \tau_{x{\bf i}} \rangle,~ 
  q_{3{\bf i}}\!=\!-\langle \tau_{z{\bf i}} \rangle.
\end{equation}
Thus, after straightforward algebra, 
the electron-phonon term in the MFA is given by
\begin{eqnarray}
 H_{\rm el-ph}^{\rm MF} &=& -2\sum_{\bf i} 
 [ E_{\rm br} \langle n_{{\bf i}} \rangle n_{{\bf i}}
  +E_{\rm JT} (\langle \tau_{x{\bf i}} \rangle \tau_{x{\bf i}}
 +\langle \tau_{z{\bf i}} \rangle \tau_{z{\bf i}})]  \nonumber \\
 && +\sum_{\bf i}[ E_{\rm br} \langle n_{\bf i} \rangle^2
  + E_{\rm JT} (\langle \tau_{x{\bf i}} \rangle^2
  + \langle \tau_{z{\bf i}} \rangle^2)],
\end{eqnarray}
where $E_{\rm br}$=$E_{\rm JT}/\beta$, as already explained.

Now let us turn our attention to the electron-electron interaction
term. 
At a first glance, it appears enough to make a similar decoupling
procedure for $H_{\rm el-el}$.
However, such a decoupling cannot be uniquely carried out, since
it will be shown below that 
$H_{\rm el-el}$ is invariant with respect to 
the choice of $e_{\rm g}$-electron orbitals
due to the local SU(2) symmetry in the orbital space.
Thus, it is necessary to find the optimal
orbital set by determining the relevant
$e_{\rm g}$-electron orbital self-consistently at each
site. For this purpose, it is convenient to use the expression 
Eq.~(\ref{eq:polar}) for $q_{2{\bf i}}$ and 
$q_{3{\bf i}}$.
Note in the MFA that the amplitude $q_{{\bf i}}$ and the phase 
$\xi_{\bf i}$ are, respectively, determined as 
\begin{equation}
  q_{{\bf i}}=\sqrt{\langle \tau_{x{\bf i}} \rangle^2
  +\langle \tau_{z{\bf i}} \rangle^2},~
  \xi_{\bf i}=\pi+\tan^{-1}(\langle \tau_{x{\bf i}} \rangle/
  \langle\tau_{z{\bf i}}\rangle),
\end{equation}
where ``$\pi$" is added to $\xi_{\bf i}$ in the MFA.
Originally, $\xi_{\bf i}$ is defined as 
$\xi_{\bf i}$=$\tan ^{-1}(q_{2{\bf i}}/q_{3{\bf i}})$,
but in the MFA, the distortions are given by Eq.~(\ref{eqs:mfqs}),
in which minus signs appear in front of
$\langle \tau_{x{\bf i}} \rangle$ and
$\langle \tau_{z{\bf i}} \rangle$.
Thus, due to these minus signs, 
the additional phase $\pi$ in $\xi_{\bf i}$ should appear in order
to maintain
consistency with the previous definition,
if $\xi_{\bf i}$ is obtained with the use of 
$\langle \tau_{x{\bf i}} \rangle$ and
$\langle \tau_{z{\bf i}} \rangle$ in the MFA.
By using the phase $\xi_{\bf i}$ determined by this procedure, 
it is convenient to transform
$c_{{\bf i}{\rm a}}$ and $c_{{\bf i}{\rm b}}$ into the
``phase-dressed" operators, 
${\tilde c}_{{\bf i}{\rm a}}$ and ${\tilde c}_{{\bf i}{\rm b}}$, as
\begin{equation}
  \label{eq:trans}
  \!  \left(
    \begin{array}{l}
      {\tilde c}_{{\bf i}{\rm a}} \\
      {\tilde c}_{{\bf i}{\rm b}}
    \end{array}
  \right) 
  \! = \! e^{i\xi_{\bf i}/2} \! 
  \left(
   \begin{array}{ll}
   \cos (\xi_{\bf i}/2)  & \sin (\xi_{\bf i}/2) \\
   -\sin (\xi_{\bf i}/2) & \cos (\xi_{\bf i}/2) 
    \end{array}
  \right) \!
  \left(
    \begin{array}{l}
      c_{{\bf i}{\rm a}} \\
      c_{{\bf i}{\rm b}}
    \end{array}
  \right),
\end{equation}
where the $2 \times 2$ matrix is SU(2) symmetric. 
Note that if $\xi_{\bf i}$ is increased by $2\pi$, the SU(2) matrix itself
changes its sign. To keep the transformation unchanged upon
a $2\pi$-rotation in $\xi_{\bf i}$,
a phase factor $e^{i\xi_{\bf i}/2}$ is needed.
In the expression for the ground state of the single JT molecule,
namely the single-site problem discussed before,
this phase factor has not been added, since the electron does not
hop around from site to site and the phases do not correlate
with each other. Namely it was enough to pay attention
to the double-valuedness of the wave function at a single site.
However, in the JT crystal in which $e_{\rm g}$ electrons move
in the periodic array of the JT centers,
the addition of this phase factor is useful to take into
account the effect of the Berry phase arising from the circular motion
of $e_{\rm g}$-electrons around the JT center,
as has been emphasized in Koizumi et al., 1998a and 1998b.
It could be possible to carry out the calculation
without including explicitly this phase factor,
but in that case, 
it is necessary to pay due attention to the inclusion of the 
effect of the Berry phase.
The qualitative importance of this effect will be explained later in 
the context of the ``band-insulating picture" for the CE-type phase 
of half-doped manganites.

Note also that the phase $\xi_{\bf i}$ determines the electron orbital
set at each site.
In the previous section, the single-site problem was discussed
and the ground-state at site ${\bf i}$ was found to be
\begin{equation}
  |``{\rm b}" \rangle =
  [-\sin (\xi_{\bf i}/2) d_{{\bf i}{\rm a}\sigma}^{\dag}
  + \cos (\xi_{\bf i}/2) d_{{\bf i}{\rm b}\sigma}^{\dag}] |0 \rangle,
\end{equation}
which is referred to as the ``b"-orbital, namely
the combination with the lowest-energy at a given site.
The excited-state or ``a"-orbital is simply obtained by requesting it to be 
orthogonal to ``b" as
\begin{equation}
  |``{\rm a}" \rangle =
  [\cos (\xi_{\bf i}/2) d_{{\bf i}{\rm a}\sigma}^{\dag}
  + \sin (\xi_{\bf i}/2) d_{{\bf i}{\rm b}\sigma}^{\dag}] |0 \rangle.
\end{equation}
For instance, at $\xi_{\bf i}$=$2\pi/3$, ``a'' and ``b'' denote the 
$d_{y^2-z^2}$- and $d_{3x^2-r^2}$-orbitals, respectively.
In Table~I, the correspondence between $\xi_{\bf i}$ and the local
orbital is summarized for several important values of $\xi_{\bf i}$.
In order to arrive to the results of the table,
remember that the original orbitals must be normalized such
that $(x^2-y^2)/\sqrt{2}$ and $(3z^2-r^2)/\sqrt{6}$ are used.
Note also that overall phase factors that may affect the orbitals
are not included in Table~I.
Furthermore, it should be noted that $d_{3x^2-r^2}$ and $d_{3y^2-r^2}$ 
never appear as the local orbital set.
Sometimes those were treated as an orthogonal orbital
set to reproduce the experimental results, but such a treatment is
an approximation, since the orbital ordering is not due to the
simple alternation of two arbitrary kinds of orbitals.

\begin{center}
  \begin{tabular}{|c|c|c|} \hline
    \makebox[15mm]{$\xi_{\bf i}$} & 
    \makebox[25mm]{``a"-orbital} & 
    \makebox[25mm]{``b"-orbital} \\ \hline
    \hline
    $0$    & $x^2-y^2$  & $3z^2-r^2$ \\ \hline 
    $\pi/3$  & $3y^2-r^2$ & $z^2-x^2$  \\ \hline
    $2\pi/3$ & $y^2-z^2$  & $3x^2-r^2$ \\ \hline
    $\pi$    & $3z^2-r^2$ & $x^2-y^2$  \\ \hline 
    $4\pi/3$ & $z^2-x^2$  & $3y^2-r^2$ \\ \hline
    $5\pi/3$ & $3x^2-r^2$ & $y^2-z^2$  \\ \hline
  \end{tabular}
\end{center}

\noindent TABLE I.~Phase $\xi_{\bf i}$ and the corresponding
$e_{\rm g}$-electron orbitals. Note that ``b" corresponds to
the lowest-energy orbital for $E_{\rm JT} \ne 0$.

\medskip

Using the above described transformations, $H_{\rm el-ph}^{\rm MF}$ and
$H_{\rm el-el}$ can be rewritten after some algebra as 
\begin{eqnarray}
 H_{\rm el-ph}^{\rm MF} &=& \sum_{\bf i} 
 \{ E_{\rm br} (-2\langle n_{{\bf i}} \rangle {\tilde n}_{\bf i}
  + \langle n_{\bf i} \rangle^2) \nonumber \\
  &+& E_{\rm JT}[2q_{\bf i}
  ({\tilde n}_{{\bf i}{\rm a}}-{\tilde n}_{{\bf i}{\rm b}})
  +q_{\bf i}^2]\},
\end{eqnarray}
and
\begin{equation}
 \label{eq:helel}
 H_{\rm el-el}= U'\sum_{\bf i}
 {\tilde n}_{{\bf i}{\rm a}}{\tilde n}_{{\bf i}{\rm b}}
 + V \sum_{\langle {\bf i},{\bf j} \rangle}
 {\tilde n}_{{\bf i}} {\tilde n}_{{\bf j}},
\end{equation}
where ${\tilde n}_{{\bf i}\gamma}$
=${\tilde c}_{{\bf i} \gamma}^{\dag} {\tilde c}_{{\bf i} \gamma}$ and
${\tilde n}_{\bf i}$
=${\tilde n}_{{\bf i}{\rm a}}$+${\tilde n}_{{\bf i}{\rm b}}$.
Note that $H_{\rm el-el}$ is invariant with respect to the choice of
$\xi_{\bf i}$.
Equation (\ref{eq:helel}) can be obtained by calculating
${\tilde c}_{{\bf i}{\rm a}}^{\dag}{\tilde c}_{{\bf i}{\rm a}}$
+${\tilde c}_{{\bf i}{\rm b}}^{\dag}{\tilde c}_{{\bf i}{\rm b}}$
using Eq.~(\ref{eq:trans}).
This immediately leads to ${\tilde n}_{\bf i}$=$n_{\bf i}$.
Then, from ${\tilde n}_{\bf i}^2$=$n_{\bf i}^2$ and recalling 
that $n_{{\bf i}\gamma}^2$=$n_{{\bf i}\gamma}$
and ${\tilde n}_{{\bf i}\gamma}^2$=${\tilde n}_{{\bf i}\gamma}$
for $\gamma$=a and b, it can be shown that
${\tilde n}_{{\bf i}{\rm a}}{\tilde n}_{{\bf i}{\rm b}}$
=$n_{{\bf i}{\rm a}}n_{{\bf i}{\rm b}}$.
Now let us apply the decoupling procedure as
\begin{equation} 
  {\tilde n}_{{\bf i}{\rm a}} {\tilde n}_{{\bf i}{\rm b}}
  \approx
  \langle {\tilde n}_{{\bf i}{\rm a}}\rangle {\tilde n}_{{\bf i}{\rm b}}
  +{\tilde n}_{{\bf i}{\rm a}}\langle{\tilde n}_{{\bf i}{\rm b}}\rangle 
  -\langle {\tilde n}_{{\bf i}{\rm a}}\rangle
  \langle {\tilde n}_{{\bf i}{\rm b}} \rangle,
\end{equation}
and use the relations 
$\langle {\tilde n}_{{\bf i}{\rm a}} \rangle$
=$(\langle n_{{\bf i}} \rangle-q_{\bf i})/2$,
$\langle {\tilde n}_{{\bf i}{\rm b}} \rangle$
=$( \langle n_{{\bf i}} \rangle +q_{\bf i})/2$,
which arise from 
$\langle {\tilde n}_{{\bf i}{\rm a}}-{\tilde n}_{{\bf i}{\rm b}} \rangle$
=$-q_{\bf i}$
and
$\langle {\tilde n}_{{\bf i}{\rm a}}+{\tilde n}_{{\bf i}{\rm b}} \rangle$
=$\langle n_{\bf i} \rangle$.
The former relation indicates that the modulation in the orbital density
is caused by the JT distortion, while the latter denotes the local charge
conservation irrespective of the choice of electron basis.
Then, the electron-electron interaction term is given in the MFA as
\begin{eqnarray}
 H_{\rm el-el}^{\rm MF}\! &=&\! (U'/4)\!\sum_{{\bf i}}
 [2\langle n_{\bf i} \rangle {\tilde n}_{\bf i}
 \!-\! \langle n_{{\bf i}} \rangle^2
 \!+\! 2q_{\bf i}({\tilde n}_{a{\bf i}}\! - \!{\tilde n}_{b{\bf i}})
 \!+\! q_{\bf i}^2 ] \nonumber \\
 &+& V \sum_{{\bf ia}} [ \langle n_{{\bf i+a}} \rangle 
 {\tilde n}_{{\bf i}} -(1/2) \langle n_{{\bf i+a}} \rangle 
 \langle n_{{\bf i}} \rangle],
\end{eqnarray}
where the vector ${\bf a}$ has the same meaning as 
in the hopping term $H_{\rm kin}$.
For instance, in two dimensions, 
it denotes ${\bf a}$=$(\pm 1, 0)$ and $(0,\pm 1)$, where the lattice
constant is taken as unity for simplicity. 
It should be noted that the type of orbital ordering would be
automatically fixed as either $x^2-y^2$ or $3z^2-r^2$, if the original
operators $c$ would be simply used for the Hartree-Fock
approximation. 
However, as it was emphasized above, the $H_{\rm el-el}$ term has a
rotational invariance in orbital space, and there is no reason to fix
the orbital only as $x^2-y^2$ or $3z^2-r^2$.
In order to discuss properly the orbital ordering, the local 
$e_{\rm g}$ electron basis, i.e., the phase $\xi_{\bf i}$ should be
determined self-consistently.

By combining $H_{\rm el-ph}^{\rm MF}$ with $H_{\rm el-el}^{\rm MF}$
and transforming ${\tilde c}_{{\bf i}{\rm a}}$ and 
${\tilde c}_{{\bf i}{\rm b}}$
into the original operators 
$c_{{\bf i}{\rm a}}$ and $c_{{\bf i}{\rm b}}$,
the mean-field Hamiltonian is finally obtained as 
\begin{eqnarray}
  \label{mfa}
  H^{\infty}_{\rm MF} &=& -\sum_{{\bf ia}\gamma \gamma'}
  t^{\bf a}_{\gamma \gamma'} 
  c_{{\bf i} \gamma}^{\dag} c_{{\bf i+a} \gamma'}
  + J_{\rm AF} \sum_{\langle {\bf i,j} \rangle}
  {\bf S}_{\bf i} \cdot {\bf S}_{\bf j} \nonumber \\ 
  &+& {\tilde E_{\rm JT}} \sum_{\bf i} 
  [-2(\langle \tau_{x{\bf i}} \rangle \tau_{x{\bf i}}
  +\langle \tau_{z{\bf i}} \rangle \tau_{z{\bf i}})
  +\langle \tau_{x{\bf i}} \rangle^2
  + \langle \tau_{z{\bf i}} \rangle^2]    \nonumber \\ 
  &+& \sum_{\bf i} [({\tilde U'}/2) \langle n_{{\bf i}} \rangle
  + V \sum_{\bf a} \langle n_{{\bf i+a}} \rangle]
  (n_{{\bf i}}- \langle n_{{\bf i}} \rangle/2),
\end{eqnarray}
where the renormalized JT energy is given by
\begin{equation}
  {\tilde E_{\rm JT}}=E_{\rm JT}+U'/4,
\end{equation}
and the renormalized inter-orbital Coulomb interaction is 
expressed as 
\begin{equation}
  {\tilde U'}=U'- 4E_{\rm br}.
\end{equation}
Physically, the former relation indicates that the JT energy is
effectively enhanced by $U'$.
Namely, the strong on-site Coulombic correlation plays the $same$ role
as that of the JT phonon, at least at the mean-field level, 
indicating that it is not necessary to include $U'$ explicitly in the
models, as has been emphasized by the present authors in several
publications (see for instance Hotta, Malvezzi and Dagotto (2000)). 
The latter equation for ${\tilde U'}$ means that the one-site 
inter-orbital Coulomb interaction is effectively reduced by the
breathing-mode phonon, since the optical-mode phonon provides an
effective attraction between electrons.
The expected positive value of ${\tilde U'}$ indicates that 
$e_{\rm g}$ electrons dislike double occupancy at the site, since the 
energy loss is proportional to the average local electron number in
the mean-field argument.
Thus, to exploit the gain due to the static JT energy and avoid 
the loss due to the on-site repulsion, 
an $e_{\rm g}$ electron will singly occupy a given site.

Now let us briefly discuss how to solve the present mean-field 
Hamiltonian on the background of the fixed $t_{\rm 2g}$-spin 
arrangement. For some fixed spin pattern, by using appropriate initial
values for the local densities $\langle n_{\bf i} \rangle^{(0)}$,
$\langle \tau_{x{\bf i}} \rangle^{(0)}$, and 
$\langle \tau_{z{\bf i}} \rangle^{(0)}$,
the mean-field Hamiltonian $H_{\rm MF}^{\infty (0)}$ is constructed,
where the superscript number $(j)$ indicates the iteration step.
By diagonalizing $H_{\rm MF}^{\infty (0)}$ on 
finite clusters, and usually using a variety of boundary conditions
depending on the problem,
the improved local densities,
$\langle n_{\bf i} \rangle^{(1)}$,
$\langle \tau_{x{\bf i}} \rangle^{(1)}$, and 
$\langle \tau_{z{\bf i}} \rangle^{(1)}$,
are obtained.
This procedure is simply repeated such that in the $j$-th iteration step,
the local densities
$\langle n_{\bf i} \rangle^{(j+1)}$,
$\langle \tau_{x{\bf i}} \rangle^{(j+1)}$, and 
$\langle \tau_{z{\bf i}} \rangle^{(j+1)}$
are obtained by using the Hamiltonian $H_{\rm MF}^{\infty (j)}$.
The iterations can be terminated if
$|\langle n_{\bf i} \rangle^{(j+1)}$$-$ 
$\langle n_{\bf i} \rangle^{(j)}|$
$<$ $\delta$,
$|\langle \tau_{x{\bf i}} \rangle^{(j)}$$-$
$\langle \tau_{x{\bf i}} \rangle^{(j-1)}|$
$<$ $\delta$,
and
$\langle \tau_{z{\bf i}} \rangle^{(j)}$$-$
$\langle \tau_{z{\bf i}} \rangle^{(j-1)}|$
$<$$\delta$
are satisfied, where $\delta$ is taken to be a small number to control
the convergence.

As for the choice of the cluster, in order to obtain the charge and 
orbital ordering pattern in the insulating phase,
it is usually enough to treat a finite-size cluster with periodic
boundary conditions. 
Note that the cluster size should be large enough to reproduce the 
periodicity in the spin, charge, and orbital ordering under
investigation. 
However, to consider the transition to the metallic state from 
the insulating phase, in principle it is necessary to treat 
an infinite cluster.
Of course, except for very special cases, it is impossible to treat
the infinite-size cluster exactly, but fortunately, in the present
MFA, it is quite effective to employ the twisted-boundary condition 
by introducing the momentum ${\bf k}$ in the Bloch phase factor
$e^{i{\bf k} \cdot {\bf N}}$ at the boundary, 
where ${\bf N}$=$(N_{\bf x}, N_{\bf y}, N_{\bf z})$, and 
$N_{\bf a}$ is the size of the cluster along the ${\bf a}$-direction.
Note that if the spin directions are changed periodically, 
an additional phase factor appears to develop
at the boundary, but this is not the case.
In the present MFA, the $t_{\rm 2g}$-spin pattern is fixed from the
outset, and the periodicity due to the spin pattern is already taken
into account in the cluster.

Finally, here comments on some of the 
assumptions employed in the present MFA are provided.
In the first approximation for $t_{\rm 2g}$-spins,
their pattern is fixed throughout the mean-field calculation
and the nearest-neighbor spin configuration is assumed to be
only FM or AFM. 
Note that this assumption does not indicate only the fully FM
phase or three-dimensional G-type AFM spin pattern, but
it can include more complicated spin patterns such as the CE-type AFM 
phase. 
However, under this assumption, several possible phases such as
the spin canted phase and the spin flux phase, in which 
neighboring spins are neither FM nor AFM, are neglected from the outset.

Unfortunately, this assumption for the fixed $t_{\rm 2g}$ spin pattern 
cannot be justified without extra tests.
Thus, it is unavoidable to confirm the assumption using other methods.
In order to perform this check, unbiased numerical calculations such
as the Monte Carlo simulations and relaxation techniques
are employed to determine the local distortions, as well as
the local spin directions.
Especially for a fixed electron number, the optimization technique
is found to work quite well in this type of problems.

Then, our strategy to complete the mean-field calculations is as
follows:
(i) For some electron density and small-size cluster,
the mean-field calculations are carried out for several fixed
configurations of $t_{\rm 2g}$ spins.
(ii) For the same electron density in the same size of cluster as in
(i), both local distortions and $t_{\rm 2g}$-spin directions are
optimized by using an appropriate computer code.
(iii) Results obtained in (i) are compared to those in (ii).
If there occurs a serious disagreement between them, go back to step
(i) and/or (ii) to do again the calculations by changing the initial
inputs.
In this retrial, by comparing the energies between the cases (i) and
(ii), the initial condition for the case with higher energy should be
replaced with that for the lower energy.
To save CPU time it is quite effective to combine analytic MFA and
numerical techniques.
(iv) After several iterations, if a satisfactory agreement between
(i) and (ii) is obtained, the MFA on a larger-size
cluster is used to improve the results in (i).
Namely, by combining the MFA and the optimization technique, it is
possible to reach physically important results in a rapid and reliable
way.

As for the assumption made regarding the use of non-cooperative phonons
in the MFA, it is also checked by comparing the non-cooperative
mean-field results with the optimized ones for cooperative
distortions. 
Note here that, due to the CPU and memory restrictions, the
optimization technique cannot treat large-size clusters. 
However, this numerical technique has the clear advantage that it is
easily extended to include the cooperative effect 
by simply changing the coordinates from $\{ Q \}$ to $\{ u \}$,
where $\{ u \}$ symbolically indicates the oxygen displacements,
while $\{ Q \}$ denotes the local distortions of the MnO$_6$ octahedron.
The effect of the cooperative phonons will be discussed separately
for several values of the hole density.

\begin{center}
  {\bf III.d Main Results: One Orbital Model}
\end{center}

\noindent{\bf Phase Diagram with Classical Localized Spins:}
\medskip

Although the one-orbital model for manganites is clearly incomplete to 
describe these compounds since, by definition, it has only one active
orbital, nevertheless it has been shown in recent calculations
that it captures part of the interesting competition between
ferromagnetic and antiferromagnetic phases in these compounds. For
this reason, and since this model is far simpler than the more 
realistic two-orbital model, it is useful to study it in detail.

A fairly detailed analysis of the phase diagram of the one-orbital
model has been recently presented, mainly using computational 
techniques.
Typical results are shown in Fig.III.d.1a-c for $D$=1, 2, and $\infty$
($D$ is spatial dimension), the first two obtained with Monte Carlo
techniques at low temperature, and the third with the dynamical
mean-field approximation in the large $J_{\rm H}$ limit varying
temperature.
There are several important features in the results which are common
in all dimensions.
At $e_{\rm g}$-density \densi=1.0, the system is antiferromagnetic (although
this is not clearly shown in Fig.III.d.1). 
The reason is that at large Hund coupling, double occupancy in the
ground state is negligible at $e_{\rm g}$-density \densi=1.0 
or lower, and at these densities
it is energetically better to have 
nearest-neighbor spins antiparallel, gaining an energy of order 
$t^2/J_{\rm H}$, rather than to align them, since in such a case the system
is basically frozen due to the Pauli principle.
On the other hand, at finite hole density, antiferromagnetism is
replaced by the tendency of holes to polarize the spin background to
improve their kinetic energy, as discussed in Section III.a.
Then, a very prominent ferromagnetic phase develops in the model as 
shown  in Fig.III.d.1. This FM tendency appears in all dimensions of 
interest, and it manifests itself in the Monte Carlo simulations
through the rapid growth with decreasing temperature, and/or increasing
number of sites, of the zero-momentum spin-spin correlation, as shown
in Fig.III.d.2a-b reproduced from Yunoki et al. (1998a). 
In real space, the results correspond to spin correlations between two
sites at a distance $d$ which do not decay to a vanishing number as $d$
grows, if there is long-range order
(see results in Dagotto et al., 1998).
In 1D, quantum fluctuations are expected to be so strong that
long-range order cannot be achieved, but in this case the spin 
correlations still can decay slowly with distance following a power law.
In practice, the tendency toward FM or AF is so strong even in 1D that 
issues of long-range order vs power-law decays are not of much
importance for studying the dominant tendencies in the model.
Nevertheless, care must be taken with these subtleties if very
accurate studies are attempted in 1D.

In 3D, long-range order can be obtained at finite temperature and
indeed it occurs in the one-orbital model. A rough estimation of the
critical Curie temperature $T_{\rm C}$ is shown in Fig.III.d.3
based on small 6$^3$ 3D clusters (from Yunoki et al., 1998a). 
$T_{\rm C}$ is of the order of just 0.1t, while other estimations predicted
a much higher value (Millis, Littlewood and Shraiman (1995)). 
More recent work has refined $T_{\rm C}$, but the order of magnitude
found in the first Monte Carlo simulations remains the same 
(see Calderon and Brey, 1998; Yi, Hur, and Yu, 1999;
Motome and Furukawa, 1999; Held and Vollhardt, 1999).
If $t$ is about 0.2eV, the $T_{\rm C}$ becomes of the order of 200K,
a value in reasonable agreement with experiments.
However, remember that this model cannot describe orbital order
properly, and thus it remains a crude approximation to manganites. 

The most novel result emerging from the computational studies of the
one-orbital model is the way in which the FM phase is reached by hole
doping of the AF phase at \densi=1.0.
As explained before, mean-field approximations by de Gennes (1960)
suggested that this interpolation should proceed through a so-called
``canted'' state in which the spin structure remains antiferromagnetic 
in two directions but develops a uniform moment along the third
direction.
For many years this canted state was assumed to be correct, and many
experiments were analyzed based on such state.
However, the computational studies showed that instead of a canted
state, an electronic  ``phase separated'' (PS) regime interpolates
between the FM and AF phases. This PS region is very prominent in the
phase diagram of Fig.III.d.1a-c in all dimensions.

As an example of how PS is obtained from the computational work, consider
Fig.III.d.4. In the Monte Carlo simulations carried out in this
context, performed in the grand-canonical ensemble, the density of
mobile $e_{\rm g}$-electrons \densi 
is an output of the calculation, the input
being the chemical potential $\mu$.
In Fig.III.d.4a, the density \densi vs. $\mu$ is shown for one dimensional 
clusters of different sizes at low temperature and large Hund
coupling, in part (b) results in two dimensions are presented, and in 
part (c) the limit $D$=$\infty$ is considered.
In all cases, a clear $discontinuity$ in the density appears at a
particular value of $\mu$, as in a first-order phase transition.
This means that there is a finite range of densities which are simply
unreachable, i.e., that they cannot be stabilized regardless of how
carefully $\mu$ is tuned.
If the chemical potential is fixed to the value where the
discontinuity occurs, frequent tunneling events among the two limiting 
densities are observed (Dagotto et al., 1998).
In the inset of Fig.III.d.4a, the spin correlations are shown for the
two densities at the extremes of the discontinuity, and they
correspond to FM and AF states.

Strictly speaking, the presence of PS means that the model has a range
of densities which cannot be accessed, and thus, those densities are
simply $unstable$. 
This is clarified better using now the canonical ensemble, where the
number of particles is fixed as an input and $\mu$ is an output.
In this context, suppose that one attempts to stabilize a density
such as \densi=0.95 (unstable in
Fig.III.d.4), by locating, say, 95 electrons into a
10$\times$10 lattice.
The ground-state of such a system will not develop a uniform
density, but instead two regions separated in space will be formed: a
large one with approximately 67 sites and 67 electrons (density 1.0)
and a smaller one with 33 sites and 28 electrons (density
$\sim$0.85). The last density is the lower value in the discontinuity
of Fig.III.d.4b in 2D, i.e., the first stable density after \densi=1.0 when
holes are introduced.
Then, whether using canonical or grand-canonical approximations, a
range of densities remains unstable.

The actual spatial separation into two macroscopic regions (FM and AF
in this case) leads to an energy problem. 
In the simulations and other mean-field approximations that produce
PS, the ``tail'' of the Coulomb interaction was not explicitly
included.
In other words, the electric charge was not properly accounted for.
Once  this long-range Coulomb interaction is introduced into the
problem, the fact that the FM and AF states involved in PS have
different densities leads to a huge energy penalization 
even considering a large dielectric constant due to polarization
(charge certainly cannot be accumulated in a macroscopic 
portion of a sample).
For this reason, it is more reasonable to expect that the PS domains
will break into smaller pieces, as sketched in Fig.III.d.5 (Moreo et
al., 1999a; see also Section III.i). 
The shape of these pieces remains to be investigated in detail since
the calculations are difficult with long-range interactions (for
results in 1D see below), but droplets or stripes appear as a serious
possibility. 
This state would now be $stable$, since it would satisfy in part the 
tendency toward phase separation and also it will avoid a macroscopic 
charge accumulation. Although detailed calculations are not available,
the common folklore is that the typical size of the clusters in the
mixed-phase state arising from the competition PS vs. $1/r$ Coulomb
will be in the $nanometer$ scale, i.e., just a few lattice spacings 
since the Mn-Mn distance is about 4$\rm \AA$.
This is the electronic ``Phase Separated'' state that one usually has
in mind as interpolating between FM and AF.
Small clusters of FM are expected to be created in the AF background,
and as the hole density grows, these clusters will increase in number
and eventually overcome the AF clusters.
For more details see also Section III.i, where the effort of other
authors in the context of PS is also described.

\medskip
\noindent{\bf Spin Incommensurability and Stripes:}

In the regime of intermediate or small $J_{\rm H}$, the one-orbital 
model does not have ferromagnetism at small hole densities, which is
reasonable since a large $J_{\rm H}$ was needed in the discussion of
Section III.a to understand the stabilization of a spin polarized
phase.
Instead, in this regime of $J_{\rm H}$ the spin sector develops
$incommensurability$ (IC), namely the peak in the Fourier transform of
the real space spin-spin correlations is neither at 0 (FM) nor at
$\pi$ (AF), but at intermediate momenta.
This feature is robust and it appears both in 1D and 2D simulations,
as well as with both classical and quantum spins (Yunoki et al., 1998a,
Dagotto et al., 1998). An example in 2D is presented in
Fig.III.d.6a.
Since a regime with IC characteristics had not been found in
experiments by the time the initial Monte Carlo simulations were carried out,
the spin IC regime was not given much importance, and its origin
remained unclear.
However, recent neutron scattering results (Adams et al., 2000; Dai et
al., 2000; Kubota et al., 2000) suggest that spin incommensurability
may appear in some compounds leading to stripe formation, similar to
that found in the cuprates.
This result induced us to further examine the numerical data obtained
in the original Monte Carlo simulations.
It turns out that the spin IC structure found in the 2D one-orbital
model has its origin in $stripes$, as shown in Fig.III.d.6b.
These structures correspond to 1D-like regions of the 2D plane that
are populated by holes, leaving undoped the area between the stripes,
similar to those structures that are
believed to occur in some high temperature
superconductors and $t$-$J$-like models
(Tranquada, 1995; Dai et al., 1998; Burges et al., 2000. See also
Emery et al., 1997; Zaanen, 1998; White and Scalapino, 1998;
Martins et al., 2000).
In fact, the results shown in Fig.III.d.6b are very similar to those
found recently by Buhler et al. (2000) in the context of the so-called
spin-fermion model for cuprates, with classical spins used for the
spins (the spin-fermion model for cuprates and the one-orbital model for
manganites only differ
in the sign of the Hund coupling). 
Stripe formation with hole density close to \densi=1.0, i.e., electronic
density close to 0.0, is natural near phase separation regimes.
Stripes have also been identified in the more realistic case of the
two orbital model (see Section III.e below). A discussion of the 
similarities and differences between the electronic phase separation
scenarios for manganites and cuprates, plus a substantial body of
references, can be found in Hotta, Malvezzi,
and Dagotto (2000). 

\medskip
\noindent{\bf Influence of $J_{\rm AF}$:}

The one-orbital model described in Section III.c included an
antiferromagnetic coupling among the localized spins that is
regulated by a
parameter $\rm J_{AF}$, which was not considered in the previous
subsections.
In principle, this number is the smallest of the couplings in the
model according to the estimations discussed in Section III.c, and one
may naively believe that its presence is not important.
However, this is incorrect as can be easily understood in the limit of
\densi=0.0 (x=1.0), which is realized in materials such as $\rm Ca Mn O_3$.
This compound is antiferromagnetic and it is widely believed that such
magnetic order is precisely caused by the coupling among the localized
spins. Then, $\rm J_{AF}$ cannot be simply neglected.
In addition, the studies shown below highlight the (unexpected)
importance of this coupling in other contexts: it has been found to be 
crucial for the stabilization of an A-type AF phase at \densi=1.0 in
the two-orbital model, and also to make stable the famous CE phase at
\densi=0.5, at least within the context of a two orbital model with
strong electron Jahn-Teller phonon coupling.
Then, it is important to understand the influence of $J_{\rm AF}$
starting with the one-orbital model.

The first numerical study that included a nonzero $J_{\rm AF}$ was
reported by Yunoki and Moreo (1998) (note that hereafter $J'$ will be an
alternative notation for $J_{\rm AF}$, as used sometimes 
in previous literature).
An interesting observation emerging from their analysis is that PS
occurs not only near \densi=1.0 but also near the other extreme of
\densi=0.0, where again a FM-AF competition exists. In this regime,
Batista et al. (2000) have shown the formation of ferromagnetic polarons
upon electron doping of the \densi=0.0 AF state. Considering several of
these polarons it is likely that 
extended structures may form, as in a phase separated state.
The 1D phase diagram at low temperature in the ($J'$, \densi)-plane
is in Fig.III.d.7.
Three AF regions and two PS regions are shown, together with a FM
regime at intermediate densities already discussed in previous
subsections.  
In addition, a novel phase exists at intermediate values of $J'$ and
\densi.
This phase has a curious spin arrangement given by a periodically
arranged pattern $\uparrow \uparrow \downarrow \downarrow$ of
localized spins, namely it has an equal number of FM and AF links, and
for this reason interpolates at constant density between FM and AF
phases (see also Garcia et al., 2000; Aliaga et al., 2000).
This phase is a precursor in 1D of the CE phase in 2D, as will be
discussed later. Calculations of the Drude weight show that this state
is insulating, as expected since it has AF links.

\medskip
\noindent{\bf Quantum Localized Spins:}

An important issue in the context discussed in this Section is whether
the approximation of using classical degrees of freedom to represent
the $t_{\rm 2g}$ spins is sufficiently accurate.
In principle, this spin should be $S$=3/2, which appears large enough
to justify the use of classical spins.
Unfortunately, it is very difficult to study quantum spins in
combination with mobile fermions, and the approximation can be
explicitly tested only in a few cases.
One of them is a 1D system, where the density matrix renormalization
group (DMRG) method and Lanczos techniques allow for a fairly accurate
characterization of the fully quantum model.
The phase diagram obtained in this context by Dagotto et al. (1998) is
reproduced in Fig.III.d.8.
Fortunately, the shape and even quantitative aspects of the diagram
(with AF, FM, IC and PS regions) are in good agreement with those
found with classical spins.
The PS regime certainly appears in the study, although finite values
of $J_{\rm H}$ are needed for its stabilization.
The study leading to Fig.III.d.8 was carried out in the canonical
ensemble, with fixed number of particles, and the possibility of PS
was analyzed by using the compressibility ($\kappa$), criterion where
a $\kappa$$<$0 corresponds to an unstable system, as it is well-known
from elementary thermodynamic considerations.
$\kappa^{-1}$ is proportional to the second derivative of the ground
state with respect to the number of particles, which can be obtained
numerically for $N$ electrons by discretizing the derivative using the 
ground state energies for $N$, $N+2$ and $N-2$ particles at the fixed 
couplings under consideration (for details see Dagotto et al., 1998).
Following this procedure, a negative compressibility was obtained,
indicative of phase separation.
Another method is to find $\mu$ from the ground state energies at
various number of electrons, and plot density vs $\mu$.
As in Monte Carlo simulations with classical spins, a discontinuity 
appears in the results in the regime of PS. It is clear that the
tendency toward these unstable regimes, or mixed-states after proper
consideration of the $1/r$ Coulomb interaction, is very robust and
independent of details in the computational studies. 
Results for the much simpler case of localized $S$=1/2 spins can also
be obtained numerically. The phase diagram (Dagotto et al., 1998) is 
still in qualitative agreement with $S$=3/2 and $\infty$, although not 
quantitatively. 
PS appears clearly in the computational studies, as well as FM and
spin IC phases.

\medskip
\noindent{\bf Influence of Long-Range Coulomb Interactions:}

As already explained before, it is expected that long-range Coulomb
interactions will break the electronic 
PS regime with two macroscopic  FM and AF
regions, into a stable state made out of small coexisting clusters of
both phases.
However, calculations are difficult in this context. One of the few
attempts was carried out by Malvezzi et  al. (1999) using a 1D
system.
On-site $U$ and nearest-neighbor $V$ Coulomb interactions were added
to the one-orbital model. The resulting phase diagram can be found in
Fig.14 of Malvezzi et al. (1999).
At $V$=0, the effect of $U$ is not much important, namely PS is found
at both extremes of densities, and in between a charge-disordered FM
phase is present, results in good agreement with those described in
previous subsections.
This is reasonable since a large Hund coupling by itself  suppresses
double occupancy even without $U$ added explicitly to the model.

However, when $V$ is switched-on, the PS regime of small hole density
is likely to be affected drastically due to the charge
accumulation.
Indeed, this regime is replaced by a charge-density wave with a peak
in the spin structure factor at a momentum different from 0 and $\pi$
(Malvezzi et al., 1999).
Holes are spread over a few lattice spacings, rather than being close
to each other as in PS. In the other extreme of many holes, the very
small electronic density makes $V$ not as important.
In between, the FM phase persists up to large value of $V$, but a
transition exists in the charge sector, separating a charge-disordered
from a charge-ordered state.
Certainly more work in this interesting model is needed to fully
clarify its properties, and extensions to 2D would be important, but
the results thus far are sufficient to confirm that PS is rapidly
destroyed by a long-range Coulomb interaction leading to nontrivial
charge density waves (Malvezzi et al., 1999).

\medskip
\noindent{\bf Tendencies Toward Electronic Phase Separation in
  t-J-like Models for Transition Metal Oxides:}

The first indications of a strong tendency toward 
electronic phase separation in
models for manganites were actually obtained by Riera, Hallberg and
Dagotto (1997) using computational techniques applied to $t$-$J$-like
models for  Ni- and Mn-oxides. 
In these models, it was simply assumed that the Hund coupling was
sufficiently large that the actual relevant degrees of freedom at low
energy are ``spins'' (of value 2 and 1, for Mn- and Ni-oxides,
respectively) and ``holes'' (with spin 3/2 and 1/2, for Mn- and
Ni-oxides, respectively).
The Hamiltonian at large $J_{\rm H}$ can be perturbatively deduced
from the quantum one-orbital model and its form is elegant, with
hole hopping which can take place with rearrangement of the spin
components of ``spin'' and ``hole''. 
Details can be found in Riera et al. (1997). The phase diagram found 
with DMRG and Lanczos methods is in Fig.III.d.9 for the special case
of one dimension.
The appearance of ferromagnetic and phase-separated regions is clear 
in this figure, and the tendency grows as the magnitude of the spin
grows.
In between PS and FM, regions with hole binding were identified that
have not been studied in detail yet. The result in Fig.III.d.9 has to
be contrasted against those found for the standard 1D $t$-$J$ model
for the cuprates (see results in Dagotto, 1994) where the PS regime
appears at unphysically large values of $J/t$, and FM was basically absent.
There is a substantial $qualitative$ difference between the results
found for Cu-oxides and those of Ni- and Mn-oxides, mainly caused by
the presence of localized spins in the last two.
This suggests that cuprates do $not$ share the same physics as other
transition metal oxides. In particular, it is already  known that
cuprates are superconductors upon hole doping, while nickelates and
manganites are not.

\begin{center}
  {\bf III.e Main Results: Two Orbital Model}
\end{center}

\noindent{\bf Phase Diagram at Density x=0.0:}
\medskip

The results of the previous Section showed that the one-orbital model
for manganites contains interesting physics, notably a 
FM-AF competition that has similarities with those found in experiments.
However, it is clear that to explain the notorious orbital order
tendency in Mn-oxides, it is crucial to use a model with two orbitals,
and in Section III.c such a model was defined for the case where there
is an electron Jahn-Teller phonon coupling and also Coulomb
interactions.
Under the assumption that both localized $t_{\rm 2g}$ spins and
phonons are classical, the model without Coulombic terms can be
studied fairly accurately using numerical and mean-field
approximations.
Results obtained with both approaches will be presented here.
For the case where Coulomb terms are included, unfortunately, 
computational studies
are difficult but mean-field approximations can still be carried out.

As in the case of one orbital, let us start with the description of
the phase diagram of the two orbital model. In this model there are
more parameters than in the previous case, and more degrees of
freedom, thus at present only a fraction of parameter space has been
investigated. Consider first the case of $e_{\rm g}$-density \densi=1.0,
which is relatively simple to study numerically since this density is
easy to stabilize in the grand canonical simulations.
It corresponds to having one electron on average per site, and in 
this respect it must be related to the physics found in hole undoped 
compounds such as $\rm LaMnO_3$. Carrying out a Monte Carlo simulation 
in the localized spins and phonons, and considering exactly the
electrons in the absence of an explicit Coulomb repulsion (as is done for
the one-orbital case), a variety of correlations have been calculated
to establish the \densi=1.0 phase diagram. 
Typical results for the spin and orbital structure factors, $S(q)$ and 
$T(q)$ respectively, at the momenta of relevance are shown in
Fig.III.e.1, obtained at a large Hund coupling equal to 8$t$, 
$J'$=0.05$t$, and a small temperature, plotted as a function of the
electron-phonon coupling $\lambda$.
Results at dimensions 1, 2 and 3 are shown.
At small $\lambda$, $S(0)$ is dominant and $T(q)$ is not active.
This signals a ferromagnetic state with disordered orbitals, namely a
standard ferromagnet (note, however, that Khomskii (2000) and Maezono
and Nagaosa (2000) believe that this state in experiments may have
complex orbital ordering).  
The result with FM tendencies dominating may naively seem strange
given the fact that for the one-orbital model at \densi=1.0 an AF
state was found. 
But here two orbitals are being considered and one electron per site
is 1/2 electron per orbital.
In this respect, \densi=1.0 with two orbitals should be similar to 
\densi=0.5 for one orbital and indeed in the last case a ferromagnetic 
state was observed (Section III.d).

Results become much more interesting as $\lambda$ grows beyond 1.
In this case, first $T({\bf Q})$ increases rapidly and dominates
(in the spin sector still $S(0)$ dominates, i.e. the system remains  
ferromagnetic in the spin channel).
The momentum {\bf Q} corresponds to $\pi$, ($\pi$,$\pi$), and
($\pi$,$\pi$,$\pi$), in 1D, 2D, and 3D, respectively. 
It denotes a $staggered$ orbital order, namely a given combination of
the a and b original orbitals is the one mainly populated in the even
sites of the cluster, while in the odd sites another orbital
combination is preferred.
These populated orbitals are not necessarily only the two
initial ones used in the definition of the Hamiltonian, in the same
way that in a Heisenberg spin system not only spins up and down in the
$z$-direction (the usual basis) are possible in mean value.
Actually, spins can order with a mean-value pointing in any direction 
depending on the model and couplings, and the same occurs with the
orbitals, which in this respect are like ``pseudospins''. A particular
combination of the original orbitals 1 and 2 could be energetically
the best in even sites and some other combination in the odd sites.

In Fig.III.e.1, as $\lambda$ increases further, a second transition was 
identified this time into a state which is staggered in the spin, and
uniform in the orbitals.
Such a state is the one in correspondence with the AF state found in
the one-orbital model, namely only one orbital matters at low energies
and the spin, as a consequence, is antiferromagnetic.
However, from the experimental point of view, the intermediate regime
between 1.0 and 2.0 is the most relevant, since staggered orbital
order is known to occur in experiments.
Then, the one orbital model envisioned to work for manganites, at least
close to \densi=1.0 due to the static Jahn-Teller distortion, actually
does not work even there since it misses the staggered orbital
order but instead assumes a uniform order.
Nevertheless the model is qualitatively interesting, as remarked
upon before. 

Why does orbital order occur here? This can be easily understood
perturbatively in the hopping $t$, following Fig.III.e.2 where a
single Mn-Mn link is used and the four possibilities (spin FM or AF, orbital
uniform or staggered) are considered.
A hopping matrix only connecting the same orbitals,
with hopping parameter $t$, is assumed for simplicity. 
The energy difference between $e_{\rm g}$ orbitals at a given site is 
$E_{\rm JT}$, which is a monotonous function of $\lambda$.
For simplicity, in the notation let us refer to orbital uniform
(staggered) as orbital ``FM'' (``AF'').
Case (a) in Fig.III.e.2 corresponds to spin FM and orbital AF: In this
case when an electron moves from orbital a on the left to the same
orbital on the right, which is the only possible hopping by assumption, 
an energy of order $E_{\rm JT}$ is lost, but kinetic energy is gained.
As in any second order perturbative calculation the energy gain is then
proportional to $t^2/E_{\rm JT}$.
In case (b), both spin and orbital FM, the electrons do not move and
the energy gain is zero (again, the nondiagonal hoppings are assumed
negligible just for simplicity).
In case (c), the spin are AF but the orbitals are FM.
This is like a one orbital model and the gain in energy is
proportional to $t^2/(2J_{\rm H})$.
Finally, in case (d) with AF in spin and orbital, both Hund and
orbital splitting energies are lost in the intermediate state, and the
overall gain becomes proportional to $t^2/(2J_{\rm H} + E_{\rm JT})$.
As a consequence, if the Hund coupling is larger than $E_{\rm JT}$,
then case (a) is the best, as it occurs at intermediate $E_{\rm JT}$ 
values in Fig.III.e.1.
However, in the opposite case (orbital splitting larger than Hund
coupling) case (c) has the lowest energy, a result also compatible to
that found in Fig.III.e.1. 
Then, the presence of orbital order can be easily understood from a
perturbative estimation, quite similarly as done by Kugel and Khomskii (1974)
in their pioneering work on orbital order. Recently, x-ray resonant
scattering studies have confirmed the orbital order in manganites
(Murakami et al., 1998a; Murakami et al., 1998b).

\medskip
\noindent{\bf A-type AF at x=0.0:}

The alert reader may have noticed that the state reported in the
previous analysis at intermediate $\lambda$'s is actually not quite the same 
state as found in experiments.
It is known that the actual state has A-type AF spin order, while in the 
analysis of Fig.III.e.1, such a state was not included.
The intermediate $\lambda$ region has FM spin in the three directions
in the 3D simulations of that figure.
Something else must be done in order to arrive at an A-type
antiferromagnet.
Recent investigations by Hotta et al. (1999) have shown that, in the 
context of the model with Jahn-Teller phonons, this missing ingredient
is $J_{\rm AF}$ itself, namely by increasing this coupling from 0.05
to larger values, a transition from a FM to an A-type AF exists (the
relevance of JT couplings at \densi=1.0 has also been remarked by 
Capone, Feinberg, and Grilli, 2000).
This can be visualized easily in Fig.III.e.3
where the energy vs. $J_{\rm AF}$ at fixed intermediate $\lambda$ and
$J_{\rm H}$ is shown.
Four regimes were identified: FM, A-AF, C-AF, and G-AF, states that 
are sketched also in that figure. The reason is simple: as $J_{\rm AF}$
grows, the tendency toward spin AF must grow since this coupling favors
such an order.
If $J_{\rm AF}$ is very large, then it is clear that a G-AF state must
be the one that lowers the energy, in agreement with the Monte Carlo 
simulations. If $J_{\rm AF}$ is small or zero, there is no reason why
spin AF will be favorable at intermediate $\lambda$ and the density
under consideration, and then the
state is ferromagnetic to improve the electronic mobility.
It should be no surprise that at intermediate $J_{\rm AF}$, the
dominant state is intermediate between the two extremes, with A-type
and C-type antiferromagnetism becoming stable in intermediate regions
of parameter space.

It is interesting to note that similar results regarding the relevance 
of $J_{\rm AF}$ to stabilize the A-type order have been found by
Koshibae et al. (1997) in a model with 
Coulomb interactions.
An analogous conclusion was found by 
Solovyev, Hamada, and Terakura (1996) and Ishihara et al. (1997).
Betouras and Fujimoto (1999), using bosonization techniques for the 
1D one-orbital model, also emphasized the importance of $J_{\rm AF}$,
similarly as did Yi, Yu, and Lee (1999) based on Monte Carlo studies
in two dimensions of the same model.
The overall conclusion is that there are clear analogies between
the strong Coulomb and strong Jahn-Teller coupling approaches, as
discussed elsewhere in this review.
Actually in the mean-field approximation presented in Section III.c it 
was shown that the influence of the Coulombic terms can be hidden in 
simple redefinitions of the electron-phonon couplings
(see also Benedetti and Zeyher, 1999).
In our opinion, both approaches (JT and Coulomb) have strong
similarities and it is not surprising that basically the same physics
is obtained in both cases.
Actually, Fig.2 of Maezono, Ishihara, and Nagaosa (1998b) showing the
energy vs. $J_{\rm AF}$ in mean-field calculations of the Coulombic
Hamiltonian without phonons is very similar to our Fig.III.e.3, aside
from overall scales.
On the other hand, 
Mizokawa and Fujimori (1995, 1996) states that the A-type AF 
is stabilized only when the Jahn-Teller distortion is included,
namely, the FM phase is stabilized in the purely Coulomb model,
based on the unrestricted Hartree-Fock calculation
for the $d$-$p$ model.

The issue of what kind of orbital order is concomitant with A-type
AF order is an important matter. This has been discussed at length by 
Hotta et al. (1999), and the final conclusion, after the introduction
of perturbations caused by the experimentally known difference in
lattice spacings between the three axes, is that the order shown in
Fig.III.e.3c minimizes the energy. 
This state has indeed been identified in recent x-ray experiments, and
it is quite remarkable that such a complex pattern of spin and orbital
degrees of freedom indeed emerges from mean-field and computational
studies. Studies by van den Brink et al. (1998) using purely Coulombic
models arrived at similar conclusions. 

\medskip
\noindent{\bf Electronic Phase Separation with Two Orbitals:}

Now let us analyze the phase diagram at densities away from 
\densi=1.0. In the case of the one-orbital model, phase separation
was very prominent in this regime. A similar situation was observed
with two orbitals, as Fig.III.e.4a-b illustrates where \densi~vs.
$\mu$ is shown at intermediate $\lambda$, $J_{\rm H}$=$\infty$, and
low temperature.
A clear discontinuity is observed, both near \densi=1.0, as well as
at low density. 
Measurements of spin and orbital correlations, as well as the Drude
weight to distinguish between metallic and insulating behavior, have
suggested the phase diagram in one dimension reproduced in
Fig.III.e.4c.
There are several phases in competition.
At \densi=1.0 the results were already described in the previous
subsection. Away from the \densi=1.0 phases, only the spin-FM
orbital-disordered survives at finite hole density, as expected due to
the mapping at small $\lambda$ into the one-orbital model with half
the density.
The other phases at $\lambda$$\geq$1.0 are not stable, but 
electronic phase
separation takes place.
The \densi$<$1.0 extreme of the PS discontinuity is given by a
spin-FM orbital-FM metallic state, which is a 1D precursor of the
metallic orbitally-ordered A-type state identified in some compounds 
precisely at densities close to 0.5.
Then, the two states that compete in the \densi$\sim$1.0 PS regime
differ in their orbital arrangement, but not in the spin sector.
This is PS triggered by the $orbital$ degrees of freedom, which is a
novel concept.
On the other hand, the PS observed at low density is very similar to 
that observed in the one-orbital model involving spin FM and AF states 
in competition.
Finally, at \densi$\sim$0.5 and large $\lambda$, charge ordering
takes place, but this phase will be discussed in more detail
later. Overall, it is quite reassuring to observe that the stable
phases in Fig.III.e.4c all have an analog in experiments.
This gives support to the models used and to the computational and
mean-field techniques employed.

In addition, since all stable regions are realistic, it is natural to
assume that the rest of the phase diagram, namely the PS regions, must 
also have an analog in experiments in the form of mixed-phase
tendencies and nanometer-size cluster formation, as discussed in the
case of the one-orbital model.
PS is very prominent in all the models studied, as long as proper
many-body techniques are employed. 
For instance, using accurate mean-field approximations, the PS
tendencies in 1D can also be properly reproduced (see section V of 
Hotta, Malvezzi and Dagotto, 2000).
It is also important that even in purely Coulombic cases (without JT
phonons), PS has been found in 1D models in some regions of parameter
space (see Hotta, Malvezzi and Dagotto, 2000), and, thus, this
phenomenon is not restricted to Jahn-Teller systems.
Kagan, Khomskii, and Kugel (2000) also reported phase separation near
x=0.5 without using JT phonons.
Guerrero and Noack (2000) reported phase separation in a
one-dimensional copper-oxide model with only Coulomb interactions.
Varelogiannis (2000) found coexistence and competition of CO, AF and
FM phases in a multicomponent mean-field-theory, without using a
particular microscopic mechanism.

It is important to remark that plenty of work still remains to be done 
in establishing the phase diagram of the two-orbitals model. Studies
in 2D carried out by our group suggest that the phase diagram is
similar to that found in 1D, but details remain to be settled. The 3D
diagram is known only in special cases.
Although the experience gained in the one-orbital model suggests that
all dimensions have similar phase diagrams, this issue remains to be
confirmed in the two-orbital case at large $\lambda$. In addition,
also note that the intermediate $\rm J_H$ regime has not been explored
and surprises may be found there, such as the stripes described for
the one-orbital case at intermediate Hund coupling. Work is in progress
in this challenging area of research.

\medskip
\noindent{\bf Charge Ordering at x=0.5 and the CE-state:}

The so-called CE-type AFM phase has been established as the ground
state of half-doped perovskite manganites in the 1950's. 
This phase is composed of zigzag FM arrays of $t_{\rm 2g}$-spins, 
which are coupled antiferromagnetically
perpendicular to the zigzag direction.
Furthermore, the checkerboard-type charge ordering in the $x$-$y$
plane, the charge stacking along the $z$-axis, and
($3x^2-r^2$/$3y^2-r^2$) orbital ordering are associated with this
phase.

Although there is little doubt that the famous CE-state of Goodenough, 
reviewed in Sec.III.a, is indeed the ground state of x=0.5
intermediate and low bandwidth manganites, only very recently such a
state has received theoretical confirmation using unbiased techniques, 
at least within some models.
In the early approach of Goodenough it was $assumed$ that the charge
was distributed in a checkerboard pattern, upon which spin and orbital
order was found. But it would be desirable to obtain the CE-state
based entirely upon a more fundamental theoretical analysis, as
the true state of minimum energy of a well-defined and realistic
Hamiltonian.
If such a calculation can be done, as a bonus one would find out which 
states compete with the CE-state in parameter space, an
issue very important in view of the mixed-phase tendencies of
Mn-oxides, which cannot be handled within the approach of Goodenough.

One may naively believe that it is as easy as introducing a huge
nearest-neighbor Coulomb repulsion $V$ to stabilize a charge-ordered
state at x=0.5, upon which the reasoning of Goodenough can be
applied. However, there are at least two problems with this approach.
First, such a large $V$ quite likely will destabilize the
ferromagnetic charge-disordered state and others supposed to be
competing with the CE-state. It may be possible to explain the
CE-state with this approach, but not others also observed at
x=0.5 in large bandwidth Mn-oxides.
Second, a large $V$ would produce a checkerboard pattern in the
$three$ directions.
However, experimentally it has been known for a long time (Wollan and
Koehler, 1955) that the charge $stacks$ along the $z$-axis, namely the
same checkerboard pattern is repeated along $z$, rather than being
shifted by one lattice spacing from plane to plane.
A dominant Coulomb interaction $V$ can not be the whole story
for x=0.5 low-bandwidth manganese oxides.

The nontrivial task of finding a CE-state with charge stacked along
the $z$-axis without the use of a huge nearest-neighbors repulsion
has been recently performed by Yunoki, Hotta, and Dagotto (2000) 
using the two-orbital model with strong electron Jahn-Teller phonon
coupling.
The calculation proceeded using an unbiased Monte Carlo simulation,
and as an output of the study, the CE-state indeed emerged as the
ground-state in some region of coupling space.
Typical results are shown in Fig.III.e.5.
In part (a) the energy at very low temperature is shown as a function
of $J_{\rm AF}$ at fixed density x=0.5, $J_{\rm H}$=$\infty$ for
simplicity, and with a robust electron-phonon coupling $\lambda$=1.5
using the two orbital model of Sec.III.c.
At small $J_{\rm AF}$, a ferromagnetic phase was found to be
stabilized, according to the Monte Carlo simulation.
Actually, at $J_{\rm AF}$=0.0 it has not been possible to stabilize a
partially AF-state at x=0.5, namely the states are always
ferromagnetic at least within the wide range of $\lambda$'s
investigated (but they can have charge and orbital order).
On the other hand, as $J_{\rm AF}$ grows, a tendency to form AF links
develops, as it happens at x=0.0. 
At large $J_{\rm AF}$ eventually the system transitions to
states that are mostly antiferromagnetic, such as the so-called
``AF(2)'' state of Fig.III.e.5b (with an up-up-down-down spin pattern
repeated along one axis, and AF coupling along the other axis), or
directly a fully AF-state in both directions. 

However, the intermediate values of $J_{\rm AF}$ are the most
interesting ones. In this case the energy of the 2D clusters 
become flat as a function of $J_{\rm AF}$ 
suggesting that the state has the same
number of FM and AF links, a property that the CE-state indeed has.
By measuring charge-correlations it was found that a checkerboard
pattern is formed particularly at intermediate and large $\lambda$'s,
as in the CE-state.
Finally, after measuring the spin and orbital correlations, it was
confirmed that indeed the complex pattern of the CE-state was fully
stabilized in the simulation. This occurs in a robust portion of the
$\lambda$-$J_{\rm AF}$ plane, as shown in Fig.III.e.5b.
The use of $J_{\rm AF}$ as the natural parameter to vary in order
to understand 
the CE-state is justified based on Fig.III.e.5b since the region of
stability of the CE-phase is elongated along the $\lambda$-axis,
meaning that its existence is not so much dependent on that coupling
but much more on $J_{\rm AF}$ itself.
It appears that some explicit tendency in the Hamiltonian toward the
formation of AF links is necessary to form the CE-state.
If this tendency is absent, a FM state if formed, while if it is too 
strong an AF-state appears. The x=0.5 CE-state, similar to the 
A-type AF at x=0.0, needs an intermediate value of $J_{\rm AF}$ for
stabilization. The stability window is finite and in this respect there
is no need to carry out a $fine$ tuning of parameters to find the CE
phase. However, it is clear that there is a balance of AF and FM
tendencies in the CE-phase that makes the state somewhat fragile.

Note that the transitions among the many states obtained when varying
$J_{\rm AF}$ are all of $first$ order, namely they correspond to
crossings of levels at zero temperature.
The first-order character of these transitions is a crucial ingredient
of the recent scenario proposed by Moreo et al. (2000) involving
mixed-phase tendencies with coexisting clusters with $equal$ density,
to be described in more detail below.
Recently, first-order transitions have also been reported in the
one-orbital model at x=0.5 by Alonso et al. (2000a, 2000b), as well as
tendencies toward phase separation. Recent progress in the development
of powerful techniques for manganite models (Alonso et al., 2000c;
Motome and Furukawa, 2000a, 2000b) will
contribute to the clarification of these issues in the near future.

\medskip
\noindent{\bf Charge Stacking:}

Let us address now the issue of charge-stacking along the $z$-axis.
For this purpose simulations using 3D clusters were carried out.
The result for the energy vs. $J_{\rm AF}$ is shown in Fig.III.e.6,
with $J_{\rm H}$=$\infty$ and $\lambda$=1.5 fixed.
The CE-state with charge-stacking was found to be the ground state 
on a wide $J_{\rm AF}$ window.
The reason that this state has lower energy than the so-called
``Wigner-crystal'' (WC) version of the CE-state, namely with the
charge spread as much as possible, is once again the influence of 
$J_{\rm AF}$. With a charge stacked arrangement, the links along the
$z$-axis can all be simultaneously antiferromagnetic, thereby minimizing the
energy. In the WC-state this is not possible.

It should be noted that this charge stacked CE-state is not
immediately destroyed when the weak nearest-neighbor repulsion $V$
is introduced to the model, as shown in the mean-field calculations by
Hotta, Malvezzi, and Dagotto (2000).
If $V$ is further increased for a realistic value of $J_{\rm AF}$,
the ground state eventually changes from the charge stacked
CE-phase to the WC version of the CE-state or the C-type AFM phase 
with WC charge ordering.
As explained above, the stability of the charge stacked phase to 
the WC version of the CE-state is due to the magnetic energy difference.
However, the competition between the charge-stacked CE-state
and the C-type AFM phase with the WC structure is not simply understood 
by the effect of $J_{\rm AF}$, since those two kinds of AFM phases
have the same magnetic energy.
In this case, the stabilization of the charge stacking originates
from the difference in the geometry of the 1D FM-path, namely a
zigzag-path for the CE-phase and a straight-line path for the C-type AFM
state.
As will be discussed later in detail, the energy for $e_{\rm g}$
electrons in the zigzag path is lower than that in the straight-line
path, and this energy difference causes the stabilization of the
charge stacking.
In short, the stability of the
charge-stacked structure at the expense of $V$ is supported
by ``the geometric energy'' as well as the magnetic energy.
Note that each energy gain is just a fraction of $t$.
Thus, in the absence of other mechanisms to understand the
charge-stacking, another consequence of this analysis is that $V$
actually must be substantially $smaller$ than naively expected,
otherwise such a charge pattern would not be stable.
In fact, estimations given by Yunoki, Hotta, and Dagotto (2000)
suggest that the manganites must have a large dielectric function at
short distances (see Arima and Tokura, 1995) to prevent the melting of 
the charge-stacked state.

Note also that the mean-field approximations by Hotta, Malvezzi, and
Dagotto (2000) have shown that on-site Coulomb interactions $U$ and
$U'$ can $also$ generate a 2D CE-state, in agreement with the
calculations by van den Brink et al. (1999).
Then, the present authors believe that strong JT and Coulomb couplings 
tend to give similar results.
This belief finds partial confirmation in the mean-field
approximations of Hotta, Malvezzi, and Dagotto (2000), where the
similarities between a strong $\lambda$ and $(U,U')$ were
investigated.
Even doing the calculation with Coulombic interactions, the influence 
of $J_{\rm AF}$ is still crucial to inducing charge-stacking (note that 
the importance of this parameter has also been recently remarked by
Mathieu, Svedlindh and Nordblad, 2000, based on experimental results).

Many other authors carried out important work in the context of the
CE-state at x=0.5.
For example, with the help of Hartree-Fock calculations, Mizokawa and
Fujimori (1997) reported the stabilization of the CE-state at x=0.5
only if Jahn-Teller distortions were incorporated into a model with
Coulomb interactions.
This state was found to be in competition with a uniform FM state, as
well as with an A-type AF-state with uniform orbital order.
In this respect the results are very similar to those found by Yunoki,
Hotta and Dagotto (2000) using Monte Carlo simulations.
In addition, using a large nearest neighbor repulsion and the
one-orbital model, charge ordering and a spin structure compatible
with the zigzag chains of the CE state was found by Lee and Min (1997)
at x=0.5.
Also Jackeli et al. (1999) obtained charge-ordering at x=0.5 using
mean-field approximations and a large $V$.
Charge-stacking was not investigated by those authors.
The CE-state in x=0.5 $\PCMO$ was also obtained
by Anisimov et al. (1997) using LSDA+U techniques.

\medskip
\noindent{\bf Topological Origin of the x=0.5 CE-state}

The fact that $\lambda$ does not play the most crucial role for the
CE-state also emerges from the ``topological'' arguments 
of Hotta et al. (2000), where at least the formation of zigzag
ferromagnetic chains with antiferromagnetic interchain coupling, 
emerges directly for $\lambda$=0 and large $J_{\rm H}$, as a 
consequence of the  ``band-insulator''character of those chains. 
Similar conclusions as those reached by Hotta et al. (2000),
were independently obtained by Solovyev and Terakura (1999),
Solovyev (2000), and by van den Brink et al. (1999).
The concept of a band insulator in this context was first described in
Hotta et al. (1998).

To understand the essential physics present in half-doped manganites,
it is convenient to consider the complicated CE-structure in
Hamiltonians simpler than those analyzed in Section III.c.
Based on the concept of ``adiabatic continuation'' for the
introduction of the JT distortion and/or the Coulombic interactions,
the following approximations will be made:
(i) $H$=$H^{\infty}$($E_{\rm JT}$=$U'$=$V$=0): In the first place,
this simple Hamiltonian is considered based on the DE mechanism, but
even in this situation, qualitative concepts can be learned for the 
stabilization of the zigzag AFM phase.
(ii) $H$=$H^{\infty}$($E_{\rm JT}$$\ne$0,$U'$=$V$=0): 
In order to consider the charge and orbital ordering,
the non-cooperative JT phonons are included in the two-orbital
DE model by using the analytic MFA.
In particular, the charge-stacked structure is correctly reproduced,
and its origin is clarified based on a ``topological'' framework.
(iii) $H$=$H^{\infty}$($E_{\rm JT}$ $\ne$0,$U'$$\ne$0,$V$$\ne$0).
Here the effect of the long-range Coulomb interaction for the
charge-stacked phase is discussed within the MFA.
(iv) $H$=$H_{\rm JT}$ both for JT and non-JT phonons.
Finally, to complete the above discussions, 
unbiased calculations for the JT model are performed 
using Monte Carlo simulations and the relaxation method.
In this subsection, a peculiar ``band-insulating'' state of the
CE-type is discussed in detail by focusing on the effect of the local
phase $\xi_{\bf i}$ for determining the orbitals.

As is well known, the CE-type antiferromagnetic 
phase is composed of a bundle of spin FM chains, each
with the zigzag geometry, and with antiferromagnetic interchain coupling.
Although the reason for the stabilization of this special zigzag
structure should be clarified further, for the time being
let us discuss what happens if this zigzag
geometry is assumed, and how it compares with a straight line.
To simplify the discussion, the limiting case of $J_{\rm H}$=$\infty$
is considered.
Namely, the $e_{\rm g}$-electrons can move only along the zigzag FM
path, since the hopping perpendicular to the zigzag direction vanishes
due to the standard DE mechanism and the antiferromagnetism between
chains, indicating that the spin degree of freedom can be effectively
neglected. 
Thus, the problem is reduced to the analysis of the $e_{\rm g}$
electron motion along the one-dimensional zigzag chain. 
However, it should be emphasized that this is still a highly
non-trivial system.

To solve the present one-body problem, a unit-cell is defined as shown 
in Fig.III.e.7, in which the hopping amplitudes change with a period of
four lattice spacings,
since the hopping direction changes as 
$\{ \cdots, x, x, y, y, \cdots \}$ along the zigzag chain,
with $t_{\mu \nu}^{\bf x}$=$-t_{\mu \nu}^{\bf y}$ for $\mu \ne \nu$
according to the values of the hopping amplitudes discussed before.
This difference in sign, i.e., the phase change, is essential for this
problem.
To make this point clear, it is useful to transform the
spinless $e_{\rm g}$-electron operators by using a unitary matrix as
(see Koizumi et al., 1998a),
\begin{equation}
  \label{eq:trans2}
  \left(
    \begin{array}{l}
      \alpha_{\bf i} \\
      \beta_{\bf i}
    \end{array}
  \right)
  = {1 \over \sqrt{2}} \left(
   \begin{array}{cc}
   1 &  i \\
   1 & -i 
    \end{array}
  \right)
  \left(
    \begin{array}{l}
      c_{{\bf i}{\rm a}} \\
      c_{{\bf i}{\rm b}}
    \end{array}
  \right).
\end{equation}
After simple algebra, $H_{\rm kin}$ is rewritten as
\begin{eqnarray}
  H_{\rm kin} &=& -t_0/2
  \sum_{\bf i,a} (\alpha_{\bf i}^{\dag}\alpha_{\bf i+a} 
  + \beta_{\bf i}^{\dag} \beta_{\bf i+a} \nonumber \\
  &+& e^{i \phi_{\bf a}}\alpha_{\bf i}^{\dag} \beta_{\bf i+a}
 +e^{-i \phi_{\bf a}}\beta_{\bf i}^{\dag} \alpha_{\bf i+a}),
\end{eqnarray}
where the phase $\phi_{\bf a}$ depends only on the hopping direction,
and it is given by
$\phi_{\bf x}$=$-\phi$, 
$\phi_{\bf y}$=$\phi$,
and $\phi_{\bf z}$=0, with $\phi$=$\pi/3$.
Note that the $e_{\rm g}$-electron picks up a phase change
when it moves between different neighboring orbitals.
In this expression, the effect of the change of the local phase
is correctly included in the Hamiltonian.

To introduce the momentum $k$ along the zigzag chain,
the Bloch's phase $e^{ \pm i k}$ is added to the hopping
term between adjacent sites.
Then, the problem is reduced to finding the eigenvalues
of an $8 \times 8$ matrix, given by
\begin{equation}
  {\hat h} \left(
    \begin{array}{c}
        \psi_{\alpha 1} \\ 
        \psi_{\beta 1} \\ 
        \psi_{\alpha 2} \\ 
        \psi_{\beta 2} \\
        \psi_{\alpha 3} \\
        \psi_{\beta 3} \\ 
        \psi_{\alpha 4} \\
        \psi_{\beta 4}
    \end{array}
  \right)
= \varepsilon_{k} 
\left(
   \begin{array}{c}
        \psi_{\alpha 1} \\ 
        \psi_{\beta 1} \\ 
        \psi_{\alpha 2} \\ 
        \psi_{\beta 2} \\
        \psi_{\alpha 3} \\
        \psi_{\beta 3} \\ 
        \psi_{\alpha 4} \\
        \psi_{\beta 4}
   \end{array}
\right),
\end{equation}
where $\psi_{\alpha j}$ and $\psi_{\beta j}$ are the basis function
for $\alpha$- and $\beta$-electrons at the $j$-site of the unit cell, 
respectively, and 
the Hamiltonian matrix ${\hat h}$ is given by
\begin{equation}
{\hat h}=
  -{t_0 \over 2} \left(
    \begin{array}{cccc}
        {\hat O} & {\hat T}_{k}^{\bf x} 
        & {\hat O} & {\hat T}_{k}^{{\bf y}*} \\
        {\hat T}_{k}^{{\bf x}*} & {\hat O} 
        & {\hat T}_{k}^{\bf x} & {\hat O} \\
        {\hat O} & {\hat T}_{k}^{{\bf x}*} 
        & {\hat O} & {\hat T}_{k}^{\bf y} \\
        {\hat T}_{k}^{\bf y} & {\hat O} & 
        {\hat T}_{k}^{{\bf y}*} & {\hat O} \\
  \end{array}
  \right).
\end{equation}
Here ${\hat O}$ is the $2 \times 2$ matrix in which  all
components are zeros, and the hopping matrix 
${\hat T}_{k}^{\bf a}$ along the ${\bf a}$-direction is
defined by
\begin{equation}
  {\hat T}_{k}^{\bf a}
  = e^{ik} \left(
    \begin{array}{cc}
        1 & e^{i\phi_{\bf a}} \\
        e^{-i\phi_{\bf a}} & 1
    \end{array}
\right),
\end{equation}
where note again that $-\phi_{\bf x}$=$\phi_{\bf y}$=$\phi$=$\pi/3$.

Although it is very easy to solve the present eigenvalue problem 
by using the computer, it is  instructive to find the solution
analytically. 
In the process of finding this solution, several important points will
be clarified.
First, note that there are two eigenfunctions of the 8$\times$8 matrix 
which have a ``localized'' character,
satisfying
\begin{equation}
   {\hat h}(\psi_{\alpha 2} - e^{-i\phi} \psi_{ \beta 2})=0,
\end{equation}
and
\begin{equation}
   {\hat h}(\psi_{\alpha 4} - e^{+i\phi} \psi_{ \beta 4})=0.
\end{equation}
As easily checked by simple algebra, those localized basis functions 
correspond to $y^2-z^2$ and $z^2-x^2$ orbitals 
at sites 2 and 4, respectively.
By orthogonality, the active orbitals are then fixed as $3x^2-r^2$
and $3y^2-r^2$ at sites 2 and 4, respectively.
This fact suggests that if some potential acts over the
 $e_{\rm g}$-electrons, the 
$(3x^2-r^2/3y^2-r^2)$-type orbital ordering immediately occurs
in such a one-dimensional zigzag path due to the standard Peierls
instability. 
This point will be discussed again later in the context of 
charge-orbital ordering due to the JT distortion.

To find the other extended eigenstates, it is quite natural to
consider active basis functions at sites 2 and 4, given by
$(\psi_{\alpha 2}+e^{-i \phi}\psi_{\beta 2})/\sqrt{2}$
and
$(\psi_{\alpha 4}+e^{+i \phi}\psi_{\beta 4})/\sqrt{2}$,
respectively.
Then, the bonding and antibonding 
combinations of those basis are constructed by including appropriate
phases such as
\begin{eqnarray}
  \Phi^{\pm}_1 &=& (1/2)[ 
  (e^{-i\phi/2}\psi_{\alpha 4}+e^{i \phi/2}\psi_{\beta 4})
  \nonumber \\ &\pm&
  (e^{i \phi/2}\psi_{\alpha 2}+e^{-i \phi/2}\psi_{\beta 2})].
\end{eqnarray}
By acting with ${\hat h}$ over $\Phi_1^{\pm}$, it is found that
two kinds of $3 \times 3$ block Hamiltonians ${\tilde h}^{\pm}$
can be constructed using new basis functions
defined as
\begin{equation}
  \Phi^{\pm}_2=
  (\psi_{\alpha 1} \pm \psi_{\beta 3})/\sqrt{2},
\end{equation}
and
\begin{equation}
  \Phi^{\pm}_3=
  (\psi_{\beta 1} \pm \psi_{\alpha 3})/\sqrt{2}.
\end{equation}
As expected, the block Hamiltonian including the ground state 
is constructed using the bonding-type basis functions 
\begin{equation}
  {\tilde h}^{+}
  \left(
    \begin{array}{c}
      \Phi_1^{+} \\
      \Phi_2^{+} \\
      \Phi_3^{+}
    \end{array}
  \right) = \varepsilon_{k}^{+}
  \left(
    \begin{array}{c}
      \Phi_1^{+} \\
      \Phi_2^{+} \\
      \Phi_3^{+}
    \end{array}
  \right),
\end{equation}
where the $3 \times 3$ matrix ${\tilde h}^{+}$ is given by
\begin{equation}
  {\tilde h}^{+} \!=\! - \sqrt{2} t_0 
  \left(
    \begin{array}{ccc}
      0 &  \cos k_{-} & \cos k_{+}\\
      \cos k_{-} & 0 & 0 \\
      \cos k_{+} & 0 & 0 \\
    \end{array}
  \right),
\end{equation}
with $k_{\pm}=k \pm \phi/2$.
By solving this eigenvalue problem, 
it can be easily shown that the eigenenergies are
\begin{equation}
  \varepsilon_k^{+} = 0, ~\pm t_0 \sqrt{2 + \cos (2k)}.
\end{equation}
Note here that the momentum $k$ is restricted to the reduced zone,
$-\pi/4 \le k \le \pi/4$.
As expected, the lowest-energy band has a minimum
at $k$=0, indicating that this block correctly includes the ground state
of the one $e_{\rm g}$-electron problem.
At a first glance, this point appears obvious, but if the effect of the
local phase is not treated correctly, an unphysical solution
easily appears, as will be discussed below.

The block Hamiltonian for the antibonding sector
is given by
\begin{equation}
  {\tilde h}^{-}
  \left(
    \begin{array}{c}
      \Phi_1^{-} \\
      \Phi_2^{-} \\
      \Phi_3^{-}
    \end{array}
  \right) = \varepsilon_{k}^{-}
  \left(
    \begin{array}{c}
      \Phi_1^{-} \\
      \Phi_2^{-} \\
      \Phi_3^{-}
    \end{array}
  \right),
\end{equation}
with
\begin{equation}
  {\tilde h}^{-} \!=\! - \sqrt{2} t_0 
  \left(
    \begin{array}{ccc}
      0 &  i\sin k_{-} & i\sin k_{+}  \\
      -i\sin k_{-} & 0 & 0 \\
      -i\sin k_{+} & 0 & 0 \\
    \end{array}
  \right).
\end{equation}
The eigenenergies are given by
\begin{equation}
  \varepsilon_k^{-} = 0, ~ \pm t_0 \sqrt{2 - \cos (2k)}.
\end{equation}
In summary, eight eigenenergies have been obtained as
\begin{equation}
  \varepsilon_k = 0, ~\pm t_0\sqrt{2+\cos(2k)}, ~
  \pm t_0\sqrt{2-\cos(2k)},
\end{equation}
where the flat band $\varepsilon_{\bf k}$=0 has four-fold degeneracy.
The band structure is shown in Fig.III.e.8.
The most remarkable feature is that the system is {\it band-insulating},
with a bandgap of value $t_0$ for quarter filling, i.e., x=0.5.
This band insulating state, without any explicit potential among the
electrons moving along the zigzag chains,
is caused by the phase difference between
$t^{\bf x}_{\mu \nu}$ and $t^{\bf y}_{\mu \nu}$.
Intuitively, such band-insulator originates in the presence of
a standing-wave state due to the interference between
two traveling waves running along the $x$- and $y$-directions.
In this interference picture, the nodes of the wavefunction can exist
on ``the corner" of the zigzag structure, and the probability
amplitude becomes larger in the ``straight" segment of the path.
Thus, even a weak potential can produce the charge and orbital
ordering based on this band-insulating phase.
Since $t_0$ is at least of the order of 1000K, this band-insulating
state is considered to be very robust.
In fact, if some potential is included into such an insulating phase,
the system maintains its ``insulating'' properties, and a modulation
in the orbital density appears.

The problem in the zigzag one-dimensional chain provided us with a
typical example to better understand the importance of the additional
factor $e^{i\xi_{\bf i}/2}$ in front of the $2 \times 2$ SU(2) unitary
matrix to generate the phase dressed operator at each site.
As clearly shown above, the ``a'' and ``b'' orbitals should be chosen
as ``a''=$y^2-z^2$ and ``b''=$3x^2-r^2$ at site 2, and ``a''=$z^2-x^2$
and ``b''=$3y^2-r^2$ at site 4, respectively. 
Namely, $\xi_2$=$2\pi/3$ and $\xi_4$=$4\pi/3$. 
The reason for these choices of $\xi_{\bf i}$ is easily understood
due to the fact that the orbital tends to polarize along the hopping
direction to maximize the overlap. 
Thus, to make the Hamiltonian simple, it is useful to fix the orbitals
at sites 2 and 4 as $\xi_2$=$2\pi/3$ and $\xi_4$=$4\pi/3$.
Here, the phase factor $e^{i\xi_{\bf i}/2}$ in the basis function is
essential to reproduce exactly the same solution as obtained in the
discussion above. 
As already mentioned, in a single-site problem, this phase factor can
be neglected, since it provides only an additional phase to the whole
wave function.
However, if the $e_{\rm g}$-electron starts moving from site to site, 
the accumulation of the phase difference between adjacent sites
does not lead just to an additional phase factor to the whole wave
function. 
In fact, if this additional phase is accidentally neglected,
the band structure will shift in momentum space as
$k$$\rightarrow$$k+\pi$,
indicating that the minimum of the lowest-energy band is not located 
at $k$=0, but at $k$=$\pi$, as already pointed out by 
Koizumi et al., 1998b. 
Of course, this can be removed by the redefinition of $k$ by including 
``the crystal momentum'', but it is not necessary to redefine $k$,
if the local phase factors are correctly included in the problem.

Now let us discuss the stabilization of the zigzag structure in the
CE-type phase. 
Although it is true that the zigzag one-dimensional FM chain has a
large band-gap, this fact does not guarantee that this band-insulating
phase is the lowest-energy state.
To prove that the CE-type AFM phase composed of these zigzag FM chains 
is truly the ground-state, at least the following three points should
be clarified:
(i) Does this zigzag structure have the lowest energy compared to other
zigzag paths with the same periodicity and compared with the straight
one-dimensional path? 
(ii) Does the periodicity with four lattice spacings produce the
$global$ ground-state? In other words, can zigzag
structures with another periodicity be the global ground-state?
(iii) Is the energy of the zigzag AFM phase lower than that of the FM
or other AFM phases ? 
All these points have been clarified in Hotta et al. (2000), and here
the essential points are discussed briefly.

The first point can be checked by directly comparing the energies for
all possible zigzag structures with the periodicity of four lattice
spacings. Due to translational invariance, there exist four types
of zigzag states which are classified by the sequence of the hopping
directions: $\{x,x,x,x\}$, $\{x,y,x,y\}$, $\{x,x,x,y\}$, and $\{x,x,y,y\}$.
For quarter-filling, by an explicit calculation it has been shown that
the zigzag pattern denoted by $\{x,x,y,y\}$ has the lowest energy among
them (see details in Hotta et al. (2000)), but here an intuitive
explanation is provided. 
The state characterized by $\{x,x,x,x\}$ is, of course, the
one-dimensional metal with a dispersion relation simply given by
$-2t_{0} \cos k$.
Note that in this straight FM chain, the active orbital at every site 
is $3x^2-r^2$, since the hopping direction is restricted to be along
the $x$-direction.
As emphasized in the above discussion, a periodic change in the
hopping amplitude produces a band-gap, indicating that the metallic
state has an energy higher than the band-insulating states.
Among them, the state with the largest bandgap will be the
ground-state.
After several calculations at quarter-filling, the zigzag state with 
$\{x,x,y,y\}$ was found to have the lowest energy, but without any calculation
this result can be deduced based on the interference effect.
As argued above, due to the interference of two traveling waves along
the $x$- and $y$-directions, a standing-wave state occurs with
nodes on the corner sites, and a large probability amplitude at the
sites included in the straight segments.
To contain two $e_{\rm g}$-electrons in the unit cell with four
lattice spacings, at least two sites in the straight segment are
needed. Moreover, the nodes are distributed with equal spacing in the
wavefunction as long as no external potential is applied.
Thus, it is clear that the lowest-energy state corresponds to the
zigzag structure with $\{x,x,y,y\}$.

As for the second point regarding the periodicity, it is quite
difficult to carry out the direct comparison among the energies for
all possible states, since there are infinite possibilities for the
combinations of hopping directions.
Instead, to mimic the periodic change of the phase $\phi_{\bf a}$ in
the hopping process, let us imagine a virtual situation in which a JT
distortion occurs in the one-dimensional $e_{\rm g}$-electron system, 
by following Koizumi at al. (1998a). 
To focus on the effect of the local phase, it is assumed that the
amplitude of the JT distortion $q_i$ is independent of the site index,
i.e., $q_i=q$, and only the phase $\xi_i$ is changed periodically.
For simplicity, the phase is uniformly twisted with the period of $M$
lattice spacings, namely, $\xi_j$=$j$$\times$$(2\pi)/M$ for
1$\le$$j$$\le$$M$.
Since the periodic change of the hopping direction is mimicked by the
phase change of the JT distortion, $t_{\mu \nu}^{\bf a}$ is simply
taken as the unit matrix $t_0 \delta_{\mu \nu}$ to avoid the
double-counting of the effect of the phase change.
If the potential amplitude is written as $v$=$2qE_{\rm JT}$, the
Hamiltonian for the present situation is given by
\begin{eqnarray}
  && H \!=\! -t_0 \sum_{\langle i,j \rangle}
  (c_{i{\rm a}}^{\dag} c_{j{\rm a}}
  \!+\!c_{i{\rm b}}^{\dag} c_{j{\rm b}} + {\rm h.c.}) \nonumber \\
  &\!+\!& v \sum_i [\sin \xi_i (c_{i{\rm a}}^{\dag} c_{i{\rm b}}
  \!+\! c_{i{\rm b}}^{\dag} c_{i{\rm a}})
  \!+\! \cos \xi_i (c_{i{\rm a}}^{\dag} c_{i{\rm a}}
  \!-\! c_{i{\rm b}}^{\dag} c_{i{\rm b}})],
\end{eqnarray}
where the spinless $e_{\rm g}$-electron operator is used since the
one-dimensional FM chain is considered here, and the potential term
for the JT distortion is neglected since it provides only a constant
energy shift in this case.
By using the transformation Eq.~(\ref{eq:trans}), this Hamiltonian is
rewritten as 
\begin{eqnarray}
  H &=& -t_0 \sum_{\langle i,j \rangle} [ e^{i(\xi_i-\xi_j)/2}
  ({\tilde c}_{i{\rm a}}^{\dag} {\tilde c}_{j{\rm a}}+
  {\tilde c}_{i{\rm b}}^{\dag} {\tilde c}_{j{\rm b}})
  + {\rm h.c.}] \nonumber \\
  &+& v \sum_i ({\tilde c}_{i{\rm a}}^{\dag} {\tilde c}_{i{\rm a}}
  -{\tilde c}_{i{\rm b}}^{\dag} {\tilde c}_{i{\rm b}}).
\end{eqnarray}
The Hamiltonian in momentum space is obtained by the Fourier transform
as 
\begin{eqnarray}
  H &=& \sum_{k} \varepsilon_{k}
  [\cos (\pi/M) ({\tilde c}_{k{\rm a}}^{\dag} {\tilde c}_{k{\rm a}} 
  +{\tilde c}_{k{\rm b}}^{\dag} {\tilde c}_{k{\rm b}}) \nonumber \\
  && +i \sin (\pi/M)
  ({\tilde c}_{k{\rm a}}^{\dag} {\tilde c}_{k{\rm b}}-
  {\tilde c}_{k{\rm b}}^{\dag} {\tilde c}_{k{\rm a}})] \nonumber \\
  && + v \sum_{k} ({\tilde c}_{k{\rm a}}^{\dag} {\tilde c}_{k{\rm a}}
  -{\tilde c}_{k{\rm b}}^{\dag} {\tilde c}_{k{\rm b}}),
\end{eqnarray}
where $\varepsilon_k$=$-2t_0 \cos k$ and the periodic boundary
condition (PBC) is imposed. 
Note that in this expression, $k$ is the generalized quasimomentum,
redefined as $k-\pi/M \rightarrow k$, to incorporate the additional
phase $\pi/M$ which appears to arise from a fictitious magnetic field 
(See Koizumi et al., 1998b). 
The eigenenergies are easily obtained by diagonalization as
\begin{eqnarray}
  E_{k}^{\pm} &=& \varepsilon_k \cos (\pi/M)
  \pm \sqrt{v^2 + \varepsilon_k^2 \sin ^2 (\pi/M)} \nonumber \\
  &=& (1/2) [
  \varepsilon_{k+\pi/M}+\varepsilon_{k-\pi/M} \nonumber \\
  && \pm \sqrt{v^2 +(\varepsilon_{k+\pi/M}-\varepsilon_{k-\pi/M})^2}].
\end{eqnarray}
Since this is just the coupling of two bands,
$\varepsilon_{k+\pi/M}$ and $\varepsilon_{k-\pi/M}$,
it is easily understood that the energy gain due to the opening of the
bandgap is the best for the filling of $n$=$2/M$.
In other words, when the periodicity $M$ is equal to $2/n$, the energy
becomes the lowest among the states considered here with several
possible periods.
Although this is just a proof in an idealized special situation, it is
believed that it captures the essence of the problem.

Here the effect of the local phase factor $e^{i \xi_{\bf i}/2}$ should
be again noted.
If this factor is dropped, the phase $\pi/M$ due to the fictitious
magnetic field disappears and the eigenenergies are given by the
coupling of $\varepsilon_{k+\pi+\pi/M}$ and
$\varepsilon_{k+\pi-\pi/M}$,
which has been also checked by the computational calculation.
This ``$\pi$" shift in momentum space appears at the boundary, 
modifying the PBC to anti-periodic BC, even if there is no intention
to use APBC.
Of course, this is avoidable when the momentum $k$ is redefined as
$k+\pi \rightarrow k$, as pointed out in Koizumi et al. (1998b).
However, it is natural that the results for PBC are obtained in
the calculation using PBC.
Thus, also from this technical viewpoint, it is recommended that the
phase factor $e^{i \xi_{\bf i}/2}$ is added for the local rotation in
the orbital space.

To show the last item of the list needed to show the stability of the CE
state (see before), it is necessary to include the effect of the
magnetic coupling between adjacent $t_{\rm 2g}$-spins.
The appearance of the AFM phase with the zigzag geometry can be
understood by the competition between the kinetic energy of
$e_{\rm g}$ electrons and the magnetic energy gain of $t_{\rm 2g}$
spins based on the double-exchange mechanism.
Namely, if $J_{\rm AF}$ is very small, for instance equal to zero, 
the FM phase best optimizes the kinetic energy of the
$e_{\rm g}$-electrons.
On the other hand, when $J_{\rm AF}$ is as large as $t_0$, the system
stabilizes a G-type AFM phase to exploit the magnetic energy of the
$t_{\rm 2g}$-spins. 
For intermediate values of $J_{\rm AF}$, the AFM phase with a zigzag
structure can appear to take advantage at least partially of 
both interactions. Namely, along the FM zigzag chain with alignment of
$t_{\rm 2g}$-spins, the $e_{\rm g}$-electrons can move easily,
optimizing the kinetic energy, and at the same time there is a
magnetic energy gain due to the antiferromagnetic coupling between
adjacent zigzag chains. 
The ``window" in $J_{\rm AF}$ in which the zigzag AFM phase is
stabilized has been found to be around $J_{\rm AF}$$\approx$0.1$t_0$
in Monte Carlo simulations and the mean-field approximation, as
discussed elsewhere in this review.

In summary, at x=0.5, the CE-type AFM phase can be stabilized 
even without the Coulombic and/or the JT phononic interactions, only 
with large Hund and finite $J_{AF}$ couplings.
Of course, those interactions are needed to reproduce the charge and
orbital ordering, but as already mentioned in the above discussion,
because of the special geometry of the one-dimensional zigzag FM
chain, it is easy to imagine that the checkerboard type
charge-ordering and $(3x^2-r^2/3y^2-r^2)$ orbital-ordering pattern
will be stabilized.
Furthermore, the charge confinement in the straight segment (sites 2
and 4 in Fig.III.e.7), will naturally lead to charge stacking along the
$z$-axis, with stability caused by the special geometry of the zigzag
structure.
Thus the complex spin-charge-orbital structure for half-doped
manganites can be understood intuitively simply from the viewpoint of
its band-insulating nature. 

\medskip
\noindent{\bf Bi-stripe structure at x $>$ 0.5:}

In the previous subsection, the discussion focused on the CE-type AFM 
phase at x=0.5.
Naively, it may be expected that similar arguments can be extended to
the regime x$>$1/2, since in the phase diagram for $\LCMO$, the AFM
phase has been found at low temperatures in the region
0.50$<$x$\alt$0.88.
Then, let us try to consider the band-insulating phase for density
x=2/3 based on $H^{\infty}$ (as defined in Sec.III.c), 
without both the JT phononic and
Coulombic interactions, since this doping is quite important for the
appearance of the bi-stripe structure, as already discussed in
previous Sections (see Mori et al. (1998)).
Following the discussion on the periodicity of the optimal zigzag
path, at x=2/3 it is enough to consider the zigzag structure with
$M$=6.
After several calculations for x=2/3, as reported by Hotta et al.
(2000), the lowest-energy state was found to be characterized by the
straight path, not the zigzag one, leading to the C-type AFM phase
which was also discussed in previous Sections (for a visual
representation of the C-type state see Fig.4 of Kajimoto et al., 1999).
At first glance, the zigzag structure, for instance the
$\{x,x,x,y,y,y\}$-type path, could be the ground-state for the same
reason, as it occurs in the case of x=0.5. 
However, while it is true that the state with such a zigzag structure
is a band-insulator, the energy gain due to the opening of the bandgap
is not always the dominant effect.
In fact, even in the case of x=0.5, the energy of the bottom of the
band for the straight path is $-2t_0$, while for the zigzag path,
it is $-\sqrt{3}t_0$. For x=1/2, the energy gain due to the gap
opening overcomes the energy difference at the bottom of the band,
leading to the band-insulating ground-state. 
However, for x=2/3 even if a band-gap opens the energy of the zigzag
structure cannot be lower than that of the metallic straight-line
phase. Intuitively, this point can be understood as follows: 
An electron can move smoothly along the one-dimensional path
if it is straight. However, if the path is zigzag, ``reflection" of
the wavefunction occurs at the corner, and then a smooth movement of
one electron is no longer possible. Thus, for small numbers of
carriers, it is natural that the ground-state is characterized by the
straight path to optimize the kinetic energy of the $e_{\rm g}$
electrons. 

However, in neutron scattering experiments a spin pattern similar 
to the CE-type AFM phase has been suggested (Radaelli et al., 1999).
In order to stabilize the zigzag AFM phase to reproduce those
experiments it is necessary to include the JT distortion effectively. 
As discussed in Hotta et al. (2000), a variety of zigzag paths could
be stabilized when the JT phonons are included.
In such a case, the classification of zigzag paths is an important
issue to understand the competing ``bi-stripe" vs. ``Wigner-crystal"
structures.
The former has been proposed by Mori et al. (1998), while the latter 
was claimed to be stable by Radaelli et al. (1999).
In the scenario by Hotta et al.(2000), the shape of the zigzag structure is
characterized by the ``winding number'' $w$ associated with the 
Berry-phase connection of an $e_{\rm g}$-electron parallel-transported 
through Jahn-Teller centers, along zigzag one-dimensional paths.
Namely, it is defined as
\begin{equation}
  \label{winding}
  w= \oint {d{\bf r} \over 2\pi} \nabla \xi.
\end{equation}
This quantity has been proven to be an integer, which is a topological
invariant (See Hotta et al., 1998). 
Note that the integral indicates an accumulation of the phase
difference along the one-dimensional FM path in the unit length $M$.
This quantity is equal to half of the number of corners included in
the unit path, which can be shown as follows. 
The orbital polarizes along the hopping direction, indicating that
$\xi_{\bf i}$=$2\pi/3$($4\pi/3$) along the 
$x$-($y$-)direction, as was pointed out above.
This is simply the double exchange mechanism in the orbital degree of
freedom.
Thus, the phase does not change in the straight segment part,
indicating that $w$=0 for the straight-line path. 
However, when an $e_{\rm g}$-electron passes a corner site, the
hopping direction is changed, indicating that the phase change occurs
at that corner.
When the same $e_{\rm g}$-electron passes the next corner, the hopping
direction is again changed.
Then, the phase change in $\xi_{\bf i}$ after moving through a couple
of corners should be $2\pi$, leading to an increase of unity in $w$. 
Thus, the total winding number is equal to half of the number of corners
included in the zigzag unit path. 
Namely, the winding number $w$ is a good label to specify the shape of
the zigzag one-dimensional FM path. 

After several attempts to include effectively the JT phonons, it was
found that the bi-stripe phase and the Wigner crystal phase
universally appear for $w$=$\rm x/(1-x)$ and $w$=1, respectively.
Note here that the winding number for the bi-stripe structure has a
remarkable dependence on x, reflecting  the fact that the distance
between adjacent bi-stripes changes with x.
This x-dependence of the modulation vector of the lattice distortion
has been observed in electron microscopy experiments 
(Mori et al., 1998).
The corresponding zigzag paths with the charge and orbital ordering 
are shown in Fig.III.e.9. In the bi-stripe structure, the charge is
confined in the short straight segment as in the case of the CE-type
structure at x=0.5.
On the other hand, in the Wigner-crystal structure, the straight
segment includes two sites, indicating that the charge prefers to
occupy either of these sites. 
Then, to minimize the JT energy and/or the Coulomb repulsion, the
$e_{\rm g}$ electrons are distributed with equal spacing. 
The corresponding spin structure is shown in Fig.III.e.10.
A difference in the zigzag geometry can produce a significant
different in the spin structure. Following the definitions for the C-
and E-type AFM structures (see Wollan and Koehler (1955) and
introductory section of this review), the bi-stripe and Wigner crystal
structure have $\rm C_{1-x}E_{x}$-type and $\rm C_{x}E_{1-x}$-type
AFM spin arrangements, respectively.
Note that at x=1/2, half of the plane is filled by the C-type,
while another half is covered by the E-type, clearly illustrating the
meaning of ``CE" in the spin structure of half-doped manganites.

As for the charge structure along the $z$-axis for x=2/3
shown in Fig.III.e.11, a remarkable feature can be observed. 
Due to the confinement of charge in the short straight segment for the
bi-stripe phase, the charge stacking is suggested from our topological
argument. 
On the other hand, in the Wigner-crystal type structure, charge is not
stacked, but it is shifted by one lattice constant to avoid the
Coulomb repulsion. 
Thus, if the charge stacking is also observed in the experiment for
x=2/3, our topological scenario suggests the bi-stripe phase as the
ground-state in the low temperature region.
To firmly establish the final ``winner" in the competition between the
bi-stripe and Wigner-crystal structure at x=2/3, more precise
experiments, as well as quantitative calculations, will be needed in the
future. 

\medskip
\noindent{\bf Charge Order at x$\le$0.5}

Regarding densities smaller than 0.5, the states at x=1/8, 1/4 and
3/8 have received considerable attention recently (see Mizokawa et
al., 1999; Korotin et al., 1999; Hotta and Dagotto, 2000).
These investigations are still in a ``fluid'' state, and the
experiments are not quite decisive yet, and for this reason, 
this issue will not be discussed in much detail here.
However, without a doubt, it is very important to clarify the
structure of charge-ordered states that may be in competition  
with the ferromagnetic states in the range in which the latter is
stable in some compounds.
``Stripes'' may emerge from this picture, as recently remarked in
experiments
(Adams et al., 2000; Dai et al., 2000; Kubota et al., 2000. See also
Vasiliu-Doloc et al., 1999) and 
calculations (Hotta, Feiguin, and Dagotto, 2000), and surely the
identification of charge/orbital arrangements at x$<$0.5 will be an
important area of investigations in the very near future.

Here a typical result for this stripe-like charge ordering is shown in
Fig.III.e.12, in which the lower-energy orbital at each site is depicted,
and its size is in proportion to the electron density occupying that
orbital. 
This pattern is theoretically obtained by the relaxation technique for
the optimization of oxygen positions, namely including the cooperative
JT effect. 
At least in the strong electron-phonon coupling region, the stripe
charge ordering along the $diagonal$ direction in the $x$-$y$ plane
becomes the global ground-state.
Note, however, that many meta-stable states can appear very close to
this ground state. 
Thus, the shape of the stripe is considered to fluctuate both in space 
and time, and in experiments it may occur that only some fragments of
this stripe can be detected. 
It should also be emphasized that the orbital ordering occurs concomitant
with this stripe charge ordering.  
In the electron-rich region, the same antiferro orbital-order exists
as that corresponding to x=0.0.
On the other hand, the pattern around the diagonal array of
electron-poor sites is quite similar to the building block of the
charge/orbital structure at x=0.5. 

If these figures are rotated by 45 degrees,
the same charge and orbital structure is found to stack
along the $b$-axis.
Namely, it is possible to cover the whole 2D plane by 
some periodic charge-orbital array along the $a$-axis
(see, for instance, the broken-line path).
If this periodic array is taken as the closed loop $C$ in 
Eq.~(\ref{winding}), the winding numbers are $w$=1, 2, and 3,
for x=1/2, 1/3, and 1/4, respectively.
Note that in this case $w$ is independent of the path along the $a$-axis.
A relation $w$=$N_{\rm c}/2$ holds only when the 1D FM path is fixed 
in the AFM spin arrangement.
The results imply a general relation $w$=$(1-x)/x$
for the charge-orbital stripe in the FM phase,
reflecting the fact that the distance between the diagonal arrays of holes
changes with x.
Our topological argument predicts 
stable charge-orbital stripes at special doping such as
x=$1/(1+w)$, with $w$ an integer.

This orbital ordering can be also interpreted as providing 
a ``$\pi$"-shift in the orbital sector, 
by analogy with the dynamical stripes found in
cuprates (see, for instance, Buhler et al.(2000)), although in copper
oxides the charge/spin stripes mainly appear along the $x$- or
$y$-directions. 
The study of the similarities and differences between
stripes in manganites and cuprates is one of the most interesting open 
problems in the study of transition metal oxides, and considerable
work is expected in the near future.

Finally, a new zigzag AFM spin configuration for x$<$0.5 is here briefly
discussed (Hotta, Feiguin, and Dagotto, 2000).
In Fig.III.e.13, a schematic view of this novel
spin-charge-orbital structure
on the 8$\times$8 lattice at x=1/4 is shown,
deduced using the numerical relaxation technique applied to
 cooperative Jahn-Teller 
phonons in the strong-coupling region. 
This structure appears to be the global ground state, but
many excited states with different spin and charge
structures are also found with small excitation energy,
suggesting that the AFM spin structure for x$<$0.5 in the layered
manganites is easily disordered due to this ``quasi-degeneracy''
in the ground state.
This result may be related to the ``spin-glass'' nature 
of the single layer manganites 
reported in experiments (see Moritomo et al. 1995).

It should be noted that the charge-orbital structure is essentially the
same as that in the 2D FM phase (see Fig.III.e.12).
This suggests the following scenario for the layered manganites:
When the temperature is decreased from the higher temperature region,
first charge ordering occurs due to the cooperative Jahn-Teller
distortions in the FM (or paramagnetic) region.
If the temperature is further decreased, the zigzag AFM spin arrangement
is stabilized, adjusting itself to the orbital structure.
Thus, the separation between the charge ordering temperature $T_{\rm CO}$
and the N\'eel temperature $T_{\rm N}$ occurs naturally in this context.
This is not surprising, since $T_{\rm CO}$ is due to the electron-lattice 
coupling, while $T_{\rm N}$ originates in the coupling $J_{\rm AF}$.
However, if the electron-phonon coupling is weak, then $T_{\rm CO}$ becomes
very low. In this case, the transition to the zigzag AFM phase may occur
prior to the charge ordering. 
As discussed above, the $e_{\rm g}$ electron hopping is confined to
one dimensional structures in the zigzag AFM environment.
Thus, in this situation, even a weak coupling electron-phonon coupling can
produce the charge-orbital ordering, as easily understood from the Peierls
instability argument.
Namely, just at the transition to the zigzag AFM phase, the charge-orbital
ordering occurs simultaneously, indicating that $T_{\rm CO}$=$T_{\rm N}$.
Note also that in the zigzag AFM phase, there is no essential difference
in the charge-orbital structures for the non-cooperative and cooperative
phonons, due to the one-dimensionality of those zigzag chains.

\begin{center}
  {\bf III.f Pseudogap in Mixed-Phase States}
\end{center}

Recent theoretical investigations suggest that the density of states
(DOS) in mixed-phase regimes of manganites 
may have ``pseudogap'' characteristics, 
namely a prominent depletion of weight at the chemical potential.
This feature is similar to that extensively discussed in copper
oxides. The calculations in the Mn-oxide context have been carried out
using both the one- and two-orbital models, with and without disorder 
(see Moreo, Yunoki and Dagotto, 1999b; Moreo et al., 2000).
Typical results are shown in Fig.III.f.1. Part (a) contains the DOS of
the one-orbital model on a 2D cluster varying the electronic density
slightly below \densi=1.0, as indicated in the caption.
At zero temperature, this density regime is unstable due to phase
separation, but at the temperature of the simulation those densities
still correspond to stable states, but with a dynamical mixture of AF
and FM features (as observed, for instance, in Monte Carlo snapshots
of the spin configurations).
A clear minimum in the DOS at the chemical potential can be observed.
Very similar results appear also in 1D simulations (Moreo, Yunoki and
Dagotto, 1999b). Part (b) contains results for two-orbitals and a
large electron-phonon coupling, this time at a fixed density and changing
temperature. Clearly a pseudogap develops in the system as a precursor
of the phase separation that is reached as the temperature is further
reduced. Similar results have been obtained in other parts of
parameter space, as long as the system is near unstable
phase-separated regimes.
A pseudogap appears also in cases where disorder is added to the
system. In Fig.III.f.1c, taken from Moreo et al. (2000), results can
be found for the case where a random on-site energy is added to the
one-orbital model.

A tentative explanation of this phenomenon for the case without
disorder was described by Moreo, Yunoki, and Dagotto (1999b), and it
is explained in Fig.III.f.2.
In part (a) a typical mixed-phase FM-AF state is sketched.
Shown are the localized spins. In the FM regions, the $e_{\rm g}$-electrons
improve their kinetic energy, and thus they prefer to be located in
those regions as shown in (b). 
The FM domains act as effective attractive potentials for electrons,
as sketched in part (c).
When other electrons are added, FM clusters are created and new
occupied levels appear below the chemical potential, creating a 
pseudogap (part (d) of Fig.III.f.2). The DOS is clearly non-rigid. 
These results are compatible with the photoemission experiments by 
Dessau et al. (1998) for bilayer manganites. Other features of the 
experiments are also reproduced such as the large width of the peaks,
and the momentum independence of the results.
This agreement adds to the notion pursued in this review that
mixed-phase states are important to understand
 the behavior of manganese oxides.
The reduction of the DOS at the chemical potential is also compatible
with the insulating characteristics of the bilayers in the regime of
the photoemission experiments.
It is conceivable that the manganites present a pseudogap regime above
their Curie and N\'eel temperatures, as rich as that found in the
cuprates. More details are given in the Discussion section.

\begin{center}
  {\bf III.g Phase Separation Caused by the Influence of
    Disorder on First-Order Transitions:}
\end{center}

Although it is frequently stated in the literature that a variety of
chemical substitutions in manganites lead to modifications in the
bandwidth due to changes in the ``average'' A-site cation radius 
$\langle r_{\rm A} \rangle$, this statement is only partially true.
Convincing analysis of data and experiments by Rodriguez-Martinez and
Attfield (1996) have shown that the disorder introduced by chemical
replacements in the A-sites is also crucially important in determining
the properties of manganites.
For instance, Rodriguez-Martinez and Attfield (1996) found that the
critical temperature $T_{\rm C}$ can be reduced by a large factor if
the variance $\sigma^2$ of the ionic radii about the mean
$\langle r_{\rm A} \rangle$ is modified, keeping 
$\langle r_{\rm A} \rangle$ constant.
Rodriguez-Martinez and Attfield (1996) actually observed that maximum
magnetoresistance effects are found in materials not only with a low
value of $\langle r_{\rm A} \rangle$ (small bandwidth) but also a
small value of $\sigma^2$.
A good example is $\PCMO$ since the Pr$^{3+}$ and Ca$^{2+}$ ions are
similar in size (1.30 $\rm \AA$ and 1.34 $\rm \AA$, respectively,
according to Tomioka and Tokura (1999)).

Disorder, as described in the previous paragraph, is
important for the phase separation scenario.
The recent experimental results showing the existence of micrometer
size coexisting clusters in 
$\rm (La_{5/8-y} Pr_y) Ca_{3/8} Mn O_3$ (LPCMO)
by Uehara et al. (1999), to be reviewed in detail later, highlights a
property of manganites that appears universal, namely the presence of
intrinsic inhomogeneities in the system, even in single crystals. 
This issue is discussed at length in various sections of this review.
In the theoretical framework described thus far, the scenario that is
the closest to predicting such inhomogeneous state is the one based 
on electronic phase separation. However, the analysis presented before
when considering the influence of long-range Coulomb interactions over
a phase separated state, led us to believe that only nanometer size
coexisting clusters are to be expected in this problem.
Those found in LPCMO are much larger, suggesting that there must be
another mechanism operative in manganites to account for their
formation.

A possible explanation of the results of Uehara et al. (1999) has been
recently proposed by Moreo et al. (2000), and it could be considered
as a form of ``disorder-induced'' or
``structural'' phase separation, rather than electronic.
The idea is based on the influence of disorder over the first-order
metal-insulator (or FM-AF) transition found in models where the
interactions are translationally invariant (without disorder),
as it was described in Sections III.d and e.
When such a transition occurs, abruptly a metal changes into an
insulator, as either concentrations or couplings are suitably
changed. 
Unless metastable states are considered, there is no reason to assume 
that in the actual stable ground-state of this system coexisting
clusters will be found, namely the state is entirely FM or AF
depending on parameters. 
However, different is the situation when disorder is considered into
the problem.
The type of disorder taken into account by Moreo et al. (2000) is
based on the influence of the different ionic radius of the various
elements that compose the manganites, as discussed at the beginning of 
this section.
Depending on the environment of A-type ions (which in LPCMO involve
La, Pr or Ca) a given Mn-O-Mn bond can be straight (180$^{\circ}$) or 
distorted with an angle less than  180$^{\circ}$. 
In the latter, the hopping across the bond under study will be less
than the ideal one. For a schematic representation of this idea 
see Fig.III.g.1.
The random character of the distribution of A
ions, leads to a concomitant random distribution of hoppings, and also 
random exchange between the localized spins $J_{\rm AF}$ since this
quantity is also influenced by the angle of the Mn-O-Mn bond.

To account for this effect, Moreo et al. (2000) studied the one- and
two-orbital models for manganites described before, including a
small random component to both the hoppings and $J_{\rm AF}$.
This small component did not influence the FM and AF phases much
away from their transition boundary, but in the vicinity of the
first-order transition its influence is important.
In fact, numerical studies show that the transition now becomes
continuous, with FM and AF clusters in coexistence in a narrow region
around the original transition point. 

Typical results are shown in Fig.III.g.2a-f, using one-dimensional
clusters as an example. In the two upper frames, the energy versus 
$J_{\rm AF}$ (or $J'$) is shown at fixed values of the other
couplings such as $J_{\rm H}$ and $\lambda$, in the absence of
disorder  and at a fixed density x=0.5. 
The abrupt change in the slope of the curves in (a) and (d) clearly
shows that the transition is indeed first-order.
This is a typical result that appears recurrently in all Monte Carlo
simulations of manganite models, namely FM and AF are so different
that the only way to change from one to the other at low temperature
is abruptly in a discontinuous transition (and spin canted 
phases have not been found in our analysis in the absence of magnetic 
fields, as possible intermediate phases between FM and AF). 
These results are drastically changed upon the application of
disorder, as shown in frames (b,c,e, and f) of Fig.III.g.2, where the
mean couplings have been fixed such that the model is located exactly
at the first-order transition of the non-disordered system.
In these frames, the nearest-neighbor spin correlations along the chain 
are shown. Clearly this correlation is positive in some portions of
the chain, while it alternates from positive to negative in
others. This alternation is compatible with an AF state, with an
elementary unit cell of spins in the configuration
 up-up-down-down, but the particular
form of the AF state is not important in the following; only its
competition with other ordered states, such as the FM one is significant.
The important point is that there are coexisting FM and AF regions.
The cluster size is regulated by the strength of the disorder, such
that the smaller the disorder, the larger the cluster size. Results such as 
those in Fig.III.g.2 have appeared in all simulations carried out in
this context, and in dimensions larger than one 
(see Moreo et al., 2000).
The conclusions appear independent of the particular type of AF
insulating state competing with the FM state, the details of the
distribution of random numbers used, and the particular type of
disorder considered which could also be in the form of a random
on-site energy in some cases (Moreo et al., 2000). 
Note that the coexisting clusters have the $same$ density, namely
these are FM and AF phases that appear at a fixed hole concentration
in the non-disordered models, for varying couplings.
Then, the problem of a large penalization due to the accumulation of
charge is not present in this context.

What is the origin of such a large cluster coexistence with equal
density? 
There are two main opposing tendencies acting in the system.
On one hand, energetically it is not convenient to create FM-AF
interfaces and from this perspective a fully homogeneous system is
preferable.
On the other hand, locally at the level of the lattice spacing
the disorder in $t$ and $J_{\rm AF}$ alter
the couplings such that the system prefers to be either on the FM or
AF phases, since these couplings fluctuate around the 
transition value. From the perspective of the disorder, 
the clusters should be as
small as possible such that the local different tendencies can be
properly accounted for. From this competition emerges the large
clusters of Fig.III.g.2, namely by creating large clusters, the number of
interfaces is kept small while the local tendencies toward one phase
or the other are partially satisfied. ``Large'' here means substantially
larger in size than the lattice spacing.
A region where accidentally the distribution of random couplings
favors the FM or AF state on average, will nucleate such a phase in
the form of a bubble.

These simple ideas can be made more elegant using the well-known
arguments by Imry and Ma (1975), which were applied originally to the
Random Field Ising Model (RFIM) (see contributions on the subject in
the book of Young (1998)), namely a model with a FM Ising interaction
among spins of a lattice in the presence of a magnetic field in the
$z$-direction which changes randomly from site to site. 
This local field is taken from a distribution of random numbers of
width 2$W$. In the context of manganites it can be imagined that the
spin up and down of the RFIM represent the two states in competition
(metal and insulator) in the real compounds.
The random field represents the local tendency to prefer either a
metal or an insulator, due to the fluctuations in the disorder of the
microscopic models.
As a function of an external uniform magnetic field, the RFIM at zero
temperature has a first-order transition at zero external field
in the absence of random fields (between all spins up and all down), 
which turns continuous as those random fields are added, quite
similar to the case described above in the FM-AF competition.
Then, the RFIM captures at least part of the physics of Mn-oxides that
emerged from the study of realistic Hamiltonians in the presence of
disorder, as shown above. For this reason it is instructive to study
this simple spin model, which can be analyzed on lattices much larger
than those that can be reached with the one or two orbital models of
Section III.c. However, note that the use of the RFIM is only to guide
the intuition, but it is not claimed that this model belongs to
exactly the same universality class as the microscopic Hamiltonians
for Mn-oxides used here. The study of universality is very complex and
has not been addressed in this context yet. Nevertheless, it is
expected that the RFIM will at least provide some intuition as to how real
manganites behave.

Typical results are shown in Fig.III.g.3. In part (a), the data
corresponding to a simulation at low temperature on a 100$\times$100 
cluster for a fixed set of random fields is shown. The clusters are
basically frozen, namely the result is representative of the
ground-state. The presence of coexisting clusters of spins up and down
is clear. Their distribution is certainly random, and their shape
fractalic, similar to that observed in experiments for LPCMO.
Upon reduction of $W$, in frame (b) results now for a 500$\times$500
cluster show that the typical size
of the clusters grow and can easily involve a few hundred lattice 
spacings. When an external field is applied, a $percolation$ among
disconnected clusters emerges. This is a very important point, in
agreement with the expectations arising from several experiments,
namely percolative characteristics should appear in real manganites to
the extend that the theoretical investigations presented in this
section are correct. 
Uehara et al. (1999) and other experimentalists intuitively concluded
that indeed percolation is important in the study of Mn-oxides, and in
the following Section
it will be shown that it plays a key role in rationalizing the d.c. resistivity
of these compounds.
Gor'kov and Kresin (1998) also briefly discussed a possible
percolation process at low temperature.

Summarizing, phase separation can be 
driven by energies other than purely electronic.
In fact it can also be triggered by the influence of
disorder on first-order transitions. In this case the competing
clusters have the same density and for this reason can be very
large. Micrometer size clusters, such as those found in the RFIM,  are
possible in this context, and have been observed 
in experiments. This result  is very
general, and should apply to a variety of compounds where two very
different ordered states are in competition at low temperatures. 

The remarkable phenomenological results of Rodriguez-Martinez and 
Attfield (1996) appear to be in qualitative agreement with the
theoretical calculations. As explained above, Moreo et al. (2000) 
found that the size of the clusters induced by disorder near a
transition, such as those
produced by chemical substitutions in real manganites, 
which would be of first-order in the clean limit, can be
controlled by the ``strength'' of that disorder. In practice this
strength is monotonically related to $\sigma^2$ (in the limit
$\sigma$=0 there is no disorder).
At small (but not vanishing) $\sigma$ or disorder in the calculations
of Moreo et al. (2000), the coexisting clusters are large. As the
disorder grows, the clusters reduce their size. 
To the extent that the size of the coexisting clusters is directly
proportional the strength of the CMR effect, then weak disorder is
associated with large magnetoresistance changes with the composition, 
magnetic fields or pressure, a somewhat counter-intuitive result since
naively strong disorder could have been expected to lead to larger
modifications in the resistivity. 

\begin{center}
  {\bf III.h Resistivity of Manganites in the Mixed-Phase Regime:}
\end{center}

One of the main lessons learned from the previous analysis of models 
for manganites is that intrinsic inhomogeneities are very important 
in this context.
It is likely that the real Mn-oxides in the CMR regime are in such 
a mixed-phase state, a conclusion that appears inevitable based on 
the huge recent experimental literature, to be reviewed in the next
Section, reporting phase separation tendencies in some form or
another in these compounds.
However, note that until recently estimations of the d.c. resistivity
$\rho_{\rm dc}$ in such a mixed-phase regime were not available. 
This was unfortunate since the interesting form of the $\rho_{\rm dc}$
vs. temperature curves, parametric with magnetic fields, is one of the
main motivations for the current huge effort in the manganite context.
However, the lack of $reliable$ estimations of $\rho_{\rm dc}$ is not 
accidental: it is notoriously difficult to calculate transport
properties in general, and even more complicated in regions of
parameter space that are expected to be microscopically inhomogeneous.
Although there have been some attempts in the literature to calculate
$\rho_{\rm dc}$, typically a variety of approximations that are not
under control have been employed. In fact, the micrometer size of some
of the coexisting clusters found in experiments strongly suggest that
a fully microscopic approach to the problem will likely fail since,
e.g., in a computational analysis it would be very difficult to study
sufficiently large clusters to account for such large scale
structures. It is clear that a more phenomenological approach is needed
in this context.

For all these reasons, recently Mayr et al. (2000) carried out a
study of $\rho_{\rm dc}$ using a $random$ $resistor$ $network$ model 
(see Kirkpatrick, 1973), and other approximations.
This model was defined on square and cubic lattices, but with a
lattice spacing much larger than the $\rm 4\AA$ distance between
nearest-neighbor Mn ions. A schematic representation is presented in
Fig.III.h.1.
Actually, the new lattice spacing is a fraction of micrometer, since
the random network tries to mimic the complicated fractalic-like
structure found experimentally.
At each link in this sort of effective lattice, randomly either a
metallic or insulating resistance was located in such a way that the
total fraction of metallic component was $p$, a number between 0 and 1.

The actual values of these resistances as a function of temperature
were taken from experiments. Mayr et al. (2000) used the
$\rho_{\rm dc}(T)$ plots obtained by Uehara et al. (1999)
corresponding to $\rm (La_{5/8-y} Pr_y) Ca_{3/8} Mn O_3$ (LPCMO), one
of the compounds that presents the coexistence of giant FM and CO
clusters at intermediate values of the Pr concentration. More
specifically, using for the insulating resistances the results of
LPCMO at y=0.42 (after the system becomes a CO state with 
increasing Pr doping) and for the metallic ones the results at y=0.0
(which correspond to a metallic state, at least below 
its Curie temperature), the results of a numerical study 
on a 100$\times$100 cluster
are shown in Fig.III.h.2 (the Kirchoff equations were solved by a
simple iterative procedure).
It is interesting to observe that, even using such a simple
phenomenological model, the results are already in reasonable
agreement with the experiments, namely, (i) at large temperature
insulating behavior is observed even for $p$ as large as 0.65 (note
that the classical percolation is expected to occur near $p=0.5$; see 
Kirkpatrick (1973)); (ii) at small temperature a (``bad'') metallic
behavior appears; and (iii) a broad peak exists in between.
Results in both 2D and 3D lead to similar conclusions. It is clear
that the experimental results for manganites can be at least partially
accounted for within the mixed-phase scenario.

The results of Fig.III.h.2 suggest a simple qualitative picture to
visualize why the resistivity in Mn-oxides has the peculiar shape it
has. The relevant state in this context should be imagined as
percolated, as sketched in Fig.III.h.3a, as predicted by the
analysis of the previous section.  
Metallic filaments from one side of the sample to the other exist 
in the system. At low temperature, conduction is through those
filaments. 
Necessarily, $\rho_{\rm dc}$ at $T$=0 must be large, in such a
percolative regime.
As temperature increases, the $\rho_{\rm dc}$ of the filaments grows as
in any metal. However, in the other limit of large or room temperature,
the resistance of the percolated metallic filament is expected to be
much larger than
that corresponding to one of the insulator paths. Actually, near room
temperature in many experimental graphs, it can be observed that
$\rho_{\rm dc}$ in the metallic and insulating regimes are quite
similar in value, even comparing
results away from the percolative region.
Then, at room temperature it is more likely that conduction will occur
through the insulating portions of the sample, rather than through the
metallic filaments. Thus, near room temperature insulating
behavior is expected. In between low and high temperature, it is
natural that $\rho_{\rm dc}$ will present a peak. Then, a simple 
``two resistances in parallel'' description appears appropriate (see 
Fig.III.h.3b). The insulating resistance behaves like any insulator,
while the metallic one starts at $T$=0 at a high value and then it
behaves like any metal. The effective resistance shown in Fig.III.h.3b 
properly reproduces the experiments at least qualitatively.

Note, however, that many experimental results suggest that
$\rho_{\rm dc}$ has an intermediate temperature peak sharper than
shown in Fig.III.h.2. In some compounds this is quite notorious, while
in others the peak is fairly broad as in Fig.III.h.2. Nevertheless, it
is important to find out alternative procedures to sharpen the
$\rho_{\rm dc}$ peak to better mimic experiments. 
One possible solution to this problem is to allow for the metallic 
fraction $p$ to vary with temperature. This is a reasonable assumption 
since it is known that the metallic portions of the sample in
mixed-phase manganites originate in the ferromagnetic arrangement of
spins that improves conduction.
The polarization of the spins deteriorates as the temperature increases,
and it is reasonable to imagine that the FM islands decrease in size as
the temperature grows.
Then, a pattern of FM clusters that are connected at low temperature
leading to a metallic behavior may become disconnected at higher
temperatures. The tendencies toward a metallic percolation decrease
with increasing temperature. Such a conjecture was studied
qualitatively by Mayr et al. (2000) using the Random Field Ising model
and the one-orbital model for Mn-oxides. In both cases, indications of
the disappearance of percolation with increasing temperature were
indeed found. Then, assuming that $p$ decreases with increasing
temperature, approximately following the magnetization, seems a
reasonable assumption.

Results with a temperature dependent $p$ are shown in Fig.III.h.4.
The actual values of $p$ are indicated, at least in part, in the
figure. Certainly the peak is now sharper than in Fig.III.h.2, as
expected, and the results indeed resemble those found in a variety of
experiments. Note that the function $p$=$p(T)$ has not been 
fine-tuned, and actually a variety of functions lead to similar
conclusions as those in Fig.III.h.4. Note also that obtaining such a
result from a purely microscopic approach would have been quite
difficult, although Mayr et al. (2000) showed that data taken on small
clusters using the one-orbital model are at least compatible with
those of the phenomenological approach. To evaluate the conductance
of these clusters, the approach of Datta (1995) and Verges (1999) were
used. Also note that calculations using a cubic cluster with either
metallic or insulating ``hopping'' (Avishai and Luck, 1992), 
to at least partially account for quantum effects, lead to results
similar to those found in Fig.III.h.4.

The success of the phenomenological approach described above leads to
an interesting prediction. In the random resistor network, it is clear
that above the peak in the resistivity, the mixed-phase character of
the system remains, even with a temperature dependent metallic
fraction $p$.
Then, it is conceivable to imagine that above the Curie temperature in
real manganites, a substantial fraction of the system should remain in
a metallic FM state (likely not percolated, but forming disconnected
clusters). 
A large variety of experiments reviewed in the next section indeed
suggest that having FM clusters above $T_{\rm C}$ is possible. As a
consequence, this has led us to conjecture that there must exist a
temperature $T^*$ at which those clusters start forming. This defines
a new temperature scale in the problem, somewhat similar to the famous
pseudogap $T^*$ scale of the high temperature superconducting
compounds. 
In fact, in mixed phase FM-AF states it is known that a pseudogap
appears in the density of states (Section III.g; Moreo et al., 1999b;
Moreo et al., 2000), thus increasing the analogy between these two
materials. In our opinion, the experimental verification that indeed
such a new scale $T^*$ exists in manganites is important to
our understanding of these compounds. In fact, 
recent results by Kim, Uehara and
Cheong (2000) for $\LCMO$ at various densities
have been interpreted as caused by small FM segments of the CE-type CO
state, appearing at hole densities smaller than x=1/2 and at high temperature.
This result is in qualitative agreement with the theoretical
analysis presented here.

The study of effective resistivities and conductances has also been 
carried out in the presence of magnetic fields (Mayr et al., 2000),
although still mainly within a phenomenological approach. From the
previous results Figs.III.h.2-4, it is clear that in the percolative
regime ``small'' changes in the system may lead to large changes in
the resistivity. For instance, if $p$ changes by only 5\% from 0.45 to
0.5 in Fig.III.h.4, $\rho_{\rm dc}$ is modified by two orders of
magnitude! It is conceivable that small magnetic fields could induce
such small changes in $p$, leading to substantial modifications in the
resistivity. Experiments by Parisi et al. (2000) indeed show a rapid
change of the fraction of the FM phase in $\rm La_{0.5} Ca_{0.5} Mn O_3$
upon the application of magnetic fields.
In addition, studies of the one-orbital model carried out
in one dimension (Mayr et al., 2000) also showed that other factors
may influence the large $\rho_{\rm dc}$ changes upon the
application of external fields.
For instance, in Fig.III.h.5, the inverse conductance $C^{-1}$ of a 64-site
chain is shown in the presence of small magnetic fields (in units of
the hopping), in the regime of FM-AF cluster coexistence, which is
achieved by the introduction of disorder where a first-order FM-AF
transition occurs, as discussed in the previous subsection.
The results of Fig.III.h.5 clearly indicate that $C^{-1}$ can indeed
change by several orders of magnitude in the presence of small fields 
even in a 1D system that certainly cannot have a percolation.
There must be some other mechanism at work in this context. 
Mayr et al. (2000) believe this alternative mechanism is caused by
small modifications in the conductivity of the $insulating$ portions
of the sample, independent of what occurs in the metallic clusters. 
It is possible that in an AF region, with zero conductivity at large
Hund coupling due to the perfect anti-alignment of the
nearest-neighbor $t_{\rm 2g}$-spins, the small fields may induce a small
canting effect that leads to a nonzero conductivity.
While this effect should be negligible if the AF phases is totally
dominating, it may become more important if small AF clusters separate FM
ones. A sort of ``valve'' effect may occur, in other words magnetic
fields can induce a small connection between metallic states leading
to a substantial change in the resistivity. 
This idea can be studied qualitatively by simply altering by a small
amount the conductivity of the insulating regions in the
random-resistor-network. 
Results are shown in Fig.III.h.5b, using the same functions $p$=$p(T)$
employed before in Fig.III.h.4. 
As anticipated, small conductivity changes lead to large
resistivity modifications, comparable to those observed in experiments 
upon the application of magnetic fields.
Although the analysis discussed above is only semi-quantitative and 
further studies in magnetic fields should actively continue in this
context, Mayr et al. (2000) have shown that in the percolative
regime two mechanisms (described above) can lead to a large MR, 
leading to at least a possible framework for describing how the famous CMR 
effect can occur.

\begin{center}
  {\bf III.i Related Theoretical Work on Electronic Phase
    Separation Applied to Manganites}
\end{center}

The possibility of ``electronic'' phase separation was already discussed by 
Nagaev (1967, 1968, 1972) well before it became a popular subject in 
the context of compounds such as the high temperature
superconductors. Its original application envisioned by Nagaev was to 
antiferromagnetic semiconductors, where the doping of electrons 
creates ferromagnetic-phase regions embedded in an AF matrix.
Nagaev (1994,1995) remarked that if the two phases have opposite
charge, the Coulombic forces will break the macroscopic clusters into
microscopic ones, typically of nanometer scale size, as remarked in this
review before. When the number
of these FM clusters is small, the system resembles a  regular array
of charge sort of a Wigner crystal, as found also in the
simulations of Malvezzi et al. (1999), and the system remains an
insulator. However, as the density grows, a transition will be found
where the clusters start overlapping, and a metal is formed. Although
it may seem tempting to assign to this transition percolative
properties, as Nagaev does, note that at least without incorporating
disorder the clusters are regularly spaced and thus the transition
does not correspond to the usual percolative ones described in
textbooks and in the previous subsection  where the random position of 
the clusters play a key role. In particular, the critical density at
which regularly spaced clusters begin overlapping triggers a process
that occurs in all clusters at the same time, different from the
notion of a percolative filament with fractalic shape which is crucial 
in percolative theories. For this reason it is unclear to these
authors to what extend electronic phase separation can describe
percolative physics in the absence of disorder. 
It appears that only when randomly distributed
clusters of two phases are stabilized, as described in Sec.III.h, can
true percolation occur.

The calculations of Nagaev (1994,1995,1996) have been carried out for
one orbital models and usually in the limit where the hopping $t$ of
the conduction electrons is much larger than the Hund coupling
(although Nagaev expects the results to qualitatively hold even in the
opposite limit $J_{\rm H}$$>$$t$). Also a low density of carriers was 
assumed, and many calculations were performed mainly for the one
electron problem (magneto polaron), and then rapidly generalized to
many electrons. The formation of lattice polarons is not included in
the approach of Nagaev.
These parameters and assumptions are reasonable for AF semiconductors,
and Nagaev (1995) argued that his results can explain a
considerable body of experimental data for EuSe and EuTe.
However, note that Mauger and Mills (1984,1985) have shown that self-trapped
FM polarons (ferrons) are $not$ stable in three dimensions. 
Instead, Mauger and Mills (1984,1985) proposed that
electrons bound to donor sites induce a ferromagnetic moment,
and they showed that
those bound magnetic polarons can account for the FM clusters observed
in EuTe. Free carriers appear ``frozen'' 
at low temperatures in these materials, and there are no ferron-like
solutions of the underlying equations in the parameter range appropriate
to Eu chalcogenides.

In addition, note that the manganites have a large $J_{\rm H}$ and a large 
density of electrons, and in principle calculations such as those
described above have to be
carried out for more realistic parameters, if the results can
indeed apply to manganites. 
These calculations are difficult without the aid of computational
techniques. In addition, it is clearly important to consider two
orbitals to address the orbital-ordering of the manganites, the
possibility of orbital phase separation, and the influence of
Jahn-Teller or Coulombic interactions that lead to charge-order AF
states. Disorder also appears to play a key role in manganites.

The unstable character of the low hole-density region of the phase
diagram corresponding to the one-orbital model for manganites has also
been analyzed by other authors using mostly analytic approximate 
techniques.
In fact, Arovas and Guinea (1998) 
found an energy convex
at small hole concentration, indicative of phase separation,
within a mean-field treatment of the
one-orbital model using the Schwinger formalism (see also 
Mishra et al. (1997), Arovas, Gomez-Santos, and Guinea (1999); Guinea,
Gomez-Santos, and Arovas (1999); Yi and Lee (1999); 
Chattopadhyay, Millis and Das Sarma (2000); 
and Yuan, Yamamoto, and Thalmeier, 2000).
Nagaev (1998) using the one-orbital model also arrived at the
conclusion that the canted AF state of the small hole density region
is unstable.
The same conclusion was obtained
in the work of Kagan, Khomskii and Mostovoy (1999)
where the dominance of  phase separation in the small hole-density
region was remarked upon, both using classical and quantum spins.
Ferromagnetic polarons embedded into an AF surrounding were also
discussed by those authors.
Polarons in electron-doped one-orbital models where also analyzed by
Batista et al. (1998). 
Nagai, Momoi and Kubo (1999) using the dynamical mean-field
approximation (exact in infinite dimension) studied the one-orbital
model with $S$=1/2 localized spins. 
Nagai, Momoi and Kubo (1999) (see also Momoi and Kubo, 1998)
identified FM, AF, and PM phases.
Regimes of phase separation were observed involving the AF and PM
phases, as well as the PM and FM ones.
A representative  density vs. chemical potential plot is shown in
Fig.III.i.1.
The results obtained by those authors are qualitatively similar to
those found using the DMRG in one-dimension (Dagotto et al., 1998) for
$S$=1/2 localized spins, and also similar to results obtained in
higher dimensions with classical localized spins
 (Yunoki et al., 1998a). 
An AF-PM phase separation was also detected in infinite dimension
calculations (Yunoki et al., 1998a), showing that not only AF-FM
coexistence is
possible. Calculations using $t$-$J$-like models, derived at large
$J_{\rm H}$ starting with the one-orbital model, also reveal phase
separation, as shown by Shen and Wang (1998). 
Overall, it can be safely concluded that using a variety of numerical
and analytical techniques, convincing evidence has accumulated that the
canted AF state of deGennes (1960) is simply not stable in models
believed to be realistic for manganese oxides. This state is replaced
by a mixed-phase or phase-separated regime.
The importance of heterogeneity in manganites was also remarked upon by von
Molnar and Coey (1998), based on an analysis of several experiments. 
Also Khomskii (1999) remarked upon the importance of phase-separation and
percolation. 

Other calculations have also shown tendencies to phase separation.
For instance, Yamanaka, Koshibae and Maekawa (1998) studied the
one-orbital model in two and three dimensions and found phase
separation between a flux and antiferromagnetic states (see also
Agterberg and Yunoki, 2000). 
Working with the one-orbital model, computational studies by Yi and Yu
(1998) arrived to the same conclusions, previously presented by Yunoki
and Moreo (1998), regarding the presence of PS at both small and large
hole density once the direct Heisenberg coupling among the 
$t_{\rm 2g}$ spins is considered.
Golosov, Norman and Levin (1998), using a mean-field approximation for
the one-band model, also found that the spin canted state was unstable,
and indications of phase separation were reported.
Schlottmann (1999) using a simple alloy-analogy model showed that the
system is unstable to phase separation.
Symmetry arguments discussed by Zhong and Wang (1999) also led to PS
at low hole doping. In the continuum model, PS has also been found (Roman
and Soto, 1998).

Even for the two-orbital model, evidence has accumulated that phase
separation is present, particularly at low- and high-density of holes. 
Besides the already described robust computational evidence for the
case where the orbital degree of freedom plays the key triggering role
for this effect (Yunoki et al., 1998b), mean-field approximations
presented by Okamoto, Ishihara and Maekawa (1999) also detected phase
separation involving two phases with the same spin characteristics
(ferromagnetic), but differing  orbital arrangement.
A representative result is reproduced in Fig.III.i.2, where the
orbital states are also shown. 

\begin{center}
  {\bf III.j On-site Coulomb Interactions and Phase Separation:}
\end{center}

What happens with phase separation when the on-site Coulomb $U$
interaction is dominant over other interactions? This question does
not have an easy answer due to the technical complications of carrying
out reliable calculations with a nonzero $U$.
In fact, the one-band Hubbard model has been studied for a long time as 
a model of high temperature superconductors and after more than 10
years of work it is still unclear whether it phase separates in
realistic regimes of parameters.
Thus, it is not surprising that similar uncertainties may arise in the
context of models for manganites. As remarked before, studies of the 
one-dimensional one-orbital model including a nonzero $U$ were carried 
out by Malvezzi et al. (1999).
In this study, a region of phase separation was identified in a similar
location as obtained in the Monte Carlo simulations without $U$
(Yunoki et al., 1998). 
Then, certainly switching on $U$ ``slowly'' starting in the phase
separated regime of the one-orbital model does not alter the presence
of this regime.
On the other hand, Shen and Wang (1999a) claimed that if $U$ is made 
larger than $J_{\rm H}$, the model does not lead to phase separation
according to their calculations (see also Gu et al., 1999). 
This issue is somewhat complicated by the well-known fact that pure
Hubbard-like models tend to present a large compressibility near
half-filling, namely the slope of the curve density vs. chemical
potential is large at that density (Dagotto, 1994).
This may already be indicative of at least a $tendency$ to phase
separation that could be triggered by small extra terms in the
Hamiltonian. 
More recently, it has been shown that a two-orbital 1D model with a 
form resembling those studied in manganites (but without localized
spins) indeed presents phase separation when studied using the DMRG
technique (Hotta, Malvezzi and Dagotto, 2000).
This is in agreement with the results of Shen and Wang (1999b) 
using the two-orbital model at both large $U$ and $J_{\rm H}$, where it was 
concluded that having $both$ couplings leads to a rich phase diagram
with phase-separated and charge-ordered states. 
It is likely that this conclusion is correct, namely phase separation
may be weak or only incipient in the purely Coulombic models, but in
order to become part of the phase diagram, the Hund coupling to
localized spins may play a key role.
More work is needed to clarify these issues. Finally, the reader
should recall the discussion of Section III.c, where at least within a
mean-field approximation it was shown that a large electron-JT phonon
coupling or large Coulombic couplings are qualitatively
equivalent. This is especially true when issues such as phase
separation induced by disorder are considered, in which the actual origin of
the two competing phases is basically irrelevant.
Note also that Motome, Nakano, and Imada (1998) have found
phase-separation in a two orbital model for manganites when a
combination of Coulombic and Jahn-Teller interactions is considered. 
Recently, Laad, Craco, and M\"uller-Hartmann (2000) have also
investigated a model including both Coulombic and JT-phononic
couplings, analyzing experiments at x=0.3 $\LSMO$.

\begin{center}
  {\bf III.k Theories based on Anderson localization:}
\end{center}

There is an alternative family of theories which relies on the
possibility of electron localization induced by two effects:
(1). off-diagonal disorder caused by the presence of an effective
$complex$ electron-hopping in the large Hund-coupling limit (see for instance
M\"uller-Hartmann and E. Dagotto (1996) and Varma (1996)), and
(2). nonmagnetic diagonal disorder due to the different charge and sizes
of the ions involved in manganese oxides, as discussed before.
Calculations in this context by Sheng et al. (1997a), using scaling
theory and a mean-field distribution for the spin orientations (one
orbital model, $J_{\rm H}$=$\infty$), were claimed to reproduce
quantitatively the magnetoresistance effect of real materials. 
Related calculations have been presented by Allub and Alascio
(1996,1997) and Aliaga, Allub and Alascio (1998).
In these calculations, the electrons are localized above $T_{\rm C}$   
due to strong disorder, while at low temperature the alignment of 
the spins
reduce the spin disorder and the electrons are delocalized.
In this framework, also Coey et al. (1995) argued that the
$e_{\rm g}$-electrons, while delocalized at the Mn-Mn scale, 
are localized at larger scales. 

There are some problems with approaches based on simple Anderson
localization. For example, the phases competing with ferromagnetism
are in general of little importance, and the mixed-phase tendencies of
manganites, which are well-established from a variety of experiments as
shown in Section IV, are not particularly relevant in this context. 
The first-order-like nature of the transitions in these compounds is
also not used.
Note also that recently Smolyaninova et al. (1999) have experimentally
shown that the
metal-insulator transition of $\LCMO$ at x=0.33 is not an Anderson
localization transition. In addition, the x=0.5 CO state, crucial in
real manganites to drive the strong CMR effect near this density,
plays no important role in this context. 
It appears also somewhat unnatural to deal
with an on-site disorder with such a large strength, typically 
$W$ $\sim$ 12 $t$ (the random energies $\epsilon_i$ are taken from
the distribution [$-W/2$,$W/2$], and t is the one-orbital hopping 
amplitude). However, it may occur that this strong disorder is 
a way to effectively mimic, in a sort of coarse-grained lattice,
the disorder induced by cluster formation, similar to the
calculation of the resistivity in Section III.h.
For instance, Sheng et al. (1997b) noticed the relation between the
$T$=0 residual resistivity and the presence of a peak in the same
quantity at $T_{\rm C}$.
However, instead of assigning the large $\rho_{\rm dc}(T=0)$ to the
percolative process described in Sec.III.h, nonmagnetic randomness was
used, and naturally a large $W$ was needed to arrive at the large
resistivities that appear near percolative transitions. 
The authors of this review believe that theories based on electron
localization ideas, although they appear at first sight not directly
related to the ubiquitous clustering tendencies of real manganites,
may effectively contain part of the answer to the manganite puzzle, 
and further work in this context should be encouraged, if possible
including in the approach a description of how localization phenomena
relates to the phase separation
character of manganites.
Steps in this direction were recently taken by Sheng et al. (1999),
in which calculations with JT phonons were carried out, and phase
separation tendencies somewhat similar to those reported by Yunoki et
al. (1998) were observed. 

\section{EXPERIMENTAL EVIDENCE OF INHOMOGENEITIES IN MANGANITES}

\begin{center}
  {\bf IV.a $\bf La_{1-x} Ca_x Mn O_3$ at density
    0.0$\leq$x$<$0.5}
\end{center}

The regime of intermediate and low hole densities of $\LCMO$, the
former being close to the AF CE-type state at x=0.5 and the latter to 
the antiferromagnetic A-type state at x=0, is complex and
interesting. In this region, the FM metallic state believed to be caused by
double-exchange is in competition with other states, notably AF ones, 
leading to the mixed-phase tendencies that are the main motive of this 
review. The special density x=0.33 in $\LCMO$ has received
considerable experimental attention, probably caused by the peak in
the Curie temperature which occurs near this hole concentration (see
phase diagram in Section II.b). 
For potential technological applications of manganites it is important
that the FM transition temperature be as high as possible, and thus
it is important to understand this particular composition. However,
although this reason for focusing efforts at x=0.33 is reasonable,
recent experimental and theoretical work showed that it is convenient
to move away from the optimal density for ferromagnetism to understand
the behavior of manganites, since many of the interesting effects in
these compounds are magnified as $T_{\rm C}$ decreases. Nevertheless,
the information gathered at the hole density x=0.33 is certainly 
important, and analyzed together with the results at other densities,
illustrates the inhomogeneous character of manganites. 

Historically, the path followed in the study of data at low and
intermediate densities of LCMO is fairly clear. Early worked focused on ideas
based on polarons, objects assumed to be usually small in size, and
simply represented as a local distortion of the homogeneous background
caused by the presence of a hole. The use of polarons was
understandable due to the absence of theoretical alternatives until a
few years ago, and it may still be quite appropriate in large regions
of parameter space.
However, recent experimental work has shifted toward the currently
more widely accepted mixed-phase picture where the ferromagnetic 
regions are not small isolated polarons but substantially larger
clusters, at least in the important region in the vicinity of 
$T_{\rm C}$ (polaronic descriptions may still be realistic well 
above $T_{\rm C}$).
Note that the various efforts reporting polarons usually employed
techniques that obtained spatially-averaged 
information, while only recently, real-space
images of the local electronic properties have been obtained that
clearly illustrates the mixed-phase character of the manganite states. Below
follows a summary of the main experimental results addressing
mixed-phase characteristics in $\LCMO$ at densities between x=0.0 and
0.5 (excluding the latter which will be analyzed separately). These
results are not presented in historical order, but are 
mainly grouped by technique. Although the list is fairly complete,
certainly it is not claimed that all reports of 
mixed-phase tendencies are described here as
other efforts in this direction may have escaped our attention. 

\medskip
\noindent{\bf Electron Microscopy:}

Among the most important experimental results that have convincingly
shown the presence of intrinsic mixed-phase tendencies in manganites
are those recently obtained by Uehara et al. (1999) in their study
of $\rm La_{5/8-y} Pr_y Ca_{3/8} Mn O_3$ using transport, magnetic,
and electron microscopy techniques (see also Kiryukhin et al., 2000). 
The results reported by those
authors for the resistivity vs. temperature at several Pr compositions
are reproduced in Fig.IV.a.1a.  Note the rapid reduction with
increasing y of the temperature at which the peak occurs, which
correlates with the Curie temperature. Note also the hysteretic
behavior of the resistivity, signalling the presence of
first-order-like characteristics in these compounds.
Another of the striking features of Fig.IV.a.1a is the presence of an
abnormally large residual resistivity at low temperatures in spite of
the fact that ${\rm d}\rho/{\rm d}T$$>$0 suggests metallic behavior.
The magnetoresistance factor shown in Fig.IV.a.1b is clearly 
large and increases rapidly as $T_{\rm C}$ is reduced. This factor is robust
even at low temperatures where the resistivity is flat, namely the
large MR effect does not happen exclusively at $T_{\rm C}$.

The results of Uehara et al. (1999) have been interpreted by those
authors as evidence of two-phase coexistence, involving a stable FM state
at small y, and a stable CO state in the large y PCMO
compound. A percolative transition in the intermediate regime of
compositions was proposed. The phase diagram is in Fig.IV.a.1c and it
contains at low temperatures and a small range of Pr densities a phase
labeled ``CO+FM'' which corresponds to the two-phase regime. In other
regions of parameter space, short-range ``s-r'' FM or CO order has been
observed. Uehara et al. (1999) substantiated their claims of phase
separation using electron microscopy studies. Working at y=0.375 and at
low temperature of 20K, coexisting domains having sizes as large as 500 nm
were found. At 120K, the clusters become nanometer in size.
Note that these low-temperature large clusters appear at odds with at
least one of the sources of inhomogeneities discussed in the
theoretical review (electronic phase separation), since $1/r$ Coulomb
interactions are expected to break large clusters into smaller ones of
nanometer size. In fact,  Uehara et al. (1999) remarked it is
reasonable to assume that the competing phases are of the same charge
density. However, the experimental results for 
$\rm La_{5/8-y} Pr_y Ca_{3/8} Mn O_3$ are in excellent agreement with 
the other proposed source of mixed-phase tendencies, namely the ideas
presented by Moreo et al. (2000), where first-order transitions are
transformed into regions of two-phase coexistence by the intrinsic
chemical disorder of the manganites (Sec.III). This effect is called
``disorder-induced phase separation''.

\medskip
\noindent{\bf Scanning Tunneling Spectroscopy:}

Another remarkable evidence of mixed-phase characteristics in
$\LCMO$ with x$\sim$0.3 has been recently reported 
by F\"ath et al. (1999) using scanning tunneling spectroscopy. With
this technique, a clear phase-separated state was observed below
$T_{\rm C}$ using thin-films. The clusters involve metallic and
insulating phases, with a size that is dependent on magnetic fields. 
F\"ath et al. (1999) believe that $T_{\rm C}$ and the associated
magnetoresistance behavior is caused by a percolation process. In
Fig.IV.a.1d, a generic spectroscopic image is shown. A coexistence of
metallic and insulating ``clouds'' can be observed, with a variety of
typical sizes involving tens to hundreds of nanometers. F\"ath et al. 
(1999) remarked that it is clear that such length scales
are not compatible with a picture of homogeneously distributed small
polarons. The authors of this review agree with that statement.

The results of F\"ath et al. (1999) suggest that small changes in the
chemical composition around $\LCMO$ at x=0.25 can lead to dramatic
changes in transport properties. This is compatible with results by
other groups. For example, Ogale et al. (1998) reported transport  
measurements applied to $\rm La_{0.75} Ca_{0.25} Mn_{1-x} Fe_x O_3$, 
i.e., with a partial replacement of Mn by Fe, the latter being in a  
Fe$^{3+}$ state. In this case, just a 4\% Fe doping (x=0.04) leads to
an instability of the low-temperature ferromagnetic metallic phase
of the x=0.0 compound toward an insulating phase. The results for the
resistivity vs temperature are shown in Fig.IV.a.2. The shape of these
curves is quite similar to the results observed in other compounds,
such as those studied by Uehara et al. (1999), and they are
suggestive of a percolative process leading eventually to a fully
insulating state as x grows. Note the similarities of these curves with
the theoretical calculations shown in Figs.III.h.2 and 4.

\medskip
\noindent{\bf Small-Angle Neutron Scattering:}

Small angle neutron scattering combined with magnetic susceptibility 
and volume thermal expansion measurements by De Teresa et al. (1997b)
(see also Ibarra and De Teresa, 1998a) applied to $\LCMO$ with x=1/3
provided evidence for 
small magnetic clusters of size 12 
$\rm \AA$ above $T_{\rm C}$.  
Although to study their data De Teresa et al. (1997b) used the simple
picture of small lattice/magneto polarons available by the time of
their analysis, by now it is apparent that individual small polarons
may not be sufficient to describe the physics of manganites near the
Curie temperature.
Nevertheless, leaving aside these interpretations, the very important
results of De Teresa et al. (1997b) clearly experimentally showed the 
presence of an inhomogeneous state above $T_{\rm C}$ early in the study of
manganese oxides.
The coexisting clusters were found to grow in size
with a magnetic field and decrease in number.
Ibarra and De Teresa (1998c), have reviewed their results and concluded
that electronic phase segregation in manganites emerges from their
data.
Even percolative characteristics were assigned by Ibarra and De Teresa
(1998a) to the metal-insulator transition, in excellent agreement
with theoretical calculations
(Moreo et al., 2000; Mayr et al., 2000).
Hints of the mixed-phase
picture (involving FM clusters larger than the size of a single ferro
polaron) are also contained in the comment on De Teresa et al.'s
results presented by Goodenough and Zhou (1997). 

Using neutron diffraction, muon-spin relaxation, and magnetic
techniques, studies of $\rm (La_{1-x} Tb_x)_{2/3} Ca_{1/3} Mn O_3$
were also reported by De Teresa et al. (1996,1997a). At low
temperatures, an evolution from the FM metallic state at x=0 to the
antiferromagnetic insulating state at x=1 was reported, involving an
intermediate regime between x=0.33 and x=0.75 with spin-glass
insulating characteristics. The phase diagram is in Fig.IV.a.3a.
Static local fields randomly oriented were identified at, e.g.,
x=0.33. No long-range ferromagnetism was found in the intermediate
density regime. In view of the recent theoretical and experimental
reports of giant cluster coexistence in several manganites, it is
natural to conjecture the presence of similar phenomena in the studies
of De Teresa et al. (1996,1997a). In fact, the plots of resistance
versus temperature (see Fig.IV.a.3b, taken from Blasco et al., 1996)
between x=0.0 and 0.5 have a shape very similar to those found in
other manganites that were described using percolative ideas, such as 
$\rm La_{5/8-y} Pr_y Ca_{3/8} Mn O_3$ (Uehara et al, 1999).

Another interesting aspect of the physics of manganites that has been
emphasized by Ibarra et al. (1995), Ibarra and De Teresa (1998c) and
others, is the presence in the paramagnetic regime above $T_{\rm C}$
of a large contribution to the volume thermal expansion that cannot be
explained by the Gr\"uneisen law.
Those authors assign this extra contribution to polaron
formation. Moreover, the results for the thermal expansion
vs. temperature corresponding to several manganites at x approximately
0.30 can be collapsed into a universal curve (Fig.57 of Ibarra and De
Teresa, 1998c) showing that the phenomenon is common to all compounds
even if they have different Curie temperatures. Above $T_{\rm C}$, there is
a coexistence of a high-volume region associated with localized
carriers and a low-volume region associated with delocalized carriers.
The spontaneous or field induced metal to insulator transition is
associated with a low-volume to high-volume transition. From this
analysis it was concluded that there are two states in close
competition and that the transition should be of first-order, 
in excellent agreement with the recent simulations of Yunoki, Hotta
and Dagotto (2000). 

The analysis of elastic neutron scattering experiments by Hennion et
al. (1998) (see also Moussa et al., 1999) has provided very useful
information on the behavior of $\LCMO$ at low values of x. 
While previous work by the same authors (Hennion et al., 1997) was
interpreted using a description in terms of simple magnetic polarons,
Hennion et al. (1998) reinterpreted their results as arising from a
liquid-like spatial distribution of magnetic $droplets$. 
The radius of these droplets was estimated to be 9 $\rm \AA$ and
their number was found to be substantially  smaller than the number of
holes (ratio droplets/holes = 1/60 for x=0.08), leading to a possible
picture of hole-rich droplets within a hole-poor medium, if spin
polarized regions are induced by carriers. 
Note the use of the word droplet instead of polaron in this context: 
polarons are usually associated with only one carrier, while droplets
can contain several. It is quite remarkable that recent analysis by
the same group (Hennion et al., 1999) of the compound $\LSMO$ at
x=0.06 has lead to very similar results: ferromagnetic clusters were
found in this ``large'' bandwidth manganite and the number of these
clusters is larger by a factor 25 than the number of holes. Hennion et
al. (1999) concluded that phase separation between hole-rich and
hole-poor regions is a general feature of the low doping state of
manganites. These authors believe that this phenomenon likely occurs
even at higher concentrations close to the metal-insulator transition.

\medskip
\noindent{\bf Neutron Scattering:}

Early in the study of manganites,
results of neutron scattering experiments on $\LCMO$ for a wide range
of compositions were interpreted by Lynn et al. (1996,1997) in terms
of a competition between ferromagnetic metallic and paramagnetic
insulating states, leading to a state consisting of two coexisting
phases. The relative fraction of these two phases was believed to
change as the temperature was reduced to $T_{\rm C}$. This occurs even
at the optimal composition for ferromagnetism close to x=1/3. 
A typical result of their measurements is presented in Fig.IV.a.4a 
where the inelastic spectrum is shown at two temperatures and small
momentum transfer, for the x=1/3 compound which has a 
$T_{\rm C}$=250K. The two peaks at nonzero energy are interpreted as
spin-waves arising from the ferromagnetic regions while the central
peak is associated with the paramagnetic phase.  
Even at temperature as low as 200K the two features can be observed.
Fernandez-Baca et al. (1998) extended the analysis of Lynn et
al. (1996) to other compounds with a similar hole concentration
x$\sim$0.33. Their conclusions are very similar, i.e., a central
component near $T_{\rm C}$ is found in all the compounds studied and
those authors concluded that ``magnetism alone cannot explain the
exotic spin dynamical properties'' of manganites (Fig.IV.a.4b contains
their results for $\PSMO$ and $\NSMO$ at x$\sim$0.3). 
Even the compound $\LSMO$ at x=0.15 and 0.30 show a similar behavior
(Fig.IV.a.4c,d), in spite of the fact that Sr-based manganites are
usually associated with more conventional behavior than Ca-based ones.

Overall, the neutron scattering experimental results are in good
qualitative agreement with the conclusions reached by other
experimental techniques, such as tunneling measurements at similar
compositions which were reviewed before, and with theoretical
calculations (already reviewed in Section III).
Lynn et al. (1996) also noticed the presence of irreversibilities in
the transitions, and they remarked that these transitions are not of
second order. These early results are also in agreement with
the more recent theoretical ideas of Yunoki, Hotta, and Dagotto (2000)
and Moreo et al. (2000) where first-order transitions are crucial for
the coexistence of giant clusters of the competing phases. 

\medskip
\noindent{\bf PDF Techniques:}

Using pair-distribution-function (PDF) analysis of neutron
powder-diffraction data, Billinge et al. (1996) studied $\LCMO$ at
small and intermediate densities x. 
They explained their results in terms of lattice polaron formation
associated with the metal-insulator transition in these materials.
Below $T_{\rm C}$, Billinge et al. (1996) believe that the polarons can
be large, dynamic, and spread over more than one atomic site.
Note, however, that these authors use a  polaronic picture due to
the presence in their data of a mixture of short and long Mn-O bonds,
implying distorted and undistorted MnO$_6$ clusters. Whether the
distorted octahedra are randomly distributed, compatible with the
polaronic theory, or gathered into larger structures, compatible with
the phase separation theory, has not been analyzed.
More recent studies by Billinge et al. (1999), using the same
technique, produced the schematic phase diagram shown in Fig.IV.a.5.
Note the light shaded region inside the FM phase: in this regime
Billinge et al. (1999) believe that localized and delocalized phases
coexist. The white region indicates the only regime where an
homogeneous FM phase was found.
This result is remarkable and it illustrates the fact that the simple
double-exchange ideas, that lead to an homogeneous FM state, are valid
in $\LCMO$ in such a narrow region of parameter space that they are of
little value to describe narrow band manganites in the important CMR
regime. As remarked before, it appears that it is the $competition$
between DE and the other tendencies dominant in manganites that
produces the interesting magneto-transport properties of these
compounds. 

\medskip
\noindent{\bf X-ray Absorption, Transport and Magnetism:}

Similar conclusions as those found by Billinge et al. (1996, 1999) 
were reached by Booth et al. (1998a, 1998b) using x-ray absorption
measurements applied to $\LCMO$ at several hole concentrations.
This technique provides information about the distribution of Mn-Mn
bond lengths and the Mn-O environment. The results, obtained at
several densities, favor a picture similar to that described in the
previous subsection, namely there are two types of carriers:
localized and delocalized. The number of delocalized holes grows
exponentially with the magnetization below $T_{\rm C}$. These results clearly
show that, even in the ferromagnetic regime, there are two types of
phases in competition. In agreement with such conclusions, the
presence of large polarons below $T_{\rm C}$ at x=0.25 was also obtained by
Lanzara et al. (1998) using x-ray techniques.
Near $T_{\rm C}$ those authors believe that small and large polarons
coexist and a microscopic phase separation picture is suitable to
describe their data.

Early work using x-ray absorption for $\LCMO$ at x=0.33 by Tyson et
al. (1996) showed the presence of a complex distribution of Mn-O bond
lengths, with results interpreted as generated by small polarons. 
Hundley et al. (1995) studied the same compound using transport
techniques and, due to the observation of  exponential behavior of
the resistivity with the magnetization, they concluded that polaron
hopping could explain their data.
As remarked before, it is not surprising that early work used polaronic
pictures to analyze their results, since by that time it was the main  
theoretical possibility available for manganites.
However, Hundley et al. (1995) already noticed that the polarons could
form superlattices or domains, a conjecture that later experimental
work contained in this section showed to describe experiments more properly.

\medskip
\noindent{\bf Nuclear Magnetic Resonance:}

The coexistence of FM and AF resonances in NMR data obtained for
$\LCMO$ at several small hole densities was reported by
Allodi et al. (1997) and Allodi, De Renzi and Guidi (1998) using
ceramic samples. 
No indications of a canted phase were observed by these authors,
compatible with the conclusions of theoretical work showing that
indeed the canted phase is unstable, at least within the models
studied in Sec.III. The NMR results showing the FM-AF coexistence
contain a peak at $\sim$260 MHz which corresponds to AF, and another
one slightly above 300 MHz which is believed to be FM in origin,
according to the analysis of Allodi et al. (1997).

A study of dynamic and static magnetic properties of $\LCMO$ in the
interval between x=0.1 and 0.2 by Troyanchuk (1992) also showed
indications of a mixed-state consisting of ferromagnetic and
antiferromagnetic clusters.
Troyanchuk (1992) remarked very early on that his data was not
consistent with the canted structure of deGennes (1960). 

\medskip
\noindent{\bf Muon Spin Relaxation:}

The observation of two time scales in $\LCMO$ at x$\sim$0.3 using
zero-field muon spin relaxation were explained by Heffner et al.
(1999) in terms of a microscopically inhomogeneous FM phase below
$T_{\rm C}$, caused by the possible overlapping of growing polarons as
the temperature is reduced. 
Heffner et al. (1999) concluded that a theoretical model mixing
disorder and coupled JT-modes with the spin degrees of freedom may be
necessary to explain their results, in agreement with the more recent
theoretical calculations presented by Moreo et al. (2000) which used a
mixture of disorder and strong JT correlations.
Evidence for spatially inhomogeneous states using muon spin relaxation
methods were also discussed by the same group in early studies
(Heffner et al., 1996) where glassy spin dynamics was observed. 
Non-homogeneous states for manganites were mentioned in that work as a
possible alternative to the polaronic picture.

\medskip
\noindent{\bf Photoemission:}

Recently, Hirai et al. (2000) applied photoemission techniques to
$\LCMO$ with x=0.3, 0.4 and 0.5, measuring the photoabsorption and
magnetic circular dichroism. Interesting systematic changes in the core
level edges of Ca$2p$, O$1s$ and Mn$2p$ were observed as temperature and
stoichiometry were varied. The results were interpreted in terms of a
phase-separated state at room temperature, slightly above the Curie
temperature. The metallic regions become larger as the temperature is
reduced. These results are in excellent agreement with several other
experiments describing the physics above $T_{\rm C}$ as caused by 
a mixed-phase state, and with the theoretical calculations reviewed
in Sec.III. Based on the results of Hirai et al. (2000), it is conceivable
that photoemission experiments may play a role as important
in manganites as they do in the cuprates.

\medskip
\noindent{\bf Hall Effect:}

Recent studies of the Hall constant of $\LCMO$ at x=0.3 by 
Chun et al. (1999b) provided evidence that the picture of
independent polarons believed in earlier studies to be valid in this
compound above $T_{\rm C}$ is actually valid $only$ for temperatures
larger than 1.4$T_{\rm C}$ i.e. well above the region of main interest
from the point of view of the CMR phenomenon. In the temperature
regime between $T_{\rm C}$ and 1.4$T_{\rm C}$,  Chun et al. (1999b)
describe their results as arising from a two-phase state, with
percolative characteristics at $T_{\rm C}$.
Once again, from this study it is clear that the insulating state of
manganites above $T_{\rm C}$ is not a simple gas of independent
lattice/spin polarons
[or bipolarons, see Alexandrov and
Bratkovsky (1999)]. This is compatible with the phenomenological
two-fluid picture of localized and itinerant carriers near $T_{\rm C}$ 
which was envisioned by Jaime et al. (1996, 1999) early in the study of
manganites, and it is expected
to apply to the x=0.3 $\LCMO$ material.

\medskip
\noindent{\bf Studies with High Pressure:}

The properties of manganites are also very sensitive to pressure, as
explained in the Introduction. 
As an example, consider the results of Zhou and Goodenough (1998)
obtained analyzing $\rm (La_{0.25} Nd_{0.75} )_{0.7} Ca_{0.3} Mn O_3$ 
as a function of pressure (see also Zhou, Archibald and Goodenough, 
1996). This compound appears to have a tolerance factor slightly below
the critical value that separates the ferromagnetic regime from the
antiferromagnetic one.
While Zhou and Goodenough (1998) emphasized in their work the giant
isotope effect that they observed in this compound upon oxygen isotope 
substitution, a very interesting feature indeed, here our description
of their results will mainly focus on the resistivity vs temperature
plots at various pressures shown in Fig.IV.a.6.
In view of the recent experimental results observed in similar
materials that are also in the region of competition between FM and AF
states, it is natural to contrast the results of Fig.IV.a.6 with those
of, e.g., Uehara et al. (1999). 
Both sets of data, one parametric with pressure at fixed Nd-density
and the other (Fig.IV.a.1a) parametric with Pr-density at ambient
pressure, are  similar and also in agreement with the theoretical
calculations  (Moreo et al., 2000; Mayr et al., 2000).
The shape of the curves Fig.IV.a.6 reveal hysteretic effects as
expected in
first-order transitions, flat resistivities at low temperatures, a
rapid change of $\rho(T=0)$ with pressure, and the typical peak in
the resistivity at finite temperature that leads to CMR effects. All
those features  exist also in Fig.IV.a.1a.

Similar pressure effects in 
$\rm (La_{0.5} Nd_{0.5} )_{2/3} Ca_{1/3} Mn O_3$ were reported by
Ibarra et al. (1998b). Those authors concluded that, at low
temperatures, insulating CO and metallic FM regions coexist, and that
this is an intrinsic feature of the material. 
The interpretation of their results appears simple: a first-order
transition smeared by the intrinsic disorder in manganites can be
reached by compositional changes or by changes in the couplings
induced by pressure. But the overall physics is similar. In view of
this interpretation, it is natural to conjecture that the material
$\rm (La_{1-x} Nd_x )_{0.7} Ca_{0.3} Mn O_3$ discussed in the previous
paragraph should also contain regions with a coexistence of giant
clusters of FM and AF phases. 
Zhou and Goodenough (1998) indeed mentioned the possibility of phase
segregation between hole-rich and hole-poor regions in the
paramagnetic state, but the low temperature regime may have
mixed-phase properties as well. 
In particular, the ``canted-spin ferromagnetism'' below $T_{\rm N}$ 
reported by Zhou and  Goodenough (1998) could be induced by phase
coexistence.

\medskip
\noindent{\bf Related Work:}

Several other studies have shed light on the behavior of
ferromagnetically-optimally doped manganites.
For instance, studies of thin-films of 
$\rm La_{0.67} (Ca_{x} Sr_{1-x})_{0.33} Mn O_3$ by Broussard, Browning
and Cestone (1999a, 1999b) showed that the value of the
magnetoresistance decreases rapidly as x is reduced from 1, namely as 
the system moves from a low to a large bandwidth manganite at a fixed
hole density of 0.33. 
This interesting material should indeed present a transition from a
mixed-phase state near $T_{\rm C}$ for x=1 (all Ca), in view of the
tunneling results of F\"ath et al. (1999) and several others, to a
more standard metal at x=0 (all Sr).
In addition, Zhao et al. (1998) found a two component signal in the
pulsed laser excitation induced conductance of $\LCMO$ at x=0.3.
The results can also be interpreted as a two-phase coexistence.
Recently, Wu et al. (2000) reported the presence of colossal
electroresistance (CER) effects in $\LCMO$ with x=0.3, an
interesting effect indeed, and they attributed its presence to phase
separation tendencies. 
Kida, Hangyo, and Tonouchi (2000a) estimated the complex dielectric
constant spectrum of $\LCMO$ with x=0.3, concluding that the results
are compatible with a mixed-phase state.
Belevtsev et al. (2000) reported studies in $\LCMO$ x=0.33 films,
where upon the application of a small dose of irradiation, large
changes in the film resistivity were obtained. This is natural in a
percolative regime, where small changes can lead to important
modifications in transport.

Complementing the previous studies, recently Smolyaninova et
al. (1999) have shown that the metal-insulator transition of $\LCMO$
at x=0.33 is $not$ an Anderson localization transition, since scaling
behavior was clearly not observed in resistivity measurements of thin
films. This important study appears to rule out simple theories based
on transitions driven by magnetic disorder, such as those proposed by
M\"uller-Hartmann and Dagotto (1996), Varma (1996), and Sheng et al.
(1997). A similar conclusion was reached by Li et al. (1997) through
the calculation of density of states with random hopping (for a more
recent density-of-states and localization study of the one-orbital
model at $J_{\rm H}$=$\infty$ see Cerovski et al., 1999).
It was observed that this randomness was not sufficient to move the
mobility edge, such that at 20 or 30\% doping there was localization.
It appears that both Anderson localization and the simple picture of a
gas of independent small polarons 
are ruled out in manganites.

\begin{center}
  {\bf IV.b $\bf {La_{1-x} Ca_x Mn O_3}$ at x$\sim$0.5} 
\end{center}

After a considerable experimental effort, the evidence for mixed-phase
FM-CO characteristics in $\LCMO$ near x=0.5 is simply overwhelming. 
The current theoretical explanation of experimental data at this
density appears simple. According to computer simulations and
mean-field approximations the FM and CO phases are separated by
first-order transitions when models without disorder are studied.
This abrupt change is due to the substantial difference between these
phases that makes it difficult a smooth transition from one
to the other. Intrinsic disorder caused by the slightly different
ionic sizes of La and Ca can induce a smearing of the first-order
transition, transforming it into a continuous transition with
percolative characteristics.
Coexistence of large clusters with equal density is possible, as
described in the theoretical section of this review.
In addition, intrinsic tendencies to electronic phase separation, which
appear even without disorder, may contribute to the cluster formation.

It is important to remark that although in this review the cases of
x$<$0.5 and x$\sim$0.5 are treated separately, it is expected that a
smooth connection between the two types of mixed-phase 
behavior exists. Hopefully, future theoretical and experimental
work will clarify how the
results at, say, x$\sim$0.3 and x$\sim$0.5 can evolve one into the other
changing the hole density.

\medskip
{\bf Experimental evidence of inhomogeneities:}

Early work by Chen and Cheong (1996) and Radaelli et al. (1997) using 
electron and x-ray diffraction experiments found the surprising
coexistence of ferromagnetism and charge ordering in a narrow
temperature window of $\LCMOhalf$.
Further studies by Mori, Chen and Cheong (1998b) showed that the x=0.5
mixture of FM and CO states arises from an inhomogeneous spatial
mixture of incommensurate charge-ordered and ferromagnetic
charge-disordered microdomains, with a size of 20-30 nm.


Papavassiliou et al. (1999a,1999b) (see also Belesi et al., 2000)
observed mixed-phase tendencies in $\LCMO$ using $^{55}$Mn NMR
techniques. 
Fig.IV.b.1a shows the NMR spectra for $\LCMO$ at several densities and
low temperature $T$=3.2K obtained by those authors. The appearance of
coexisting peaks at x=0.1, 0.25, and 0.5 is clear from the figure,
and these peaks correspond to either FM metal, FM insulator, or AF
states according to the discussion presented in Papavassiliou et
al. (1999a,1999b). 
The results at x=0.5 are in agreement with previous results reported
by the same group (Papavassiliou et al., 1997). The revised phase
diagram proposed by those authors is shown in Fig.IV.b.1b. In
agreement with the conclusions of other groups, already reviewed in
the previous subsection, the region in the vicinity of $T_{\rm C}$ 
corresponds to a mixed-phase regime. The same occurs at low
temperatures in the region between the CO and FM states of x=0.5.
The coexistence of FM and AF phases was also observed in $\LCMOhalf$ 
by Allodi et al. (1998) using similar NMR techniques.
First order characteristics in the FM-AF transition were found,
including an absence of critical behavior. Their spectra is shown in
Fig.IV.b.1c. As in Fig.IV.b.1a, a clear two signal spectra is observed
in the vicinity of x=0.5 and low temperatures.
The presence of mixed-phase characteristics in NMR data was also
observed by Dho, Kim and Lee (1999a, 1999b) in their studies of
$\LCMO$. Their results apply mainly near the phase boundaries of the
ferromagnetic regime at a fixed temperature, or near $T_{\rm C}$ at a
fixed density between 0.2 and 0.5.

It can be safely concluded, overall, that the NMR results described here
are in general agreement,
and also in agreement with the phase
separation scenario which predicts that all around the FM metallic
phase in the temperature-density plane  there are regions of
mixed-phase characteristics due to the competition between
metallic and insulating
states.


Magnetization, resistivity, and specific heat data analyzed by 
Roy et al. (1998, 1999, 2000a, 2000b) led to the conclusion that in a
narrow region of hole densities centered at x=0.5, two types of carriers
coexisted: localized and free. 
The evidence for a rapid change from the FM to the CO phases as x was
varied is clear (see Fig.3 of Roy et al., 1998). This is compatible
with the fact that La$^{3+}$ and Ca$^{2+}$ have a very similar ionic
radius and, as a consequence, the disorder introduced by their mixing
is ``weak''. 
The theoretical scenario described before (Moreo et al., 2000)
suggests that, at zero temperature and for weak disorder, the density
window with large cluster coexistence should be narrow
(conversely in this region  large cluster sizes are expected).
It is also to be expected that the magnetoresistance effect for low
values of magnetic fields will appear only in the same narrow region
of densities. Actually, Roy et al. (1999) showed that at x=0.55, a field
of 9 Tesla is not enough to destabilize the charge-ordered state.
Very recently, Roy, Mitchell and Schiffer (2000a) studied, among other
quantities, the resistivity vs temperature for  magnetic fields up to
9 Tesla. The result is reproduced in Fig.IV.b.2. 
This figure clearly resembles results found by Uehara et al. (1999)
in their study of $\rm La_{5/8-y} Pr_y Ca_{3/8} Mn O_3$
(see Fig.IV.a.1a) varying the Pr concentration. In both cases, the
curves are similar to those that appear in the percolative process studied 
by Mayr et al. (2000).
Percolation between the CO and FM states appears to occur similarly
both by changing chemical compositions and also as a function of magnetic
fields, a very interesting result. 
Phase separation in x=0.5 polycrystalline
samples obtained under different thermal treatments was also reported
by Levy et al. (2000a). J. L\'opez et al. (2000) also found results
compatible with FM droplets immersed in a CO background. Kallias et
al. (1999) using magnetization and
M\"ossbauer measurements also reported coexisting FM
and AF components in x=0.5 $\LCMO$.

It is also important
to remark that experimentally it is very difficult to make reproducible
$\LCMO$ samples with x$\sim$0.5 (see for instance
Roy et al., 1998, 1999, 2000a, 2000b). Samples with the same nominal Ca
content can actually present completely different behavior. This is 
compatible with a phase separated state 
at this density, which is expected to be
very sensitive to small chemical changes. Another result compatible with
phase separation can be found in the magnetization curves (Fig.IV.b.2),
which are well below the expected saturation value for a ferromagnet,
even in several Tesla fields where the magnetization is not increasing
rapidly.


Neutron powder diffraction studies by Radaelli et al. (1995) of
$\LCMO$ at x=0.5 revealed peak broadening effects that were explained
assuming multiple phases simultaneously present in the sample.
Rhyne et al. (1998) studied $\LCMO$ with x=0.47 using elastic and
inelastic neutron scattering. Coexisting ferromagnetic and
antiferromagnetic phases were found at low temperatures. Similar
conclusions were also reached by Dai et al.(1996).
Further confirmation that near x=0.5 in $\LCMO$ the material has
mixed-phase characteristics has been recently provided by neutron
powder diffraction measurements by Huang et al. (1999). 
Discontinuous features in the results discussed by those authors also
indicate that the competing phases are likely separated by first-order
transitions in the absence of intrinsic disorder, as found in the
theoretical calculations (Yunoki, Hotta, and Dagotto, 2000).
Infrared absorption studies by Calvani et al. (1998) were also
described in terms of a phase separation scenario.

It is also very interesting to test materials with the hole density
x=0.5 but allowing for slight deviations away from the $\LCMOhalf$
chemical composition. Among these investigations are the 
transport and x-ray experiments on 
$\rm R_{1/2} Ca_{1/2} Mn_{0.97} Cr_{0.03}  O_3$, with 
R=La, Nd, Sm and Eu, carried out by Moritomo et al. (1999).
Their study allowed for a systematic analysis of the charge-ordered
state when the ionic radius of the  rare-earth ion was changed. Due to
the small presence of Cr, this material with R=La has a purely
ferromagnetic state while the other rare-earths leads to a CE-type CO
state. The main result obtained by Moritomo et al. (1999) is quite
relevant to the subject of this review and is summarized in
Fig.IV.b.3a. Moritomo et al. (1999) concluded that the region between
the CO and FM phases has mixed-phase characteristics involving the two
competing states. This hypothesis was confirmed by the use of electron
microscopy, which showed microdomains of size 20 to 50 nm, a result 
similar to those observed by other authors in other compounds.
Then, once again, mixed-phase tendencies are clear in materials with
x=0.5 (see also Oshima et al., 2000). Results for the
Fe-doped x=0.5 LCMO compound by Levy et al. (2000b) 
likely can be rationalized in a similar way.

Moreover, the study of Cr-doped compounds at many Ca densities shows
that this type of doping with impurities has an effect similar to
that of a magnetic field, namely a small Cr percentage is enough to
destabilize the CO-state into a FM-state. This result is surprising,
since impurities are usually associated with a tendency to
localize charge, and they are not expected to generate a
metallic state. In Fig.IV.b.4a-c, the phase diagrams presented by
Katsufuji et al. (1999) for three compounds are shown to illustrate
this point. In $\PCMO$, Cr-doping destabilizes the CO-state in a wide
range of densities, as a magnetic field does, while for $\LCMO$ and
$\NSMO$, it is effective only near x=0.5. The resistivity plots in
Fig.IV.b.4d show that the shape of the curves are very similar to all
the previous ones analyzed in this review, indicative of a percolative
process.

Results similar to those of Moritomo et al. (1999) were obtained using 
transport techniques by Mallik et al. (1998) studying 
$\rm La_{0.5} Ca_{0.5-x} Ba_x Mn O_3$ with x between 0, where the
sample they used is in a charge-ordered insulating state, and x=0.5,
where a ferromagnetic metallic compound is obtained.
The resistivity vs temperature at several compositions is shown in
Fig.IV.b.3b. The results certainly resemble those obtained by Uehara
et al.(1999) and other authors, especially regarding the presence of a
flat resistivity  in a substantial low temperature range and a rapid
variation of $\rho(T=0)$ with Ba concentration. Mallik et al. (1998)
observed that the difference in ionic sizes between Ca and Ba plays a
crucial role in understanding the properties of this compound.
They also found  hysteretic behavior and first-order characteristics
in their results, results all compatible with the theoretical scenario
described before
(Yunoki, Hotta, and Dagotto, 2000; Moreo et al., 2000).
The critical concentration for percolation in Fig.IV.b.3b appears to
be near x=0.1, where the $T_{\rm C}$ was found to be the smallest in
this compound. 

Finally, it is also interesting to remark that magnetic field
dependent optical conductivity studies by Jung et al. (1999) applied
to $\NSMO$ at x=0.5 have also found indications of a percolative
transition in the melting of the charge ordered state.

\begin{center}
  {\bf IV.c Electron-Doped Manganites}
\end{center}

Neutron scattering studies of $\BCMO$ single crystals in the range
between x=0.74 and x=0.82 were presented by Bao et al. (1997). 
It is expected that $\BCMO$ will have properties very similar to those 
of $\LCMO$ in the range of densities studied by those authors, and for 
this reason the analysis of a Bi-based compound is discussed in this 
subsection.
One of the most interesting results reported by Bao et al. (1997) is 
the presence of ferromagnetic correlations at high temperatures, which 
are replaced by antiferromagnetic ones as the temperature is reduced.
Fig.IV.c.1a taken from Bao et al. (1997) show the intensity of the FM
and AF peaks as a function of temperature at x=0.82. It is clear from
the figure that in the intermediate regime, roughly between 150K and
200K, there is a coexistence of FM and AF features, as in a
mixed-phase state. In a related study, Bao et al. (1998) concluded
that manganites have only two important generic states: metallic
ferromagnetic and localized antiferromagnetic. This is in agreement
with theoretical results, although certainly combinations such
as charge-ordered ferromagnetic states are also possible at least in 2D
(Yunoki, Hotta, and Dagotto, 2000).
Subsequent studies of $\BCMO$ single-crystals performed by 
Liu et al. (1998) reported optical reflectivity results in the same
compositional range (i.e. between x=0.74 and 0.82). The main result of
this effort is reproduced in Fig.IV.c.1b. Liu et al. (1998) concluded
that in the intermediate range $T_{\rm N}$$<$T$<$$T_{\rm CO}$ the
coexistence of a polaron-like response together with a charge-gap
structure signifies two-phase behavior characterized by domains of
both FM and AF spin correlations. Recently, studies of $\BCMO$ at
x=0.81 and 0.82 were interpreted in terms of spin or charge 
``chiral'' fluctuations
(Yoon et al., 2000), showing that exotic physics may occur in
this electron doped compound. 

The range of hole densities above 0.8 for $\BCMO$ was analyzed by
Chiba et al. (1996) using magnetic and transport techniques.  
They observed that large magnetoresistance effects are found even at
low $T_{\rm C}$, which is compatible with a mixed-phase state in the
ferromagnetic regime, quite different from a spin-canted state.
Actually, it is important to remark that there are previous studies of 
the electron-doped materials (not reviewed here) that have labeled the 
small x region as ``spin-canted'' due to the observation of coexisting 
FM and AF features. 
The conclusions of those papers may need revision in view of the new
results described in this section.

Studies of $\rm Ca_{1-y} Sm_y Mn O_3$ by Maignan et al. (1998), using
magnetic and transport techniques in the range from y=0.0 to y=0.12,
reported results compatible with a ``cluster glass'' (see also Martin
et al., 1999). As y increases from zero, the system rapidly becomes
ferromagnetic and metallic. 
However, those authors remark that no true long-range order exists,
and thus the FM state is unusual. The resistivity is shown 
in Fig.IV.c.2a. The metallic character at y=0.0 and high temperature
is caused by oxygen deficiency and should not be considered as really
representing the electron undoped compound, which is actually
antiferromagnetic (G-type).

More recently, a careful and systematic study of 
$\rm Ca_{1-x} La_x Mn O_3$ has been carried out by Neumeier and Cohn 
(2000) using magnetic and transport techniques. These authors
concluded that the addition of electrons to the x=0.0
antiferromagnetic state promotes phase segregation.
Representative magnetization versus temperature data are shown in
Fig.IV.c.2b. The saturated moment and conductivity versus density are
reproduced in Fig.IV.c.2c.
Neumeier and Cohn (2000) reported multiple magnetic phases emerging 
from the analysis of their data, and remarked that the long-accepted
existence of canted AF is supplanted by phase coexistence.

In addition, recent NMR studies of $\LCMO$ for x=0.65 at low
temperature by Kapusta et al. (2000) reported the existence of
electronic phase separation, with FM regions detected over a CO/AF
background. This interesting result leads us to believe that it may be
possible that the widely accepted phase diagram of $\LCMO$ may still
need further revision, since a phase with coexisting FM and AF
features may exist at low temperature and x around 0.65, with a shape
similar to the ``canted state'' that appears in the phase diagram of 
$\PCMO$ and the bilayer compounds (see Figs.II.c.1 and IV.f.1).
This conjecture could be tested experimentally with NMR techniques.

\begin{center}
  {\bf IV.d Large Bandwidth Manganites and Inhomogeneities:
    The case of $\bf {La_{1-x} Sr_x Mn O_3}$}
\end{center}

A compound as much scrutinized as the Ca-based manganites of the
previous sections is the Sr-based $\LSMO$, which has a larger
bandwidth. In spite of this property, the $\LSMO$ material
presents a very complex phase diagram, especially at low Sr-density,
with a behavior in many respects qualitatively similar to that of the
Ca-based compound. The main experimental evidence that leads to this
conclusion is reviewed below. In the other regime of large densities, 
the Sr-based material is metallic $both$ at low and high temperatures
(see phase diagram Fig.II.a.1b) and its magnetoresistance effect is
relatively small. 
In this density regime, studies using mainly dynamical mean-field
approaches ($D$=$\infty$) have provided evidence that the simple
double exchange ideas are enough to understand the main properties of
$\LSMO$ (Furukawa, 1994, 1995a, 1995b, 1995c, 1998), especially
concerning the interplay between ferromagnetism and transport.
This is a reasonable conclusion, and illustrates the fact that
materials whose couplings and densities locate them in
parameter space far away from insulating instabilities tend to present
canonical properties. 
A review of the status of the theoretical approach based on the
double-exchange ideas and its application to large bandwidth
manganites has been recently presented (Furukawa, 1998).
Additional results for the FM Kondo model have been discussed by Zang
et al. (1997), and several other authors.
However, it must be kept in mind that the more canonical, and
governed by double-exchange, the behavior of a compound is, the smaller
is the magnetoresistance effect.
For this reason, in the description of experimental results for
$\LSMO$ the effort is here mainly focused into the low-density regime
where effects other than canonical double-exchange seem to dominate in
this material.

\medskip
\noindent{\bf $\bf {La_{1-x} Sr_x Mn O_3}$ at low density:}

Among the first papers to report inhomogeneities in Sr-based
manganites are those based on atomic pair-density-functional (PDF) 
techniques. In particular, Louca et al. (1997) studied $\LSMO$ in a
wide range of densities between x=0.0 and x=0.4, and interpreted their 
results as indicative of small one-site polarons in the paramagnetic
insulating phase. Those authors found that the local atomic structure
deviates significantly from the average.
At lower temperatures their polarons increase in size, typically
involving three sites according to their analysis. 
These effects were found even in the metallic phase.
Based on such results, Louca et al. (1997) questioned the at-that-time
prevailing homogeneous picture of the metallic state of manganites, 
and based their analysis mainly on a small polaron picture rather than
large droplets or phase separation ideas. Nevertheless they
envisioned that increasing the density of polarons would lead to larger
structures, and in more recent work (Louca and Egami, 1999) they also
presented microscopic separation of charge-rich and charge-poor regions
as a possible scenario to describe their results.
In addition, they conjectured that the conductivity could be
determined by some kind of dynamic percolative mechanism, which is
the current prevailing view
(see also Egami, 1996, and Egami et al., 1997).
The possible percolative nature of the metal-insulator transition
close to x$_{\rm c}$=0.16 in $\LSMO$ was also proposed
by Egami and Louca (1998). Tendencies toward a two-phase regime in
low hole-density doped (La,Sr)-based manganites were also reported by
Demin, Koroleva and Balbashov (1999) using a variety of techniques.

Recently, Endoh et al. (1999a, 1999b) and Nojiri et al. (1999), using
transport and resonant x-ray scattering, have studied in detail the
region near x$\sim$1/8 of $\LSMO$. Interesting results were
observed in this regime, especially a first-order transition from a
ferromagnetic metal to a ferromagnetic insulator. 
This ferromagnetic insulator was reported in previous work by Yamada
et al. (1996) using neutron scattering techniques. Those authors
interpreted their results using a state with charge ordering, which
they refer to as ``polaron ordering'' with polarons involving only one
site [note, however, that other authors could not reproduce
Yamada et al. (1996)'s results. See Vasiliu-Doloc et al. (1998a)].
Endoh et al. (1999a,1999b) reported huge changes in resistivity upon
the application of a magnetic field close to the above metal-insulator 
transition in this compound. Regions with phase-separation
characteristics were identified by Endoh et al. (1999b). The key
difference between the two competing states is the orbital ordering,
as revealed by the x-ray experiments. The reported phase diagram is in
Fig.IV.d.1a. Similar conclusions were reached by Paraskevopoulos et
al. (2000a) and previously by Zhou et al. (1997) through measurements
of resistivity and thermoelectric power. 
The last authors reported a dynamic phase segregation into hole-rich
and hole-poor phases in the region of x=0.12 between the
charge-ordered transition temperature and the Curie temperature.
Their phase diagram resembles that of Endoh et al. (1999b)
(see Fig.IV.d.1b). Overall, these experimental results are in good
agreement with mean-field calculations using purely Coulombic models
(Endoh et al., 1999a) and with Monte Carlo simulations using JT
phonons (Yunoki et al., 1998b).
In both cases, phase separation triggered by the $orbital$ degree of
freedom, instead of the spin, were found. It is clear once again that 
simple Double-Exchange ideas or even the  proposal of small
polarons are not sufficient to explain the physics of manganites,
particularly in the most interesting regions of parameter space where
the CMR effect occurs.

The results of Endoh et al. (1999a,1999b) and Nojiri et al. (1999)
have characteristics similar to those of the theoretical scenario described in
Sec.III, namely a competition between two states which are
sufficiently different to generate a first-order transition between
them. The results of Moreo et al. (2000) suggest that the small ionic
radii differences between La$^{3+}$ and Sr$^{2+}$ induces weak
disorder that affects the first-order transition, inducing a narrow
region of coexistence of cluster of both phases.
Percolative properties are predicted in this regime based on the
results of Moreo et al. (2000). It would be quite interesting to
search for such properties in x$\sim$1/8 $\LSMO$ experiments.

In fact the theoretical calculations are already in qualitative
agreement with a recent experimental effort. 
Independent of the previously described results by Endoh et
al. (1999a,1999b) and Nojiri et al. (1999), Kiryukhin et al. (1999)
studied x=1/8 $\LSMO$ using synchrotron x-ray scattering. At low
temperatures, they observed an x-ray-induced transition from a
charge-ordered phase to a charge-disordered state. These results are
qualitatively similar to those reported by Kiryukhin et al. (1997)
applied to $\PCMO$. 
Kiryukhin et al.(1999) suggest that their results can be explained
within a phase separation scenario with charge-ordered regions as
large as 500 $\rm \AA$, sizes similar to those observed in
half-doped $\LCMO$, as described before in this review 
(see also Baran et al., 1999).
Wagner et al. (1999), using transport and magnetic techniques applied 
to x=1/8 $\LSMO$, also found evidence of a first-order transition
as a function of temperature. The possibility of phase separation was
briefly mentioned in that work. 
Finally, the optical conductivity spectra obtained by Jung et al.(1998)
in their study of x=1/8 $\LSMO$ has also been explained in terms of a
phase separated picture by comparing results with those of 
Yunoki et al. (1998b), which were obtained at temperatures such that
dynamical clustering was present in the Monte Carlo simulations. It is
interesting to remark that a large number of optical experiments have
been analyzed in the near past as arising from coexisting metallic
(Drude) peaks and mid-infrared bands that were usually assigned to 
polaronic features (see for instance Kaplan et al., 1996).
In view of the novel experimental evidence pointing toward coexisting
metallic and insulating clusters, even in optimal regimes for FM such
as x=0.33 in $\LCMO$, the previous optical conductivity may admit
other interpretations perhaps replacing polarons by larger droplets.
Finally, note that recent optical studies at x=0.175 by Takenaka,
Sawaki and Sugai (1999) have been interpreted as arising from a FM
metallic phase below $T_{\rm C}$ which can have either coherent or
incoherent characteristics, and a mixture of them is possible.
The anomalous metallic state of Sr-doped manganites has been
theoretically addressed recently by Ferrari and Rozenberg (1999) using
Dynamical Mean Field calculations. Motome and Imada (1999) and
Nakano, Motome, and Imada (2000)
also studied this material and concluded that to reproduce the small
Drude weight of experiments a mixture of strong Coulomb and
electron-phonon(JT) interactions is needed. 

For completeness, some remarks about related compounds are  
here included. For instance, $\rm La Mn O_{3+\delta}$ was studied
(see Ritter et al., 1997, and Ibarra and De Teresa, 
1998b) and at $\delta$$\sim$0.15 a large magnetoresistance effect was
observed. 
The magnetic and transport properties of 
$\rm La_{1-\delta} Mn O_3$ were analyzed by De Brion et al. (1998).
In their study, they concluded that a canted state was observed, but
magnetization measurements cannot distinguish between FM-AF phase
separation and spin canting.
In fact, recent studies by Loshkareva et al. (2000, 1999) of optical,
electrical, and magnetic properties of the same compound and x=0.1 
$\LSMO$ were interpreted in terms of phase-separation. 
In addition, ``cluster-glass'' features were reported for this compound 
by Ghivelder et al. (1999). On the other hand, susceptibility, magnetization,
MR and ultrasonic studies of $\LSMO$ at low doping x$<$0.1
by Paraskevopoulos et
al. (2000b) where interpreted as compatible with a canted state, rather
than a phase-separated state (see also Pimenov et al., 2000, and
Mukhin et al., 2000). 
However, those authors remark that the canting
does not arise from DE interactions because the carriers are localized 
near the Sr-ions. These trapped holes can polarize the Mn ions in their
vicinity leading to FM clusters in a PM matrix. This interesting proposal
merits theoretical studies. It is safe to conclude that at very low 
hole density in $\LSMO$ it is still unclear what kind of state dominates
the low temperature behavior, namely whether it is
homogeneous (canted) or inhomogeneous as predicted by phase separation
scenarios. 

\medskip
\noindent{\bf $\bf {La_{1-x} Sr_x Mn O_3}$ at intermediate density:} 

Although some features of $\LSMO$ at intermediate densities are
well-described by the double-exchange ideas, experiments have revealed 
mixed-phase tendencies in this region if the study is carried out
close to instabilities of the FM metallic phase. 
For instance, working at x=0.17 in $\LSMO$, Darling et al. (1998)
reported measurements of the elastic moduli using resonant ultrasound 
spectroscopy. Those authors noticed that their results suggest the
existence of very small microstructures in their single crystals.
Studies by Tkachuk et al. (1998) of 
$\rm La_{0.83} Sr_{0.17} Mn_{0.98} Fe_{0.02} O_3$ also led to the
conclusion that the paramagnetic phase contains ferromagnetic
clusters. Recent ESR studies by Ivanshin et al. (2000) have also
contributed interesting information to the study of $\LSMO$ at hole 
densities between x=0.00 and 0.20. 
Small-Angle polarized neutron scattering measurements by 
Viret et al. (1998) for $\LSMO$ at x=0.25 indicated the presence of 
nanometer size inhomogeneities of magnetic origin in the vicinity of
the Curie temperature. Approximately at this density occurs the 
metal-insulator transition above $T_{\rm C}$, and as a consequence,
mixed-phase features as observed in $\LCMO$ (which at all densities
presents an insulating state above $T_{\rm C}$) are to be expected below
x=0.25. Machida, Moritomo, and Nakamura (1998) studied the absorption
spectra of thin-films of $\rm R_{0.6} Sr_{0.4} Mn O_3$ 
with R=Sm, $\rm (La_{0.5} Nd_{0.5})$, $\rm (Nd_{0.5} Sm_{0.5})$, 
and $\rm (Nd_{0.25} Sm_{0.75})$. They concluded that cluster states
were formed in these compounds. 

\medskip
\noindent{\bf Sr-based Compounds at High Hole Density: The cases of 
  $\bf{Pr_{1-x} Sr_x Mn O_3}$ and $\bf {Nd_{1-x} Sr_x Mn O_3}$}

The antiferromagnetic manganite $\PSMO$ at x=0.5 has been recently
studied using NMR techniques by Allodi et al. (1999). 
This material has a magnetic field induced transition to a
ferromagnetic state and a CMR effect. 
The NMR results show that the transition proceeds through the
nucleation of microscopic ferromagnetic domains, with percolative 
characteristics. Allodi et al. (1999) believe that the size of the
clusters in coexistence is on the nanometer scale, to be compared with 
the micrometer scale found in other manganites.

Kajimoto et al. (1999) studied $\NSMO$ in a range of densities from
x=0.49 to x=0.75 using neutron diffraction techniques. 
Four states were observed: FM metallic, CE-type insulating, A-type
metallic, and a C-type AF insulator. The latter may be charge-ordered.
At x$\sim$0.5, Kajimoto et al. (1999) reported a possible mixed-phase 
state involving the CE-type and A-type orderings. Other groups arrived
at similar conclusions:
Woodward et al. (1999) found coexisting macroscopic FM, A-type and
CE-type phases, while Fukumoto et al. (1999) reported microscopic scale 
electronic phase separation in this compound. 
All these results are compatible with the recent theoretical work of
Moreo et al. (2000) and Yunoki et al. (2000), since computer
simulations of models with JT phonons at x=0.5 have found first-order
transitions separating the many possible states in manganites,
including one between the A-type to CE-type states. The addition of 
weak disorder would smear this sharp first-order transition into a
rapid crossover. CMR effects are to be expected in this regime.

\begin{center}
  {\bf IV.e $\bf {Pr_{1-x} Ca_x Mn O_3}$ }
\end{center}

It is interesting to observe that the low-bandwidth compound $\PCMO$
with x=0.30 undergoes an unusual insulator-metal transition when it is
exposed to an x-ray beam. Without x-rays, the material is in a
charge-ordered insulator state below 200K.
However, below 40K, x-rays convert the insulating state into a metallic
state which persists when the x-ray beam is switched off
(Kiryukhin et al., 1997; Cox et al., 1998).
A similar transition occurs upon the application of a magnetic
field. The authors of these experiments interpreted their results as
arising from a phase-segregation phenomenon induced by the x-rays,
with ferromagnetic droplets coalescing into larger aggregates.
Note that x=0.30 is at the border between the CO-state and a
FM-insulating state in this compound, and thus unusual behavior is to
be expected in such a regime. Recently, transport, optical and specific
heat results at x=0.28 by Hemberger et al. (2000a, 2000b) have been
interpreted as a percolative metal-insulator transition induced by a
magnetic field, with coexisting metallic and insulating clusters
below 100K at zero external field. 
Using neutron diffraction techniques applied to x=0.3 
PCMO, Katano, Fernandez-Baca and Yamada (2000) recently found evidence
of a phase-separated state with percolative characteristics in the
metal-insulator transition induced by magnetic fields.

Recent analysis, again using x-rays, of the related material 
$\rm Pr_{1-x} (Ca_{1-y} Sr_y)_{x} Mn O_3$ showed that the
metal-insulator transition present in this compound is not caused by a
conventional change in the electron density, but by a change in the
couplings of the system which affect the mobility of the carriers
(Casa et al., 1999). It is believed that the x-rays can help
connecting adjacent preformed metallic clusters which originally are
separated by an insulating barrier. 
In other words, the picture is similar to that of the
percolation process described in other manganites and also 
in the theoretical analysis of the influence of a magnetic field on,
e.g., the random field Ising model as a toy model for cluster
coexistence near first-order transitions (Moreo et al., 2000).

Studies of thermal relaxation effects by Anane et al. (1999a) applied 
to $\PCMO$ with x=0.33 are also in agreement with a mixed-phase
tendency and percolative characteristics description of this
compound. Anane et al. (1999a) focused their effort into the
hysteresis region that separates the metallic and insulating phases
upon the application of a magnetic field.
More recently, Anane et al. (1999b) studied the low frequency
electrical noise for the same compound, at similar temperatures and
fields. Their conclusion is once again that mixed-phase behavior and
percolation are characteristics of this material. 
More recently, Raquet et al. (2000), studying $\LCMO$ (x=0.33), observed 
a giant and random telegraph noise in the resistance fluctuations of
this compound. They attribute the origin of this effect to a dynamic 
mixed-phase percolative conduction process involving two phases with
different conductivities and magnetizations.
These important experimental results are compatible with the
theoretical expectations described earlier: if it were possible to 
switch off the intrinsic disorder of manganites, the transition would 
be first-order with more standard hysteresis effects
(Yunoki et al., 2000, Moreo et al., 2000). 
But the influence of intrinsic disorder produces a distribution of
critical fields which causes mixed-phase characteristics, which
themselves induce colossal relaxation effects.

Oxygen isotope substitution on a material at the verge of a
metal-insulator transition, such as 
$\rm (La_{0.25} Pr_{0.75})_{0.7} Ca_{0.3} Mn O_3$, 
leads to indications of phase segregation involving AF-insulating and
FM-metallic phases according to neutron powder diffraction studies by
Balagurov et al. (1999)
(see also Babushkina et al., 1998; Voloshin et al., 2000).
The results for the resistivity versus temperature shown in those
papers are quite similar to those observed in other materials where
percolation seems to occur.
Then, once again it is observed that near a metal-insulator transition 
it is easy to alter the balance by small changes in the composition. 

Finally, neutron scattering studies of $\PCMO$ by Kajimoto et
al. (1998) have shown that in the temperature regime between 
$T_{\rm CO}$ and $T_{\rm N}$, ferromagnetic spin fluctuations have
been observed. In addition, antiferromagnetic fluctuations appear to
be present also in the same temperature regime (see Fig.2 of Kajimoto
et al. (1998)), and thus a coexistence of FM and AF correlations exist
in a finite window of temperatures. 
This result is similar to that observed in the same temperature window 
$T_{\rm N}$$<$T$<$$T_{\rm CO}$ for $\BCMO$ with large x 
(see Bao et al. (1997), Liu et al. (1998)), and adds to the mixed-phase
tendencies of these compounds.
Very recently, neutron diffraction and inelastic neutron scattering
results by Radaelli et al. (2000) obtained in $\PCMO$ (x=0.30)
indicated mesoscopic and microscopic phase segregation at different
temperatures and magnetic fields.

\begin{center}
  {\bf IV.f Mixed-Phase Tendencies in Bilayered Manganites:}
\end{center}

Early neutron scattering experiments by Perring et al. (1997) reported
the presence of long-lived antiferromagnetic clusters coexisting with
ferromagnetic critical fluctuations in $\bilayered(1.8)$, which has a
nominal hole density of x=0.4.  
Figure IV.f.1a contains the intensity of their signal vs momenta. The
peaks at 0.0, 1.0 and 2.0 in the horizontal axis correspond to
ferromagnetism. The relatively small peak at 0.5 corresponds to an  
antiferromagnetic signal. 
In view of their results, Perring et al. (1997) concluded that a 
simple mean-field approach where a given typical site interacts with  
other typical sites cannot be valid in the bilayered material, a
conclusion that the authors of this review fully agree with.
Note, however, that other authors disagree with the mixed-phase
interpretation of the neutron results [see Millis (1998), and the
reply contained in Perring, Aeppli, and Tokura (1998)] and with the
data itself [Osborn et al. (1998) believes that the AF signal is
smaller than it appears in Fig.IV.f.1a, although they agree with
the notion that FM and AF interactions are finely balanced in this
compound]. Nevertheless, regardless of the actual intensities and in
view of the overwhelming amount of experimental information pointing
toward mixed-phase tendencies in 3D manganites, these authors believe
that Perring et al. (1997) have provided reasonable evidence that
bilayers could also support mixed-phase states.

Kubota et al. (1999a) studying the x=0.5 bilayered manganite, concluded 
that here the CE-type insulating and the A-type metallic phases
coexist. Battle et al. (1996a, 1996b) and Argyriou et al. (2000)
arrived at similar conclusions. This is qualitatively compatible with
the Monte Carlo simulation results described in Sec.III that showed
first-order transitions between many phases in the limit of a large
electron-phonon coupling. In particular, in Section III it was shown,
based on theoretical calculations, that the A-type and CE-type 
phases are in competition, and their states cross as a function of the 
$t_{\rm 2g}$ spin coupling $J_{\rm AF}$
(Yunoki, Hotta, and Dagotto, 2000).
Weak disorder transforms the first-order transition into a second 
order one with cluster coexistence in the vicinity of the critical  
point. This is an interesting detail that deserves to be reemphasized: 
the phenomenon of mixed-phase formation and percolation is expected to 
occur whenever a first-order transition separates two competing states, 
and whenever some sort of disorder affects the system. There is $no$
need for one of the phases to be the 3D FM metallic state, which
usually appears prominently in materials that show the CMR effect in
manganites.
This also shows that the DE mechanism is not needed to have a large
magnetoresistance. This is in agreement with the conclusions of the
work by Hur et al. (1998), where CMR effects for x=0.3 bilayered
manganites were presented even $without$ long-range ferromagnetism.
Hur et al. (1998) discussed the possibility of nonhomogeneous states
at low temperature. Chauvet et al. (1998), using ESR techniques applied
to the x=0.325 bilayered system, also arrived at the conclusion that
polarons or mixed-phase tendencies are possible in this compound.

Based on powder neutron-diffraction studies for bilayered manganites
in a wide range of densities, Kubota et al. (1999b, 1999c) reported
the phase diagram shown in Fig.IV.f.1b (see also Hirota et al., 1998.
For results at x larger than 0.5 see Ling et al., 2000). 
The AFM-I and II phases are A-type AF phases with different spin
periodicities along the direction perpendicular to the FM planes.
The FM-I and II phases are ferromagnetic states with the spins
pointing in different directions
(for more details see Kubota et al. (1999b)).
For our purposes, the region of main interest is the one labelled as 
``Canted AFM'' which arises from the coexistence of AF and FM features
in the neutron-diffraction signal. However, as repeatedly stressed in
this review, a canted state is indistinguishable from a mixed FM-AF
phase if the experimental techniques used average over the sample (see
also Battle et al., 1999, and reply by Hirota et al., 1999).
Further work, such as NMR studies, is needed to address the canted
vs. mixed-phase microscopic nature of this state. Such a study would
be important for clarifying these matters. Since the
neutron-scattering peaks observed by Kubota et al. (1999b, 1999c) are
sharp, the FM and AF clusters, if they exist, will be very large
as in other manganites that have shown a giant cluster coexistence.
The resistivity vs temperature of x=0.40 and 0.45 already show
features (Kubota, 1999d) somewhat similar to those that appeared in
related experiments, namely dirty metallic behavior at low temperature
with a $\rho$($T$$\sim$0) increasing as x grows toward 0.5
(insulating phase). Very recently, Tokunaga et al. (2000) have observed with
magneto-optical measurements  a spatial variation of the
magnetization in the region of ``spin canting''. Those authors produced
clear images of the x=0.45 bilayer compound, and also of $\PCMO$ at
x=0.30, showing domains with typical length scale exceeding one
micrometer. 
Tokunaga et al. (2000) concluded that phase separation occurs
in the region that neutron scattering experiments labeled before as spin
canted,
in excellent agreement with the theoretical calculations (on the other
hand, above $T_{\rm C}$ Osborn et al. 1998 reported the presence of canted
spin correlations).
In addition, Zhou, Goodenough, and Mitchell (1998) also believe that the
x=0.4 compound has polaron formation that condenses into clusters as
the temperature is reduced. Also Vasiliu-Doloc et al. (1999), using 
x-ray and neutron scattering measurements for the x=0.4 bilayered
manganite, concluded that there are polarons above $T_{\rm C}$ (see also
Argyriou et al., 1999).
More recently, Campbell et al. (2000) found indications of micro-phase
separation on the x=0.4 bilayer compound based on neutron scattering results.
Chun et al. (2000) reported a spin-glass behavior at x=0.4 which is
interpreted as caused by FM-AF phase-separation tendencies.

The x=0.4 low-temperature phase of double-layer manganites, which
appears to be a metal according to Figs.II.e.2-3, can be transformed
into a charge-ordered state by chemical substitution using
$\rm (La_{1-z} Nd_z )_{1.2} Sr_{1.8} Mn_2 O_7$.
Data for several z's are shown in Fig.IV.f.2a.
The shape of the $\rho_{\rm dc}$ vs temperature curves resemble results
found for other materials where clear indications of inhomogeneities
were found using electron microscopy techniques.
These authors believe that Fig.IV.f.2a may be indicative of a
percolative transition between the FM and CO states at low
temperature, where clusters of one phase grow in a background of the
other until a percolation occurs. Moreover, recent theoretical work in
this context (Moreo et al., 2000) allows for CMR effects involving 
two insulators, since apparently the most important feature of the
compounds that present these effects is (i) a first-order-like
transition between the competing phases and (ii) the presence of
intrinsic disorder in the material. Thus, it is very interesting to
note that in the bilayer system $\rm Sr_{2-x} Nd_{1+x} Mn_2 O_7$ with x=0.0
and 0.10 a colossal MR effect has also been reported involving two
insulators (Battle et al., 1996),  showing that it is not necessary to
have a double-exchange induced ferromagnetic metallic phase to observe
this effect, as remarked before.

Layered electron-doped compounds are also known. 
In fact, Raychaudhuri et al. (1998) reported transport, magnetic and
specific heat studies of $\rm La_{2.3-x} Y_x Ca_{0.7} Mn_2 O_7$ with
x=0.0, 0.3, and 0.5. 
For x=0.0 the material is a FM insulator. As x grows, a transition
to a metallic state at low temperature was observed. 
The resistivity vs. temperature results are reproduced in Fig.IV.f.2b.
The similarities with the behavior of other materials is clear.
Raychaudhuri et al. (1998) concluded that the x=0.0 compound may
correspond to a FM-AF mixture involving unconnected ferromagnetic
clusters embedded in an antiferromagnetic matrix.


Additional, although indirect, evidence for mixed-phase tendencies in 
bilayer compounds can be obtained from photoemission experiments. 
In fact, the first set of high-energy resolution angle-resolved
photoemission (ARPES) measurements in the context of manganites was 
reported by Dessau et al. (1998) and the compound used was precisely 
$\bilayer$ with x=0.4 (high resolution photoemission results for
$\LSMO$ and $\LCMO$ were previously reported by Park et al., 1996.
Dessau and Shen (1999) also presented results for $\LSMO$). 
In this experiment it was observed that the low temperature
ferromagnetic state was very different from a prototypical metal.
Its resistivity is unusually high, the width of the ARPES features are 
anomalously broad, and they do not sharpen as they approach the Fermi 
momentum. Single Fermi-liquid-like quasiparticles cannot be used to
describe these features. 
In addition, the centroids of the experimental peaks never approach 
closer than approximately 0.65 eV to the Fermi energy. This implies
that, even in the expected ``metallic'' regime, the density of states
at the Fermi energy is very small. Dessau et al. (1998) refers to
these results as the formation of a ``pseudogap'' (see Fig.IV.f.3). 
Those authors found that the effect is present both in the FM and
paramagnetic regimes, namely below and above $T_{\rm C}$. The pseudogap
affects the entire Fermi surface, i.e., there is no important momentum 
dependence in its value, making it unlikely that it is caused by
charge, spin, or orbital ordering. 
Dessau et al. (1998) and Dessau and Shen (1999) argued that the origin
of this pseudogap cannot simply be a Mott-Hubbard effect since the
density is x=0.4. The effect cannot arise from the simple DE mechanism 
which does not predict a pseudogap, and also cannot be caused by
Anderson localization due to disorder, which is not expected to
significantly affect the density of states.
In other words, it is not the mobility that appears to lead to large
resistivities but the lack of states at the Fermi energy.
Recent photoemission studies for bilayers and $\LSMO$ with x=0.18 led
to similar conclusions (Saitoh et al., 1999).

These ARPES results are in qualitative agreement with recent 
calculations by Moreo, Yunoki and Dagotto (1999b) 
and Moreo et al. (2000), described 
in detail elsewhere in this review, where a pseudogap in the density
of states was shown to appear naturally in mixed-phase regimes, either
those created by electronic phase separation or by the influence of
disorder on first-order transitions that leads to giant cluster
formation.
In both cases the conductivity was shown to be very small in these
regimes (Moreo, Yunoki and Dagotto, 1999b), and a pseudogap appears in
the density of states.
It is possible that the low temperature region of the x=0.4 bilayer can
be described in terms of a percolative process, and its reported 
``spin canted'' character is simply caused by mixing AF and FM
phases. This rationalization also explains the large value of the
resistivity even at low temperature.

The photoemission results are consistent with scanning tunneling
microscopy data (Biswas et al., 1998), gathered for single crystals
and thin-films of hole-doped manganites. This study showed a rapid
variation in the density of states for temperatures near the Curie
temperature, such that below $T_{\rm C}$ a finite density of states is
observed at the Fermi energy while above $T_{\rm C}$ a hard gap opens
up. This result suggests that the presence of a gap or pseudogap is not 
just a feature of bilayers, but it appears in other manganites as
well. In addition, the work of Biswas et al. (1998) suggest that the
insulating behavior above $T_{\rm C}$ is caused by a depletion in the
density of states, rather than by a change in the mobility. 
As in the photoemission work just described, it appears that Anderson
localization is not the reason for the insulating behavior, since this
mechanism is not expected to induce a gap in the density of states.

\begin{center}
  {\bf IV.g Mixed-Phase Tendencies in Single-Layered Manganites:}
\end{center}

Bao et al. (1996) reported the presence of macroscopic 
phase separation in the planar manganite $\rm Sr_{2-x} La_x Mn O_4$
in the range between x=0.0 and x=0.38. At x=0.0 the material is a
2D AF insulator, with no carriers in the $e_{\rm g}$-band. As x grows, 
carriers are introduced and they polarize the $t_{\rm 2g}$-spins
leading to spin polaron formation, as in other compounds at low
electronic density. These polarons attract each other and form
macroscopic ferromagnetic regions. This result is in agreement with
the theoretical discussion of Sec.III where it was found, both for one 
and two orbital models, that the region of small density of 
$e_{\rm g}$-electrons has phase separation characteristics.
The conclusions of Bao et al. (1996) are also in excellent agreement
with the studies discussed in this review in the context of $\LCMO$ at
large hole density concentration.

\begin{center}
  {\bf IV.h Possible Mixed-Phase Tendencies in Non-Manganite
    Compounds:}
\end{center}

There are several other non-manganite
 compounds that present a competition between 
FM and AF regions, states which in clean systems should be separated 
by first-order transitions, at least according to theoretical
calculations.
One of these compounds is $\rm {\bf La_{1-y} Y_y Ti O_3}$. As y is 
varied, the average bandwidth $W$ of the mobile electrons changes, and
experiments have shown that a FM-AF transition appears 
(Tokura et al., 1993). 
This material may be a candidate for percolative FM-AF transitions, as
in the manganites (see also Hays et al., 1999, for results on 
$\rm {\bf La_{1-x} Sr_x Ti O_3}$ with phase-separation
characteristics).
Also $\rm {\bf Tb_2 Pd Si_3}$ and $\rm {\bf Dy_2 Pd Si_3}$ present
properties that have been interpreted as indicative of magnetic
polaron formation (Mallik, Sampathkumaran and Paulose, 1998).
Large MR effects have been found in $\rm {\bf Gd_2 Pd Si_3}$ by
Saha et al. (1999).
In addition, simply replacing Mn by Co has been shown to lead to
physics somewhat similar to that found in manganites. For instance,
results obtained for $\rm {\bf La_{1-x} Sr_x Co O_3}$ using a variety of
techniques have been interpreted as mixed-phase or cluster-glass
states (see Caciuffo et al. (1999), Nam et al. (1999), and references
therein).
Also $\rm {\bf Se_{1-x} Te_x Cu O_3}$ presents a FM-AF competition
with spin-glass-like features (Subramanian, Ramirez and Marshall,
1999), resembling the mixed-phase states discussed in this review.
First-order FM-AF transitions have also been reported in 
$\rm {\bf Ce Fe_2}$ based pseudobinary systems 
(Manekar, Roy and Chaddah, 2000).
Even results obtained in films of vinylidene fluoride with
trifluoroethylene (Borca et al., 1999) have been interpreted in terms
of a compressibility phase transition similar to those discussed by 
Moreo et al. (1999a), reviewed in Sec.III. In addition, 
$\rm {\bf Ni S_{2-x} Se_x}$ also presents some characteristics similar
to those of the materials described here, namely a metal-insulator 
transition which is expected to be of first-order, random disorder
introduced by Se substitution, and an antiferromagnetic state (see
Husmann et al., 1996; Matsuura et al., 2000; and references therein). 

Very recently, some ruthenates have been shown to present
characteristics similar to those of electron-doped $\rm Ca Mn O_3$ (as
discussed for example by Neumeier and Cohn, 2000), including a
tendency to phase separation. Transport and magnetic results by 
Cao et al. (2000) indicate that in the region between x=0.0 and x=0.1
of $\rm {\bf Ca_{2-x} La_x Ru O_4}$, the material changes rapidly from
an antiferromagnetic insulator to a ferromagnetic metal.
The behavior of the magnetic susceptibility vs temperature 
is shown in Fig.IV.h.1a. The shape of the M vs H curve (Fig.IV.h.1b)
is quite significant.
On one hand, at finite density x there appears to be a finite moment
as the field is removed, characteristic of FM samples. On the other
hand, the linear behavior with H is indicative of AF behavior, namely
for antiferromagnetically ordered spins the canting that occurs in the 
presence of a magnetic field leads to a linearly growing moment.
A mixed-phase FM-AF is probably the cause of this behavior.
The curve resistivity versus temperature
(also shown in Cao et al., 2000) indeed appears to have percolative
characteristics, as found in many manganites.
Also perovskites such as $\rm {\bf Ca Fe_{1-x} Co_x O_3}$ have an
interesting competition between AF and FM states as x is varied.
In Fig.IV.h.2 the resistivity in the range of Co densities where the
transition occurs is shown, reproduced from Kawasaki et al. (1998).
The similarities with other results described in this review are
clear.

It is also important to mention here the large MR found in the 
{\it  pyrochlore} compound $\rm {\bf Tl_{2-x} Sc_x Mn_2 O_7}$,
although it is believed that its origin maybe different from the
analogous effect found in manganites (Ramirez and Subramanian, 1997 
and references therein. See also Shimakawa, Kubo, and Manako, 1996;
Cheong et al., 1996). 
The behavior of the resistivity with temperature, parametric with the
Sc concentration and magnetic fields is shown in Fig.IV.h.3.
The similarities with the analogous plots for the manganites presented
in previous sections is clear. More work should be devoted to clarify
the possible connection between pyrochlore physics and the ideas
discussed in this review.

Diluted magnetic semiconductors also present characteristics of
phase-separated states. Ohno (1998) has recently reviewed part of the
work in this context. The physics of magneto-polarons has also been
reviewed before by Kasuya and Yanase (1968). 
The reader should consult these publications and others  to find more
references and details about this vast area of research. 
Diluted semiconductors have mobile carriers and localized moments in 
interaction. At low temperatures the spins are ferromagnetically
aligned and the charge appears localized. It is believed that at these
temperatures large regions of parallel spins are formed. The cluster
sizes are of about 100 $\rm \AA$, a large number indeed
(see Ohno et al., 1992).
At a relatively small polaron density, their overlap will be
substantial. Important experimental work in this context applied to
$\rm {\bf Eu_{1-x} Gd_x Se}$ can be found in von Molnar and Methfessel
(1967). The resistivity vs temperature at several magnetic fields of
$\rm Eu Se$ is shown in Fig.IV.h.4, reproduced from Shapira et
al. (1974). The similarity with results for manganites is clear.

Other compounds of this family present interesting FM-AF competitions.
For instance, the phase diagram of $\rm {\bf Eu B_{6-x} C_x}$
presented by Tarascon et al. (1981) contains an intermediate region
labeled with a question mark between the FM and AF phases. This
intermediate phase should be analyzed in more detail.
Already Tarascon et al. (1981) favored an interpretation of this
unusual region based on mixed-phase states. 
Recently, two magnetically similar but electronically inequivalent
phases were detected with NMR applied to EuB$_6$ by
Gavilano et al. (1998).
Also Gavilano, Hunziker and Ott (1995) reported a two component NMR
signal in {\bf CeAl$_3$}, signalling inhomogeneities in the
material. Clearly other compounds seem to present physics very similar
to that found in manganites, at least regarding the FM-AF
competition. The diluted magnetic semiconductors have been
rationalized in the past as having physics caused by magneto-polaron
formation. However, larger clusters, inhomogeneities, and percolative
processes may matter in these compounds as much as in
manganites. Actually, optical experiments by Yoon et al. (1998) have
already shown the existence of strong similarities between manganites
and EuB$_6$.  More recently, Snow et al. (2000) presented inelastic
light scattering measurements of EuO and $\rm {\bf Eu_{1-x} La_x B_6}$,
as a function of doping, magnetic fields, and temperature. A variety of
distinct regimes were observed, including a magnetic polaron regime
above the Curie temperature and a mixed FM/AF regime at La density x larger
than 0.05. These Eu-based systems do not have strong electron-lattice 
effects associated with Jahn-Teller modes. Then, the existence of 
physical properties
very similar to those of manganites show that the key feature leading
to such behavior
is the competition between different tendencies, rather than the origin
and detailed properties of those competing phases.
It is clear that further experimental work should be
devoted to clarify these interesting issues. The authors of this
review firmly believe that mixed-phase tendencies and percolation are
not only interesting properties of manganites, but  should be
present in a large variety of other compounds as well.

\section{\bf Discussion, Open Questions, and Conclusions}

In this review, the main results gathered in recent years in the
context of theoretical studies of models for manganites have been
discussed. In addition, the main experiments that have helped
clarify the physics of these interesting compounds have also been
reviewed. Several aspects of the problem are by now widely accepted,
while others still need further work to be confirmed. Intrinsic
inhomogeneities exist in models and experiments and seem to play
a key role in these compounds. 

\vskip0.5cm

Among the issues related with inhomogeneities that after a
considerable effort appear well-established are the following:

\noindent {\bf (1)} 
Work carried out by several groups using a variety of techniques
have shown that electronic phase separation is a dominant feature of 
models for manganites, particularly in the limits of small and large 
hole doping. This type of phase separation leads to nanometer size
coexisting clusters once the long-range Coulombic repulsion is
incorporated into the models.

\noindent {\bf (2)} 
Working at constant density, the transitions between metallic
(typically FM) and insulating (typically CO/AF) states are of 
{\bf first} order at zero temperature. No counter-example has been
found to this statement thus far.

\noindent {\bf (3)}
A second form of phase separation has been recently discussed.
It is produced by the influence of disorder on the first-order 
metal-insulator transitions described in the previous item. A simple
intuitive explanation is given in Fig.V.1. If couplings are fixed such
that one is exactly at the first-order 
transition in the absence of disorder, 
the system is ``confused'' and does not know whether to
be metallic or insulating (at zero disorder). On the other hand, if
the couplings are the same, but the strength of disorder is large in such
a way that it becomes dominating, then tiny clusters of the two competing
phases are formed with the lattice spacing as the typical length scale.
For nonzero but weak disorder, an intermediate situation develops where
fluctuations in the disorder pin either one phase or the other in large
regions of space. 

This form of phase separation is even more promising than the
electronic one for explaining the physics of manganites for a variety of
reasons: (i) it involves phases with the same density, thus there are
no constraints on the size of the coexisting clusters which can be as
large as a micrometer in scale, as found in experiments. 
(ii) The clusters are randomly distributed and have fractalic shapes,
leading naturally to {\bf percolative} transitions 
from one competing phase to
the other, as couplings or densities are
varied. This is in agreement with many experiments that have reported
percolative features in manganites.
(iii) The resistivity obtained in this context is similar to that
found in experiments, as sketched in Fig.V.2: Near the critical amount
of metallic fraction for percolation, at room temperature the charge
conduction can occur through the insulating regions since their
resistivity at that temperature is very similar to that of the
metallic state. Thus, the system behaves as an insulator. However, at
low temperatures, the insulator regions have a huge resistivity and,
thus, conduction is through the percolative metallic filaments which
have a large intrinsic resistivity. The system behaves as a bad metal,
and $\rho_{\rm dc}(T=0)$ can be very large.
(iv) Finally, it is expected that in a percolative regime there must
be a high sensitivity to magnetic fields and other naively ``small''  
perturbations, since tiny changes in the metallic fraction can induce
large conductivity modifications. This provides the best explanation
of the CMR effect of which these authors are aware.

\noindent {\bf (4)} 
The experimental evidence for inhomogeneities in manganites is by now
simply overwhelming. Dozens of groups, using a variety of techniques, 
have converged to such a conclusion. 
It is clear that homogeneous descriptions of manganites in the region
of interest for the CMR effect are incorrect.
These inhomogeneities appear even above the Curie temperature. In
fact, the present authors believe that a new scale of temperature
$T^*$ should be introduced, as very roughly sketched in Fig.V.3.
There must be a temperature window where coexisting clusters exist
above the temperatures where truly long-range order develops. Part of
the clusters can be metallic, and their percolation may induce
long-range order as temperature decreases.  The region below $T^*$ can
be as interesting as that observed in high temperature
superconductors, at temperatures higher than the critical values. It
is likely that it contains pseudogap characteristics, due to its low
conductivity in low bandwidth manganites. The search for a
characterization of $T^*$ should be pursued actively in experiments.

\noindent {\bf (5)} 
The famous CE-state of half-doped manganites has been shown to be
stable in mean-field and computational studies of models for
manganites. Although such a state was postulated a long time ago, it is
only recently that it has emerged from unbiased studies. The simplest
view to understand the  CE-state is based on a ``band insulating''
picture: it has been shown that in a zigzag FM chain a gap opens at
x=0.5, reducing the energy compared with straight chains. Thus,
elegant geometrical arguments are by now available to understand the
origin of the naively quite complicated CE-state of manganites. Its
stabilization can be rationalized based simply on models of
non-interacting spinless fermions in 1D geometries.
In addition, theoretical  studies have allowed one to analyze the
properties of the states competing with the CE at x=0.5. In order to
arrive at the CE-state, the use of a strong long-range Coulomb
interaction to induce the staggered charge pattern is not correct,
since by this procedure the experimentally observed charge-stacking
along the $z$-axis could not be reproduced, and in addition the
metallic regimes at x=0.5 found in some manganites would not be
stable. Manganese oxides are in the subtle regime where many different
tendencies are in competition.

\noindent {\bf (6)} 
Contrary to what was naively believed until recently, studies with
strong electron Jahn-Teller phonon coupling or with strong on-site
Coulomb interactions lead to quite similar phase diagrams. The reason
is that both  interactions disfavor double occupancy of a given
orbital. Thus, if the goal is to understand the CMR effect, the
important issue is not whether the material is Jahn-Teller or Coulomb
dominated, but how the metallic and insulating phases, of whatever
origin, compete. Calculations with Jahn-Teller phonons are the
simplest in practice, and they have led to phase diagrams that contain
most of the known phases found experimentally for manganites, such as
the A-type AF insulating state at x=0, the A-type AF metallic state at
x=0.5, the CE-state at x=0.5, etc.
Such an agreement theory-experiment is quite remarkable and encouraging.

\noindent {\bf (7)} 
Also contrary to naive expectations, the smallest parameter in
realistic Hamiltonians for Mn-oxides, namely ``$J_{\rm AF}$'' between
localized $t_{\rm 2g}$ spins, plays an important role in stabilizing
the experimentally observed phases of manganites, including the
CE-state. Modifications of this coupling due to disorder are as
important as those in the hopping amplitudes for $e_{\rm g}$ electron
movement.

\vskip0.5cm

In short, it appears that some of the theories proposed in early
studies for manganites can already be shown to be incorrect.
This includes (i) simple Double Exchange ideas where the high
resistivity above $T_{\rm C}$ is caused by the disordered character of
the localized spins that reduce the conductivity in the $e_{\rm g}$
band. This is not enough to produce an insulating state above
$T_{\rm C}$, and does not address the notorious inhomogeneities found
in experiments. It may be valid in some large bandwidth compounds away
from the region of competition between metal and insulator.
(ii) Anderson localization also appears unlikely to explain the
experimental data. An unphysically large value of the disorder strength is
needed for this to work at high temperature, the pseudogap found in
photoemission experiments cannot be rationalized in this context where
the density of states is not affected by disorder, and large
inhomogeneities, once again, cannot be addressed in this
framework. However, note that once a percolative picture is accepted
for manganites, then some sort of localization in such a fractalic
environment is possible. 
(iii) Polaronic ideas can explain part of the experimental data at
least at high temperatures, far from the Curie temperature. However,
the region where CMR is maximized cannot be described by a simple gas
of heavy polarons or bipolarons (see experimental results in Section IV). 
There is
no reason in the polaronic framework for the creation of micrometer
size coexisting clusters in these compounds. Actually, note that 
theories based on small polarons and phase separation do not differ 
only on subtle points if the phase separation involves microdomains.
It may happen that nanometer phase separation leads to physics similar
to that created by polaronic states, but certainly not when much
larger clusters are formed. 

As a conclusion, it is clear that the present prevailing paradigm for
manganites relies on a phase-separated view of the dominant state, as
suggested by dozens of experiments and also by theoretical
calculations once powerful many-body techniques are used to study
realistic models.

\vskip0.6cm

Although considerable progress has been achieved in recent years in
the analysis of manganites, both in theoretical and experimental
aspects, there are still a large number of issues that require further
work. Here a partial list of {\bf open questions} is included.

\noindent {\bf (a)} 
The phase separation scenario needs further experimental
confirmation. Are there counterexamples of compounds where CMR
occurs but the system appears homogeneous? 

\noindent {\bf (b)} 
On the theory front, a phase-separated percolative state is an
important challenge to our computational abilities. Is it possible to
produce simple formulas with a small number of parameters that
experimentalists can use in order to fit their, e.g., transport data?
The large effort needed to reproduce the zero magnetic field
resistivity vs. temperature results (reviewed here) suggests that this
will be a hard task.

\noindent {\bf (c)}
It is believed that at zero temperature the metal-insulator transition
is of first-order and upon the introduction of disorder it becomes
continuous, with percolative characteristics. A very important study
that remains to be carried out is the analysis of the influence of
temperature on those results. These authors believe that the
generation of a ``Quantum Critical Point'' (QCP) is likely in this
context, and preliminary results support this view 
(Burgy et al., 2000). The idea is
sketched in Fig.V.4. Without disorder (part (a)), the first-order
transition survives the introduction of temperature, namely in a
finite temperature window the transition between the very different FM
and AF states remains first-order. However, introducing disorder (part
(b)), a QCP can be generated since the continuous zero temperature
transition is unlikely to survive at finite temperature at fixed
couplings. The presence of such QCP would be a conceptually important
aspect of the competition between FM and AF phases in manganites. 
Experimental results showing that the generation of such QCP is possible
have already been presented (Tokura, 2000).

\noindent {\bf (d)} 
There is not much reliable theoretical work carried out in the
presence of magnetic fields addressing directly the CMR effect. The
reason is that calculations of resistivity are notoriously difficult,
and in addition, the recent developments suggest that percolative
properties are important in manganites, complicating the theoretical
analysis. Nevertheless, the present authors believe that a very simple
view of the CMR effect could be as follows. It is known that the
metallic and insulating phases are separated by first-order
transitions. Then, when energy is plotted vs the parameter ``g'' that
transforms one phase into the other (it could be a coupling in the
Hamiltonian or the hole density), a level crossing occurs at zero
temperature, as sketched in Fig.V.5. In the vicinity of the transition
point, a small magnetic field can produce a rapid destabilization of
the insulating phase in favor of the metallic phase. This can occur
only in a small window of densities and couplings if realistic (small)
magnetic fields are used. At present it is unknown how disorder, and
the percolation phenomena it induces, will affect these sketchy results.
In addition, there are compounds such as $\PCMO$ that present CMR in a
large density window, suggesting that the simple picture of Fig.V.5
can be a good starting point, but is incomplete. Thus, quantitative
calculations addressing the CMR effect are still needed.

\noindent {\bf (e)} 
Does a spin-canted  phase ever appear in simple models with competing 
FM and AF phases in the absence of magnetic fields? Are the regions
labeled as spin-canted in some experiments truly homogeneous or
mixed-states?

\noindent {\bf (f)} If the prediction of a phase-separated state in
the CMR regime of manganites is experimentally fully confirmed, what
are the differences between that state and a canonical ``spin-glass''?
Both share complexity and complicated time dependences, but are they
in the same class? Stated in more exciting terms, can the
phase-separated regime of manganites be considered a ``new'' state 
of matter in any respect?

\noindent {\bf (g)} 
Considerable progress has been achieved in understanding the x=0 and
x=0.5 charge/orbital/spin order states of manganites. But little is
known about the ordered states at intermediate densities, 
both in theory and experiments. Are
there stripes in manganites at those intermediate hole densities as
recently suggested by experimental and theoretical work? 

\vskip0.6cm

Summarizing, the study of manganites continues challenging our
understanding of transition metal oxides. While considerable progress
has been achieved in recent years, much work remains to be done. In
particular, a full understanding of the famous CMR effect is still
lacking, although evidence is accumulating that it may be caused by
intrinsic tendencies toward inhomogeneities in Mn-oxides and other
compounds.  Work in this challenging area of research should continue
at its present fast pace.


\section*{Acknowledgement}

The authors would like to thank our many collaborators that have
helped us in recent years in the development of the phase separation
scenario for manganites. 
Particularly, the key contributions of Seiji Yunoki are here acknowledged. 
It is remarked that a considerable portion of the subsection entitled 
``Monte Carlo Simulations'' has been originally prepared by Seiji Yunoki.
We also thank C. Buhler, J. Burgy, S. Capponi, A. Feiguin, N. Furukawa,
K. Hallberg, J. Hu, H. Koizumi, A. Malvezzi, M. Mayr, D. Poilblanc,
J. Riera, Y. Takada, J. A. Verges, for their help in these projects.

We are also very grateful to S. L. Cooper, T. Egami, J. P. Hill, 
J. Lynn, D. Mills, J. Neumeier, A. Pimenov, and P. Schiffer, for
their valuable comments on early drafts of the present review.

E.D. and A.M. are supported by grant NSF-DMR-9814350, 
the National High Magnetic Field Laboratory (NHMFL), and the
Center for Materials Research and Technology (MARTECH).
T.H. has been supported by the Ministry of Education, Science, Sports, 
and Culture (ESSC) of Japan during his stay in the National High
Magnetic Field Laboratory, Florida State University.
T.H. is also supported by the Grant-in-Aid for Encouragement of Young 
Scientists under the contact No. 12740230 from the Ministry of ESSC.

\begin{center}
  {\bf REFERENCES}
\end{center}

\parskip4pt

\noindent C. P. Adams, J. W. Lynn, Y. M. Mukovskii, A. A. Arsenov,
and D. A. Shulyatev, 2000, preprint, to appear in Phys. Rev. Letters.

\noindent D. Agterberg and S. Yunoki, 2000, preprint.

\noindent T. Akimoto, Y. Maruyama, Y. Moritomo, A. Nakamura, 
K. Hirota, K. Ohoyama, and M. Ohashi, Phys. Rev. B{\bf 57}, R5594
(1998).

\noindent A. S. Alexandrov and A. M. Bratkovsky, 1999,
Phys. Rev. Lett. {\bf 82}, 141 (1999).

\noindent R. Aliaga, R. Allub, and B. Alascio, 1998, preprint,
cond-mat/9804248.

\noindent H. Aliaga, K. Hallberg, B. Alascio, and B. Normand, 2000,
preprint.

\noindent P. B. Allen and V. Perebeinos, Phys. Rev. B{\bf 60}, 10747
(1999).

\noindent G. Allodi, R. De Renzi, G. Guidi, F. Licci, and
M. W. Pieper, Phys. Rev. B{\bf 56}, 6036 (1997).

\noindent G. Allodi, R. De Renzi, and G. Guidi, 
Phys. Rev. B{\bf 57}, 1024 (1998).

\noindent G. Allodi, R. De Renzi, F. Licci, and M. W. Pieper,
Phys. Rev. Lett. {\bf 81}, 4736 (1998).

\noindent G. Allodi, R. De Renzi, M. Solzi, K. Kamenev,
G. Balakrishnan, and M. W. Pieper, 1999, preprint, cond-mat/9911164.

\noindent R. Allub and B. Alascio, Solid State Commun. {\bf 99}, 613
(1996).

\noindent R. Allub and B. Alascio, Phys. Rev. B{\bf 55}, 14113
(1997).

\noindent J. L. Alonso, L. A. Fern\'andez, F. Guinea, V. Laliena, and 
V. Mart\'in-Mayor, 2000a, preprint, cond-mat/0003472.

\noindent J. L. Alonso, L. A. Fern\'andez, F. Guinea, V. Laliena, and 
V. Mart\'in-Mayor, 2000b, preprint, cond-mat/0007438.

\noindent J. L. Alonso, L. A. Fern\'andez, F. Guinea, V. Laliena, and 
V. Mart\'in-Mayor, 2000c, preprint, cond-mat/0007450.

\noindent A. Anane, J. P. Renard, L. Reversat, C. Dupas, P. Veillet,
M. Viret, L. Pinsard, and A. Revcolevschi, Phys. Rev. B{\bf 59}, 77
(1999).

\noindent A. Anane, B. Raquet, S. von Molnar, L. Pinsard-Godart,
and A. Revcolevschi, 1999, preprint, cond-mat/9910204.

\noindent P. W. Anderson and H. Hasegawa, Phys. Rev. {\bf 100}, 675
(1955).

\noindent V. I. Anisimov, I. S. Elfimov, M. A. Korotin, and K. Terakura,
Phys. Rev. B{\bf 55}, 15494 (1997).

\noindent D. N. Argyriou, H. N. Bordallo, J. F. Mitchell,
J. D. Jorgensen, and G. F. Strouse, Phys. Rev. B{\bf 60}, 6200 (1999).

\noindent D. N. Argyriou, H. N. Bordallo, B. J. Campbell,
A. K. Cheetham, D. E. Cox, J. S. Gardner, K. Hanif, A. dos Santos, and
G. F. Strouse, Phys. Rev. B{\bf 61}, 15269 (2000).

\noindent T. Arima, Y. Tokura, and J. B. Torrance,
Phys. Rev. B{\bf 48}, 17006 (1993).

\noindent T. Arima and Y. Tokura, J. Phys. Soc. Jpn. {\bf 64}, 2488
(1995).

\noindent D. Arovas, and F. Guinea,  Phys. Rev. B{\bf 58}, 9150 (1998). 

\noindent D. Arovas, G. G\'omez-Santos,
and F. Guinea, Phys. Rev. B{\bf 59}, 13569 (1999).

\noindent J. P. Attfield, A. L. Kharlanov, and J. A. McAllister,
Nature {\bf 394}, 157 (1998). 

\noindent Y. Avishai and J. M. Luck,
Phys. Rev. B{\bf 45}, 1074 (1992).

\noindent N. A. Babushkina, L. M. Belova, V. I. Ozhogin,
O. Yu. Gorbenko, A. R. Kaul, A. A. Bosak, D. I. Khomskii, and
K. I. Kugel, J. Appl. Phys. {\bf 83}, 7369 (1998).

\noindent A. M. Balagurov, V. Yu. Pomjakushin, D. V. Sheptyakov,
V. L. Aksenov, N. A. Babushkina, L. M. Belova, A. N. Taldenkov,
A. V. Inyushkin, P. Fischer, M. Gutmann, L. Keller, O. Yu. Gorbenko, 
and A. R. Kaul, Phys. Rev. B{\bf 60}, 383 (1999).

\noindent W. Bao, C. H. Chen, S. A. Carter, and S.-W. Cheong,
Solid State Commun. {\bf 98}, 55 (1996).

\noindent W. Bao, J. D. Axe, C. H. Chen, and S.-W. Cheong, 
Phys. Rev. Lett. {\bf 78}, 543 (1997).

\noindent W. Bao, J. D. Axe, C. H. Chen, S.-W. Cheong, P. Schiffer,
and M. Roy, Physica B {\bf 241-243}, 418 (1998).

\noindent M. Baran, S. Gnatchenko, O. Gorbenko, A. Kaul, R. Szymczak, 
and H. Szymczak, Phys. Rev. B{\bf 60}, 9244 (1999).

\noindent C. D. Batista, J. M. Eroles, M. Avignon, and B. Alascio,
preprint, cond-mat/9807361.

\noindent C. D. Batista, J. M. Eroles, M. Avignon, and B. Alascio,
preprint, cond-mat/0008367.

\noindent P. D. Battle, S. J. Blundell, M. A. Green, W. Hayes, M.
Honold, A. K. Klehe, N. S. Laskey, J. E. Millburn, L. Murphy,
M. J. Rosseinsky, N. A. Samarin, J. Singleton, N. E. Sluchanko,
S. P. Sullivan, and J. F. Vente, J. Phys.: Condens. Matter {\bf 8},
L427 (1996).

\noindent P. D. Battle, M. A. Green, N. S. Laskey, J. E. Millburn,
M. J. Rosseinsky, S. P. Sullivan, and J. F. Vente, 
Chem. Comm., 767 (1996). 

\noindent P. D. Battle, M. A. Green, N. S. Laskey, J. E. Millburn,
P. G. Radaelli, M. J. Rosseinsky, S. P. Sullivan, and J. F. Vente,
Phys. Rev. B{\bf 54}, 15967 (1996).

\noindent P. D. Battle, M. J. Rosseinsky, and P. G. Radaelli, 
J. Phys. Soc. Jpn. {\bf 68}, 1462 (1999).

\noindent M. Belesi, G. Papavassiliou, M. Fardis, G. Kallias, and 
C. Dimitropoulos, 2000, preprint, cond-mat/0004332.

\noindent B. I. Belevtsev, V. B. Krasovitsky, V. V. Bobkov,
D. G. Naugle, K. D. D. Rathnayaka, and A. Parasiris, 2000, preprint,
cond-mat/0001372.

\noindent P. Benedetti and R. Zeyher,
Phys. Rev. B{\bf 59}, 9923 (1999).

\noindent J. J. Betouras and S. Fujimoto,
Phys. Rev. B{\bf 59}, 529 (1999).

\noindent S. J. L. Billinge, R. G. DiFrancesco, G. H. Kwei, 
J. J. Neumeier, and J. D. Thompson,
Phys. Rev. Lett. {\bf 77}, 715 (1996).

\noindent S. J. L. Billinge, Th. Proffen, V. Petkov, J. L. Sarrao, 
and S. Kycia, 1999, preprint, cond-mat/9907329.

\noindent A. Biswas, S. Elizabeth, A. K. Raychaudhuri and H. L. Bhat, 
1998, preprint, cond-mat/9806084.


\noindent J. Blasco, J. Garcia, J. M. de Teresa, M. R. Ibarra, 
P. A. Algarabel and C. Marquina, J. Phys.: Condens. Matter {\bf 8},
7427 (1996).

\noindent A. Bocquet, T. Mizokawa, T. Saitoh, H. Namatame, and
A. Fujimori, Phys. Rev. B{\bf 46}, 3771 (1992).

\noindent C. H. Booth, F. Bridges, G. H. Kwei, J. M. Lawrence,
A. L. Cornelius, and J. J. Neumeier, 
Phys. Rev. Lett. {\bf 80}, 853 (1998).

\noindent C. H. Booth, F. Bridges, G. H. Kwei, J. M. Lawrence,
A. L. Cornelius, and J. J. Neumeier, 
Phys. Rev. Lett.B{\bf 57}, 10440 (1998).

\noindent C. N. Borca, S. Adenwalla, J. Choi, P. T. Sprunger, 
S. Ducharme, L. Robertson, S. P. Palto, J. Liu, M. Poulsen, 
V. M. Fridkin, H. You and P. A. Dowben
 Phys. Rev. Lett. {\bf 83}, 4562 (1999). 


\noindent P. Bourges {\it et al.}, Science {\bf 288}, 1234 (2000).

\noindent S. de Brion, F. Ciorcas, G. Chouteau, P. Lejay, P. Radaelli, 
and C. Chaillout, 1999 preprint, cond-mat/9803024.

\noindent P. R. Broussard, V. M. Browning, and V. C. Cestone,
1999 preprint, cond-mat/9901189.

\noindent P. R. Broussard, S. B. Qadri, V. M. Browning, and
V. C. Cestone, 1999 preprint, cond-mat/9902020.

\noindent C. Buhler, S. Yunoki and A. Moreo, 
Phys. Rev. Lett. {\bf 84}, 2690 (2000).

\noindent J. Burgy et al., 2000, in preparation.

\noindent R. Caciuffo, D. Rinaldi, G. Barucca, J. Mira, J. Rivas,
M. A. Senaris-Rodriguez, P. G. Radaelli, D. Fiorani, and
J. B. Goodenough, Phys. Rev. B{\bf 59}, 1068 (1999).

\noindent M. J. Calderon, and L. Brey, 
Phys. Rev. B{\bf 58}, 3286 (1998).

\noindent M. Calderon, J. Verg\'es, and L. Brey,
Phys. Rev. B{\bf 59}, 4170 (1999). 

\noindent P. Calvani, G. De Marzi, P. Dore, S. Lupi, P. Maselli,
F. D'Amore, and S. Gagliardi,
Phys. Rev. Lett. {\bf 81}, 4504 (1998).

\noindent B. J. Campbell et al., 2000, unpublished.

\noindent G. Cao, S. McCall, V. Dobrosavljevic, C. S. Alexander,
J. E. Crow, and R. P. Guertin, 2000 preprint, to appear in PRB.

\noindent M. Capone, D. Feinberg, and M. Grilli, 2000 preprint,
cond-mat/0001243.

\noindent D. Casa, V. Kiryukhin, O. A. Saleh, B. Keimer, J. P. Hill,
Y. Tomioka, and Y. Tokura, Europhys. Lett. {\bf 47}, 90 (1999).

\noindent C. Castellani, C. R. Natoli, and J. Ranninger,
Phys. Rev. {\bf 18}, 4945 (1978).

\noindent V. Z. Cerovski, S. D. Mahanti, T. A. Kaplan, and
A. Taraphder, Phys. Rev. B{\bf 59}, 13977 (1999).

\noindent A. Chattopadhyay, A. J. Millis, and S. Das Sarma,
2000 preprint, cond-mat/0004151.

\noindent O. Chauvet, G. Goglio, P. Molinie, B. Corraze, and
L. Brohan, Phys. Rev. Lett. {\bf 81}, 1102 (1998).

\noindent C. H. Chen and S.-W. Cheong, 
Phys. Rev. Lett. {\bf 76}, 4042 (1996).

\noindent Z. Y. Chen, A. Biswas, I. Zuti\'c, T. Wu, S. B. Ogale,
A. Orozco, R. L. Greene, and T. Venkatesan, 2000, cond-mat/0007353.

\noindent S.-W. Cheong, H. Y. Hwang, B. Batlogg, and L. W. Rupp, Jr., 
Solid State Comm. {\bf 98}, 163 (1996).

\noindent H. Chiba, M. Kikuchi, K. Kusaba, Y. Muraoka, and Y. Syono,
Solid State Commun. {\bf 99}, 499 (1996).

\noindent S.-W. Cheong and H. Y. Hwang, contribution to {\it Colossal 
Magnetoresistance Oxides} edited by Y. Tokura (Gordon \& Breach, 
Monographs in Condensed Matter Science, London, 1999).

\noindent S. H. Chun, M. B. Salamon, P. D. Han, Y. Lyanda-Geller,
P. M. Goldbart, preprint 1999a, cond-mat/9904332.

\noindent S. H. Chun, M. B. Salamon, Y. Tomioka, and Y. Tokura,
preprint 1999b, cond-mat/9906198.

\noindent S. H. Chun, Y. Lyanda-Geller, M. B. Salamon,
R. Suryanarayanan, G. Dhalenne, and A. Revcolevschi, 
preprint 2000, cond-mat/0007249.

\noindent J. M. D. Coey, M. Viret, L. Ranno, and K. Ounadjela,
Phys. Rev. Lett. {\bf 75}, 3910 (1995).

\noindent J. M. D. Coey, M. Viret, and S. von Molnar, 
{\it Mixed-valence Manganites}, Adv. Phys. 1998, in press.

\noindent D. Coffey, K. Bedell, and S. Trugman,
Phys. Rev. B{\bf 42}, 6509 (1990).

\noindent D. E. Cox, P. G. Radaelli, M. Marezio, 
et al., Phys. Rev. B{\bf 57}, 3305 (1998).

\noindent E. Dagotto, Rev. Mod. Phys. {\bf 66}, 763 (1994).

\noindent E. Dagotto and T. M. Rice, Science {\bf 271}, 618 (1996).

\noindent E. Dagotto, S. Yunoki, A. L. Malvezzi, A. Moreo, J. Hu,
S. Capponi, D. Poilblanc, and N. Furukawa, 
Phys. Rev. {\bf 58 B}, 6414 (1998).

\noindent P. Dai, J. A. Fernandez-Baca, B. C. Chakoumakos,
J. W. Cable, S. E. Nagler, P. Schiffer, N. Kalechofsky, M. Roy,
Y.-K. Tsui, P. McGinn, S. Einloth, and A. P. Ramirez, 
preprint 1996 (unpublished).

\noindent P. Dai {\it et al.}, 
Phys. Rev. Lett. {\bf 80}, 1738 (1998).

\noindent P. Dai, J. A. Fernandez-Baca, N. Wakabayashi, E. W. Plummer, 
Y. Tomioka, and Y. Tokura, 2000, Phys. Rev. Lett. {\bf 85}, 2553 (2000).

\noindent S. Datta,
{\it Electronic Transport in Mesoscopic Systems}, Cambridge University 
Press, Cambridge, 1995.

\noindent P. G. deGennes, Phys. Rev. {\bf 118}, 141 (1960).

\noindent R. V. Demin, L. I. Koroleva, and A. M. Balbashov, JETP
Letters {\bf 70}, 314 (1999), and references therein.

\noindent D. S. Dessau, T. Saitoh, C.-H. Park, Z.-X. Shen,
P. Villella, N. Hamada, Y. Moritomo, and Y. Tokura, 
Phys. Rev. Lett. {\bf 81}, 192 (1998).

\noindent D. S. Dessau and Z.-X. Shen, contribution to {\it Colossal
  Magnetoresistance Oxides} edited by Y. Tokura (Gordon \& Breach, 
Monographs in Condensed Matter Science, London, 1999).

\noindent J. M. De Teresa, M. R. Ibarra, J. Garcia, J. Blasco,
C. Ritter, P. A. Algarabel, C. Marquina, and A. del Moral, 
Phys. Rev. Lett. {\bf 76}, 3392 (1996).

\noindent J. M. De Teresa, C. Ritter, M. R. Ibarra, P. A. Algarabel, 
J. Garcia-Mu\~noz, J. Blasco, J. Garcia, and C. Marquina,
Phys. Rev. B{\bf 56}, 3317 (1997a).

\noindent J. M. De Teresa, M. R. Ibarra, P. A. Algarabel, C. Ritter,
C. Marquina, J. Blasco, J. Garcia, A. del Moral, and Z. Arnold, 
Nature {\bf 386}, 256 (1997b).

\noindent J. Dho, I. Kim, S. Lee, K. H. Kim, H. J. Lee, J. H. Jung, 
and T. W. Noh, Phys. Rev. B{\bf 59}, 492 (1999).

\noindent J. Dho, I. Kim, and S. Lee, preprint 1999.

\noindent L. Dworin and A. Narath,
Phys. Rev. Lett. {\bf 25}, 1287 (1970).

\noindent T. Egami, J. of Low Temp. Physics {\bf 105}, 791 (1996).

\noindent T. Egami, D. Louca, and R. J. McQueeney, 
J. of Superconductivity {\bf 10}, 323 (1997).

\noindent T. Egami and D. Louca, proceedings of the Euroconference on
``Polarons: Condensation, Pairing, Magnetism'', Erice, Sicily, June
1998, to be published in J. Superconductivity.

\noindent J. B. Elemans, B. Van Laar, K. R. Van Der Keen, and 
B. Loopstra, J. Solid State Chem. {\bf 3}, 238 (1971).

\noindent V. Emery, S. A. Kivelson, and O. Zachar, 
Phys. Rev. B{\bf 56}, 6120 (1997).

\noindent Y. Endoh, H. Nojiri, K. Kaneko, K. Hirota, T. Fukuda,
H. Kimura, Y. Murakami, S. Ishihara, S. Maekawa, S. Okamoto, and
M. Motokawa, J. Mater. Sci. Eng. B {\bf 56}, 1 (1999) (see also 
cond-mat/9812404).

\noindent Y. Endoh, K. Hirota, S. Ishihara, S. Okamoto, Y. Murakami, 
A. Nishizawa, T. Fukuda, H. Kimura, H. Nojiri, K. Kaneko, and
S. Maekawa, Phys. Rev. Lett. {\bf 82}, 4328 (1999).

\noindent M. V. Eremin and V. N. Kalinenkov, Sov. Phys. Solid State  
{\bf 20}, 2051 (1978).

\noindent M. V. Eremin and V. N. Kalinenkov, Sov. Phys. Solid State 
{\bf 23}, 828 (1981).

\noindent M. F\"ath, S. Freisem, A. A. Menovsky, Y. Tomioka, J. Aarts,
and J. A. Mydosh, Science {\bf 285}, 1540 (1999).

\noindent J. A. Fernandez-Baca, P. Dai, H. Y. Hwang, C. Kloc, and
S.-W. Cheong, Phys. Rev. Lett. {\bf 80}, 4012 (1998).

\noindent V. Ferrari and M. J. Rozenberg, 
1999 preprint, cond-mat/9906131.

\noindent A. Fert and I. A. Campbell, 
J. Phys. F{\bf 6}, 849 (1976). 

\noindent R. Fr\'esard and G. Kotliar, 
Phys. Rev. B{\bf 56}, 12909 (1997). 

\noindent H. Fujishiro et al., 
J. Phys. Soc. Jpn. {\bf 67}, 1799 (1998). 

\noindent N. Fukumoto, S. Mori, N. Yamamoto, Y. Moritomo,
T. Katsufuji, C. H. Chen, and S-W. Cheong, 
Phys. Rev. B{\bf 60}, 12963 (1999). 

\noindent N. Furukawa, J. Phys. Soc. Jpn. {\bf 63}, 3214 (1994).

\noindent N. Furukawa, J. Phys. Soc. Jpn. {\bf 64}, 2734 (1995).

\noindent N. Furukawa, J. Phys. Soc. Jpn. {\bf 64}, 2754 (1995).

\noindent N. Furukawa, J. Phys. Soc. Jpn. {\bf 64}, 3164 (1995).

\noindent N. Furukawa, Y. Moritomo, K. Hirota, and Y. Endoh, 1998
preprint, cond-mat/9808076.

\noindent N. Furukawa, 1998 preprint, cond-mat/9812066.

\noindent D. J. Garcia, K. Hallberg, C. D. Batista, M. Avignon, and 
B. Alascio, 2000, preprint, cond-mat/9912227, to appear in Phys. Rev. Letters.

\noindent J. Gavilano, J. Hunzifer, and H. R. Ott,
Phys. Rev. B{\bf 52}, R13106 (1995).

\noindent J. L. Gavilano, B. Ambrosini, P. Vonlanthen, H. R. Ott, 
D. P. Young, and Z. Fisk, Phys. Rev. Lett. {\bf 81}, 5648 (1998).

\noindent M. Gerloch and R. C. Slade, {\it Ligand-Field Parameters},
(Cambridge, London, 1973).

\noindent L. Ghivelder, I. Abrego Castillo, M. A. Gusmao,
J. A. Alonso, and L. F. Cohen, 1999 preprint, cond-mat/9904232.

\noindent D. I. Golosov, M. R. Norman, and K. Levin, 1998 preprint, 
cond-mat/9805238.

\noindent J. Goodenough, Phys. Rev. {\bf 100}, 564 (1955).

\noindent John B. Goodenough, {\it Magnetism and the Chemical Bond},
(Interscience, New York, 1963).

\noindent J. B. Goodenough and J.-S. Zhou,
Nature {\bf 386}, 229 (1997).

\noindent L. P. Gor'kov and V. Z. Kresin,
JETP Letters {\bf 67}, 985 (1998). 

\noindent J. S. Griffith, 
{\it The Theory of Transition-Metal Ions}, (Cambridge, 1961).

\noindent R. Y. Gu, Z. D. Wang, S.-Q. Shen, and D. Y. Xing, 1999
preprint, cond-mat/9905152.

\noindent M. Guerrero, and R. M. Noack, preprint, cond-mat/0004265.

\noindent F. Guinea, G. G\'omez-Santos, and D. P. Arovas,
1999 preprint, cond-mat/9907184.

\noindent W. A. Harrison, 
{\it Electronic Structure and The Properties of Solids}, 
(Dover Publications, New York, 1989).

\noindent C. C. Hays, J.-S. Zhou, J. T. Markert, and J. B. Goodenough, 
Phys. Rev. B{\bf 60}, 10367 (1999).

\noindent R. H. Heffner, L. P. Le, M. F. Hundley, J. J. Neumeier,
G. M. Luke, K. Kojima, B. Nachumi, Y. J. Uemura, D. E. MacLaughlin,
and S.-W. Cheong, Phys. Rev. Lett. {\bf 77}, 1869 (1996).

\noindent R. H. Heffner, J. E. Sonier, D. E. MacLaughlin,
G. J. Nieuwenhuys, G. Ehlers, F. Mezei, S.-W. Cheong, J. S. Gardner,
and H. R\"oder, cond-mat/9910064.

\noindent K. Held and D. Vollhardt, preprint, cond-mat/9909311.

\noindent J. S. Helman and B. Abeles, 
Phys. Rev. Lett. {\bf 37}, 1429 (1976).

\noindent J. Hemberger, M. Paraskevoupolos, J. Sichelschmidt, 
M. Brando, R. Wehn, F. Mayr, K. Pucher, P. Lunkenheimer, and A. Loidl, 2000a, 
preprint.

\noindent J. Hemberger, 2000b
talk given at the NATO Advanced Research 
Workshop, Bled, Slovenia, April 2000.

\noindent M. Hennion, F. Moussa, G. Biotteau, J. Rodriguez Carvajal,
L. Pinsard, and A. Revcolevschi, Phys. Rev. Lett. {\bf 81}, 1957
(1998).

\noindent M. Hennion, F. Moussa, J. Rodriguez Carvajal,
L. Pinsard, and A. Revcolevschi, Phys. Rev. B{\bf 56}, R497 (1997).

\noindent M. Hennion, F. Moussa, G. Biotteau, J. Rodriguez Carvajal,
L. Pinsard, and A. Revcolevschi, preprint 1999, cond-mat/9910361.


\noindent Y. Hirai, B. H. Frazer, M. L. Schneider, S. Rast,
M. Onellion, W. L. O'Brien, S. Roy, A. Ignatov, and N. Ali, 
preprint, 2000.

\noindent K. Hirota, Y. Moritomo, H. Fujioka, M. Kubota, H. Yoshizawa, 
and Y. Endoh, J. Phys. Soc. Jpn. {\bf 67}, 3380 (1998).

\noindent K. Hirota, Y. Moritomo, H. Fujioka, M. Kubota, H. Yoshizawa,
and Y. Endoh, J. Phys.
Soc. Jpn. {\bf 68}, 1463 (1999).

\noindent T. Holstein, Annals of Physics {\bf 8}, 343 (1959).

\noindent T. Hotta, Y. Takada, and H. Koizumi,
Int. J. Mod. Phys. B{\bf 12}, 3437 (1998).

\noindent T. Hotta, S. Yunoki, M. Mayr, and E. Dagotto, 
Phys. Rev. B{\bf 60}, R15009 (1999).

\noindent T. Hotta, Y. Takada, H. Koizumi, and E. Dagotto,
Phys. Rev. Lett. {\bf 84}, 2477 (2000).

\noindent T. Hotta and E. Dagotto,
Phys. Rev. B{\bf 61}, R11879 (2000).

\noindent T. Hotta, A. Malvezzi, and E. Dagotto, 2000,
Phys. Rev. B {\bf 62}, 9432 (2000).

\noindent T. Hotta, A. Feiguin, and E. Dagotto, 2000,
preprint.

\noindent Q. Huang, J. W. Lynn, R. W. Erwin, A. Santoro, D. C. Dender, 
V. N. Smolyaninova, K. Ghosh, and R. L. Greene, 1999, preprint.

\noindent M. F. Hundley, M. Hawley, R. H. Heffner, Q. X. Jia,
J. J. Neumeier, J. Tesmer, J. D. Thompson, and X. D. Wu, Appl. Phys.
Lett. {\bf 67}, 860 (1995).

\noindent A. Husmann, D. S. Jin, Y. V. Zastavker, T. F. Rosenbaum,
X. Yao, and J. M. Honig, Science {\bf 274}, 1874 (1996).

\noindent H. Y. Hwang, S-W. Cheong, P. G. Radaelli, M. Marezio, and
B. Batlogg, 1995a, Phys. Rev. Lett. {\bf 75}, 914 (1995).

\noindent H. Y. Hwang, T. T. M. Palstra, S-W. Cheong, and B. Batlogg,
1995b, Phys. Rev. B{\bf 52}, 15046 (1995).

\noindent M. R. Ibarra, P. A. Algarabel, C. Marquina, J. Blasco
and J. Garcia, Phys. Rev. B{\bf 75}, 3541 (1995).

\noindent M. R. Ibarra and J. M. De Teresa,
J. of Mag. Mag. Mat. {\bf 177-181}, 846 (1998a).

\noindent M. R. Ibarra, G.-M. Zhao, J. M. De Teresa, B. Garcia-Landa,
Z. Arnold, C. Marquina, P. A. Algarabel, H. Keller, and C. Ritter,
Phys. Rev. B{\bf 57}, 7446 (1998b).

\noindent M. R. Ibarra and J. M. De Teresa, contribution to
{\it Colossal Magnetoresistance, Charge Ordering and Related
Properties of Manganese Oxides}, edited by C. N. R. Rao and B. Raveau,
World Scientific, 1998c.

\noindent M. N. Iliev, M. V. Abrashev, H.-G. Lee, Y. Y. Sun,
C. Thomsen, R. L. Meng, and C. W. Chu,
Phys. Rev. B{\bf 57}, 2872 (1998).

\noindent Y. Imry and S. K. Ma, 
Phys. Rev. Lett. {\bf 35}, 1399 (1975). 

\noindent S. Ishihara, M. Yamanaka, and N. Nagaosa,
Phys. Rev. B{\bf 56}, 686 (1997).

\noindent S. Ishihara, J. Inoue, and S. Maekawa, 
Phys. Rev. B{\bf 55}, 8280 (1997).

\noindent V. A. Ivanshin, J. Deisenhofer, H.-A. Krug von Nidda, 
A. Loidl, A. A. Mukhin, A. M. Balbashov, and M. V. Eremin, 2000,
Phys. Rev. B{\bf 61}, 6213 (2000).

\noindent G. Jackeli, N. B. Perkins, and N. M. Plakida,
1999 preprint, cond-mat/9910391.

\noindent H. A. Jahn and E. Teller,
Proc. Roy. Soc. London A {\bf 161}, 220 (1937).

\noindent M. Jaime, M. B. Salamon, M. Rubinstein, R. E. Treece,
J. S. Horwitz, and D. B. Chrisey, 
Phys. Rev. B{\bf 54}, 11914 (1996).

\noindent M. Jaime, P. Lin, S. H. Chun, M. B. Salamon, P. Dorsey,
and M. Rubinstein, Phys. Rev. B{\bf 60}, 1028 (1999).

\noindent  S. Jin, T. H. Tiefel, M. McCormack, R. A. Fastnacht,
R. Ramesh, and L. H. Chen, Science {\bf 264}, 413 (1994).

\noindent Z. Jirak, S. Krupicka, Z. Simsa, M. Dlouha, and Z. Vratislav, 
J. Mag. Mag. Mat. {\bf 53}, 153 (1985).

\noindent G. H. Jonker and J. H. Van Santen, 
Physica (Utrecht) {\bf 16}, 337 (1950). 

\noindent J. H. Jung, K. H. Kim, H. J. Lee, J. S. Ahn, N. J. Hur,
T. W. Noh, M. S. Kim, and J.-G. Park, 
Phys. Rev. B{\bf 59}, 3793 (1999).

\noindent J. H. Jung, H. J. Lee, T. W. Noh, E. J. Choi, Y. Moritomo,
Y. J. Wang, and X. Wei, preprint, cond-mat/9912451.

\noindent M. Yu. Kagan, D. I. Khomskii, and M. Mostovoy,
Eur. Phys. J. B{\bf 12}, 217 (1999).

\noindent M. Yu. Kagan, D. I. Khomskii, and K. I. Kugel, preprint,
cond-mat/0001245.

\noindent R. Kajimoto, 
T. Kakeshita, Y. Oohara, H. Yoshizawa, Y. Tomioka, and Y. Tokura,
Phys. Rev. B{\bf 58}, R11837 (1998).

\noindent R. Kajimoto, H. Yoshizawa, H. Kawano, H. Kuwahara,
Y. Tokura, K. Ohoyama, and M. Ohashi, 
Phys. Rev. B{\bf 60}, 9506 (1999).

\noindent G. Kallias, M. Pissas, E. Devlin, A. Simopoulos, and
D. Niarchos, Phys. Rev. B{\bf 59}, 1272 (1999).

\noindent J. Kanamori, J. Appl. Phys. Suppl. {\bf 31}, 14S (1960).

\noindent J. Kanamori, Prog. Theor. Phys. {\bf 30}, 275 (1963).

\noindent T. Kanki, H. Tanaka, and T. Kawai, 
Solid State Comm. {\bf 114}, 267 (2000).

\noindent S. G. Kaplan, M. Quijada, H. Drew, D. Tanner, G. Xiong,
R. Ramesh, C. Kwon, and T. Venkatesan, 
Phys. Rev. Lett. {\bf 77}, 2081 (1996).

\noindent T. Kaplan and S. Mahanti, editors, 
{\it Physics of Manganites}, 
Kluwer Academic/Plenum Publ. New York, 1999.

\noindent Cz. Kapusta, P. C. Riedi, M. Sikora, and M. R. Ibarra,
Phys. Rev. Lett. {\bf 84}, 4216 (2000).

\noindent T. Kasuya and A. Yanase, 
Rev. Mod. Phys. {\bf 40}, 684 (1968).

\noindent S. Katano, J. A. Fernandez-Baca, and Y. Yamada, Physica B {\bf
276-278}, 786 (2000).

\noindent T. Katsufuji, S.-W. Cheong, S. Mori and C.-H. Chen, 
J. of the Phys. Soc. of Jpn. {\bf 68}, 1090 (1999).

\noindent H. Kawano, R. Kajimoto, H. Yoshizawa, Y. Tomioka,
H. Kuwahara, and Y. Tokura, Phys. Rev. Lett. {\bf 78}, 4253 (1997). 

\noindent S. Kawasaki, M. Takano, R. Kanno, T. Takeda, and
A. Fujimori, J. Phys. Soc. Jpn. {\bf 67}, 1529 (1998). 

\noindent D. Khomskii, preprint, cond-mat/9909349.

\noindent D. Khomskii, preprint, cond-mat/0004034.

\noindent N. Kida, M. Hangyo, and M. Tonouchi, 2000a, preprint,
cond-mat/0005461.

\noindent N. Kida and M. Tonouchi, 2000b, preprint, cond-mat/0008298.

\noindent K. H. Kim, M. Uehara, and S-W. Cheong, 2000, preprint,
cond-mat/0004467.

\noindent S. Kirkpatrick, Rev. Mod. Phys. {\bf 45}, 574 (1973).

\noindent V. Kiryukhin, D. Casa, J. P. Hill, B. Keimer, A. Vigliante, 
Y. Tomioka, and Y. Tokura, Nature {\bf 386}, 813 (1997).

\noindent V. Kiryukhin, Y. J. Wang, F. C. Chou, M. A. Kastner, and 
R. J. Birgeneau, Phys. Rev. B{\bf 59}, R6581 (1999).

\noindent V. Kiryukhin, B. G. Kim, V. Podzorov, S-W. Cheong, T. Y. Koo,
J. P. Hill, I. Moon, and Y. H. Jeong, 2000, preprint, cond-mat/0007295.

\noindent H. Koizumi, T. Hotta, Y. Takada,
Phys. Rev. Lett. {\bf 80}, 4518 (1998a).

\noindent H. Koizumi, T. Hotta, Y. Takada,
Phys. Rev. Lett. {\bf 81}, 3803 (1998b).

\noindent M. Korotin, T. Fujiwara, and V. Anisimov, preprint,
cond-mat/9912456.

\noindent W. Koshibae, Y. Kawamura, S. Ishihara, S. Okamoto, J. Inoue,
and S. Maekawa, J. Phys. Soc. Jpn. {\bf 66}, 957 (1997).

\noindent K. Kubo and N. Ohata, J. Phys. Soc. Jpn. {\bf 33}, 21 (1972).

\noindent M. Kubota, H. Yoshizawa, Y. Moritomo, H. Fujioka, K. Hirota, 
and Y. Endoh, 1999a, J. Phys. Soc. Jpn. {\bf 68}, 2202 (1999).

\noindent M. Kubota, H. Fujioka, K. Ohoyama, K. Hirota, Y. Moritomo,
H. Yoshizawa, and Y. Endoh, 1999b, J. Phys. Chem. Solids {\bf 60}, 1161 (1999).

\noindent M. Kubota, H. Fujioka, K. Hirota, K. Ohoyama, Y. Moritomo,
H. Yoshizawa, and Y. Endoh, 1999c, preprint, cond-mat/9902288.

\noindent M. Kubota, 1999d, private communication.

\noindent M. Kubota, Y. Oohara, H. Yoshizawa, H. Fujioka, K. Shimizu,
K. Hirota, Y. Moritomo, and Y. Endoh, 2000, preprint.

\noindent J. Kuei and R. T. Scalettar, Phys. Rev. B{\bf 55}, 14968 (1997).

\noindent K. I. Kugel and D. I. Khomskii, 
Sov. Phys.-JETP {\bf 37}, 725 (1974).

\noindent K. Kumagai, A. Iwai, Y. Tomioka, H. Kuwahara, Y. Tokura, 
and A. Yakubovskii, Phys. Rev. B{\bf 59}, 97 (1999).

\noindent R. M. Kusters, J. Singleton, D. A. Keen, R. McGreevy, 
and W. Hayes, Physica B{\bf 155}, 362 (1989).

\noindent H. Kuwahara, Y. Tomioka, A. Asamitsu, Y. Moritomo, and
Y. Tokura, Science {\bf 270}, 961 (1995).

\noindent M. S. Laad, L. Craco, and E. M\"uller-Hartmann, 2000,
preprint, cond-mat/0007184.

\noindent A. Lanzara, N. L. Saini, M. Brunelli, F. Natali,
A. Bianconi, P. G. Radaelli, and S.-W. Cheong,
Phys. Rev. Lett. {\bf 81}, 878 (1998).

\noindent J. D. Lee and B. I. Min, 
Phys. Rev. B{\bf 55}, R14713 (1997).

\noindent P. Levy, F. Parisi, G. Polla, D. Vega, G. Leyva, H. Lanza, 
R. S. Freitas, and L. Ghivelder, 2000a, Phys. Rev. B{\bf 62}, 6437 (2000).

\noindent P. Levy, L. Granja, E. Indelicato, D. Vega, G. Polla and
F. Parisi, 2000b, preprint, cond-mat/0008236.

\noindent J. Q. Li, M. Uehara, C. Tsuruta, Y. Matsui, and Z. X. Zhao,
Phys. Rev. Lett. {\bf 82}, 2386 (1999). 

\noindent Q. Li, J. Zang, A. R. Bishop, and C. M. Soukoulis, 
Phys. Rev. B{\bf 56}, 4541 (1997).

\noindent C. D. Ling, J. E. Millburn, J. F. Mitchell, D. N. Argyriou,
J. Linton, and H. N. Bordallo, 2000, preprint.

\noindent H. L. Liu, S. L. Cooper, and S.-W. Cheong, 
Phys. Rev. Lett. {\bf 81}, 4684 (1998).

\noindent J. L\'opez, P. N. Lisboa-Filho, W. A. C. Passos,
W. A. Ortiz, and F. M. Araujo-Moreira, preprint, cond-mat/0004460.

\noindent N. N. Loshkareva, Yu. P. Sukhorukov, E. A. Neifel'd,
V. E. Arkhipov, A. V. Korolev, V. S. Gaviko, E. V. Panfilova,
V. P. Dyakina, Ya. M. Mukovskii, and D. A. Shulyatev, Journal of
Experimental and Theoretical Physics {\bf 90}, 389 (2000).

\noindent N. N. Loshkareva, Yu. P. Sukhorukov, N. I. Solin,
S. V. Naumov, Ya. M. Mukovskii, and N. I. Lobachevskaya, proceedings
of MSM'99 Moscow International Symposium on Magnetism, Moscow, June
1999. 

\noindent D. Louca, T. Egami, E. L. Brosha, H. R\"oder, 
and A. R. Bishop, Phys. Rev. B{\bf 56}, R8475 (1997).

\noindent D. Louca and T. Egami, Phys. Rev. B{\bf 59}, 6193 (1999).

\noindent V. M. Loktev and Yu. G. Pogorelov, Low Temperature Physics
{\bf 26}, 171 (2000).

\noindent Y. Lyanda-Geller, P. M. Goldbart, S. H. Chun, and 
M. B. Salamon, preprint 1999, cond-mat/9904331.

\noindent J. W. Lynn, R. W. Erwin, J. A. Borchers, Q. Huang,
A. Santoro, J. L. Peng, and Z. Y. Li, 
Phys. Rev. Lett. {\bf 76}, 4046 (1996).

\noindent J. W. Lynn, R. W. Erwin, J. A. Borchers, A. Santoro,
Q. Huang, J.-L. Peng and R. L. Greene, 
J. Appl. Phys. {\bf 81}, 5488 (1997).

\noindent A. Machida, Y. Moritomo, and A. Nakamura,
Phys. Rev. B{\bf 58}, R4281 (1998).

\noindent R. Maezono, S. Ishihara, and N. Nagaosa, 
Phys. Rev. B{\bf 57},
R13993 (1998).

\noindent R. Maezono, S. Ishihara, and N. Nagaosa, 
Phys. Rev. B{\bf 58}, 11583 (1998).

\noindent R. Maezono and N. Nagaosa, preprint, 2000.

\noindent A. Maignan, C. Martin, F. Damay, B. Raveau, and
J. Hejtmanek, Phys. Rev. B{\bf 58}, 2758 (1998).

\noindent R. Mallik, E. V. Sampathkumaran and P. L. Paulose, 1998 
preprint, cond-mat/9811387.

\noindent R. Mallik, E. S. Reddy, P. L. Paulose, S. Majumdar, and
E. V. Sampathkumaran, preprint 1998, cond-mat/9811351.

\noindent A. L. Malvezzi, S. Yunoki, and E. Dagotto,
Phys. Rev. {\bf B 59}, 7033 (1999).

\noindent M. Manekar, S. B. Roy, and P. Chaddah, preprint,
cond-mat/0005399.

\noindent C. Martin, A. Maignan, M. Hervieu, and B. Raveau,
Phys. Rev. B{\bf 60}, 12191 (1999).

\noindent G. Martins, C. Gazza, J. C. Xavier, A. Feiguin, and
E. Dagotto, Phys. Rev. Lett. {\bf 84}, 5844 (2000).

\noindent R. Mathieu, P. Svedlindh, and P. Nordblad, 2000, preprint,
cond-mat/0007154. 

\noindent G. Matsumoto, 1970a, J. Phys. Soc. Jpn. {\bf 29}, 606
(1970).

\noindent G. Matsumoto, 1970b, J. Phys. Soc. Jpn. {\bf 29}, 615
(1970). 

\noindent M. Matsuura, H. Hiraka, K. Yamada, and Y. Endoh,
preprint, cond-mat/0006185.

\noindent A. Mauger and D. L. Mills, 1984,
Phys. Rev. Lett. {\bf 53}, 1594 (1984).

\noindent A. Mauger and D. L. Mills, 1985,
Phys. Rev. B{\bf 31}, 8024 (1985).

\noindent M. Mayr, A. Moreo, J. Verg\'es, J. Arispe, A. Feiguin, and 
E. Dagotto, 2000, preprint.

\noindent A. Millis, B. I. Shraiman, and P. B. Littlewood,
Phys. Rev. Lett. {\bf 74}, 5144 (1995).

\noindent A. J. Millis, B. I. Shraiman, and R. Mueller,
Phys. Rev. Lett. {\bf 77}, 175 (1996)

\noindent A. J. Millis, B. I. Shraiman, and R. Mueller,
Phys. Rev. B{\bf 54}, 5389 (1996).

\noindent A. J. Millis, R. Mueller, and B. I. Shraiman,
Phys. Rev. B{\bf 54}, 5405 (1996).

\noindent A. J. Millis, Nature {\bf 392}, 147 (1998).

\noindent A. J. Millis, Phys. Rev. Lett. {\bf 80}, 4358 (1998). 

\noindent S. Mishra, S. Satpathy, F. Aryasetiawan, and 
O. Gunnarson, Phys. Rev. B{\bf 55}, 2725 (1997).

\noindent T. Mizokawa and A. Fujimori, Phys. Rev. B{\bf 51}, R12880
(1995). 

\noindent T. Mizokawa and A. Fujimori, Phys. Rev. B{\bf 54}, 5368
(1996). 

\noindent T. Mizokawa and A. Fujimori, Phys. Rev. B{\bf 56}, R493
(1997). 

\noindent T. Mizokawa, D. I. Khomskii, and G. A. Sawatzky, preprint,
cond-mat/9912021. 

\noindent T. Momoi and K. Kubo, Phys. Rev. B{\bf 58}, R567 (1998). 

\noindent H. A. Mook {\it et al.}, Nature {\bf 395}, 580 (1998).

\noindent A. Moreo, S. Yunoki and E. Dagotto, Science {\bf 283}, 2034
(1999a). 

\noindent A. Moreo, S. Yunoki, and E. Dagotto, 
Phys. Rev. Lett. {\bf 83}, 2773 (1999b).

\noindent A. Moreo, M. Mayr, A. Feiguin, S. Yunoki and E. Dagotto, 
Phys. Rev. Lett. {\bf 84}, 5568 (2000).

\noindent S. Mori, C. H. Chen, and S.-W. Cheong, 1998a, 
Nature {\bf 392}, 473.

\noindent S. Mori, C. H. Chen, and S.-W. Cheong, 1998b,
Phys. Rev. Lett. {\bf 81}, 3972.

\noindent Y. Moritomo, Y. Tomioka, A. Asamitsu, Y. Tokura, and 
Y. Matsui, Phys. Rev. B{\bf 51}, 3297 (1995).

\noindent Y. Moritomo, A. Asamitsu, H. Kuwahara, and Y. Tokura,
Nature {\bf 380}, 141 (1996).

\noindent Y. Moritomo, H. Kuwahara, Y. Tomioka, and Y. Tokura,
Phys. Rev. B{\bf 55}, 7549 (1997).

\noindent Y. Moritomo, T. Akimoto, A. Nakamura, K. Ohoyama, and 
M. Ohashi, Phys. Rev. B{\bf 58}, 5544 (1998).

\noindent Y. Moritomo, A. Machida, S. Mori, N. Yamamoto, and
A. Nakamura, Phys. Rev. B{\bf 60}, 9220 (1999).

\noindent Y. Moritomo, Phys. Rev. B{\bf 60}, 10374 (1999).

\noindent Y. Motome, H. Nakano, and M. Imada, 
1998 preprint, cond-mat/9811221.

\noindent Y. Motome, and M. Imada, Phys. Rev. B{\bf 60}, 7921 (1999).

\noindent Y. Motome and N. Furukawa, 
J. Phys. Soc. Jpn. {\bf 68}, 3853 (1999).

\noindent Y. Motome and N. Furukawa, 2000a, preprint, cond-mat/0007407.

\noindent Y. Motome and N. Furukawa, 2000b, preprint, cond-mat/0007408.

\noindent F. Moussa, M. Hennion, G. Biotteau, J. Rodriguez Carvajal,
L. Pinsard, and A. Revcolevschi, Phys. Rev. B{\bf 60}, 12299 (1999).

\noindent A. A. Mukhin, V. Yu. Ivanov, V. D. Travkin, A. Pimenov, A. Loidl,
and A. M. Balbashov, 2000, Europhs. Lett. {\bf 49}, 514 (2000).

\noindent E. M\"uller-Hartmann and E. Dagotto, 
Phys. Rev. B{\bf 54}, R6819 (1996).

\noindent Y. Murakami, H. Kawada, H. Kawata, M. Tanaka, T. Arima,
Y. Moritomo, and Y. Tokura, Phys. Rev. Lett. {\bf 80}, 1932 (1998a).

\noindent Y. Murakami et al., Phys. Rev. Lett. {\bf 81}, 582 (1998b). 

\noindent E. L. Nagaev, JETP Lett. {\bf 6}, 18 (1967).

\noindent E. L. Nagaev, Sov. Phys. Lett. {\bf 27}, 122 (1968).

\noindent E. L. Nagaev, JETP Lett. {\bf 16}, 394 (1972).

\noindent E. L. Nagaev, phys. stat. sol. (b) {\bf 186}, 9 (1994).

\noindent E. L. Nagaev, Physics-Uspekhi {\bf 38}, 497 (1995).

\noindent E. L. Nagaev, Physics-Uspekhi {\bf 39}, 781 (1996).

\noindent E. L. Nagaev, Phys. Rev. B{\bf 58}, 2415 (1998).

\noindent K. Nagai, T. Momoi, and K. Kubo, 1999 preprint,
cond-mat/9911091.

\noindent H. Nakano, Y. Motome, and M. Imada, preprint 2000,
cond-mat/0004232.

\noindent D. N. H. Nam, K. Jonason, P. Nordblad, N. V. Khiem,
and N. X. Phuc, Phys. Rev. B{\bf 59}, 4189 (1999).


\noindent J. J. Neumeier, M. F. Hundley, J. D. Thompson, and 
R. H. Heffner, Phys. Rev. B{\bf 52}, R7006 (1995).

\noindent J. J. Neumeier and J. L. Cohn, Phys. Rev. B{\bf 61}, 14319
(2000).

\noindent H. Nojiri, K. Kaneko, M. Motokawa, K. Hirota, 
Y. Endoh, and K. Takahashi, Phys. Rev. B{\bf 60}, 4142 (1999).

\noindent S. B. Ogale, R. Shreekala, R. Bathe, S. K. Date,
S. I. Patil, B. Hannoyer, F. Petit, and G. Marest,
Phys. Rev. B{\bf 57}, 7841 (1998). 

\noindent H. Ohno, H. Munekata, T. Penney, S. von Molnar, and 
L. L. Chang, Phys. Rev. Lett. {\bf 68}, 2664 (1992).

\noindent H. Ohno, Science {\bf 281}, 951 (1998).

\noindent S. Okamoto, S. Ishihara, and S. Maekawa, 
Phys. Rev. B{\bf 61}, 451 (2000).

\noindent Y. Okimoto, T. Katsufuji, T. Ishikawa, A. Urushibara, 
T. Arima, and Y. Tokura, Phys. Rev. Lett. {\bf 75}, 109 (1995).

\noindent R. Osborn, S. Rosenkranz, D. N. Argyriou, L. Vasiliu-Doloc, 
J. W. Lynn, S. K. Sinha, J. F. Mitchell, K. E. Gray, and S. D. Bader,
Phys. Rev. Lett. {\bf 81}, 3964 (1998).

\noindent H. Oshima, Y. Ishihara, M. Nakamura, and K. Miyano,
preprint, 2000.

\noindent G. Papavassiliou, M. Fardis, F. Milia, A. Simopoulos, 
G. Kallias, M. Pissas, D. Niarchos, N. Ioannidis, C. Dimitropoulos,
and J. Dolinsek, Phys. Rev. B{\bf 55}, 15000 (1997).

\noindent G. Papavassiliou, M. Fardis, M. Belesi, M. Pissas,
I. Panagiotopoulos, G. Kallias, D. Niarchos, C. Dimitropoulos, and  
J. Dolinsek, Phys. Rev. B{\bf 59}, 6390 (1999).

\noindent G. Papavassiliou, M. Fardis, M. Belesi, T. Maris,
G. Kallias, M. Pissas, C. Dimitropoulos, and J. Dolinsek, 
preprint 1999. 

\noindent M. Paraskevopoulos, F. Mayr, C. Hartinger, A. Pimenov,
J. Hemberger, P. Lunkenheimer, A. Loidl,  A. A. Mukhin,
V. Yu. Ivanov, and A. M. Balbashov, 2000a, 
J. Mag. Mag. Mat. {\bf 211}, 118 (2000).

\noindent M. Paraskevopoulos, F. Mayr, J. Hemberger, A. Loidl,
R. Heichele, D. Maurer, V. M\"uller, A. A. Mukhin,
and A. M. Balbashov, 2000b, J. Phys.: Condens. Matter {\bf 12}, 3993 (2000).

\noindent F. Parisi, P. Levy, G. Polla, and D. Vega, 2000,
cond-mat/0008080. 

\noindent J.-H. Park, C. T. Chen, S.-W. Cheong, W. Bao, G. Meigs,
V. Chakarian, and Y. U. Idzerda, Phys. Rev. Lett. {\bf 76}, 4215
(1996).

\noindent T. G. Perring, G. Aeppli, Y. Moritomo, and Y. Tokura,
Phys. Rev. Lett. {\bf 78}, 3197 (1997).

\noindent T. G. Perring, G. Aeppli, Y. Tokura, 
Phys. Rev. Lett. {\bf 80}, 4359 (1998).

\noindent A. Pimenov, M. Biberacher, D. Ivannikov, A. Loidl, V. Yu. Ivanov,
A. A. Mukhin, and A. M. Balbashov, Phys. Rev. B{\bf 62}, 5685 (2000).

\noindent W. H. Press, S. A. Teukolsky, W. T. Vitterling, and
B. P. Flannery, {\it Numerical Recipes},
(Cambridge University Press, New York, 1986).


\noindent M. Quijada, J. Cerne, J. R. Simpson, H. D. Drew, K. H. Ahn,
A. J. Millis, R. Shreekala, R. Ramesh, M. Rajeswari, and
T. Venkatesan, Phys. Rev. B{\bf 58}, 16093 (1998).

\noindent P. G. Radaelli, D. E. Cox, M. Marezio, S.-W. Cheong, 
P. E. Schiffer, and A. P. Ramirez,
Phys. Rev. Lett. {\bf 75}, 4488 (1995). 

\noindent P. G. Radaelli, D. E. Cox, M. Marezio, and S.-W. Cheong,
Phys. Rev. B{\bf 55}, 3015 (1997).

\noindent P. G. Radaelli, D. E. Cox, L. Capogna, S.-W. Cheong, and
M. Marezio, Phys. Rev. B{\bf 59}, 14440 (1999).

\noindent P. G. Radaelli, R. M. Ibberson, D. N. Argyriou, H. Casalta, 
K. H. Andersen, S-W. Cheong, and J. F. Mitchell, 2000, preprint,
cond-mat/0006190.

\noindent A. P. Ramirez, P. Schiffer, S.-W. Cheong, C. H. Chen,
W. Bao, T. T. M. Palstra, P. L. Gammel, D. J. Bishop, and B. Zegarski, 
Phys. Rev. Lett. {\bf 76}, 3188 (1996).

\noindent A. P. Ramirez and M. A. Subramanian, 
Science {\bf 277}, 546 (1997).

\noindent A. P. Ramirez,
J. Phys.: Condens. Matter {\bf 9}, 8171 (1997).

\noindent C. A. Ramos, H. R. Salva, R. D. Sanchez, M. Tovar,
F. Rivadulla, J. Mira, J. Rivas, A. M. Lopez-Quintela, L. Hueso,
M. Saint-Paul, P. Lejay, and Y. Tokura, preprint, 2000,
proceedings of the SCM 2000 conference, Recife, Brazil.

\noindent C. N. R. Rao, P. Ganguly, K. K. Singh, and R. A. Mohan
Ram, J. Solid State Chem. {\bf 72}, 14 (1988).

\noindent C. N. R. Rao and B. Raveau, editors, {it Colossal
  Magnetoresistance, Charge Ordering and Related Properties of
  Manganese Oxides}, World Scientific, Singapore, 1998.

\noindent B. Raquet, A. Anane, S. Wirth, P. Xiong, and S. von Moln\'ar,
2000 preprint.

\noindent P. Raychaudhuri, C. Mitra, A. Paramekanti, R. Pinto,
A. K. Nigam, and S. K. Dhar,
J. Phys.: Condens. Matter {\bf 10}, L191 (1998).

\noindent J. J. Rhyne, H. Kaiser, H. Luo, G. Xiao, and M. L. Gardel, 
J. of Applied Physics {\bf 83}, 7339 (1998).

\noindent J. Riera, K. Hallberg, and E. Dagotto, 
Phys. Rev. Lett. {\bf 79}, 713 (1997).

\noindent C. Ritter, M. R. Ibarra, J. M. De Teresa, P. A. Algarabel, 
C. Marquina, J. Blasco, J. Garcia, S. Oseroff, and S-W. Cheong,
Phys. Rev. B{\bf 56}, 8902 (1997).

\noindent H. R\"oder, J. Zang, and A. R. Bishop, 
Phys. Rev. Lett. {\bf 76}, 1356 (1996).

\noindent H. R\"oder, R. R. P. Singh, and J. Zang, 
Phys. Rev. B{\bf 56}, 5084 (1997).

\noindent L. M. Rodriguez-Martinez and J. P. Attfield, 
Phys. Rev. B{\bf 54}, R15622 (1996).

\noindent J. M. Rom\'an and J. Soto, 1998 preprint, cond-mat/9810389. 

\noindent M. Roy, J. F. Mitchell, A. P. Ramirez, and P. Schiffer,
Phys. Rev. B{\bf 58}, 5185 (1998).

\noindent M. Roy, J. F. Mitchell, A. P. Ramirez, and P. Schiffer,
J. Physics, Condens. Matter {\bf 11}, 4843 (1999).

\noindent M. Roy, J. F. Mitchell, S. J. Potashnik, and P. Schiffer, 
2000, J. Magn. Magn. Mat. {\bf 218}, 191 (2000).

\noindent M. Roy, J. F. Mitchell, and P. Schiffer, 2000, 
J. of Appl. Phys. {\bf 87}, 5831 (2000).

\noindent S. R. Saha, H. Sugawara, T. D. Matsuda, H. Sato, R. Mallik,
and E. V. Sampathkumaran, Phys. Rev. B{\bf 60}, 12162 (1999).

\noindent T. Saitoh, A. Bocquet, T. Mizokawa, H. Namatame,
A. Fujimori, M. Abbate, Y. Takeda, and M. Takano, 
Phys. Rev. B{\bf 51}, 13942 (1995).

\noindent T. Saitoh, D. S. Dessau, Y. Moritomo, T. Kimura,
Y. Tokura, and N. Hamada, preprint, cond-mat/9911189.

\noindent S. Satpathy, Z. S. Popovic, and F. R. Vukajlovic,
Phys. Rev. Lett. {\bf 76}, 960 (1996)

\noindent P. Schiffer, A. P. Ramirez, W. Bao, and S.-W. Cheong,
Phys. Rev. Lett. {\bf 75}, 3336 (1995).

\noindent P. Schlottmann, Phys. Rev. B{\bf 59}, 11484 (1999).

\noindent C. W. Searle and S. T. Wang, 
Can. J. Phys. {\bf 47}, 2703 (1969).

\noindent Y. Shapira, S. Foner, and N. Oliveira Jr., 
Phys. Rev. B{\bf 10}, 4765 (1974). 

\noindent S.-Q. Shen, and Z. D. Wang, 
Phys. Rev. B{\bf 58}, R8877 (1998).

\noindent S.-Q. Shen, and Z. D. Wang,
1999 preprint, cond-mat/9904420. 

\noindent S.-Q. Shen, and Z. D. Wang,
1999 preprint, cond-mat/9906126.

\noindent L. Sheng, D. Y. Xing, D. N. Sheng, and C. S. Ting,
Phys. Rev. Lett. {\bf 79}, 1710 (1997); 
Phys. Rev. B{\bf 56}, R7053 (1997).

\noindent L. Sheng, D. N. Sheng, and C. S. Ting, 
Phys. Rev. B{\bf 59}, 13550 (1999).

\noindent Y. Shimakawa, Y. Kubo and T. Manako, 
Nature {\bf 379}, 53 (1996).

\noindent V. N. Smolyaninova, X. C. Xie, F. C. Zhang, M. Rajeswari, 
R. L. Greene and S. Das Sarma, 1999 preprint, cond-mat/9903238.

\noindent C. S. Snow, S. L. Cooper, D. P. Young, and Z. Fisk, 
2000, preprint.

\noindent I. Solovyev, N. Hamada, and K. Terakura,
Phys. Rev. Lett. {\bf 76}, 4825 (1996).

\noindent I. Solovyev and K. Terakura, 
Phys. Rev. Lett. B{\bf 83}, 2825 (1999).

\noindent I. Solovyev, 2000, preprint.

\noindent B. J. Sternlieb, J. P. Hill, U. C. Wildgruber, G. M. Luke, 
B. Nachumi, Y. Moritomo, and Y. Tokura, Phys. Rev. Lett. {\bf 76}, 
2169 (1996).


\noindent M. A. Subramanian, A. P. Ramirez, and W. J. Marshall,
Phys. Rev. Lett. {\bf 82}, 1558 (1999).

\noindent K. Takenaka, Y. Sawaki, and S. Sugai, 
Phys. Rev. B{\bf 60}, 13011 (1999).

\noindent H. Tang, M. Plihal, and D. L. Mills, 
J. Magn. Magn. Mat. {\bf 187}, 23 (1998).

\noindent J. M. Tarascon, J. L. Soubeyroux, J. Etourneau, R. Georges, 
J. M. D. Coey, and O. Massenet, 
Solid State Commun. {\bf 37}, 133 (1981).

\noindent A. Tkachuk, K. Rogacki, D. E. Brown, B. Dabrowski,
A. J. Fedro, C. W. Kimball, B. Pyles, X. Xiong, D. Rosenmann, and
B. D. Dunlap, Phys. Rev. B{\bf 57}, 8509 (1998).

\noindent M. Tokunaga, Y. Tokunaga, M. Yasugaki, and T. Tamegai,
2000 preprint, submitted to Physica B.

\noindent Y. Tokura et al., 
J. Phys. Soc. Jpn. {\bf 63}, 3931 (1994).

\noindent Y. Tokura et al., 
Phys. Rev. Lett. {\bf 70}, 2126 (1993).

\noindent Y. Tokura, Y. Tomioka, H. Kuwahara, A. Asamitsu,
Y. Moritomo, and M. Kasai, J. Appl. Phys. {\bf 79}, 5288 (1996). 

\noindent Y. Tokura, Physica B{\bf 237-238}, 1 (1997).

\noindent Y. Tokura, {Fundamental Features of Colossal
  Magnetoresistive Manganese Oxides},  contribution to {\it Colossal
  Magnetoresistance Oxides} edited by Y. Tokura (Gordon \& Breach,
Monographs in Condensed Matter Science, London, 1999). 

\noindent Y. Tokura, talk given at 
SCM 2000 conference, Recife, Brazil, August 2000.

\noindent Y. Tomioka, A. Asamitsu, Y. Moritomo, H. Kuwahara,
and Y. Tokura, Phys. Rev. Lett. {\bf 74}, 5108 (1995).

\noindent Y. Tomioka, A. Asamitsu, H. Kuwahara, Y. Moritomo, and 
Y. Tokura, Phys. Rev. B{\bf 53}, R1689 (1996).

\noindent Y. Tomioka, A. Asamitsu, H. Kuwahara, and Y. Tokura,
J. Phys. Soc. Jpn. {\bf 66}, 302 (1997).

\noindent Y. Tomioka and Y. Tokura, 1999, {\it Metal-Insulator
  Phenomena Relevant to Charge/Orbital-Ordering in Perovskite-Type
  Manganese Oxides}, preprint.

\noindent J. M. Tranquada {\it et al.}, Nature {\bf 375}, 561 (1995). 

\noindent I. O. Troyanchuk, Sov. Phys. JETP {\bf 75}, 132 (1992).

\noindent T. A. Tyson, J. Mustre de Leon, S. R. Conradson,
A. R. Bishop, J. J. Neumeier, H. R\"oder, and J. Zang,
Phys. Rev. B{\bf 53}, 13985 (1996).

\noindent M. Uehara, S. Mori, C. H. Chen, and S.-W. Cheong,
Nature {\bf 399}, 560 (1999).

\noindent A. Urushibara, Y. Moritomo, T. Arima, A. Asamitsu, G. Kido, 
and Y. Tokura, Phys. Rev. B{\bf 51}, 14103 (1995).

\noindent G. Varelogiannis, preprint, cond-mat/0003107.

\noindent J. van den Brink, P. Horsch, F. Mack, and A. M. Ole\'s,
1998 preprint, cond-mat/9812123.

\noindent J. van den Brink, G. Khaliullin, and D. Khomskii, 1999,
Phys. Rev. Lett. {\bf 83}, 5118 (1999).

\noindent C. Varma, 1996, Phys. Rev. B{\bf 54}, 7328 (1996).

\noindent L. Vasiliu-Doloc, J. W. Lynn, A. H. Moudden, 
A. M. de Leon-Guevara, and A. Revcolevschi, 1998a,
Phys. Rev. B{\bf 58}, 14913 (1998).

\noindent L. Vasiliu-Doloc, J. W. Lynn, Y. M. Mukovskii, A. A. Arsenov,
and D. A. Shulyatev, 1998b, J. of Appl. Phys. {\bf 83}, 7342 (1998).

\noindent L. Vasiliu-Doloc, S. Rosenkranz, R. Osborn, S. K. Sinha,
J. W. Lynn, J. Mesot, O. H. Seeck, G. Preosti, A. J. Fedro, and
J. F. Mitchell, 1999, Phys. Rev. Lett. {\bf 83}, 4393 (1999).

\noindent A. Venimadhav, M. S. Hedge, V. Prasad, and S. V. Subramanyam,
2000, cond-mat/0006388.

\noindent J. A. Verg\'es, cond-mat/9905235. 

\noindent I. F. Voloshin, A.  V. Kalinov, S. E. Savel'ev,
L. M. Fisher, N. A. Babushkina, L. M. Belova, D. I. Khomskii, and
K. I. Kugel, JETP Letters {\bf 71}, 106 (2000).

\noindent R. von Helmolt, J. Wecker, B. Holzapfel, L. Schultz, and 
K. Samwer, Phys. Rev. Lett. {\bf 71}, 2331 (1993).

\noindent S. von Molnar and S. Methfessel, J. Appl.Phys. {\bf 38},
959 (1967).

\noindent S. von Molnar and J. M. D. Coey, Current Opinion in Solid
State \& Materials Science 1998, {\bf 3}:171-174.

\noindent P. Wagner, I. Gordon, S. Mangin, V. V. Moshchalkov, and
Y. Bruynseraede, 1999 preprint, cond-mat/9908374.

\noindent S. White and D. Scalapino, 
Phys. Rev. Lett. {\bf 80}, 1272 (1998).

\noindent E. O. Wollan and W. C. Koehler, 
Phys. Rev. {\bf 100}, 545 (1955).

\noindent P. M. Woodward, D. E. Cox, T. Vogt, C. N. R. Rao, and
A. K. Cheetham, preprint, submitted to Chemistry of Materials.

\noindent T. Wu, S. B. Ogale, J. E. Garrison, B. Nagaraj, Z. Chen,
R. L. Greene, R. Ramesh, T. Venkatesan, and A. J. Millis,
preprint, 2000.

\noindent G. C. Xiong, Q. Li, H. L. Ju, S. N. Mao, L. Senapati,
X. X. Xi, R. L. Greene, and T. Venkatesan, Appl. Phys. Lett. {\bf 66},
1427 (1995).

\noindent Y. Yamada, O. Hino, S. Nohdo, R. Kanao, T. Inami, and
S. Katano, Phys. Rev. Lett. {\bf 77}, 904 (1996).

\noindent M. Yamanaka, W. Koshibae, and S. Maekawa,
Phys. Rev. Lett. {\bf 81}, 5604 (1998).

\noindent H. Yi and J. Yu, Phys. Rev. B{\bf 58}, 11123 (1998).

\noindent H. Yi and S.-I. Lee, Phys. Rev. B{\bf 60}, 6250 (1999).

\noindent H. Yi, J. Yu, and S.-I. Lee, 1999 preprint,
cond-mat/9910152. 

\noindent H. Yi, N. H. Hur, and J. Yu, 1999 preprint,
cond-mat/9910153. 

\noindent S. Yoon, H. L. Liu, G. Schollerer, S. L. Cooper, P. D. Han,
D. A. Payne, S.-W. Cheong, and Z. Fisk, 
Phys. Rev. B{\bf 58}, 2795 (1998). 

\noindent S. Yoon, M. R\"ubhausen, S. L. Cooper, K. H. Kim, and 
S-W. Cheong, preprint, cond-mat/0003250, to appear in PRL, Oct. 2000.

\noindent K. Yoshida, {\it Theory of Magnetism}, 
(Springer-Verlag, Berlin, 1996). 

\noindent H. Yoshizawa, H. Kawano, Y. Tomioka, and Y. Tokura,
Phys. Rev. B{\bf 52}, R13145 (1995). 

\noindent H. Yoshizawa, H. Kawano, J. A. Fernandez-Baca, H. Kuwahara,
and Y. Tokura, Phys. Rev. B{\bf 58}, R571 (1998).

\noindent A. P. Young, editor, 1998, 
{\it Spin Glasses and Random  Fields}, World Scientific.

\noindent Q. Yuan, T. Yamamoto, and P. Thalmeier, 2000, preprint,
cond-mat/0008296.

\noindent S. Yunoki, J. Hu, A. Malvezzi, A. Moreo, N. Furukawa, and
E. Dagotto, Phys. Rev. Lett. {\bf 80}, 845 (1998).

\noindent S. Yunoki, A. Moreo, and E. Dagotto, 
Phys. Rev. Lett. {\bf 81}, 5612 (1998).

\noindent S. Yunoki and A. Moreo. Phys. Rev. B{\bf 58}, 6403 (1998). 

\noindent S. Yunoki, T. Hotta, and E. Dagotto, 
Phys. Rev. Lett. {\bf 84}, 3714 (2000).

\noindent J. Zaanen, J. Phys. Chem. Solids {\bf 59}, 1769 (1998).

\noindent J. Zang, A. R. Bishop, and H. R\"oder, 
Phys. Rev. B{\bf 53}, R8840 (1996).

\noindent J. Zang, S. A. Trugman, A. R. Bishop, and H. R\"oder,
Phys. Rev. B{\bf 56}, 11839 (1997).


\noindent C. Zener, Phys. Rev. {\bf 81}, 440 (1951a).

\noindent C. Zener, Phys. Rev. {\bf 82}, 403 (1951b).

\noindent G.-M. Zhao, K. Conder, H. Keller, and K. A. M\"uller,
Nature {\bf 381}, 676 (1996).

\noindent G.-M. Zhao, K. Conder, H. Keller, and K. A. M\"uller,
Phys. Rev. B{\bf 60}, 11914 (1999).

\noindent G.-M. Zhao, preprint, cond-mat/0001390.

\noindent Y. G. Zhao, J. J. Li, R. Shreekala, H. D. Drew, C. L. Chen, 
W. L. Cao, C. H. Lee, M. Rajeswari, S. B. Ogale, R. Ramesh,
G. Baskaran, and T. Venkatesan,
Phys. Rev. Lett. {\bf 81}, 1310 (1998). 

\noindent F. Zhong and Z. D. Wang, 
Phys. Rev. B{\bf 60}, 11883 (1999).

\noindent J.-S. Zhou, W. Archibald, and J. B. Goodenough,
Nature {\bf 381}, 770 (1996).

\noindent J.-S. Zhou, J. B. Goodenough, and J. F. Mitchell,
Phys. Rev. B{\bf 58}, R579 (1998).

\noindent J.-S. Zhou and J. B. Goodenough,
Phys. Rev. Lett. {\bf 80}, 2665 (1998).

\noindent M. v. Zimmermann, J. P. Hill, D. Gibbs, M. Blume, D. Casa, 
B. Keimer, Y. Murakami, Y. Tomioka, and Y. Tokura, 1999,
Phys. Rev. Lett. {\bf 83}, 4872 (1999).

\noindent M. v. Zimmermann, C. S. Nelson,
 J. P. Hill, D. Gibbs, M. Blume, D. Casa, 
B. Keimer, Y. Murakami, C.-C. Kao,
C. Venkataraman, T. Gog,
Y. Tomioka, and Y. Tokura, 2000, preprint,
cond-mat/0007231.

\noindent L.-J. Zou, Q.-Q. Zheng, and H. Q. Lin, 1998 preprint,
cond-mat/9806015.

\noindent S. Zvyagin, C. Saylor, G. Martins, L.-C. Brunel, K. Kamenev, 
G. Balakrishnan, and D. M. K. Paul, preprint, 2000.

\begin{center}
 {\bf FIGURE CAPTIONS}
\end{center}

\begin{description}


\item{I.1}
(a) Arrangement of ions in the perovskite structure of
manganites (from Tokura, 1999).
(b) Possible magnetic structures and their labels. The circles
represent the position of Mn ions, and the sign that of their spin
projections along the $z$-axis. The G-type
is the familiar antiferromagnetic arrangement in the three directions,
while B is the familiar ferromagnetic arrangement. Taken from Wollan
and Koehler (1955).

\item{I.2}
(a) Temperature dependence of the resistivity of a
$\rm La_{2/3} Ba_{1/3} Mn O_x$ thin film at zero and five Teslas, 
taken from von Helmolt et al. (1993). (a) and (b) are results
as-deposited and after a subsequent annealing, respectively.
For more details see von Helmolt et al. (1993).
(b) Resistivity of $\rm Nd_{0.5} Pb_{0.5} Mn O_3$
as a function of temperature and magnetic fields taken
from Kusters et al. (1989). The inset is the magnetoresistance
at the indicated temperatures.

\item{I.3}
Field splitting of the five-fold degenerate atomic 3$d$ levels into
lower $t_{\rm 2g}$ and higher $e_{\rm g}$ levels.
The particular Jahn-Teller distortion sketched in the figure further
lifts each degeneracy as shown. Figure taken from Tokura (1999).


\item{II.a.1} 
(a) Temperature dependence of resistivity for various single crystals
of $\LSMO$. Arrows indicate the Curie temperature. 
The open triangles indicate anomalies due to structural transitions.
For more details see Urushibara et al. (1995) from where this figure
is reproduced.
(b) Phase diagram of $\LSMO$ (courtesy of Y. Tokura and Y. Tomioka)
prepared with data from Urushihara et al. (1995) and Fujishiro et al.
(1998). The AFM phase at large x is an A-type AF metal with uniform
orbital order. PM, PI, FM, FI, and CI denote paramagnetic metal,
paramagnetic insulator, FM metal, FM insulator, and spin-canted
insulator states, respectively. $T_{\rm C}$ is the Curie temperature
and $T_{\rm N}$ is the N\'eel temperature. A more detailed version of
this phase diagram is shown below in Fig.IV.d.1, with emphasis on the
small hole-density region which presents tendencies to charge-ordering
and mixed-phase states.

\item{II.a.2} 
(a) Temperature dependence of the resistivity in magnetic fields
corresponding to $\LSMO$ at x=0.175 (from Tokura et al., 1994).
(b) Temperature dependence of the total infrared-absorption
spectral weight $N_{\rm eff}$ (open circles) and Drude weight $D$
(closed circles) for a single crystal of $\LSMO$ (x=0.175). 
The solid line is the square of the normalized ferromagnetic
magnetization $(M/M_{\rm s})^2$, with $M_{\rm s}$ the saturated
magnetization. Results reproduced from Tokura (1999).


\item{II.b.1}
The magnetization, resistivity, and magnetoresistance of 
$\LCMO$ (x=0.25), as a function of temperature at various fields.
The inset is $\rho$ at low temperatures. 
Reproduced from Schiffer et al. (1995).
  
\item{II.b.2}
Resistivity vs temperature at three hydrostatic pressures for $\LCMO$
(x=0.21). In the inset the pressure dependence of $T_{\rm C}$ and the
activation energy $e_{\rm g}$ are plotted.
For details see Neumeier et al. (1995), from where these results have
been reproduced.

\item{II.b.3}
Phase diagram of $\LCMO$, constructed from measurements of macroscopic
quantities such as the resistivity and magnetic susceptibility,
reproduced from Cheong and Hwang (1999).
FM: Ferromagnetic Metal, FI: Ferromagnetic Insulator,
AF: Antiferromagnetism, CAF: Canted AF, and CO: Charge/Orbital
Ordering. FI and/or CAF could be a spatially inhomogeneous states with
FM and AF coexistence.

\item{II.b.4}
The charge and orbital ordering configurations for $\LCMO$ with x=0,
1/2, and 2/3. Open circles are Mn$^{4+}$ and the lobes show the
orbital ordering of the $e_{\rm g}$-electrons of Mn$^{3+}$.
Figure reproduced from Cheong and Hwang (1999).

\item{II.b.5}
Resistivity at 300K and 100K vs Ca concentration for $\LCMO$
reproduced from Cheong and Hwang (1999).


\item{II.c.1} Phase diagram of $\PCMO$. PI and FI denote the
paramagnetic insulating and ferromagnetic insulating states, 
respectively. For hole density between 0.3 and 0.5, the 
antiferromagnetic insulating (AFI) state exists in the
charge/orbital-ordered insulating (COI) phase. The canted
antiferromagnetic insulating (CAFI) state, which may be a mixed FM-AF
state,  also has been identified between
x=0.3 and 0.4. Reproduced from Tomioka and Tokura (1999).

\item{II.c.2} Temperature dependence of the resistivity
of $\PCMO$ with x=0.3 at various magnetic fields.
The inset is the phase diagram in the temperature-magnetic field
plane. The hatched region has hysteresis. Results reproduced from
Tomioka and Tokura (1999).

\item{II.c.3} Temperature dependence of resistivity for
$\PCMO$ at x=0.3 under the various pressures indicated. Reproduced
from Moritomo et al. (1997).

\item{II.c.4} Temperature dependence of the resistivity
corresponding to $\PCMO$ at the hole concentrations and
magnetic fields indicated. Reproduced from Tomioka et al. (1996).

\item{II.c.5} The charge/orbital-ordered state of $\PCMO$ at
several hole concentrations, plotted on the magnetic 
field-temperature plane. The hatched area indicates the
hysteresis region. For more details see Tomioka and Tokura (1999).

\item{II.c.6} (a) Temperature dependence of the resistivity for
$\rm Pr_{0.65} (Ca_{1-y} Sr_y)_{0.35} Mn O_3$ crystals with
varying y. (b) Resistivity vs temperature of
$\rm Pr_{0.65} (Ca_{1-y} Sr_y)_{0.35} Mn O_3$ (y=0.2) for
several magnetic fields. Reproduced from Tomioka and Tokura (1999).


\item{II.d.1} Phase diagram of $\NSMO$, reproduced from 
Kajimoto et al. (1999).The notation is standard.

\item{II.d.2} The charge-ordered phase of various 
compounds $\rm (RE)_{1/2} (AE)_{1/2} Mn O_3$ plotted on the 
magnetic field-temperature plane. The hatched area indicates the
hysteresis region. Reproduced from Tomioka and Tokura (1999).

\item{II.d.3} (a) Average ionic radius at x=0.5 corresponding
to a mixture of a trivalent ion (upper abscissa) and a divalent ion
(lower abscissa). (b) Critical Curie temperature $T_{\rm C}$ and 
charge/orbital ordering transition $T_{\rm CO}$ for various
trivalent-divalent ion combinations. Reproduced from Tomioka and
Tokura (1999).


\item{II.e.1} 
Schematic crystal structure of four representatives of the
Ruddlesden-Popper series of manganese oxides (taken from Tokura, 1999).

\item{II.e.2} Temperature dependence of the resistivity in the
n=1 (single layer), n=2 (double layer) and n=$\infty$ (cubic)
representatives of the Ruddlesden-Popper series of manganese oxides.
The hole concentration is x=0.4. Results along the layers and
perpendicular to them are shown for n=1 and 2. Reproduced from
Moritomo et al. (1996).

\item{II.e.3} Temperature dependence of the resistivity for single
crystals of the n=2 compound at x=0.4 (from Moritomo et al., 1996),
with an external field parallel to the layer.


\item{II.f.1} Phase diagram corresponding to the single layer
compound $\rm La_{1-x} Sr_{1+x} Mn O_4$. AF, SG, and CO stand for
the antiferromagnetic, spin-glass, and charge-ordering phases, 
respectively. Solid lines are a guide to the eye. Reproduced from
Moritomo et al. (1995).


\item{II.g.1} (a) Phase diagram of temperature versus tolerance factor 
for the system $\rm A_{0.7} A'_{0.3} Mn O_3$, where A is a trivalent
rare earth ion and $\rm A'$ is a divalent alkali earth ion. Open 
and closed symbols denote $T_{\rm C}$ measured from the magnetization
and resistivity, respectively. For more details see Cheong and Hwang
(1999), from where this figure is reproduced. A very similar figure
appeared in Hwang et al. (1995a).
(b) Top panel: log $\rho(T)$ in 0 and 5 Teslas for a series of
samples of $\rm La_{0.7-y} A'_y Ca_{0.3} Mn O_3$, with $\rm A'$
mainly Pr but also Y. Bottom panel: MR factor. For details see
Hwang et al. (1995a).


\item{III.a.1} (a) Sketch of the Double Exchange mechanism
which involves two Mn ions and one O ion.
(b) The mobility of $e_{\rm g}$-electrons improves if the localized
spins are polarized.
(c) Spin canted state which appears as the interpolation between
FM and AF states in some mean-field approximations.
For more details see the text.

\item{III.a.2} Generation of antiferromagnetic (a) or
ferromagnetic (b) effective interactions between the spins of Mn ions
mediated by oxygen,
depending on the orientation of the Mn orbitals. For details see text.

\item{III.c.1} MnO$_6$ octahedron at site${\bf i}$. 
The labeling for oxygen ions is shown. 

\item{III.d.1}
Phase diagram of the one-orbital model with classical spins (and
without $J_{\rm AF}$ coupling).
(a) are results obtained with Monte Carlo methods 
at low temperature in 1D (Yunoki et al., 1998a; Dagotto
et al., 1998). FM, PS, and IC, denote ferromagnetic, phase-separated,
and spin incommensurate phases, respectively. Although not shown
explicitly, the \densi=1.0 axis is antiferromagnetic. The dashed lines
correspond to results obtained using quantum localized spins. For more
details see Yunoki et al. (1998a) and Dagotto et al. (1998).
(b) Similar to (a) but in 2D. The grey region denotes the possible 
location of the PS-IC transition at low Hund coupling, which is
difficult to determine.
Details can be found in Yunoki et al. (1998a).
(c) Results obtained in the infinite dimension limit and at large Hund 
coupling varying the temperature (here in units of the half-width
$W$ of the density of states). Two regions with PS were identified, as
well as a paramagnetic PM regime.
For details see Yunoki et al. (1998a).

\item{III.d.2} 
Spin-spin correlations of the classical spins at zero momentum
S({\bf q}=0) vs temperature T (units of t) obtained with the Monte Carlo
technique, taken from Yunoki et al. (1998a).
Density, Hund coupling, and lattice sizes are shown. (a) and (b) 
correspond to one and two dimensions, respectively.
Closed shells and open boundary conditions were used in (a) and (b),
respectively. For details see Dagotto et al. (1998).

\item{III.d.3} 
Rough estimation of the Curie temperature $T_{\rm C}$ in 3D and in
the limit $J_{\rm H}$=$\infty$, as reported by Yunoki et al. (1998a).
Other calculations discussed in the text produce results in reasonable
agreement with these Monte Carlo simulations (see Motome and Furukawa,
1999).

\item{III.d.4}
Density of $e_{\rm g}$ electrons vs chemical potential $\mu$.
The coupling is $J_{\rm H}$=8$t$ in (a) and (b) and 4$W$ in (c) ($W$
is the half-width of the density of states). Temperatures and lattice
sizes are indicated.
(a) Results in 1D with PBC. The inset contains the spin correlations
at the electronic densities 1.00 and 0.72, that approximately limit
the density discontinuity. (b) Same as (a) but in 2D. (c) Same as (a)
but in D=$\infty$. Results reproduced from Yunoki et al. (1998a).

\item{III.d.5}
Schematic representation of a macroscopic phase-separated state (A),
as well as possible charge inhomogeneous states stabilized by the
long-range Coulomb interaction (spherical droplets in (B), stripes 
in (C)). Reproduced from Moreo et al. (1999). Similar conclusions have 
been reached before in the context of phase separation applied to
models of high temperature superconductors.

\item{III.d.6}
(a) Spin-spin correlation S({\bf q}) vs momentum, for 2D clusters.
Couplings, temperature, and densities are indicated. The cluster is
6$\times$6. Reproduced from Dagotto et al. (1998).
(b) Snapshot obtained with Monte Carlo techniques applied to 
the one-orbital model using an 8$\times$8 cluster, $J_{\rm H}$=2.0 and 
\densi=0.75, illustrating the existence of stripes.
The area of the circles are proportional to the electronic density at
each site. The arrows are proportional to the value of the z-component
of the spin. Result courtesy of C. Buhler, using a program prepared by
S. Yunoki (unpublished).

\item{III.d.7} 
Phase diagram of the one-orbital model for manganites including an
antiferromagnetic Heisenberg coupling among the localized spins, here
denoted by $J'$ (while in other parts of the text it is referred
to as $J_{\rm AF}$). The Hund coupling is fixed to 8 and t=1. Two PS
regions are indicated, three AF regimes, and one FM phase. 
The ``I'' insulating phase is described in more detail in the
text. Reproduced from Yunoki and Moreo (1998).

\item{III.d.8}
Phase diagram of the one-orbital model with $S$=3/2 localized 
$t_{\rm 2g}$-spins, obtained with the DMRG and Lanczos methods applied 
to the chains of finite length L indicated. The notation is as in
previous figures. Results reproduced from Dagotto et al. (1998), where
more details can be found.

\item{III.d.9}
Phase diagram of $t$-$J$-like models in 1D corresponding to (a)
nickelates and (b) manganites. J is the coupling between Heisenberg
spins at each site in the large Hund coupling limit and x is the hole
density. PS, B, and FM, denote phase-separated, hole binding, and
ferromagnetic phases.  
For the meaning of the various symbols used to find the boundaries of
the phases see Riera et al. (1997). DMRG and Lanczos techniques were
used for this result.

\item{III.e.1}
(a) T(q) and S(q), orbital and spin structure factors vs $\lambda$,
working with the two-orbital model at \densi=1.0, low temperature,
$J_{\rm H}$=8, $J'$=0.05, and in 1D chains. The hopping set 
$t_{\rm aa}$=$t_{\rm bb}$=2$t_{\rm ab}$=2$t_{\rm ba}$ was used, but
qualitatively the results are similar for other hoppings.
(b) Same as (a) but using a 4$\times$4 cluster and realistic hoppings 
(Section III.c).
(c) Same as (a) but for a 4$^3$ cluster and realistic hoppings, and
using $J_{\rm H}$=$\infty$. Results reproduced from Yunoki et
al. (1998b), where more details can be found.

\item{III.e.2}
Magnetic and orbital structures discussed in the text to justify the 
Monte Carlo results for the two-orbital model at electronic density
1.0.

\item{III.e.3}
(a) Total energy vs $J'$ on a 2$^3$ cluster at low temperature
with $J_{\rm H}$=8$t$ and $\lambda$=1.5. The results were obtained
using Monte Carlo and relaxational techniques, with excellent
agreement among them. (b) The four spin arrangements are also shown.
(c) Orbital order corresponding to the A-type AF state.
For more details the reader should consult Hotta et al. (1999).

\item{III.e.4}
(a) \densi vs $\mu$ at the couplings and temperature indicated on a
L=22 site chain. The discontinuities characteristic of phase
separation are clearly shown. (b) Same as (a) but in 2D at the
parameters indicated.
The two sets of points are obtained by increasing and decreasing
$\mu$, forming a hysteresis loop. (c) Phase diagram of the two
orbitals model in 1D, $J_{\rm H}$=8, $J'$=0.05, and using the hopping
set $t_{\rm aa}$=$t_{\rm bb}$=2$t_{\rm ab}$=2$t_{\rm ba}$. 
The notation has been explained in the text. For more details see
Yunoki et al. (1998b), from where this figure was reproduced.

\item{III.e.5}
(a) Monte Carlo energy per site vs $J_{\rm AF}$ at density x=0.5,
$\lambda$=1.5, low temperature $T$=1/100, and $J_{\rm H}$=$\infty$,
using the two orbital model in 2D with Jahn-Teller phonons
(non-cooperative ones).
FM, CE, and AF states were identified measuring charge, spin, and 
orbital correlations. ``AF(2)'' denotes a state with
spins $\uparrow \uparrow \downarrow \downarrow$ in one direction,
and antiferromagnetically coupled in the other. The clusters used are 
indicated.
(b) Phase diagram in the plane $\lambda$-$J_{\rm AF}$ at x=0.5,
obtained numerically using up to 8$\times$8 clusters. All transitions
are of first-order. The notation is the standard one (CD = charge
disorder, CO = charge order, OO = orbital order, OD = orbital
disorder). Results reproduced from Yunoki, Hotta and Dagotto (2000),
where more details can be found.

\item{III.e.6}
Monte Carlo energy per site vs $J_{\rm AF}$ obtained working on a
$4^3$ cube ($\lambda$=1.5, $J_{\rm H}$=$\infty$, $T$=1/100). Results
with both cooperative and non-cooperative phonons are shown, taken
from Yunoki et al. (2000). The state of relevance here is the
``CE (CS)'' one, which is the CE-state with charge-stacking.

\item{III.e.7}
The unit cell for the zigzag FM chain in the CE-type AFM phase at
x=0.5. Note that the hopping direction is changed periodically as 
$\{ \cdots,x,x,y,y,\cdots\}$.

\item{III.e.8}
Band structure for the zigzag 1D chain (solid curve) in the reduced
zone $-\pi/4$$\leq$$k$$\leq$$\pi/4$.
For reference, the band structure $-2t_0 \cos k$ for the straight 1D
path is also shown (broken curve). 
Note that the line at zero energy indicates the four-fold degenerate
flat-band present for the zigzag 1D path.

\item{III.e.9}
(a) Path with $w$=$1$ at x=$1/2$. Charge and orbital densities are
calculated in the MFA for $E_{\rm JT}$=$2t$. At each site, the orbital
shape is shown with its size in proportion to the orbital density.
(b) The BS-structure path with $w$=$2$ at x=$2/3$. 
(c) The BS-structure path with $w$=$3$ at x=$3/4$. 
(d) The WC-structure path with $w$=$1$ at x=$2/3$. 
(e) The WC-structure path with $w$=$1$ at x=$3/4$.

\item{III.e.10}
(a) C- and E-type unit cell (Wollan and Koehler, 1955).
(b) The spin structure in the $a$-$b$ plane at x=1/2.
Open and solid circle denote the spin up and down, respectively.
The thick line indicates the zigzag FM path.
The open and shaded squares denote the C- and E-type 
unit cells.
At x=1/2, C-type unit cell occupies half of the 1D plane,
clearly indicating the ``CE'' type phase.
(c) The spin structure at x=2/3 for Wigner-crystal type
phase. Note that 66\% of the 2D lattice is occupied by C-type
unit cell. Thus, it is called ``C$_2$E''-type AFM phase.
(d) The spin structure at x=2/3 for bi-stripe type
phase. Note that 33\% of the 2D lattice is occupied by C-type
unit cell. Thus, it is called ``CE$_2$''-type AFM phase.

\item{III.e.11}
Schematic figures for spin, charge, and orbital ordering for (a) WC
and (b) BS structures at x=2/3. The open and solid symbols indicate
the spin up and down, respectively. The FM 1D path is denoted by the
thick line. The empty sites denote Mn$^{4+}$ ions, while the robes
indicate the Mn$^{3+}$ ions in which $3x^2-r^2$ or $3y^2-r^2$ orbitals
are occupied.

\item{III.e.12}
Orbital densities in the FM phase for (a)x=1/2, (b)1/3, and (c)1/4.
The charge density in the lower-energy orbital is shown, and the size
of the orbital is in proportion to this density. The broken line
indicates one of the periodic paths to cover the whole 2D plane.

\item{III.e.13}
Schematic representation of the spin-charge-orbital structure at x=1/4 in
the zigzag AFM phase at low temperature and large electron-phonon coupling.
The symbol convention is the same as in Fig.III.e.11.
This figure was obtained using numerical techniques,
and $cooperative$ phonons, for $J_{\rm H}$=$\infty$ and $J_{\rm AF}$=$0.1t$.
For the non-cooperative phonons, basically the same pattern
can be obtained.

\item{III.f.1}
(a) DOS of the one-orbital model on a 10$\times$10 cluster
at $J_{\rm H}$=$\infty$ and temperature $T$=1/30 (hopping $t$=1).
The four lines from the top correspond to densities 0.90, 0.92, 0.94,
and 0.97. The inset has results at \densi=0.86, a marginally stable
density at $T$=0. 
(b) DOS of the two-orbital model on a 20-site chain, working 
at \densi=0.7, $J_{\rm H}$=8, and $\lambda$=1.5. Starting from the 
top at $\omega$-$\mu$=0, the three lines represent temperatures
1/5, 1/10, and 1/20, respectively.
Here the hopping along $x$ between orbitals a is the unit of energy.
Both, (a) and (b) are taken from Moreo, Yunoki, and Dagotto (1999b).
(c) DOS using a 20-site chain of the one-orbital model
at $T$=1/75, $J_{\rm H}$=8, \densi=0.87, and at a chemical potential
such that the system is phase-separated in the absence of
disorder. $W$ regulates the strength of the disorder, 
as explained in Moreo et al. (2000) from where this figure was taken.

\item{III.f.2}
Schematic explanation of pseudogap formation at low electronic
density (taken from Moreo, Yunoki, and Dagotto, 1999b). In (a) 
a typical Monte Carlo configuration of localized spins is shown.
In (b), the corresponding electronic density is shown. In (c),
the effective potential felt by electrons is presented. A populated
cluster band (thick line) is formed. In (d), the resulting DOS is
shown. Figure taken from Moreo, Yunoki, and Dagotto (1999b).


\item{III.g.1}
Schematic representation of the influence of the A-site ionic size on 
the hopping amplitude ``$t$'' between two Mn ions.

\item{III.g.2}
Results that illustrate the generation of ``giant'' coexisting
clusters in models for manganites (taken from Moreo et al., 2000).
(a-c) are Monte Carlo results for the two-orbital model with 
$\langle n \rangle$=0.5, $T$=1/100, $J_{\rm H}$=$\infty$, 
$\lambda$=1.2, $t$=1, PBC, and using a chain with L=20 sites.
(a) is the energy per site vs $J_{\rm AF}/t$ for the non-disordered
model, with level crossing at 0.21.
(b) MC averaged nearest-neighbor $t_{\rm 2g}$-spins correlations 
vs position along the chain (denoted by i) for one set of random
hoppings $t^{\alpha}_{\rm ab}$ and $J_{\rm AF}$ couplings 
($J_{\rm AF}/t$ at every site is between 0.21-$\delta$ and
0.21+$\delta$ with $\delta$=0.01). FM and AF regions are shown. For
more details see Moreo et al. (2000). 
(c) Same as (b) but with $\delta$=0.05.
(d-f): results for the one-orbital model with
$\langle n \rangle$=0.5, $T$=1/70, $J_{\rm H}$=$\infty$, $t$=1, open
boundary conditions, and L=64 (chain). (d) is energy per site vs 
$J_{\rm AF}$ for the non-disordered model, showing the FM-AF states
level crossing at $J_{\rm AF}$$\sim$0.14. 
(e) are the MC averaged nearest-neighbor $t_{\rm 2g}$-spin
correlations vs position for one distribution of random hoppings and
$t_{\rm 2g}$ exchanges, such that $J_{\rm AF}/t$ is 
between 0.14-$\delta$ and 0.14+$\delta$ with $\delta$=0.01.
(f) Same as (e) but with $\delta$=0.03.

\item{III.g.3}
Results of a Monte Carlo simulation of the Random Field Ising Model 
at $T$=0.4 ($J$=1), with PBC, taken from Moreo et al. (2000).
The dark (white) small squares represent spins up (down).
At $T$=0.4 the thermal fluctuations appear negligible, and the results
shown are those of the lowest energy configuration.
(a) was obtained for a random field with strength $W$=3 taken from a
box distribution [$-W$,$W$], external field $H_{\rm ext}$=0,  using a 
100$\times$100 cluster, and one set of random fields $\{h_{\bf i}\}$.
(b) Results using a 500$\times$500 cluster with $W$=1.2 and for one
fixed configuration of random fields. 
The dark regions are spins up in the $H_{\rm ext}$=0 case, the grey
regions are spins down at zero field that have flipped to up at 
$H_{\rm ext}$=0.16, while the white regions have spins down with and
without the field.
The percolative-like features of the giant clusters are apparent in
the zero field results. Special places are arrow-marked where narrow
spin-down regions have flipped linking spin-up domains. For more
details see Moreo et al. (2000).

\item{III.h.1}
Schematic representation of the random resistor network
approximation. On the left is a sketch of the real system with
metallic and insulating regions. On the right is the resistor network 
where dark (light) resistances represent the insulator (metal). ``a''
is the Mn-Mn lattice spacing, while L is the actual lattice spacing of
the resistor network.

\item{III.h.2}
Net resistivity $\rho_{\rm dc}$ of a 100$\times$100 random resistor
network cluster vs temperature, at the indicated metallic fractions
$p$ (result taken from Mayr et al., 2000). Inset: Results for a 20$^3$
cluster with (from the top) $p$=0.0, 0.25, 0.3, 0.4 and 0.5. 
In both cases, averages over 40 resistance configurations were
made. The $p$=1 and 0 limits are from the experiments  
corresponding to LPCMO (see Uehara et al., 1999). 
Results on 200$\times$200 clusters (not shown) indicate that size
effects are negligible. 

\item{III.h.3}
(a) Schematic representation of the mixed-phase state near
percolation. The arrows indicate conduction either through the
insulating or metallic regions depending on temperature (see text). 
(b) Two-resistances in parallel model for Mn-oxides. The (schematic)
plot for the effective resistance $R_{\rm eff}$ vs $T$ arises from the 
parallel connection of metallic (percolative) $R^{\rm per}_M$ and
insulating $R_{\rm I}$ resistances.
Figure taken from Mayr et al. (2000). 

\item{III.h.4}
Net resistivity $\rho_{\rm dc}$ of the 100$\times$100
random-resistor-network used in the previous figure,
but with a metallic fraction $p$ changing with $T$.
Representative values of $p$ are indicated.
Results averaged over 40 resistance configurations are shown (taken
from Mayr et al., 2000). 

\item{III.h.5}
(a) Inverse conductivity of the half-doped one-orbital model on a
64-site chain in the regime of coexisting clusters, with 
$J_{\rm H}$=$\infty$, AF coupling among localized spins 
$J'$=0.14, $t$=1, and $\Delta$=0.03, varying a magnetic field as
indicated. The data shown corresponds to a particular
disorder configuration, but results with other configurations are
similar. 
(b) Effective resistivity of a 100$\times$100 network of
resistances. Results at $\Delta \sigma$=0.0 (full circles, open
triangles, and open squares starting at $T$=0 with $p$=0.45, 0.5 and
0.7, respectively) are the same as found in Fig.III.h.4.
Full triangles, inverse open triangles, and diamonds, correspond to
the same metallic fractions, but with a small addition to the
insulating conductivity ($\Delta \sigma$=0.1 
($\Omega$cm$)^{-1}$), to simulate the effect of magnetic fields (see
text). Results taken from Mayr et al. (2000).

\item{III.i.1}Electron density versus chemical potential in the ground
state of the one-orbital model with $S$=1/2, and a large Hund
coupling. AF, P, and F, denote antiferromagnetic, paramagnetic, and
ferromagnetic states, respectively. The result is taken from Nagai,
Momoi, and Kubo, 1999, where more details can be found.

\item{III.i.2}  Phase diagram at zero temperature in the plane
of AF interaction $J_{\rm AF}$ and hole concentration x, using a
two-orbital model with Coulomb interactions. F$_1$ and F$_2$
are the ferromagnetic phases with different types of orbital
ordering (indicated). 
PS(F$_1$/F$_2$) is the phase separated state between the
F$_1$ and F$_2$ phases. Results taken from Okamoto, Ishihara, and 
Maekawa (2000) where more details, including couplings, can be found. 


\item{IV.a.1} Transport and magnetic properties of 
$\rm La_{5/8-y} Pr_y Ca_{3/8} Mn O_3$ as a function of temperature and
y, reproduced from Uehara et al. (1999). (a) contains the temperature
dependence of the resistivity. Both cooling (solid lines) and heating
(dotted lines) curves are shown. (b) Magnetoresistance of
representative specimens at 4 kOe.
(c) Phase diagram of $\rm La_{5/8-y} Pr_y Ca_{3/8} Mn O_3$ 
as a function of the ionic radius of (La,Pr,Ca). 
$T_{\rm C}$ and $T_{\rm CO}$ are shown as filled circles (or
triangles) and open circles, respectively. For more details, 
see Uehara et al. (1999) from where this figure was taken.
(d) Generic spectroscopic images reported by F\"ath et al. (1999)
using scanning tunneling spectroscopy applied to a thin-film of
$\LCMO$ with x close to 0.3, and the temperature just below $T_{\rm C}$. 
The size of each frame is 0.61 $\mu$m by 0.61 $\mu$m. From 
left to right and top to bottom the magnetic fields are 0, 0.3, 1,
3, 5, and 9T. The light (dark) regions are insulating (metallic).

\item{IV.a.2} Resistivity vs temperature at several densities for 
$\rm La_{0.75} Ca_{0.25} Mn_{1-x} Fe_x O_3$, 
taken from Ogale et al. (1998).
The results for the undoped sample are shown on an expanded 
scale (right) still using $\Omega$-cm as unit.

\item{IV.a.3} (a) Electronic and magnetic phase diagram of 
$\rm (La_{1-x} Tb_x)_{2/3} Ca_{1/3} Mn O_3$ as a function of x, 
reproduced from  De Teresa et al. (1997a), where more details can be
found. SGI is a ``spin-glass'' insulating state.
(b) Resistance versus temperature corresponding to 
$\rm (La_{1-x} Tb_x)_{2/3} Ca_{1/3} Mn O_3$ at the densities
indicated, reproduced from Blasco et al. (1996). The similarities with
analogous plots for other manganites described by a percolative
process are clear. 

\item{IV.a.4} (a) Inelastic spectrum at the two temperatures indicated
and for q=0.07 $\rm \AA^{-1}$, reported by Lynn et al. (1996) in their
study of $\LCMO$ at x=0.33. The left and right peaks are associated
with spin-waves in FM portions of the sample, while the central peak
is attributed to paramagnetic regions.
(b) Similar as (a) but for $\PSMO$ (x=0.37) and $\NSMO$ (x=0.30) at
the temperatures and momenta indicated (reproduced from 
Fernandez-Baca et al., 1998).
(c,d) Similar as (a) but for $\LSMO$ at the compositions,
temperatures, and momenta indicated. (c) is reproduced from Vasiliu-Doloc
et al. (1998a), while (d) is from Vasiliu-Doloc et al. (1998b).

\item{IV.a.5} Schematic phase diagram of $\LCMO$, from 
Billinge et al. (1999). The solid lines are transport and magnetic
transitions taken from Ramirez et al. (1996). The notation is
standard. The small insets are PDF peaks
(for details see Billinge et al., 1999). The dark shaded regions are
claim to contain fully localized polaronic phases. The light shaded
region denotes coexistence of localized and delocalized phases, while
the white region is a FM homogeneous phases. The boundaries between 
the three regimes are diffuse and continuous, and are only suggestive.

\item{IV.a.6} Resistivity versus temperature for 
$\rm (La_{0.25} Nd_{0.75} )_{0.7} Ca_{0.3} Mn O_3$ reproduced from 
Zhou and Goodenough (1998). Pressures are indicated.


\item{IV.b.1}
(a) $^{55}$Mn NMR spectra of $\LCMO$ at $T$=3.2K for the densities 
shown, reproduced from Papavassiliou et al. (1999b). Coexistence of
features corresponding to two phases appear in the data.
(b) Revised temperature-density phase diagram proposed by
Papavassiliou et al. (1999b). The circles denote the NMR results
presented in that reference. The notation is standard.
(c) $^{55}$Mn NMR spectra at $T$=1.3K of $\LCMO$ with x=0.5 
in zero and applied field. FM lines are marked with filled symbols,
while AF ones are marked with open symbols. Figure reproduced from
Allodi et al. (1998).

\item{IV.b.2} 
Upper panel: Resistivity versus temperature on cooling (solid) and
warming (open) at fields between 0 and 9T, in steps of 1T starting
from the top. The material is $\LCMOhalf$ and the results are
reproduced from Roy, Mitchell and Schiffer (2000a).
Bottom panel: magnetization versus temperature for fields between 1T
and 7T in steps of 2T, starting from the bottom.
The inset illustrates two distinct features in the resistivity
associated with the coexistence of two states. For more details see
Roy, Mitchell and Schiffer (2000a).

\item{IV.b.3} (a) Phase diagram of a 3\% Cr-doped manganites,
$\rm R_{1/2} Ca_{1/2} (Mn_{0.97} Cr_{0.03}) O_3$, against averaged
ionic radius $r_R$ of the rare-earth ion. Closed circles and squares
are Curie temperatures and critical temperatures for the
charge-ordering transition, respectively. PS is the region of phase
separation. Open symbols represent the data for the Cr-undoped
compounds. Figure reproduced from Moritomo et al. (1999). 
(b) Electrical resistivity versus temperature for the compounds 
$\rm La_{0.5} Ca_{0.5-x}  Ba_x Mn O_3$, reproduced from
Mallik et al. (1998).

\item{IV.b.4} Phase diagrams of 
(a) $\rm Pr_{1-x} Ca_x Mn_{0.97} Cr_{0.03} O_3$, 
(b) $\rm La_{1-x} Ca_x Mn_{0.97} Cr_{0.03} O_3$, and
(c) $\rm Nd_{1-x} Sr_x Mn_{0.97} Cr_{0.03} O_3$, taken from
Katsufuji et al. (1999). The grey regions are the FM phases
in the absence of Cr, while the dark regions are FM metallic
phases stabilized by Cr doping. The rest of the notation is
standard.
(d) Resistivity vs temperature for 
$\rm Pr_{1-x} Ca_x Mn_{0.97} Cr_{0.03} O_3$. The inset contains
results at x=0.5 with (y=0.03) and without (y=0.0) Cr. Results
taken from Katsufuji et al. (1999).


\item{IV.c.1} (a) Neutron scattering results of Bao et al. (1997)
corresponding to $\BCMO$ at x=0.82. The solid circles correspond to 
the AF response, while the open circles are the FM response. The dotted
line is the background. A region of FM-AF coexistence is observed.
For more details the reader should consult Bao et al. (1997).
(b) Real part of the optical conductivity at the three temperatures
indicated, from Liu, Cooper and Cheong (1998) where the details of
the fitting results (dashed and dot-dashed lines) are explained.
The upper inset contains the temperature dependence of the energy gap 
(filed squares) and the polaron oscillator strength (open circles).
The lower inset is the effective number of carriers. The peak B
evolves into a clean charge-gap as T decreases, while A corresponds
to polarons.

\item{IV.c.2} (a) Temperature dependence of the resistivity of
$\rm Ca_{1-x} Sm_x Mn O_3$ for several values of x (shown).
For more details see Maignan et al. (1998).
(b) Magnetization $M$ versus temperature of 
$\rm Ca_{1-x} La_x Mn O_3$ (x shown). In the inset $M$ vs the
magnetic field $H$ is plotted. 
(c) Upper panel: Magnetic saturation moment at 5K versus x. Region I
is a G-type AF with local ferrimagnetism. Region II has local FM
regions and G-type AF.
Region III contains C-type and G-type AF, as well as local FM. Region
IV is a C-type AF. Lower panel: Electrical conductivity at T=5K versus x.
All the results are taken from Neumeier and Cohn (2000).


\item{IV.d.1} (a) Magnetic and structural phase diagram of $\LSMO$
determined by neutron diffraction data, reproduced from 
Endoh et al. (1999a). The notation is standard. Note that at
densities roughly between 0.10 and 0.15, a FM metallic phase
can be identified in a narrow temperature region upon changing
the temperature.
(b) Phase diagram of $\LSMO$ according to Zhou and Goodenough (1997).
Most of the notation is standard. The FMP region corresponds to
ferromagnetic polarons in the analysis of Zhou and Goodenough (1997),
where more details can be found.


\item{IV.f.1} (a) Intensity of neutron scattering experiments
by Perring et al. (1997)
performed on $\bilayer$ with x=0.4. The main figure shows 
the dependence with Q$_x$, while the inset contains 
a Q$_z$ dependence (for details the reader should consult the
original reference). At 150K and 0.5 in the horizontal axis,
a weak peak is observed corresponding to AF correlations,
while the most dominant peaks denote ferromagnetism.
(b) Magnetic phase diagram of $\bilayer$ reproduced from
Kubota et al. (1999a). Most of the notation is standard, but
a more detailed  explanation of the various phases can
be found in the text or in the original reference. Note the
prominent ``Canted AFM'' phase, which the authors of this review
believe may have mixed-phase characteristics.

\item{IV.f.2} 
(a) In-plane resistivity component $\rho_{\rm ab}$ 
of $\rm (La_{1-z} Nd_z)_{1.2} Sr_{1.8} Mn_2 O_7$ (single
crystals). The arrows indicate the Curie temperature.
Reproduced from Moritomo et al. (1997).
(b) Resistivity of the electron-doped manganite
$\rm La_{2.3-x} Y_x Ca_{0.7} Mn_2 O_7$ versus temperature
for x=0.0, 0.3, and 0.5, reproduced from Raychaudhuri et al. (1998).

\item{IV.f.3} Low temperature (10K) ARPES spectra corresponding
to $\bilayered(1.8)$ along various high symmetry directions. Results
reproduced from Dessau et al. (1998).


\item{IV.h.1} (a) Magnetic susceptibility defined as $M/H$
($M$=magnetization, $H$=magnetic field) vs temperature $T$ for the 
densities indicated of $\rm Ca_{2-x} La_x Ru O_4$
(from Cao et al., 2000). Inset: Magnetization vs temperature. (b)
Magnetization M as a function of magnetic field for the densities
indicated.

\item{IV.h.2} Resistivity (R) vs. temperature for
$\rm Sr Fe_{1-x} Co_x O_3$ and $\rm Ca Fe_{1-x} Co_x O_3$, reproduced 
from Kawasaki et al. (1998). For the meaning of the arrows the reader 
should consult the original reference.

\item{IV.h.3} Resistivity vs. temperature of 
$\rm Tl_{2-x} Sc_x Mn_2 O_7$ for various values of x. The upper,
middle, and lower curves for each x correspond to applied fields of
$H$= 0, 3, and 6 T, respectively.
Result reproduced from Ramirez and Subramanian (1997).

\item{IV.h.4} Resistivity vs. temperature of $\rm EuSe$
for several magnetic fields. The inset contains the zero field
resistivity versus temperature in a different scale. Results
reproduced from Shapira et al. (1974).

\item{V.1}
Sketch of the competition metal-insulator in the presence of
disorder, leading to equal-density coexisting large clusters
in the ``disorder-induced'' phase separation scenario.

\item{V.2}
Sketch of the expected resistivity vs temperature in the percolative
picture. For more details see text.

\item{V.3}
Illustration of a conjectured new temperature scale $T^*$ in manganites.
Above the ordering temperatures $T_{\rm CO}$, $T_{\rm N}$, and 
$T_{\rm C}$, a region with coexisting clusters could exist, in view of
the theoretical ideas described in this review and the many
experiments that are in agreement. 
It is possible that this region may have pseudogap characteristics, as
in the high temperature superconductors. The sketch shown here tries to
roughly mimic the phase diagram of LCMO. The doping independence of
$T^*$ in the figure is just to simplify the discussion. Actually, a
strong hole density dependence of $T^*$ is possible.

\item{V.4}
Illustration of how a Quantum Critical Point can be generated
in models for manganites. In (a) the first-order FM-AF transition
is shown as a function of temperature, without disorder ($\Delta$=0). 
In (b), the expected behavior with disorder is shown. In both cases
``g'' is a coupling or hole density that allows the system to change
from a metal to an insulator, and the disorder under discussion
involves adding a random component to ``g''.

\item{V.5}
Simple rationalization of the CMR effect based on a first-order
transition metal-insulator. In this context CMR can only occur in a
narrow window of couplings and densities. Sketched is the ground-state
energy vs a parameter ``g'' that causes the transition from metal to
insulator (coupling or density). The FM phase is shown with and
without a magnetic field ``h''.

\end{description}

\end{multicols}
\end{document}